\documentclass[12pt]{combine}

\usepackage{amsmath}
\usepackage{graphicx}

\usepackage{subfigure}

\usepackage{multicol}
\usepackage{tabularx}
\usepackage{amssymb}
\usepackage{mathrsfs}
\usepackage{xfrac}
\usepackage{textcomp}
\usepackage{bm}
\usepackage{cite}
\usepackage{color}
\usepackage{setspace} 
\usepackage{float}

\usepackage{epsfig}

\title{Proceedings of the third French-Ukrainian workshop on the
instrumentation developments for HEP}
\date{October 15-16, 2015, LAL, Orsay, France}
\author{***}

\begin{document}
\pagestyle{combine}

\maketitle 

\newpage
\import{first/picture}
\newpage

\tableofcontents 
\clearpage

\coltoctitle{Foreword}
\label{Foreword}
\newpage
\pagestyle{plain}
\normalsize
 
\noindent
{\Large {\bf Foreword}} 

 \vskip 1.5cm
 
\noindent
The reports collected in these proceedings have been presented in
the third French-Ukrainian workshop on the instrumentation
developments for high-energy physics held at LAL, Orsay on October
15-16. Joint developments between French and Ukrainian
laboratories and universities as well as new proposals have been
discussed. Workshop hosted about 40 presentations from more than
50 participants from the two countries. The program was focused on
developments for accelerator and beam monitoring, detector
developments, experimental platforms, joint developments for
large-scale high-energy physics projects, neutrino physics and
medical applications. Organisation of the workshop was possible
thanks to the support from Science and Technology Centre in
Ukraine (STCU), National Centre for Scientific Research (CNRS) and
Linear Accelerator Laboratory (LAL). The workshop was conducted in
the scope of the IDEATE International Associated Laboratory (LIA).

\vskip 0.5cm

\begin{flushright}
Sergey Barsuk
 \end{flushright}


\clearpage

\coltoctitle{Program}
\label{Program} 
\newpage
\pagestyle{plain}
\normalsize
 
\noindent
 {\Large{\bf Program}} 

\vskip 0.7cm

\begin{center}
 15th October 2015
\end{center}

\noindent
 09:15~~~Achille Stocchi, Introduction

\vskip 0.3cm

\noindent
  {\bf R\&D on accelerators: beam diagnostics}

\vskip 0.2cm

\noindent
 09:30~~~Nikolai F. Shulga (KIPT and KNU): Quantum
effects at radiation by ultrarelativistic electrons in long
undulator and at scattering in ultrathin crystals

 \vskip 0.3cm
 \noindent
 10:00~~~Sergii P. Fomin (KIPT), Computer simulation of high energy charged particle passage through aligned crystals

\vskip 0.3cm \noindent
 10:15~~~Oleksiy Fomin (KIPT and KNU), Qausireflection of relativistic charged particle beam by a thin
crystal

\vskip 0.3cm

 \noindent
 10:25~~~Sergii V. Trofimenko (KIPT), Ionization loss and
transition radiation processes by high-energy 'half-bare' electron

 \vskip 0.3cm
 \noindent
 10:40~~~Angeles Faus-Golfe (LAL), Electron beam uses and applications in ThomX
 \vskip 0.5cm

\noindent
 {\bf R\&D on accelerators: beam diagnostics}
 \vskip 0.2cm
\noindent
 11:10~~~Daniil Brzhechko (TSNUK), UA9 - slow extraction
 \vskip 0.3cm
\noindent
 11:20~~~Vasyl Drozd (TSNUK), Installation and test of the
Fiber Beam Loss Monitor at PHIL
 \vskip 0.3cm
\noindent
 11:30~~~Nicolas Delerue (LAL), Smith-Purcell radiation

 \vskip 0.3cm
 \noindent
 11:40~~~Alexander V. Shchagin (KIPT), Production and applications of
X-rays emitted by relativistic particles

 \vskip 0.3cm
 \noindent
 11:55~~~Valery M. Pugatch (KINR), Hybrid and Metal Microdetector
Systems for measuring spatial and time distribution of the charged
particles and X-rays beams

 \vskip 0.3cm
 \noindent
 12:10~~~Oleksii Kovalchuk (KINR), Design
of the Metal Microstrip Detector for the beam profile studies at
PHIL

 \vskip 0.3cm
 \noindent
 12:20~~~Oleksandre Okhrimenko (KINR), LHCb RMS status and operation at 13 TeV

 \vskip 0.3cm
 \noindent
 12:30~~~Discussion

 \vskip 0.5cm
 \noindent
 {\bf R\&D on accelerators: lasers, RF and simulations}

 \vskip 0.3cm
 \noindent
 14:00~~~V.I. Maslov, I.P. Levchuk, I.N. Onishchenko (KIPT),
 Focusing of relativistic electron bunches by nonresonant wakefield excited in plasma

 \vskip 0.3cm
 \noindent
 14:15    Nicolas Delerue (LAL), New acceleration techniques 14:30
St\'ephane Berry (CEA), XFEL village

 \vskip 0.3cm
 \noindent
  14:50 Discussion

 \vskip 0.5cm
 \noindent
 {\bf Nuclear Physics}
 \vskip 0.2cm
 \noindent
 15:05~~~Eric Voutier (IPNO), Prospects for a new measurement of
the proton radius

 \vskip 0.3cm
 \noindent
 15:25~~~Georgi Georgiev (CSNSM), ALTO

 \vskip 0.3cm
 \noindent
 15:45~~~Giulia Hull (IPNO), R\&D on scintillating materials for Nuclear
Physics

 \vskip 0.3cm
 \noindent
  16:00 Discussion

 \vskip 0.5cm
 \noindent
 {\bf Medical applications}

 \vskip 0.2cm
 \noindent
  16:30~~~Marie Jacquet (LAL)

 \vskip 0.3cm
 \noindent
  16:45~~~Sara Spadola (IMNC), Optimization of an intraoperative beta imaging probe dedicated to tumor margins control during radioguided cancer surgery

 \vskip 0.3cm
 \noindent
 17:00~~~Marc-Antoine Verdier (IMNC), Intrinsic Performances of the MAGICS Compact Gamma-Camera

 \vskip 0.3cm
 \noindent
 17:15~~~Slava Sharyy (IRFU), Development of the innovative detectors for PET at IRFU

 \vskip 0.3cm
 \noindent
  17:30~~~Olga Kochebina (IRFU), CaLIPSO-PET simulation with
GATE/GEANT4

 \vskip 0.3cm
 \noindent
 17:40~~~Samuel Meyroneinc (CPO): CPO

 \vskip 0.3cm
 \noindent
 18:00 Discussion

 \vskip 0.3cm
 \noindent
 18:30   Visit of CPO

 \vskip 0.5cm

\begin{center}
 16th October 2015
\end{center}

 \vskip 0.2cm
 \noindent
 {\bf R\&D on detectors}

 \vskip 0.3cm
 \noindent
 09:15~~~Slava Kubytsky (LAL), Diamond detectors

  \vskip 0.3cm
 \noindent
 09:30~~~Andrey Nagai (LAL), SiPM

  \vskip 0.3cm
 \noindent
  09:45    Leonid Burmistrov (LAL), UA9 - CPFM

 \vskip 0.3cm
 \noindent
  10:00~~~(Oleg Bezshyyko, Andrii Chaus), Micromegas/InGrid

  \vskip 0.3cm
 \noindent
 10:15~~~Giovanni Calderini (LPNHE), Si for ATLAS upgrade

  \vskip 0.3cm
 \noindent
 10:30~~~Nikolai I. Maslov (KIPT), Radiation resistance of Si detectors research using electrons and
 bremsstrahlung of KhIPT linacs

  \vskip 0.3cm
 \noindent
 11:00~~~Dmytro Kharchenko (IAP-Sumy), Modeling Radiation Defects Dynamics and Microstructure
 Transformations in Irradiated Solids

  \vskip 0.3cm
 \noindent
 11:15~~~Gabriel Charles (IPNO), ALERT: A Low Energy Recoil Tracker

  \vskip 0.3cm
 \noindent
 11:30~~~Denys Poda (CSNSM), Molybdenum containing scintillating bolometers for double
 beta decay search (LUMINEU program)

  \vskip 0.3cm
 \noindent
 11:45    Discussion

 \vskip 0.5cm
 \noindent
  {\bf Experimental platforms}

 \vskip 0.2cm
 \noindent
 12:00~~~Hugues Monard (LAL), ThomX

 \vskip 0.3cm
 \noindent
 12:15 Pierre Lepercq (LAL), PHIL

 \vskip 0.3cm
 \noindent
 {\bf Medical applications}

 \vskip 0.2cm
 \noindent
 14:00~~~Slava Sharyy (IRFU), Development of the innovative detectors for PET at IRFU

 \vskip 0.3cm
 \noindent
 14:15~~~Olga Kochebina (IRFU), CaLIPSO-PET simulation with GATE/GEANT4

 \vskip 0.5cm
 \noindent
 {\bf Experimental platforms}

 \vskip 0.2cm
 \noindent
 14:25~~~Oleg Bezshyyko (TSNUK), Leetech

 \vskip 0.3cm
 \noindent
 14:35~~~Sophie Kazamias, Laserix

 \vskip 0.3cm
 \noindent
 14:50~~~Visit of Laserix

 \vskip 0.5cm
 \noindent
 {\bf Developments for HEP projects}

 \vskip 0.2cm
 \noindent
  15:50~~~Maksym Teklishyn (GSI), Charmonium description using decays to hadronic states with LHCb,

 \vskip 0.3cm
 \noindent
  16:05~~~Stephane Monteil (LPC-Clermont), FCC

 \vskip 0.3cm
 \noindent
  16:20~~~Emi Kou
(LAL), Flavour Physics at SuperBelle

 \vskip 0.3cm
 \noindent
 16:35~~~Francois Le Diberder (LAL), Flavour Physics at ILC

 \vskip 0.3cm
 \noindent
  16:50~~~Discussion

 \vskip 0.5cm
 \noindent
 {\bf Neutrino}

 \vskip 0.2cm
 \noindent
  17:05~~~Mathieu Bongrand (LAL), SOLID

 \vskip 0.3cm
 \noindent
 17:20~~~Fedor Danevich (KINR), Instrumentation for double beta decay experiments

 \vskip 0.3cm
 \noindent
 17:35~~~Discussion

 \vskip 0.3cm
 \noindent
 17:50~~~LIA discussion

\begin{papers} 

\coltocauthor{S.N.~Shul'ga, N.F.~Shul'ga, S.~Barsuk, I.~Chaikovska, R.~Chehab} 
\coltoctitle{On classical and quantum effects at scattering of fast charged particles in ultrathin crystal} 

\begin{center}

{\Large {\bf On classical and quantum effects at scattering of fast charged particles in ultrathin crystal}}

\vskip 0.5cm

{\bf S.N.~Shul'ga$^{a}$, N.F.~Shul'ga$^{a,b,}$, S.~Barsuk$^{c}$, I.~Chaikovska$^{c}$, R.~Chehab$^{c}$}

\vskip 0.3cm

$^{a}${\it National Science Center ``Kharkov Institute of Physics and Technology", Kharkiv 61108, Ukraine}

$^{b}${\it Karazin Kharkiv National University, Kharkiv 61022, Ukraine}

$^{c}${\it  LAL, IN2P3-CNRS and Universit\'e Paris Sud, 91898 Orsay Cedex, France}

\end{center}

\vskip 0.4cm

\begin{abstract}
\noindent Classical and quantum properties of scattering of charged particles in ultrathin crystals are considered. A comparison is made of these two ways of study of scattering process. In the classical consideration we remark the appearance of sharp maxima that is referred to the manifestation of the rainbow scatering phenomenon and in quantum case we show the sharp maxima that arise from the interference of single electrons on numerous crystal planes, that can be expressed in the terms of reciprocal lattice vectors. We show that for some parameters quantum predictions substantially differ from the classical ones. Estimated is the influence of the beam divergence on the possibility of experimental observation of the studied effects.
\end{abstract}

\noindent

\vskip 0.5cm


\vskip 0.3cm


\section{Introduction}

The motion of a fast charged particle near direction of one of its
planes or axes can be considered as a motion in the field of
continuous planes or strings. These are the cases of particle
channelling or over-barrier motion. A number of theoretical and
experimental studies have been made devoted to these phenomena
(see, e.g.,
\cite{Lindh1965,Gemmell1974,Kumakh1986,BazZhev1987,AkhiezShulga_HighEn1996,RArtDh98}
and references therein). In order to describe the effects of
interaction of a charged particle with medium we must get first of
all the characteristics of its motion.

Interesting phenomena may happen just at the beginning of such
motion, in the transitional area before the particle has completed
several oscillations inherent in channelling or above-barrier
motion in this case. Such a regime is realized in crystals thin
enough, with the thicknesses that vary from less than tenths of
micron (hundreds of \AA) for MeV particles to several tens of
microns for hundreds of GeV particles (the characteristic
dimension of such an area depends on the particle energy as a
square-root function). In our study we will call these ultrathin
crystals. In this work we will mostly consider few-MeV charged
particles, so our range of crystal thicknesses spreads from about
hundreds up to thousands of {\AA}ngstr\"oms. In the last years the
technologies were developed to produce such crystals, and these
crystals have been used for channelling experiments
\cite{Guidi,Motapoth2,MotapothNesk_PRB2012,Motapoth1,Motapoth3}.
In the paper we will propose the experimental study of angular
distributions of electrons scattered by an unltrathin crystal.

The problem of obtaining characteristics of the motion of a
charged particle in these conditions may be resolved both by means
of quantum and classical theories, at that higher is the particle
energy, more the quantum and classical solutions match one other.

In this work we will stress on the energies low enough so as
quantum effects become essential in the particle motion but still
high enough so that the crystal thickness required for observing
these effects was reachable. For electrons we can propose the
study at energies 4\,Mev for mostly quantum motion and 50\,MeV
where some comparison of quantum results with the classical ones
begins to be reasonable. The conditions, necessary for the study
of phenomena considered in this paper, can be met with use of
modern technologies of creating ultrathin crystals and the
experiments can be realized on the base of accelerators PhIL and
ThomX in the laboratory LAL in Orsay, France.

In this paper we consider the case of planar scattering only as
the one which reveals the essence of the nature of the processes,
two-dimensional case of axial scattering being mostly only a
generalization of it (although, some phenomena as, for example,
dynamical chaos, will be only possible at axial scattering). Our
observations require low beam divergence, so one should aspire to
get it low for experiments. It is possible to improve the
divergence by squeezing the beam in the direction perpendicular to
crystal planes using magnetic field: simultaneously the beam will
stretch out along the planes but this would create no problem in
our one-dimensional study.

Here we will consider the case of parallel incidence of a particle
relatively a crystal plane. The incidence under a small angle
relatively plane reveals other interesting effects and is a
subject of a separate study.

\section{Classical scattering}

We can consider the interaction of a fast particle with matter
both within the classical and the quantum theory. The classical
theory of scattering is based upon the definition of the particle
trajectory in external field. At motion of a particle along
crystal planes its trajectory in transversal direciton is defined
as a solution of the differential equation of motion
\cite{Lindh1965,Gemmell1974,Kumakh1986,BazZhev1987,AkhiezShulga_HighEn1996},

\begin{equation}
\label{eq_clasrho}
\ddot{x}  =  - \frac{\,\,{c^2}}
{E_\parallel}\frac{\partial}{\partial x} U\left( x\right),
\end{equation}

\noindent
where $x$ is the coordinate of the particle in the plane
of transverse motion, ${E_\parallel} = c\sqrt{p_\parallel^2
+{m^2}{c^2}}$ and $U(x)$ is the potential energy of a particle in
the continuous potential of crystal planes. In this article we
will neglect such phenomena as multiple scattering or radiation
energy losses of a particle by assuming them to be small enough,
that is the consequence of a small crystal thickness, therefore
small particle path in the field of atomic forces.

The continuous potential of a crystal plane is obtained as the
average of the fields of atoms along it with taking into account
of random deviations of atom positions relatively their places in
the lattice caused by heat oscillations
\cite{AkhiezShulga_HighEn1996}. As a model of a solitary atom
potential we took the Moli\`ere potential that is widely used to
describe atomic electric forces. In order to obtain the potential
of entire crystal along the chosen direction we must summarize all
non-negligible contributions of the neighbouring planes.

So, for both positively and negatively charged particles (PCP and
NCP) in a crystal the continuous plane potential is a series of
periodically placed potential wells and potential barriers. We
must turn the potential upside down making the wells become
barriers and vice versa if the sign of particle charge changes to
the opposite (see Figure \ref{fig_pot110}).

In this case the potential in the neighbourhood of the bottom of
the well for both PCP and NCP can be approximated by a parabola

\begin{equation}\label{pot_fit}
U(x)=b (x - x_0)^2 + d,
\end{equation}

\noindent
where the parameters $x_0$, $b$, $d$ may be found using
a fitting procedure. The solution of equation of motion (1) in the
case of particles moving in such parabolic potential (in the case
of parallel incidence) is a set of harmonic curves
$x\!=\!x_0\,\cos{\omega t}$, where $\omega\!=\!\sqrt{2c^2
\left|b\right|/E_\parallel}$, therefore the spatial period of
oscillations is

\begin{equation}\label{T_osc}
T=\pi \beta\sqrt{2E_\parallel/\left|b\right|}\,,
\end{equation}

\noindent
 where $\beta=v/c$. We can define $T_0$ as the basic
oscillation period that corresponds to a particle entering in the
crystal in immediate proximity to the well bottom $x_0$. As we
move away from $x_0$, the form of real potential deviates from the
parabolic one, therefore the oscillation periods deviate from
$T_0$. We can compare such a behaviour of NCP and PCP by analysing
the difference between the real continuous potential and its fit
by parabolas.

\nopagebreak
\begin{figure}[htb]
\begin{center}
 \mbox{\epsfig{figure=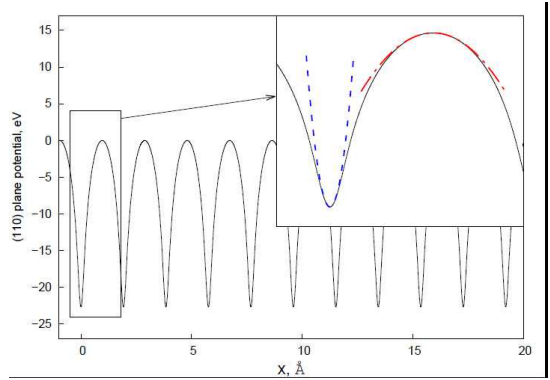,height=5.0cm}}
  \caption{Continuous potential of the plane (110) of Si crystal (solid line) and its approximations by quadratic functions of coordinates near extrema (blue dashed line -- approximation of the potential well for negative particles, red dash-and-dot line -- the one for positive particles, in this case is turned upside down)}
 \label{fig_pot110}
 \end{center}
 \end{figure}

As far as in some vicinity of the well bottom the real potential
remains close enough to its quadratic approximation, some part of
particles falling into a crystal are ``focused" after scattering
if the crystal thickness is a small integer number of half-periods
$T_{\onehf}=T_0/2$ of the particle oscillations within the
approximated parabolic potential (2). As the thickness increases,
the focusing effect weakens because of increasing difference of
real coordinates of the oscillation nodes for different impact
parameters. Obviously, the strongest focusing is observed at the
first half-oscillation, $L=T_{\onehf}$. We can see in
Figure\,\ref{fig_pot110} that the bottom of the potential well for
PCP is approximated by a parabola much better than for NCP. It
means that PCP will be focused more strongly and that the focusing
effect will persist for a larger number of periods than for NCP
(moreover, as stated below, the oscillation period for PCP is much
larger than for NCP, so the thickness where the focusing effect
can be observed is substantially  larger for PCP than for NCP that
is caused by these two factors).

For positively $(^+)$ and negatively $(^-)$ charged particles the
parameters $b^\pm$ in the fit (2) of the potential of the (110)
plane of Si crystal are
$\left|b^+\right|=17.01\,\texttt{eV/\AA$^2$}$ and
$\left|b^-\right|=407.6\,\texttt{eV/\AA$^2$}$. Therefore, we have
\begin{equation}\label{T_osc_numbers}
\begin{array}{l}
T_{\onehf}^+=0.5385 \beta\sqrt{E_\parallel[eV]}\texttt{\AA},
\\
T_{\onehf}^- = \;\;\; 0.11 \beta\sqrt{E_\parallel[eV]}\texttt{\AA}.
\end{array}
\end{equation}
So, for this crystal plane the period for positively charged
particles is almost 5 times larger than for negatively charged
particles, that is only explained by the geometry of potential.

From Figure\,\ref{fig_pot110} we see that, as far as we go away
from the well bottom, the real potential curve for NCP becomes
wider than its parabolic approximation and the one for PCP becomes
narrower. This fact causes different symmetry of the scattering
pictures for NCP and PCP in the neighbourhood of thicknesses
$L=n\!\cdot\!T_{\onehf}$, where $n$ is an integer number, what is
observed at comparison of both graphs of Figure
\ref{fig_caust_4MeV}: we see that because of this the caustic lines for PCP cross among themselves, and the number of crossing caustics increases with thickness, that all not being observed for NCP.

\begin{figure*}
\begin{center}
\resizebox{0.4\textwidth}{!}{\includegraphics{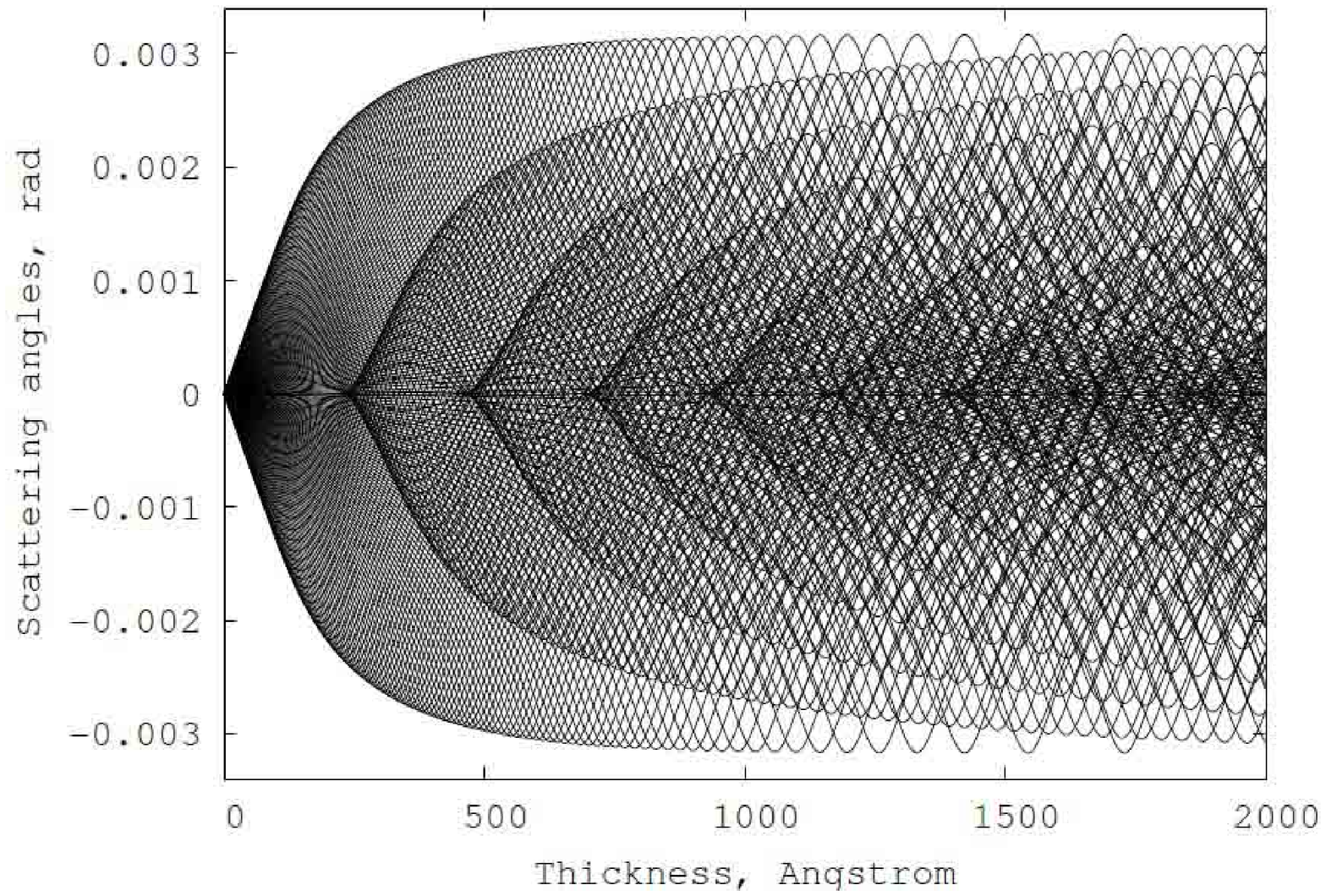}}
\resizebox{0.4\textwidth}{!}{\includegraphics{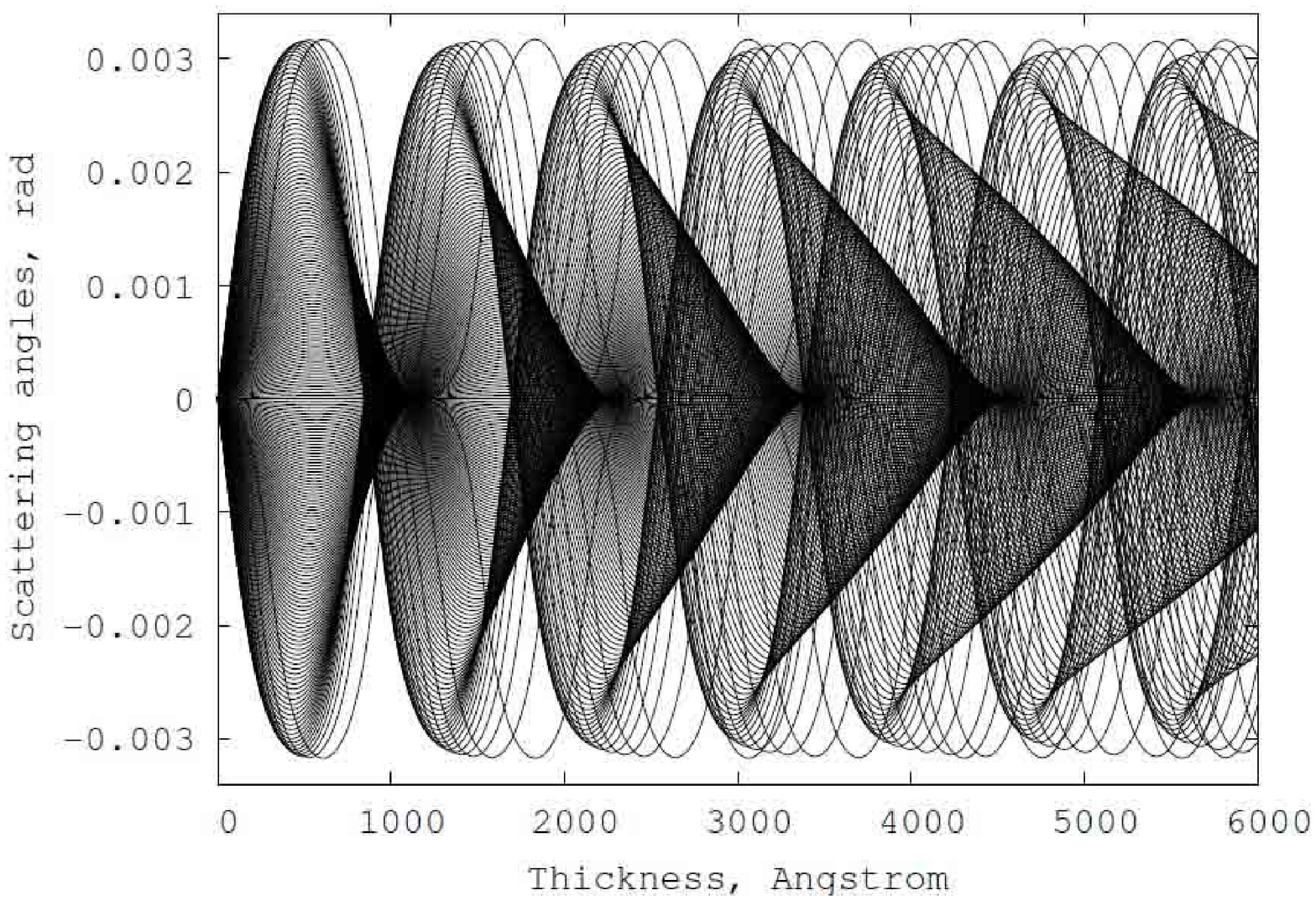}}
\caption{Classical scattering angles for different impact parameters of 4\,MeV electrons incident onto a 2000\,\AA~Si crystal parallel to (110) plane (left) and for different impact parameters of 4\,MeV positrons incident onto a 6000\,\AA~Si crystal parallel to (110) plane (right)}
 \label{fig_caust_4MeV}
\end{center}
\end{figure*}

Figure \ref{fig_caust_4MeV} is a set of scattering angles of fast charged particles with a large number (200) of different impact parameters uniformly distributed throughout the full interval between planes. The maximal angular amplitude of oscillations in these graphs (along vertical axis) corresponds to the critical channelling angle,
\begin{equation}
\psi_c = \sqrt{2U_{\rm max}/E_\parallel}.
\label{eq_psicrit}
\end{equation}
The difference in these pictures is only caused by the asymmetry
of planar potential relatively turnover upside down that is
connected with the change of sign of the particle charge. We can
see that entire scattering picture for PCP even changes its entire
angular dimensions at first half-periods of oscillations, while in
the case of NCP it quickly reaches its maximal value and then only
changes its internal structure. Near each ``focusing point'' we
see a caustic line enveloping the curves with similar impact
parameters, coming out from this focusing point. As the thickness
exceeds the ``focusing point'' the angular density having there a
sharp maximum begins to bifurcate, and the two shown up maxima
branch off in opposite directions, as observed in the plot of
angular distributions of scattered electrons. The sections of the
graphs of Figure \,\ref{fig_caust_4MeV} at constant thickness are
proportional to the density of trajectories. By using these
sections one can build the angular distributions of electrons
scattered by crystal planes (see Figure\,\ref{fig_clas_ele_div}).
Near the caustic lines, we have a strong increase of the density
of lines from one side and abrupt decrease from the other. The
angular distribution (Figure\,\ref{fig_clas_ele_div}) it is
manifested as sharp maxima at corresponding angles, the angular
distance between maxima being spread with the increase of
thickness. We treat the presence of such sharp maxima as an
appearance of the rainbow scattering phenomenon
\cite{Nussenzv1992}. It, applied to axial channelling, is studied
in works
\cite{KrausNesk_PRB1986,Nesk_PRB1986,PetroNesk_PRA2013,NeskPero_PRL1987,MotapothNesk_PRB2012,Motapoth1}.

\begin{figure*}
\begin{center}
\resizebox{0.38\textwidth}{!}{\includegraphics{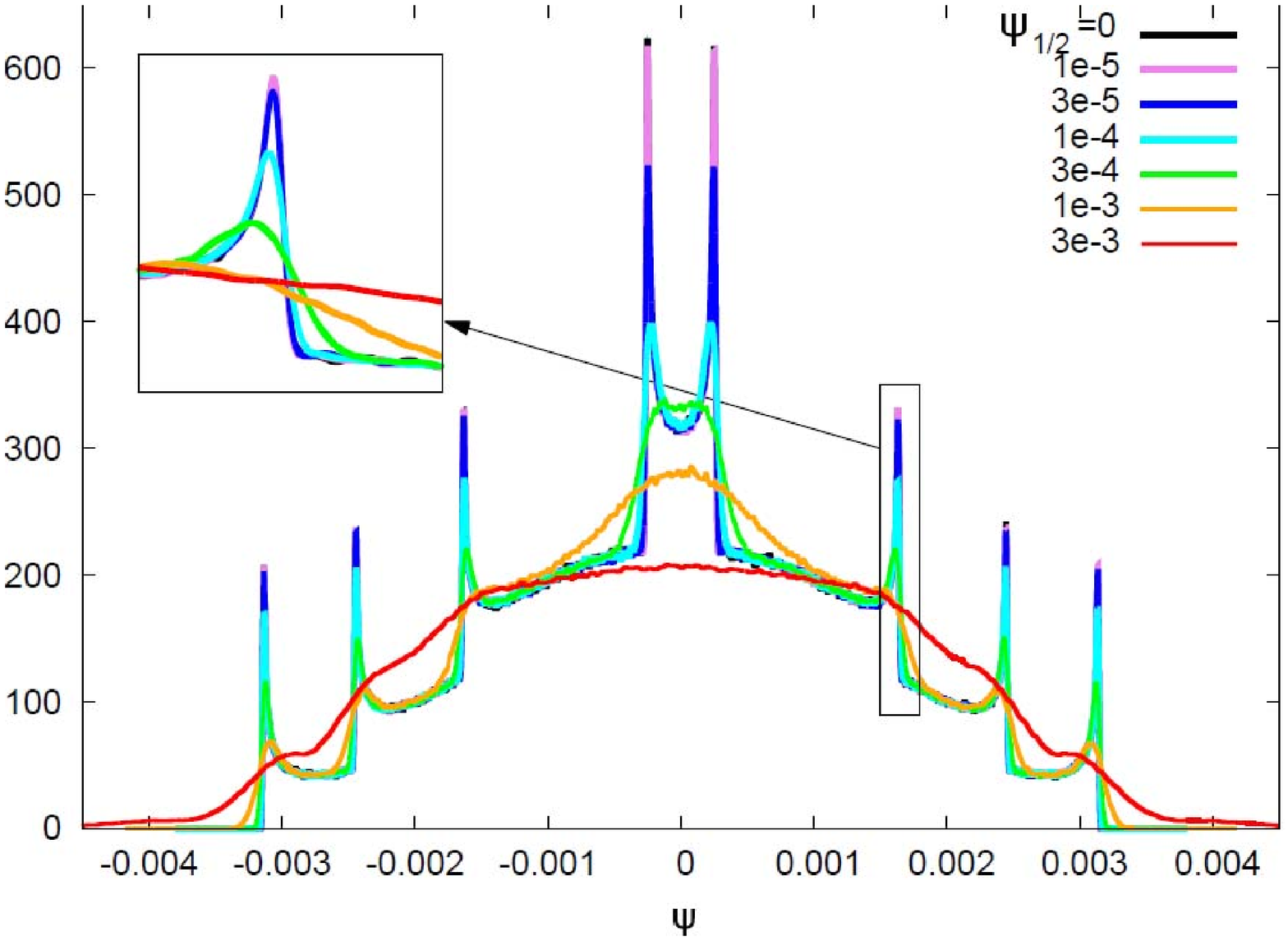}}
\resizebox{0.38\textwidth}{!}{\includegraphics{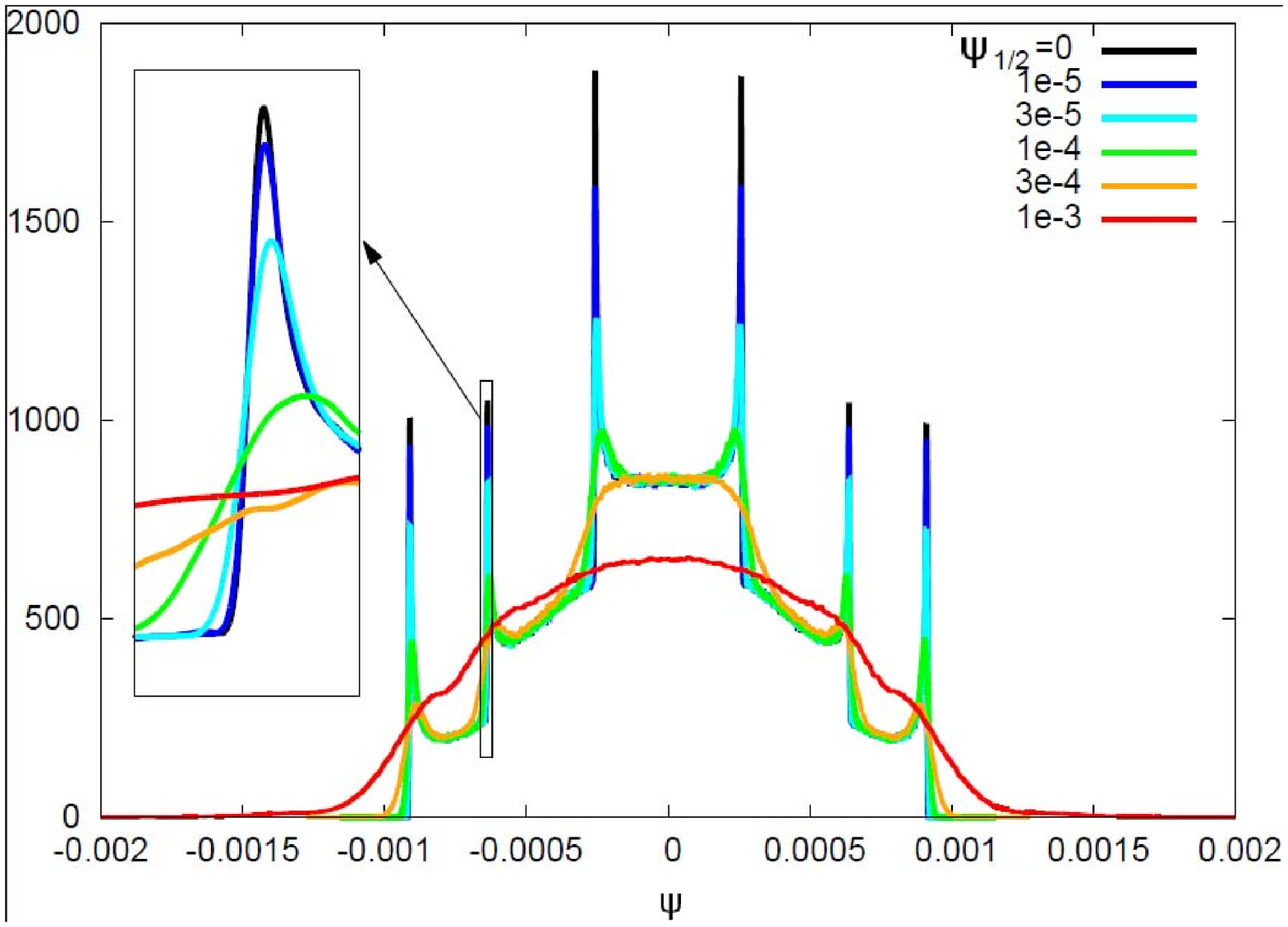}}
\caption{Classical simulation for scattering with different
initial divergences for beams of 4\,MeV electrons by a 750\,\AA Si
crystal (left) and 50\,MeV electrons by a 2000\,\AA~Si crystal
(right), both along the (110) plane}
 \label{fig_clas_ele_div}
\end{center}
\end{figure*}

A real beam is not parallel, and, as far as we consider the
angular properties of scattering, we must account for this fact.
In order to do this, we performed a simulation of the scattering
picture of charged particles of a beam with different initial
angular spread (Figure\,\ref{fig_clas_ele_div}). The ideal
situation of a parallel beam is also considered in order to see a
``pure'' picture. We can see that, as far as the initial beam
spread increases, the rainbow lines disappear being smeared. We
consider the angular distribution of the particles in the initial
beam as gaussian, and characterize the angular spread using the
parameter ``half-width-at-half-maximum" that means that at the
angle corresponding to this parameter we have half of the maximal
beam intensity (which is observed in center). We will designate
this parameter as $\psi_{\onehf}$.

By performing a simulation and using such parameters for the
description of a beam, we will see that in order to see the
rainbow effect in classical study at scattering of, e.g., 4\,MeV
electrons in 750\,\AA~Si crystal the beam angular spread must be
at least not larger than $\psi_{\onehf} \sim
3{\cdot}10^{-4}-1{\cdot}10^{-3}$\,rad. The analogous condition for
50\,MeV electrons in the 2000\,\AA~Si crystal gives us
$\psi_{\onehf}\sim 1{\cdot}10^{-4}-3{\cdot}10^{-4}$\,rad, being in
both cases about $0.1-0.3\;\psi_c$.

So, as we can see, the conditions imposed on the beam quality in
order to see the sharp maxima in classical study are strong
enough, although we find them to be reachable.

\section{Quantum scattering}

In order to realize the quantum study we need to describe the
initial beam as a wave packet, instead of the set of point-like
particles as in classical case, and to study its development with
time. As in the classical case, we must take into account that the
initial beam has some spatial dimensions and angular spread. The
analysis of the wave function will give us the information about
the wave packet motion.

Within the quantum theory, a fast particle moving in a certain
direction can be presented as a plane wave. A beam as a set of
particles is therefore a set of plane waves, their directions of
motion being distributed according to the laws of distribution of
particles in the beam. We, however, find useful to describe
mathematically the single particle wave function as a Gaussian
wave packet
\begin{equation}
\Psi \left( x,t=0 \right) =\frac{1}{\sqrt {\sigma \sqrt{\pi\,}\,}}
\exp \left(- \frac{x^2}{2\sigma^2}+ i\frac{{p_x x}}{\hbar }\right),
\end{equation}
where the parameter $\sigma$ corresponds to the wave packet that
covers a large number of neighbouring planes, hence, through the
uncertainty relations, follows a very low angular divergence of
such a wave packet.

In order to find the evolution of quantum state of the system with
time we used the action of the time evolution operator onto the
wave function. Such a way, with purpose to study the bound states
levels, has been used in the works
\cite{FeitFleck,Dabag1988,Dabag1988T1,Dabag1988T2,KozShuCherk2010,ShuSyshchNer2013}.

The essence of our way of finding the evolution of wave packet is
following: the change of the wave function with a time step
$\delta t$ is obtained as a result of action of the time evolution
operator onto its last step value:
\begin{equation}
\Psi \left( {x,t + \delta t} \right) = \exp \left( { - \frac{i}{\hbar }\delta t\,\hat H} \right)\Psi \left( {x,t} \right).
\label{psi_dt}
\end{equation}
But, we must take into account that the Hamiltonian of transverse motion is a sum of two \textit{non-commutating} terms
\begin{equation}
\hat H = -\frac{{{\hbar^2 c^2}}} {2{E_\parallel}} \frac{d^2}{dx^2}  + eU\left( {x} \right),
\label{hamilt}
\end{equation}
that calls forth that we cannot present the exponent (7) of the
hamiltonian (8) as a simple consequent product of exponents. This
does not let us take $\delta t$ as large as desired that would be
in the case of absence of the potential, and we need to look for
some approximation in order to get valuable results.  In order to
perform the expansion of the exponent in (7) in series in terms of
$\delta t$ we may use the Zassenhaus product formula. So, with
precision up to terms proportional to $(\delta t)^3$, we have:
\begin{equation}
\begin{array}{l}
\exp \left( { - \frac{i}{\hbar}\,\delta t\hat H} \right) \approx
\\
\exp \left(-\frac{i}{2\hbar}eU\left(x \right)\delta t\right)
\exp \left(i\frac{\hbar c^2\delta t}{2E_\parallel} \frac{d^2}{dx^2}\right)
\exp {\left(-\frac{i}{2\hbar}eU\left(x\right)\delta t\right)}.
\end{array}
\label{}
\end{equation}
In order to deal with the differential operator as an exponent
index and not to calculate higher order derivatives we may take
use of Fourier series formalism in which taking the derivative is
reduced to the multiplication of each Fourier series term by a
number. This procedure is exposed in the works \cite{FeitFleck,
KozShuCherk2010, ShuSyshchNer2013}, and so on.

Once we have the wave function in position space we can take a
Fourier transform in order to get it in momentum space. Therefore,
by taking the square of its absolute value we obtain the angular
distribution of scattered particles in quantum case. So, the
probability for the particle scattering in the interval
$[\psi,\psi+d\psi]$ is
\begin{equation}
dw(\psi)=\left|\int{\Psi\left(x,t\!=\!L/v\right)e^{-ip_{\parallel}\psi x/\hbar}dx}\right|^2 \frac{p_\parallel d\psi}{2\pi\hbar}.
\label{}
\end{equation}

In our study the wave functions describing single particles
correspond to almost plane waves with the divergence of about
$\psi_{\onehf}\sim0.001\psi_c$. In Figure\,\ref{fig_quant_SolEle}
we can see that the diffraction picture of single electrons in
crystal is a row of $\delta$-like maxima. The set of equidistant
narrow maxima can be explained as the expansion in reciprocal
lattice vectors. We can represent the particles as waves with the
wavelength equal to the de Broglie length
$\lambdabar=\hbar/p_\parallel$, therefore the angles corresponding
to the maxima must satisfy the relation
\begin{equation}\label{eq_reciprlat}
\psi_n=g_n/p_\parallel,
\end{equation}
where $g_n=2\pi\hbar n/d_x$ is the reciprocal lattice vector, $n$
-- integer number and $d_x$ is the distance between the crystl
planes. By the other words, we consider the crystal as a
diffraction grating and the particles scattering -- as a
scattering of de Broglie waves on this grating, so the sharp
maxima present on Figure\,\ref{fig_quant_SolEle} we explain as the
manifestation of the interference of electrons with themselves at
scattering on different planes. Higher is the particle energy,
more densely the allowed angles for particle scattering are
situated. As far as the distance between the peaks is proportional
to $1/E$ (11) and the channelling angle $\psi_c\propto
1/\sqrt{E}$, for higher energies we have the number of quantum
peaks inside the scattering range increasing proportional to the
square root of ${E}$. We can estimate the number of peaks:
$N_p\approx \psi_c/\psi_T=\frac{\sqrt{2U_{\rm max}}d}{2\pi
c\hbar}\sqrt{E}$. For the $(110)$ plane of Si crystal it is
approximately $N_p\sim\sqrt{E[MeV]}$.

\begin{figure*}
\begin{center}
\resizebox{0.38\textwidth}{!}{\includegraphics{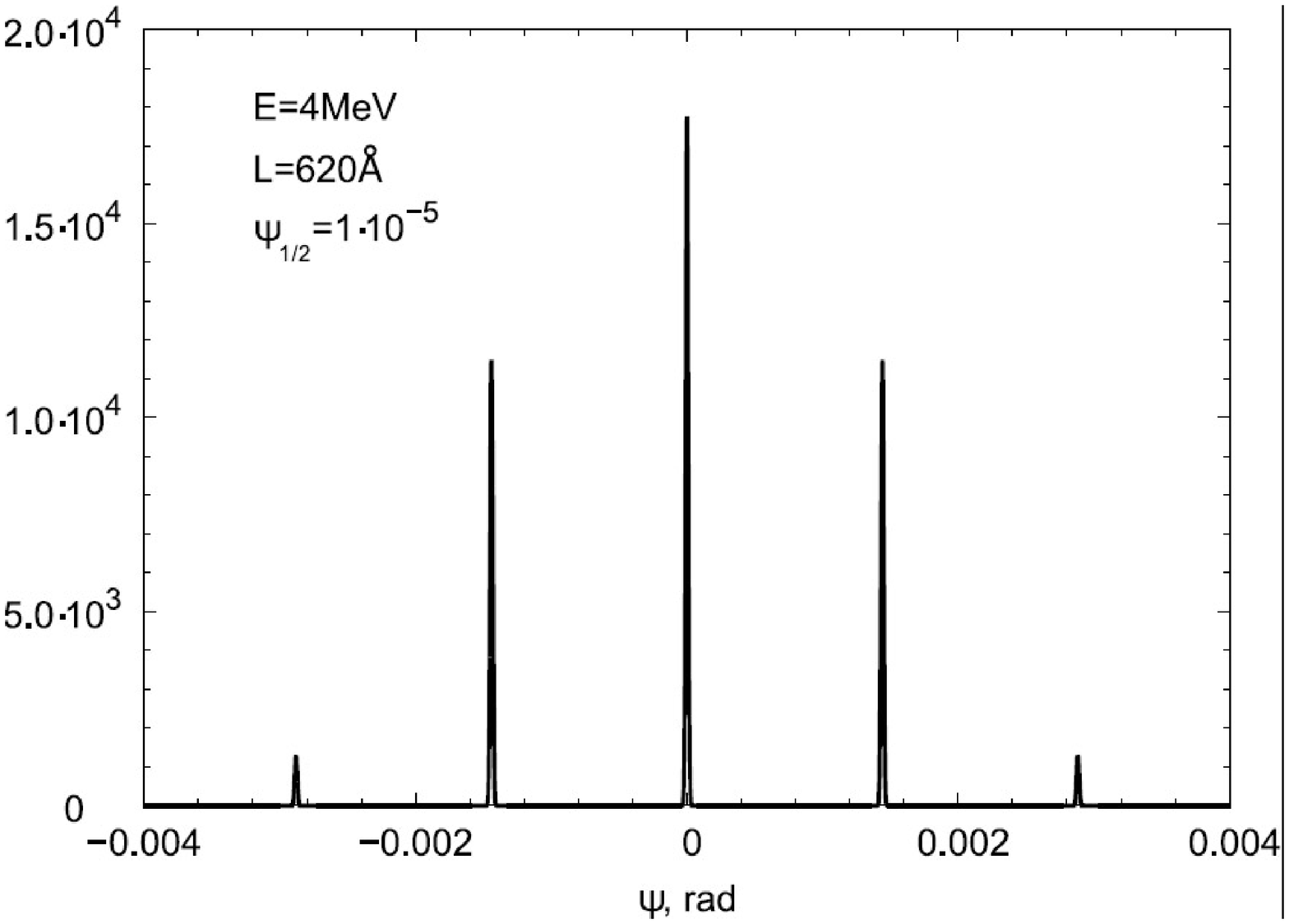}}
\resizebox{0.38\textwidth}{!}{\includegraphics{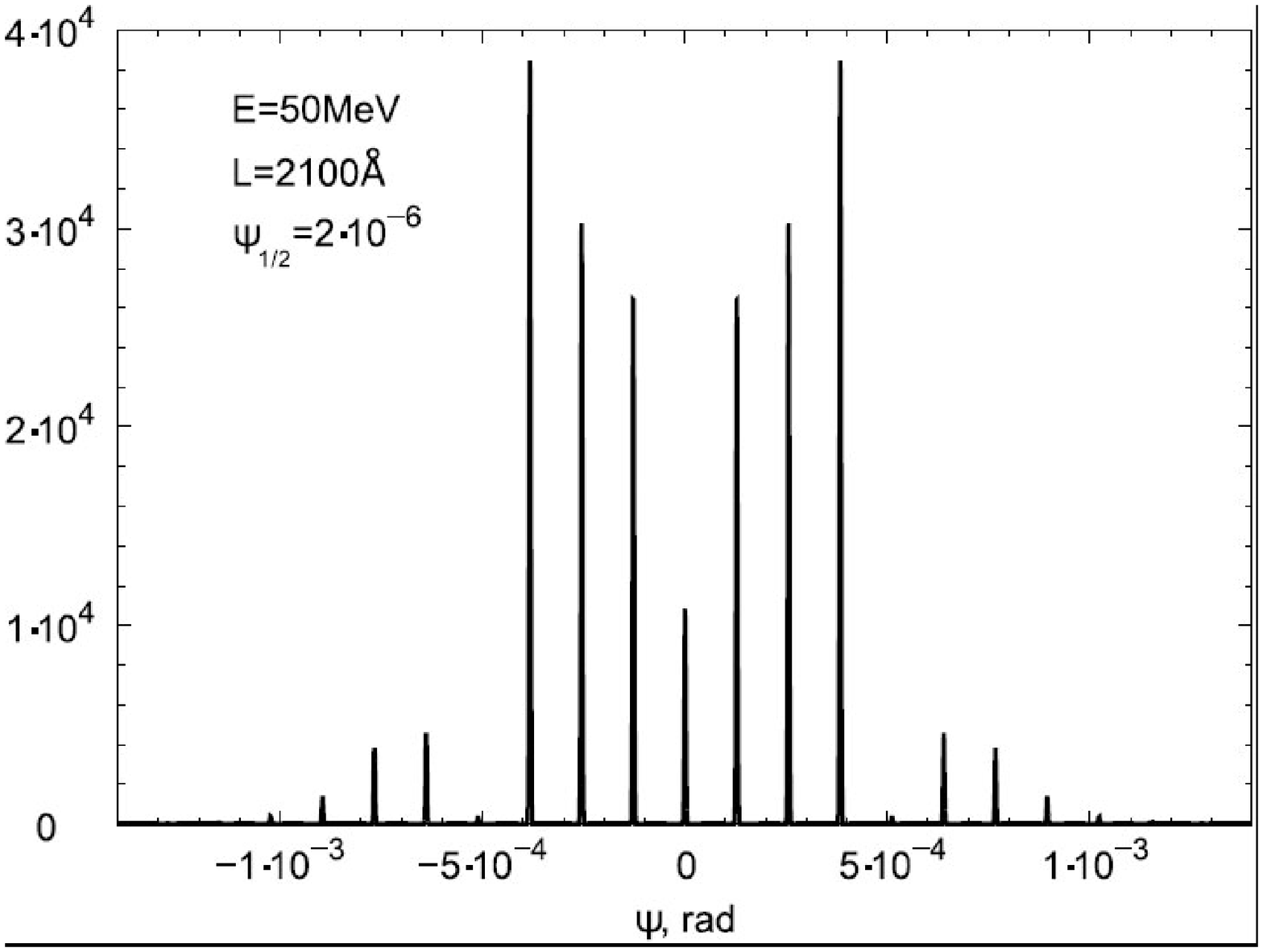}}
\caption{Diffraction of single electrons on (110) planes of Si
crystal}
 \label{fig_quant_SolEle}
\end{center}
\end{figure*}

We get the results for scattering in quantum case by averaging
contributions of solitary wave packets of single charged particles
by summing up their contributions with Gauss distribution function
that modulates the beam divergence:
\begin{equation}\label{eq_beamavrg}
\overline{w_{beam}(\psi)}=\frac{1}{\sigma_b\sqrt{\pi}} \int{e^{-\psi_i^2/\sigma_b^2}}\,w(\psi_i,\psi)\,d\psi_i,
\end{equation}
where $\overline{w_{beam}(\psi)}$ states for the density of
probability of scattering of the beam incident as a whole parallel
to the crystal planes in the direction $\psi$, and
$w(\psi_i,\psi)$ is the density of probability that the particle
incident at the angle $\psi_i$ to the planes is scattered in the
direction $\psi$.

\nopagebreak
\begin{figure}[htb]
\begin{center}
 \mbox{\epsfig{figure=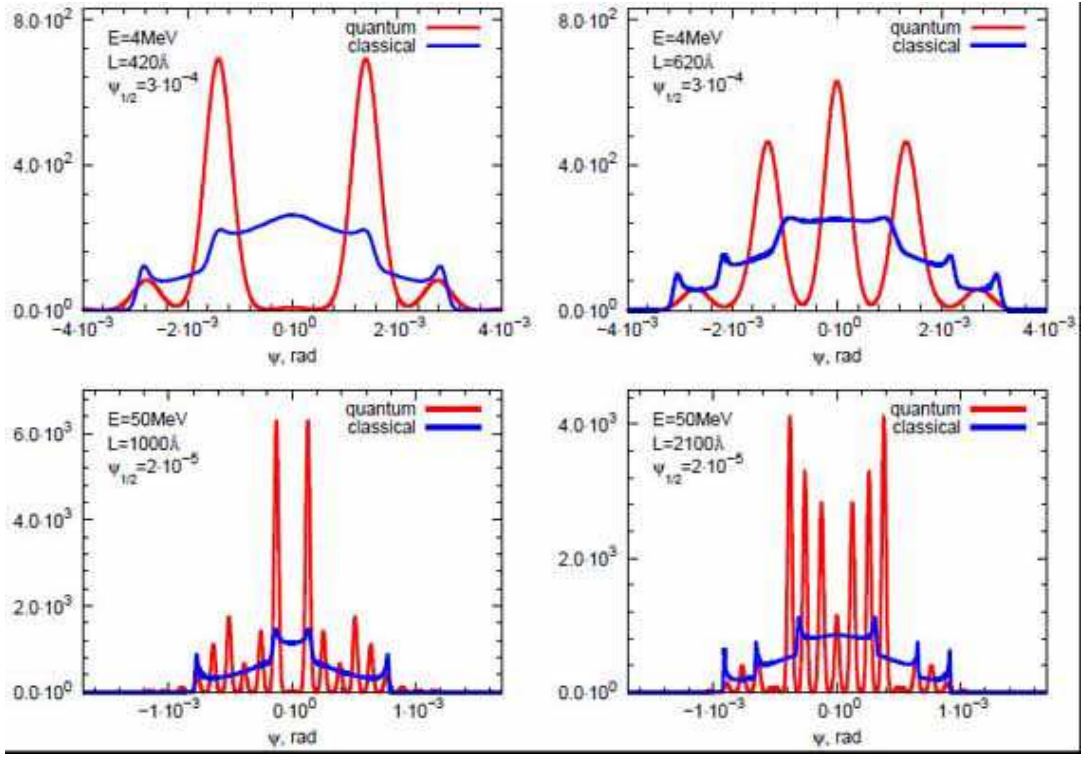,height=10.0cm}}
 \caption{Quantum and classical simulations of angular distribution
of 4\,MeV (above) and 50\,MeV (below) electrons  scattered in a Si
crystal incident parallel to the ${(}110{)}$ plane}
 \label{fig_QuCl_different}
 \end{center}
 \end{figure}

It is possible to compare quantum results for angular
distributions with the classical ones at the same parameters. In
Figure~\ref{fig_QuCl_different} we present quantum angular
distributions obtained by the method described here and the
classical ones as solutions of the classical equation of motion
(1). We see that, lower is the particle energy, greater is the
difference between the classical and quantum pictures.
Particularly for low energies, the classical and quantum pictures
may be substantially different. Besides, the averaging over a
diverging beam in quantum case can make neighbouring maxima flow
together, make them disappear or displace them. For example, on
the picture for 420\,\AA~we see that the central part of the
angular distribution of scattering, being elevated in classical
distribution,  is strongly depressed in quantum case. From the
other hand, for higher energies, the sharp maxima observed in
classical pictures have some manifestation in quantum case: for
the angles that correspond to them the quantum peaks are higher,
and for GeV energies they flow together so as quantum picture
approaches the classical one.

As written above, the positions of quantum and classical maxima
have different nature, so the positions of quantum ones do not
depend on the crystal thickness (at the absence of strong
influence of the beam divergence), whereas the classical ones
migrate towards outside of the scattering figure with the increase
of thickness, so any coincidence of classical and quantum maxima
is accidental. It can be observed at one crystal thickness and not
be observed for another one.

\section{Conclusion}

In this paper we propose an idea for experiment on planar
scattering of fast charged particles in ultrathin crystal, in
order to observe quantum and classical effects that can manifest
themselves in scattering picture. If the quality of beam and
crystal is good enough and the resolution of detectors is high
enough we expect the observing of a series of spots of different
brightness that, by varying the initial parameters may be referred
to manifestation of quantum or classical nature of processes at
interaction of charged particles with ultrathin crystal.

The obtained results for angular distributions of the electrons
scattered by a ultrathin crystal show that, for Si crystal with
thickness about several hundreds of \AA~the quantum effects at
scattering can be essential, that are connected with the
interference effect of single electrons on a set of crystal
planes. This effect is particilarly bright for the electrons
energies about a few MeV. As the plots on Figure~5 show, the
electron beam must have parameters attainable on the PhIl and
ThomX facilities in the laboratory LAL in Orsay, France.

We did not include in this paper the study of the levels of
transversal energy of particles that they occupy in the potential
wells of crystal potential. These energy levels are to be observed
by using other technical measures, such that will let us register
the photons irradiated at interaction of charged particles with
crystal, so such observations could reply the question about the
mechanisms of arising of these levels and their nature, whether
they are connected only with the potential well form or also with
the reciprocal influence of neighbouring crystal planes and
interference of charged particles on them. These questions could
be answered by performing, in addition and in connection with the
study proposed in this paper, of another study of radiation
arising at interaction of charged particles with ultrathin
crystal.

\section{Acknowledgement}

Research conducted in the scope of the IDEATE International
Associated Laboratory (LIA). The work was done with partial
support of the NAS of Ukraine, project $\Phi$-12 and of Ministry
of Education and Science of Ukraine (project no.~0115U000473). One
of the authors (S.N.\,Shul'ga) is grateful to the Laboratoire de
l'Acc\'el\'erateur Lin\'eaire where the essential part of this
work was done and its researchers for the hospitality and fruitful
discussions.

\coltoctitle{Transition radiation by low-energy relativistic "half-bare" electron}
\coltocauthor{N.F.~Shul'ga, S.V.~Trofymenko, S.Ya.~Barsuk, O.A.~Bezshyyko}

\begin{center}
{\Large {\bf Transition radiation by low-energy relativistic \lq half-bare\rq~electron}}

\vskip 0.5cm

{\bf N.F.~Shul'ga~$^{a,b}$, S.V.~Trofymenko$^{a,b}$, S.Ya. Barsuk$^{c}$, O.A.~Bezshyyko$^{d}$}

\vskip 0.3cm

$^{a}${\it Akhiezer Institute for Theoretical Physics of National Science Center \lq Kharkov Institute of Physics and Technology \rq, Kharkov, Ukraine}

$^{b}${\it Karazin Kharkov National University, Kharkov, Ukraine}

$^{c}${\it LAL, Universit\'e Paris-Sud, CNRS/IN2P3, Orsay, France}

$^{d}${\it Taras Shevchenko National University of Kyiv, Kyiv, Ukraine
}

\end{center}

\vskip 0.4cm

\begin{abstract}
\noindent The problem of transition radiation generated by relativistic electron which has non-equilibrium electromagnetic field around it at its impinging upon conducting plate is considered. The conditions under which such non-equilibrium (\lq halh-bare') state of electron significantly influences upon characteristics of the particle transition radiation are discussed. It is proposed to investigate experimentally such influence for incident electrons at low energies (several MeV).
\end{abstract}

\noindent

\vskip 0.4cm

\noindent PACS numbers: 41.60.Dk

\vskip 0.4cm

\noindent  Keywords: Transition radiation, \lq half-bare' electron

\section{Introduction}

When a charged particle traverses the interface between two media with different electromagnetic properties transition radiation (TR) is generated. In usual statement of the problem about TR of a particle traversing single interface the radiation characteristics depend on physical properties of the media, the particle energy and the angle of its incidence upon interface. It is usually assumed that the particle, before impinging upon interface, moves uniformly and rectilinearly during long period of time and has equilibrium coulomb field around it. This is however not the case for particles which change their state of motion (accelerate) before traversing the boundary between media. Such situation, for instance, takes place if a particle before impinging upon the interface is deflected from the initial direction of motion (by external field or bent crystal) (Fig. 1a).

\nopagebreak
\begin{figure}[htb]
\begin{center}
 \mbox{\epsfig{figure=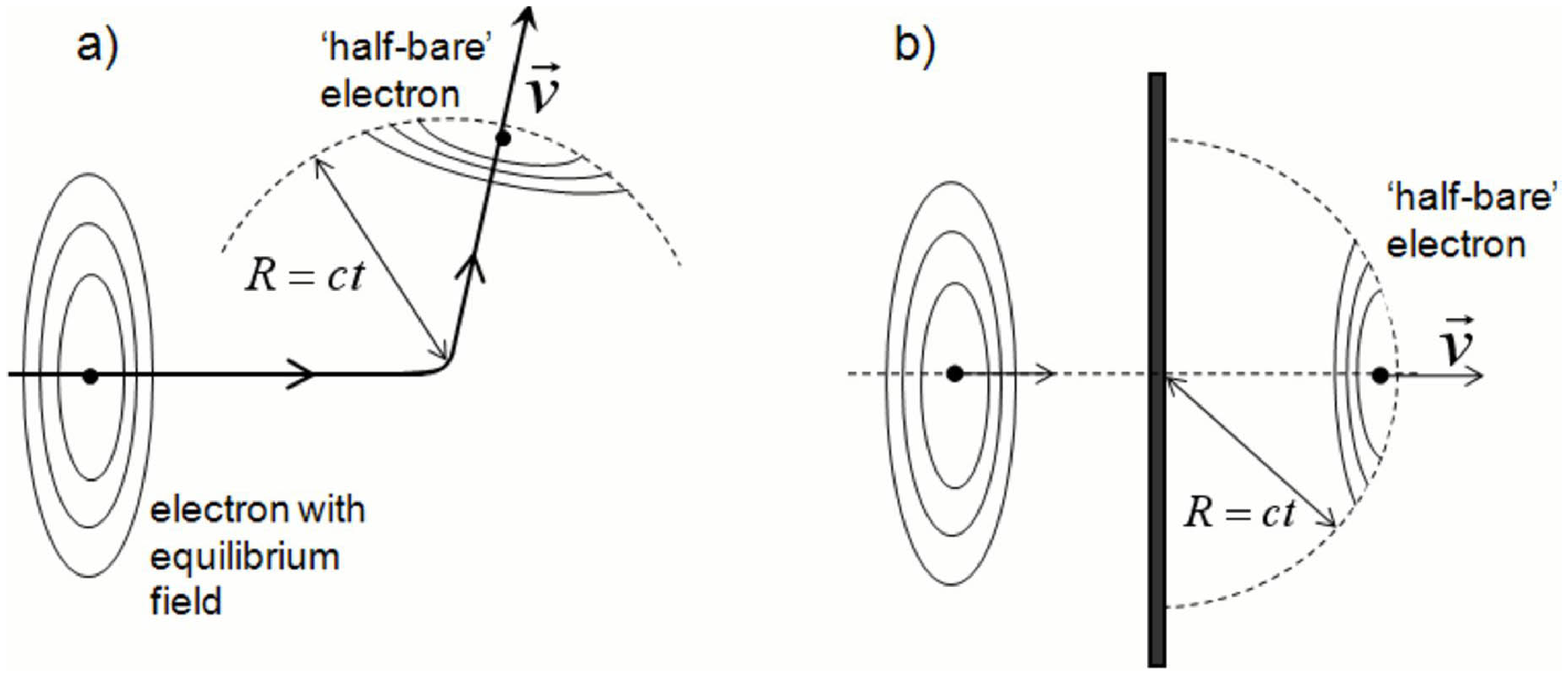,height=5.0cm}}
 \caption{Schematic structure of electromagnetic fields around the particle after its deflection to a large angle (a) and after traversal of a conducting plate (b).}
 \end{center}
 \end{figure}

According to \cite{Feinberg,Akhiezer}, by the moment of time $t$ after deflection the electromagnetic field inside the sphere of radius $R=ct$ ($c$ is the speed of light) with the center in the deflection point will coincide with coulomb field of the particle moving in final direction while outside the sphere (the region of space which the signal about deflection has not yet reached) it is still coulomb field of the particle moving in the initial direction. During some period of time after deflection the field around the deflected particle is non-equilibrium. The spectrum of virtual photons which constitute the field of scattered particle lacks significant part of frequencies comparing to corresponding spectrum of equilibrium coulomb field. The restoration of the field around the particle after scattering occurs in such way that each single-frequency component of this field (Fourier component or virtual photon) regenerates only on distance from the deflection point known as coherence length: $l_{C} \sim \gamma^{2}\lambda$, where $\gamma$ is the particle Lorenz-factor and $\lambda$ is the wavelength of the considered component. The deflected (scattered) electron with such incomplete electromagnetic field around it is known as \lq half-bare' electron. Here we use the notion of deflection \lq point' which is valid if consider the Fourier components of the fields produced in this process for which $l_C>>a$, where $a$ is the characteristic linear size of the region in which deflection occurs. The condition $R>>a$ should be fulfilled as well.

The manifestations of such non-equilibrium state of high-energy (multi-GeV) electron were previously theoretically considered \cite{Landau,Migdal,Ternovsky,Shulga} and experimentally observed \cite{Anthony,Klein,Thomsen1,Thomsen2} in bremsstrahlung emitted by the particle during its multiple scattering on the atoms while moving through substance. In this case the \lq half-bare' state, in which electron appeared after scatterings, resulted in suppression of radiation intensity in certain frequency range (Landau-Pomeranchuk-Migdal \cite{Landau,Migdal} and Ternovsky-Shul'ga-Fomin \cite{Ternovsky,Shulga} effects).

In the present paper we propose to study the influence of \lq half-bare' state of electron upon its millimeter wavelength TR characteristics which is generated at impinging of such electron upon thin conducting plate situated in vacuum at several MeV energy of incident electrons (which corresponds to the energy of electrons delivered by photoinjector PHIL at LAL Orsay). The plate could be placed in the final direction of motion of deflected electron in the vicinity of the deflection area. However calculations \cite{Trofymenko1} show that the considerable change of TR characteristics (due to \lq half-bareness' of the particle) takes place in the case when relatively large deflection angle ($\chi>>1/\gamma$) is obtained within the area much smaller than the coherence length $l_{C}$ for the considered radiation wavelength. At low energies it is technically difficult to realize such condition and another way of obtaining electrons in \lq half-bare' state is needed. The considerations made in \cite{Trofymenko2} show that the structure of electromagnetic field around the electron analogous to the one after the particle deflection to a large angle could be obtained by electron traversal of thin conducting plate (Fig. 1b). Hence, characteristics of TR generated by preliminarily deflected electron should be analogous to the ones of TR emitted by electron which before impinging upon the second plate \lq undresses' traversing the first one and the latter process is proposed to be used for investigation of \lq half-bare' electron TR at low particle energies.

\section{Transition radiation by low-energy \lq half-bare' electron at oblique incidence upon conducting plate}

Let us note that the study of radiation in millimeter wavelength region generated during high-energy electron traversal of a system of two conducting plates (Fig. 2) was previously made in the work \cite{Shibata}. In this work the coherent TR generated by 150 MeV bunched electron beam which traversed a radiator system consisting of an upstream aluminum foil and a downstream inclined aluminum mirror was investigated. In the considered experiment radiation suppression (comparing to radiation on a single mirror in the absence of the foil) and its gradual growth with the increase of distance between the mirror and the foil was observed. It is necessary to note that the foil and the mirror of rather small transversal (with respect to the direction of the electron velocity) size were used and in most part of the measurements this size was smaller than characteristic transversal size of electromagnetic fields of impinging electrons. This makes the characteristics of generated radiation influenced by diffraction of electromagnetic field on the plates edges and different from the corresponding characteristics of TR from an infinite target. However, as follows from the paper, it was the theory of TR generated on the plates of infinite size which was used to interpret the experimental data in \cite{Shibata}.

\nopagebreak
\begin{figure}[htb]
\begin{center}
 \mbox{\epsfig{figure=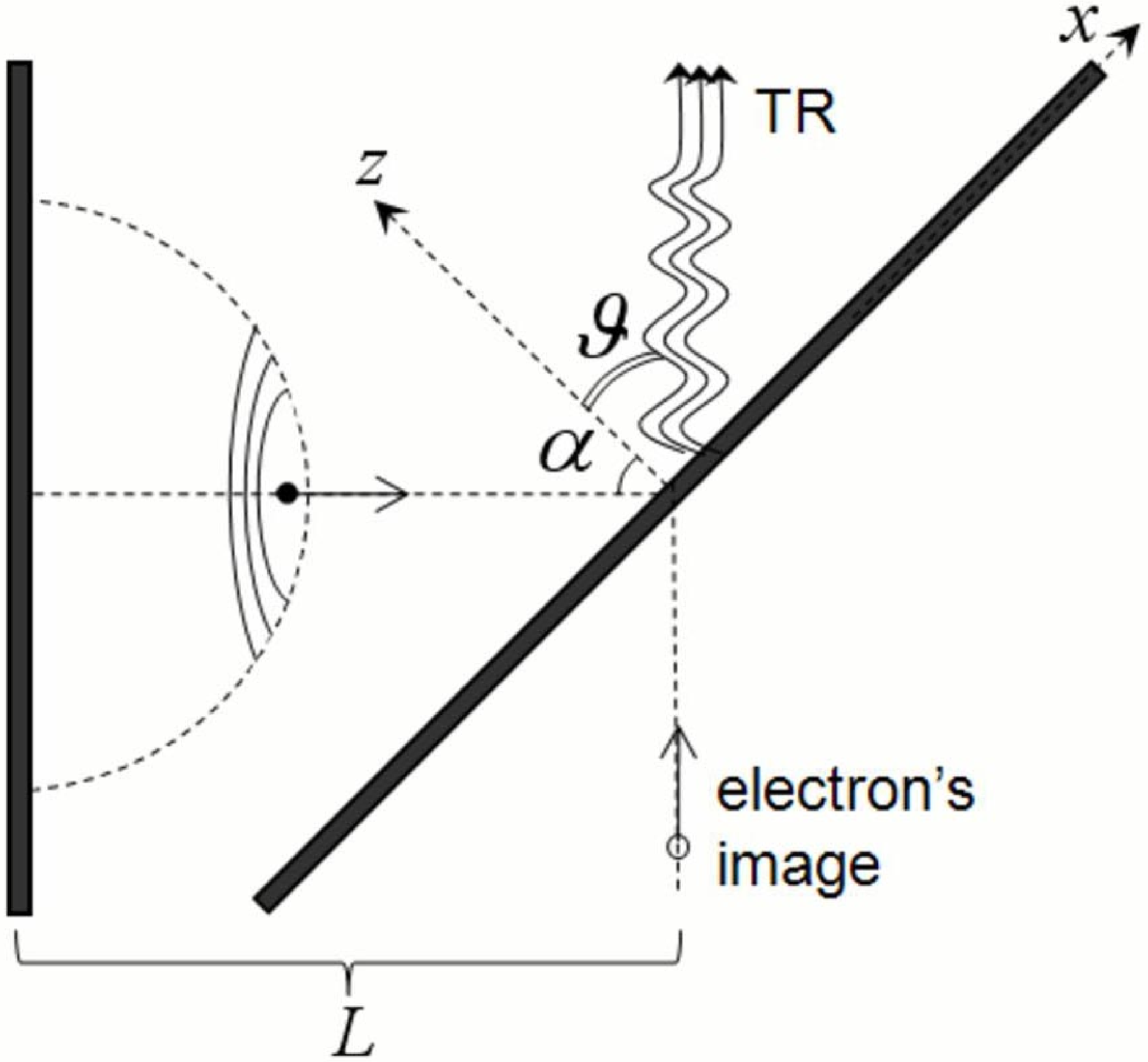,height=5.0cm}}
 \caption{TR by \lq half-bare' electron on the inclined plate (for $\alpha=\pi/4$).}
 \end{center}
 \end{figure}

In the present paper we propose to investigate the peculiarities of \lq half-bare' electron TR in the millimeter wavelength range at low energies (several MeV). At such energies the characteristic transversal  size of the particle's own field (for the given wavelength it is defined by expression $l_{T}\sim \gamma\lambda/2\pi$) is quite small and the condition of \lq infinite plate' could be easily realized. Such condition means that the linear transversal dimensions of both TR-radiating plates should be larger than $l_{T}$ and the approximation of infinite size of the plates is valid for theoretical description of measured radiation characteristics. The diffraction of the electron's field on the plates' edges is absent. Moreover under the considered conditions the radiation formation length $l_{C}$ is rather small (however macroscopic) and it is possible to make the measurements not only for distances (between the plates) much smaller than $l_{C}$ (as in \cite{Shibata}) but on distances comparable and larger than this value as well. As we show, in this case it is possible to observe new features of \lq half-bare' electron TR such as enhancement of radiation intensity comparing to the one emitted by electron with equilibrium field and oscillation of TR intensity with the change of distance between the plates.

The consideration of TR in millimeter wavelength region in our case is also preferable since it allows to make measurements in the wave zone of radiation process (place the detector on distance much larger than $l_C$ from the plates) and not to use additional equipment (parabolic focusing mirrors, for instance) to eliminate the effect of pre-wave zone in radiation measurements \cite{Castellano,Naumenko}. Among the vices of the proposed experimental conditions it is necessary to note the lower intensity of TR comparing to the case of electrons of higher energies.

Let us consider backward (to the half-space from which the electron comes to the plate) transition radiation generated at impinging upon thin conducting plate of \lq half-bare' electron which \lq undresses' during preliminary traversal of another conducting plate (Fig. 2). As was noted the TR characteristics in the considered case should be similar to TR characteristics in the case of electron traversal of the plate after being deflected to a large angle. In the considered case the second plate (on which backward TR occurs) should have some inclination relatively to the first one (vertical) so that the backward TR could pass the first plate and arrive at the detectors.

The necessity to use the inclined plate in the considered case at rather low energies (the calculations is made for 3.5 MeV electrons) causes significant perturbation of the \lq elegant' azimuthally symmetrical structure of TR calculated in high-energy approximation (presented in \cite{Shibata}). Nevertheless in the considered case the \lq half-bare' state of electron still significantly influences upon its TR characteristics and could be experimentally registered.

In order to calculate the considered TR spectral-angular density we use the method of images which is valid also in relativistic case if consider radiation from infinite ideally conducting plates \cite{Trofymenko2,Ginzburg}. In this method the radiation emitted at \lq half-bare' electron impinging upon the inclined plate is considered as radiation produced by electron and its image inside the inclined plate during their motion towards the point of electron traversal of the plate and abrupt stop at this point. Moreover the \lq half-bareness' (as a result of electron traversal of another plate) is taken into account by taking the moment of time at which this motion begins as $t_{0}=-L/v$, where $L$ is the length of electron's path between the plates and $v$ is the particle velocity. The calculations give the following expression for TR spectral-angular density of \lq half-bare' electron in this case:

\begin{equation}
\frac{\pi^{2}c}{e^{2}}\frac{d^2\textrm{\Large $\varepsilon$}}{d\omega do}=\left (\frac{\bf n\times\textrm{\mathversion{bold}$\beta_{1}$}}{1-\bf n \textrm{\mathversion{bold}$\beta_{1}$}}\right )^{2}I_{1}   +   \left (\frac{\bf n\times\textrm{\mathversion{bold}$\beta_{2}$}}{1-\bf n \textrm{\mathversion{bold}$\beta_{2}$}}\right )^{2}I_{2}     -       \frac{\textrm{\mathversion{bold}$\beta_{1}$}\textrm{\mathversion{bold}$\beta_{2}$}-(\bf n\textrm{\mathversion{bold}$\beta_{1}$})(\bf n\textrm{\mathversion{bold}$\beta_{2}$})}{(1-\bf n \textrm{\mathversion{bold}$\beta_{1}$})(1-\bf n \textrm{\mathversion{bold}$\beta_{2}$})}I_{12},
\label{one}
\end{equation}
which differs from the corresponding expression for electron with equilibrium field by the presence of interference factors

$
$

$
I_{1}=\sin^{2}\left[\textrm{\Large $\frac{\omega l}{2v}$}(1-\beta_{x}\sin\theta \cos\phi+\beta_{z}\cos\theta)\right ] ,
$

$
$

$
I_{2}=\sin^{2}\left[\textrm{\Large $\frac{\omega l}{2v}$}(1-\beta_{x}\sin\theta \cos\phi-\beta_{z}\cos\theta)\right ] ,
$

$
$

$
I_{12}=\sin^{2}\left[\textrm{\Large $\frac{\omega l}{2v}$}(1-\beta_{x}\sin\theta \cos\phi+\beta_{z}\cos\theta)\right ]    +    \sin^{2}\left[\textrm{\Large $\frac{\omega l}{2v}$}(1-\beta_{x}\sin\theta \cos\phi-\beta_{z}\cos\theta)\right ]- $

$
$

$
-\sin^{2}\left[\textrm{\Large $\frac{\omega l}{2v}$}\beta_{z}\cos\theta\right ].
$

$
$

In order to obtain the expression for TR by electron with equilibrium field (see, for example, \cite{Bass}) we have to put $I_{1}$ and $I_{2}$ equal to $1/4$ while $I_{12}$ equal to $1/2$).

The quantity $\phi$ in (\ref{one}) is the azimuthal angle in the plane of the inclined plate (the angle between $x$-axis and projection of radiation direction vector $\bf n$ on the plate surface). $\theta$ is the angle between $\bf n$ and $z$-axis. $\textrm{\mathversion{bold}$\beta_{1}$}$ and $\textrm{\mathversion{bold}$\beta_{2}$}$ are the dimensionless velocities of electron and its \lq image' which moves symmetrically to electron with respect to the inclined plate. They have the same magnitude $v/c$. Also $\beta_x=(v/c)\sin \alpha$, $\beta_z=(v/c)\cos \alpha$ and

$
$

$
\bf n\textrm{\mathversion{bold}$\beta_{1}$}=$$\beta_x\sin\theta\cos\phi-\beta_z\cos\theta$$,
$

$
$

$
\bf n\textrm{\mathversion{bold}$\beta_{2}$}=$$\beta_x\sin\theta\cos\phi+\beta_z\cos\theta$$,
$

$
$

$
(\bf n\times\textrm{\mathversion{bold}$\beta_{1,2}$})^2=$$v^2/c^2-(\bf n\textrm{\mathversion{bold}$\beta_{1,2}$})^2$$
$

$
$

For measurements in the wavelength region $\lambda<5~mm$ ($\gamma\lambda/2\pi<0.56~cm$) a plate (vertical) with transversal size $2\times2~cm$ may be used (it should be larger than the transversal size of the electron's coulomb field $2l_T\sim\gamma\lambda/\pi$). If we take $\alpha=\pi/4$ (the inclined plate in this case should have then $\sqrt{2}$ times larger dimension along the $x$-axis in order to have the same transversal cross-section) the minimal distance $L$ of the electron's path between the plates which can be achieved is $1~cm$. The expression (1) is valid for arbitrary energies of incident electron and at sufficiently high-energies coincides with expression used in \cite{Shibata}.

\section{The peculiarities of \lq half-bare' electron TR}

Changing the distance $L$, the observation angle $\theta$ and the measured TR wavelength $\lambda$ it is possible to investigate the difference between the \lq half-bare' electron TR characteristics and the ones of electron with equilibrium field. According to (\ref{one}), this difference can manifest itself both in radiation suppression and enhancement, change of angular distribution and dependence of TR intensity on the radiated wavelength (which is absent for TR by electron with equilibrium field).

Fig. 3 shows that the intensity of TR by \lq half-bare' electron in the vicinity of the radiation maxima for electron with equilibrium field can be either suppressed or enhanced depending on the value of the parameter $\delta=L/l_C$. Namely, if the condition $\delta<<1$ is fulfilled the radiation will be suppressed (thick solid curve). Otherwise enhancement may take place. In any case the considered TR has wider angular distribution (around the direction of the electron image velocity) than the radiation by electron with equilibrium field. Here we use the denomination $\textrm{\Large $\varepsilon$}_{\omega,o}=(4\pi^2 c/e^2) d^2\textrm{\Large $\varepsilon$}/d\omega do$ for the dimensionless quantity proportional to TR spectral-angular density.

\nopagebreak
\begin{figure}[htb]
\begin{center}
 \mbox{\epsfig{figure=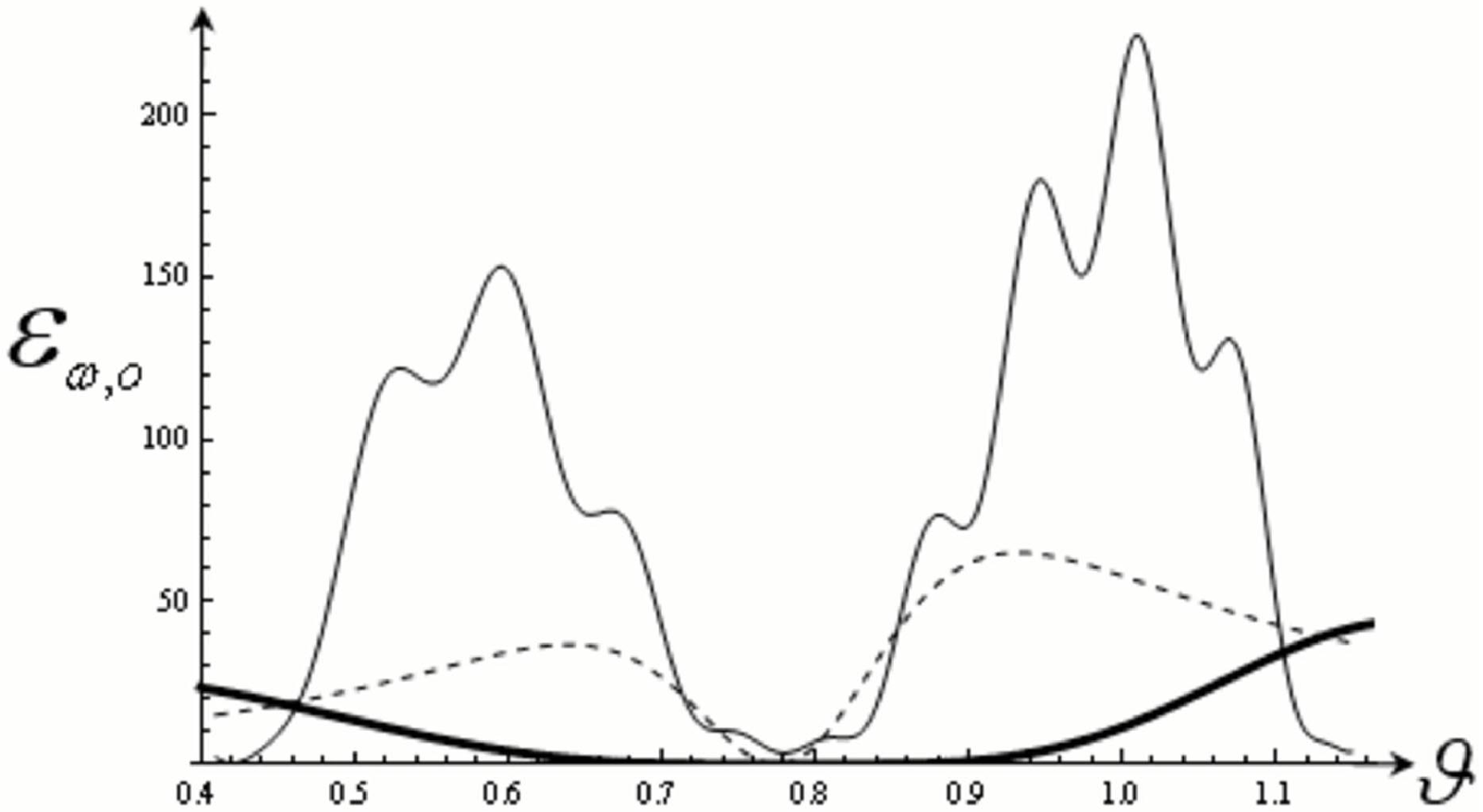,height=5.0cm}}
 \caption{TR angular distribution for $\phi=0$. Dashed line -- radiation by electron with equilibrium field; thin solid line - radiation by \lq half-bare' electron for $\lambda=3~mm$, $L=4~cm$ ($\delta\approx 0.27$); thick solid line -- radiation by 'half-bare' electron for  $\lambda=5~mm$, $L=1~cm$ ($\delta\approx 0.04$).}
 \end{center}
 \end{figure}

Fig. 4 represents the examples of dependence of TR intensity on distance $L$ between the plates (on the magnitude of the electron \lq half-bareness') for different radiation wavelengths.

\nopagebreak
\begin{figure}[htb]
\begin{center}
 \mbox{\epsfig{figure=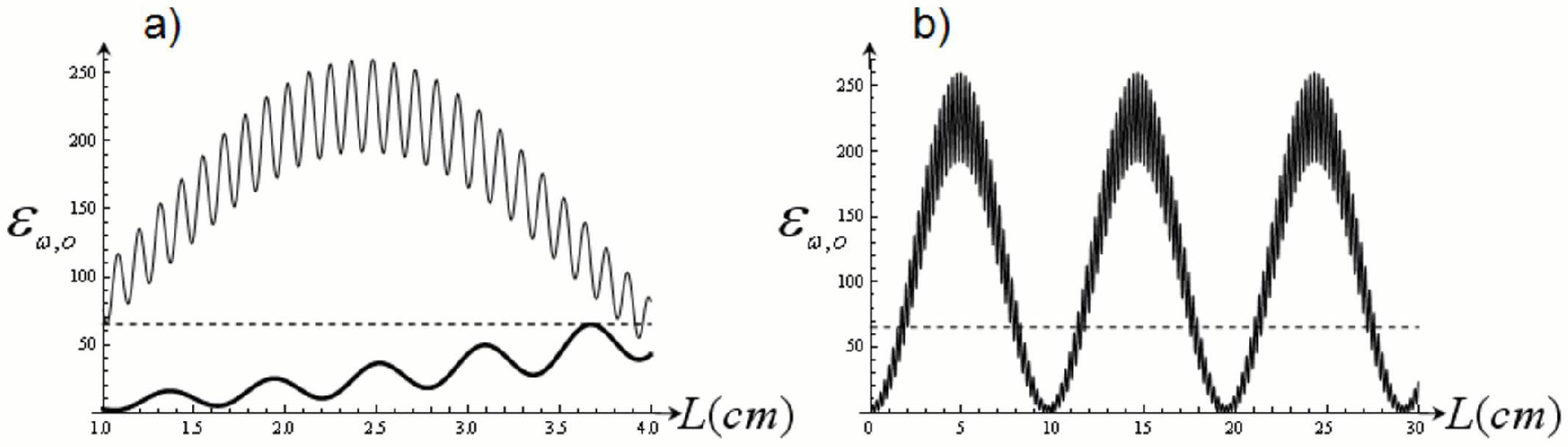,height=5.0cm}}
 \caption{TR dependence on $L$ for $\phi=0$ and $\theta=\alpha+1/\gamma$ (TR maximum for electron with equilibrium field). Dashed lines -- radiation by electron with equilibrium field. a) solid lines -- radiation by \lq half-bare' electron for $\lambda=1~mm$ (thin line) and for $\lambda=5~mm$ (thick line). b)~$\lambda=2~mm$ for much larger range of $L$.}
 \end{center}
 \end{figure}

Again for $\delta<<1$ (the electron's field is significantly suppressed) the radiation is suppressed (thick line on Fig. 4a). For larger values of this parameter radiation is enhanced and further radiation dependence on increasing $L$ has the type of oscillations (Fig. 4b). The possibility to observe at least several such oscillations depends on spectral and angular resolution of the used detector.

Let us note that finite spectral and angular resolutions of the detector should blur small amplitude quick oscillations which exist on all the presented figures and are caused by the first and the third terms in (\ref{one}).

As far as the absolute value of the radiation signal from single bunch is concerned, for the angle of distribution maximum for electron with equilibrium field $\theta=\alpha+1/\gamma$ it can be estimated as follows (the estimation is made for the case of total coherence of radiation by all $10^9$ particles of the bunch):

\begin{center}
$
\textrm{\Large $\frac{d^2\textrm{\Large $\varepsilon$}}{d\omega do}$}\sim 10~ \textrm{\Large $\frac{eV\cdot s}{sr}$}.
$
\end{center}

If the detector used for radiation registration has, for instance, the detection area $2\times 2~mm$ and the distance between the detector position and the inclined plate is about 1 meter the total amount of energy radiated by a single bunch in the wavelength range $2~mm<\lambda<2.1~mm$ which detector registers can be estimated as

\begin{center}
$
\Delta\textrm{\Large $\varepsilon$}\sim 0.5~MeV.
$
\end{center}

\section{Some aspects of radiation detection technique}

The detection technique required for the studies considered above should be efficient in the millimeter-wave range. Currently these detectors are being rapidly developed for a number of applications in science and technology -- wide-band wireless applications and communications, automotive radars, safety control, military use, mm-wave astrophysics. Therefore a variety of equipment (generators and detecting systems) for millimeter-wave range is available at the market of industrial and scientific instruments. During last years both branches progress -- room-temperature sensors with accompanying electronics (cheaper approach) and cryogenic detector technologies (maximum of sensitivity, it is used primarily for astrophysics). Individual detectors achieved photon noise limited performance for ground-based imaging in the 1990s. The photon noise from only astrophysical sources achievable in space with a cold telescope is of about $10^{-18}~W/\sqrt{Hz}$, varies somewhat with wavelength, and is within actually demonstrated sensitivities. In this decade imaging will be driven not by cryogenic detector sensitivity but by array formats. Detector arrays have doubled in format every 20 months over the past 10 years. In order to choose a given instrument, we plan a full simulation and bench tests in order to motivate using cryogenic sensitive detectors (or arrays); to determine the main sources of noise; estimate the expected levels of signal in millimeter-wave range.

\section{Conclusions}

The non-equilibrium state of electromagnetic field around relativistic electron can significantly influence upon characteristics of transition radiation by such electron. This effect can be observed, for instance, after particle deflection from initial direction of motion or during electron traversal of conducting foil. At several MeV energies the second process is more promising for experimental investigations of peculiarities of transition radiation by such \lq half-bare' electron. The calculations show that such properties are well pronounced in the millimeter region of radiated wavelengths. Both the radiation intensity suppression and enhancement as well as change of its angular distribution comparing to the case of radiation by electron with equilibrium field could be studied in this case. The results of investigation of transition radiation generated by \lq half-bare' particles might be valuable in connection with application of transition radiation based techniques for particle beams analysis. It is relevant, for instance, for deflected beams which consist of particles extracted from from primary linear beams or accelerator storage rings. Such beams may represent the example of beams of \lq half-bare' particles.

\section{Acknowledgements}

The research is conducted in the scope of the IDEATE International Associated Laboratory (LIA). The work was partially supported by the project F12 of NAS of Ukraine, the project no. 0115U000473 of the Ministry of Education and Science of Ukraine and by the project of NAS of Ukraine for young researchers (contract no. M63/56-2015).

\coltoctitle{An Improvement to Phase Reconstruction Techniques applied to Smith-Purcell Radiation Measurements with Noise}
\coltocauthor{N.~Delerue, O.~Bezshyyko, V.~Khodnevych }

\begin{center}
{\Large {\bf An Improvement to Phase Reconstruction Techniques applied to Smith-Purcell Radiation Measurements with Noise}}

\vskip 0.5cm

{\bf N.~Delerue$^{a}$, O.~Bezshyyko$^{b}$, V.~Khodnevych$^{b}$}

\vskip 0.3cm

$^{a}${\it LAL, Univ. Paris-Sud, CNRS/IN2P3, Universit\'e Paris-Saclay, Orsay, France}

$^{b}${\it Taras Shevchenko National University of Kyiv, Ukraine}

\end{center}

\vskip 0.4cm

\begin{abstract}
\noindent We recently reported on an implementation of phase noise reconstruction technique. In further work we found how to further improve our result when there is a strong noise background. We report on this here.
\end{abstract}

\section{Introduction}

We recently published a paper comparing different phase reconstruction techniques~\cite{reco_paper} and making an extensive study of the. After finalizing this paper we realized that it is possible to improve further the performance achieve by such reconstruction techniques when the signal is modified by a strong random noise. Here we report only the improvements to that paper and we invite the interested reader to refer to the original publication~\cite{reco_paper}.

\section{High frequency extrapolation}

In~\cite{reco_paper} we used a high frequency extrapolation method inspired from~\cite{LaiS}. This method relied heavily on the value of the last point of the spectrum to determine the high frequency extrapolation coefficient. Although this performs well in ideal world simulations this can be significantly affected by noise on this data point.

 Further analysis of simulated cases where this method did not perform well by giving incorrect profile FWHM led us to understand that we would get better results by replacing the value of the last data point by an average of the last 3 data points. The figure~\ref{noise} shows that using this method the error on $\Delta_{FWHM}$ is significantly decreased as it is often dominated by the wings of the profile. The $\chi^2$ of the distributions is only marginally affected by the high frequency extrapolation. Interestingly, although this method was meant to address mostly high noise measurements we see that the performance are also significantly improved when the gain is lower.

\begin{figure}[htpb]
    \centering
        \includegraphics[width=7cm ]{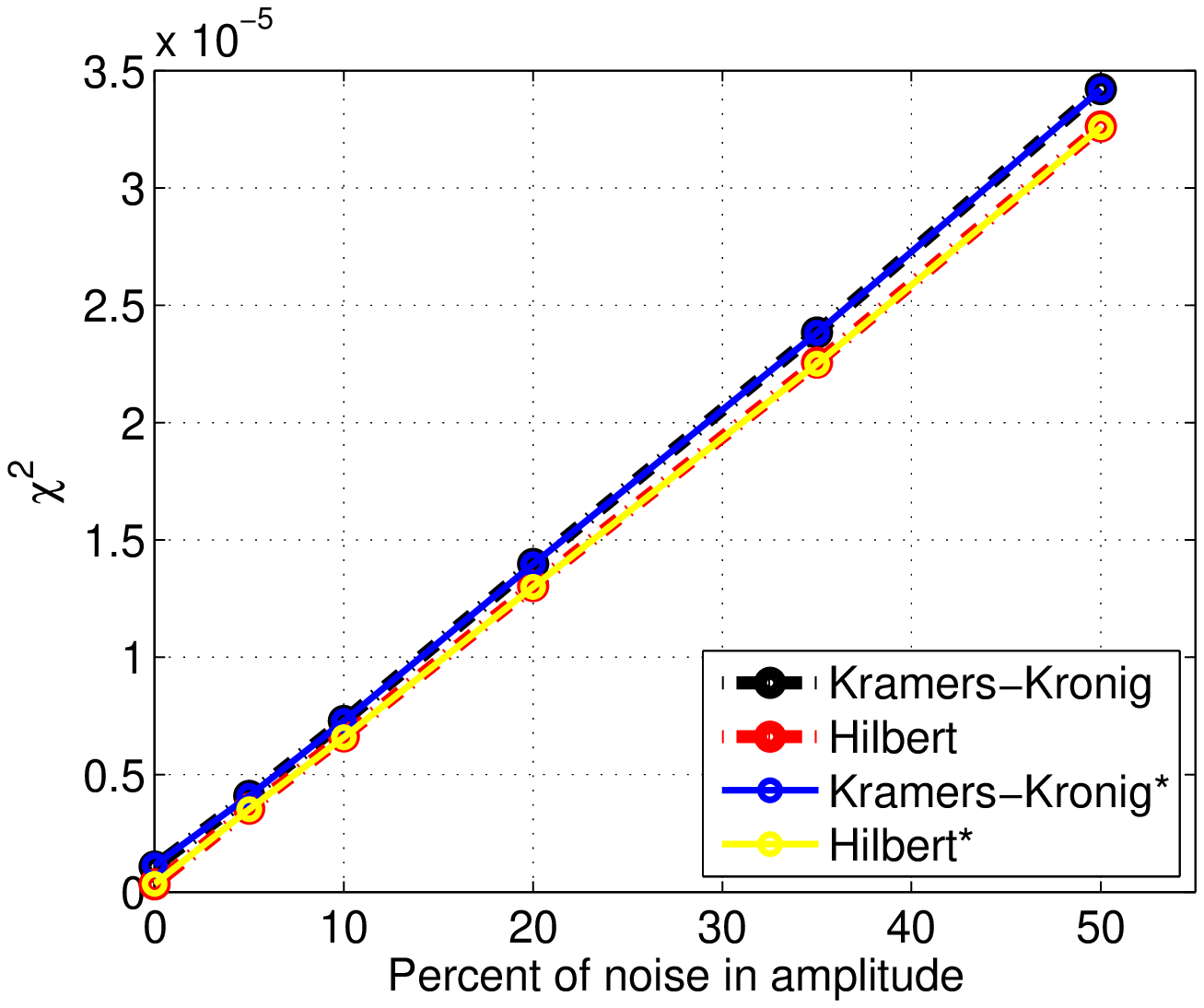}
        \includegraphics[width=7cm ]{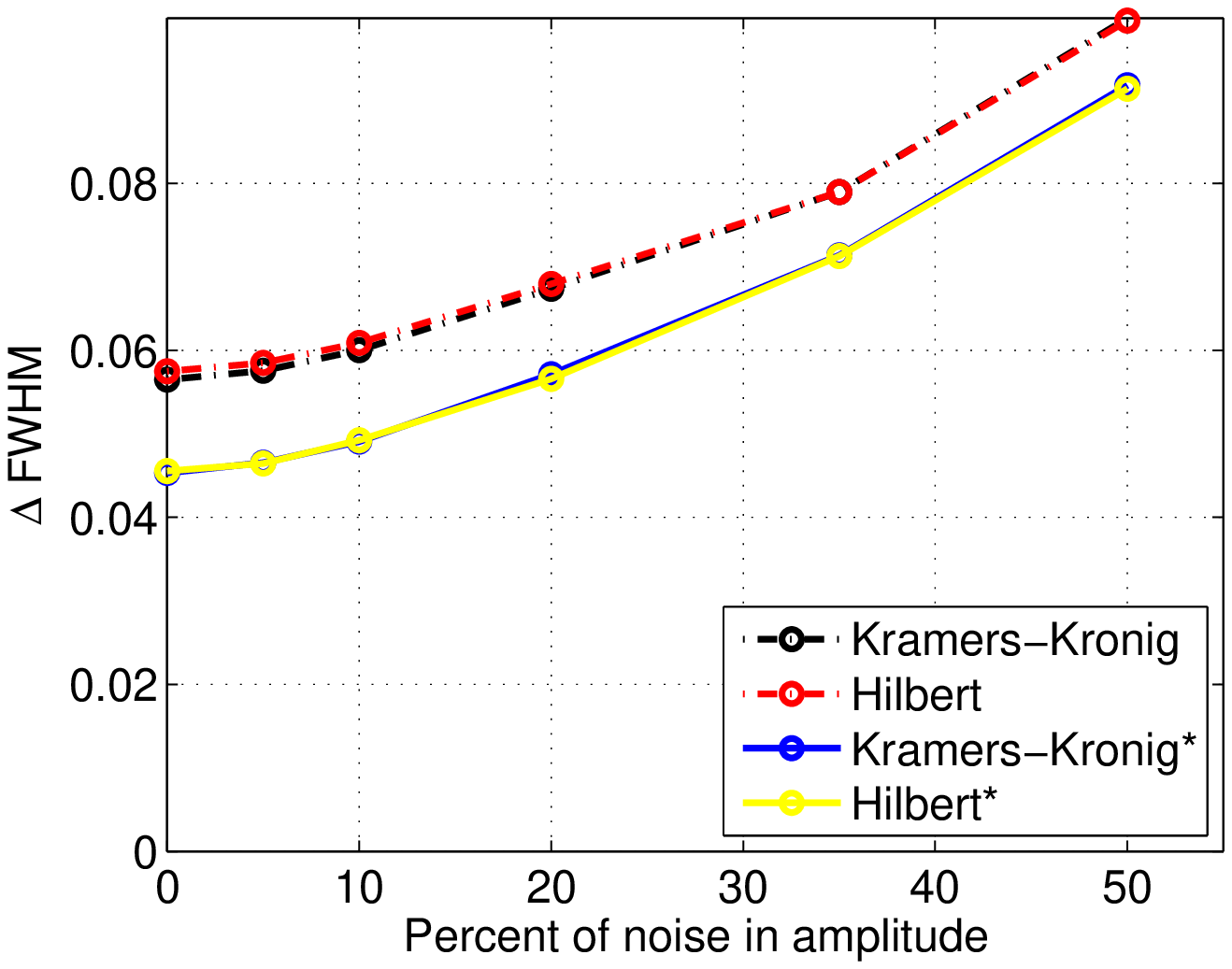}
  \caption{Mean $\chi^2$ and $\Delta_{FWXM}$  as function of noise amplitude. The black and red lines correspond to the method published in~\cite{reco_paper} whereas the data in yellow and blue use the average of the last 3 data point to estimate the high frequency extrapolation coefficients. For each data points 1000 simulations were performed.}
   \label{noise}
\end{figure}

We have also estimated the performance of our reconstruction method:  On a computer Intel Core-i5-3317U at 1.7 GHz computing the reconstruction using the Hilbert transform take 0.025 seconds. Calculating the reconstruction using the Kramers-Kronig method take 155.021 seconds.

\section{Discussion}

We have found the by modifying the high frequency extrapolation of our phase reconstruction method we significantly reduce the error on the  FWHM of the reconstructed profile.

\coltoctitle{Comparison of Smith-Purcell radiation models}
\coltocauthor{M.~Malovytsia, N.~Delerue}

\begin{center}
{\Large {\bf Comparison of Smith-Purcell radiation models}}

\vskip 0.5cm

{\bf Maksym~Malovytsia$^{a,b}$, Nicolas~Delerue$^{b}$}

\vskip 0.3cm

$^{a}${\it V.N.~Karazin Kharkiv National University, School of Physics and Technology, Department of Nuclear and Medical Physics, Kharkiv, Ukraine}

$^{b}${\it LAL, Univ. Paris-Sud, CNRS/IN2P3, Univ. Paris-Saclay, Orsay, France}

\end{center}

\vskip 0.4cm

\begin{abstract}
Smith-Purcell radiation is used in several applications including
the measurement of the longitudinal profile of electron bunches. A
correct reconstruction of such profile requires a good
understanding of the underlaying model. We have compared the two
leading models  of Smith-Purcell radiation and shown that they are
in good agreement. We have also studied the interference effects
that modulate the signal in the near-field zone.
\end{abstract}

\section{Introduction}

Smith-Purcell (SP) radiation \cite{SPRSmithPurcell} has been
discovered sixty years ago. This phenomena occurs, when charged
particles pass near periodical conductive grating. The wavelength
of the radiation produced  is given by a simple equation:

\begin{equation}\label{eq:diff_length}
            \lambda=\frac{d}{n}\left(\frac{1}{\beta}-cos{\theta}\right)
\end{equation}

\noindent where $ \lambda $ is the wavelength of the radiation, $
d $ is the period of the grating, $ n $ is the order of the
radiation and $ \theta $ is the polar observation angle (see Fig.
\ref{fig:geom1}).

There are several ideas of applications based on SP radiation,
including non-invasive bunch profile measurement methods
\cite{NIMA_SP_2014}.

Several models \cite{compsprKarlovetsPotylitsyn} of SP radiation
intensity distribution were proposed. Though there are experiment
for each of these models, that agrees with these theory a
comparison of these models is important to discriminate them
experimentally.

It has also been shown in \cite{sprpwKarlovetsPotylitsyn} that the
predicted intensities should be affected by interference effects.

\begin{figure}[htb]
\begin{center}
 \mbox{\epsfig{figure=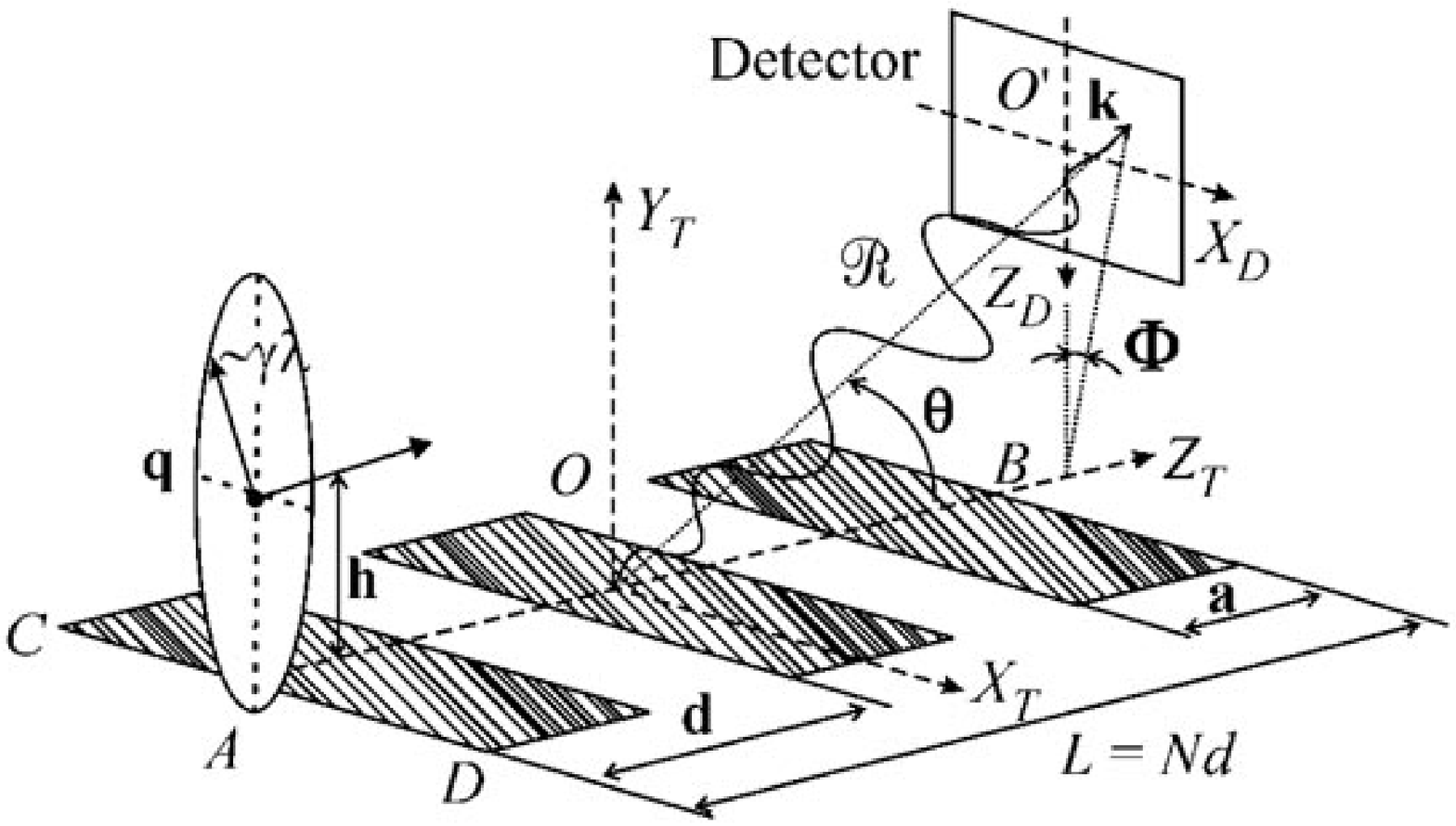,height=5.0cm}}
 \caption{Disposition of the grating, detector, bunch and radiation. Flat grating \cite{sprpwKarlovetsPotylitsyn}}
 \label{fig:geom1}
 \end{center}
 \end{figure}

\begin{figure}[htb]
\begin{center}
 \mbox{\epsfig{figure=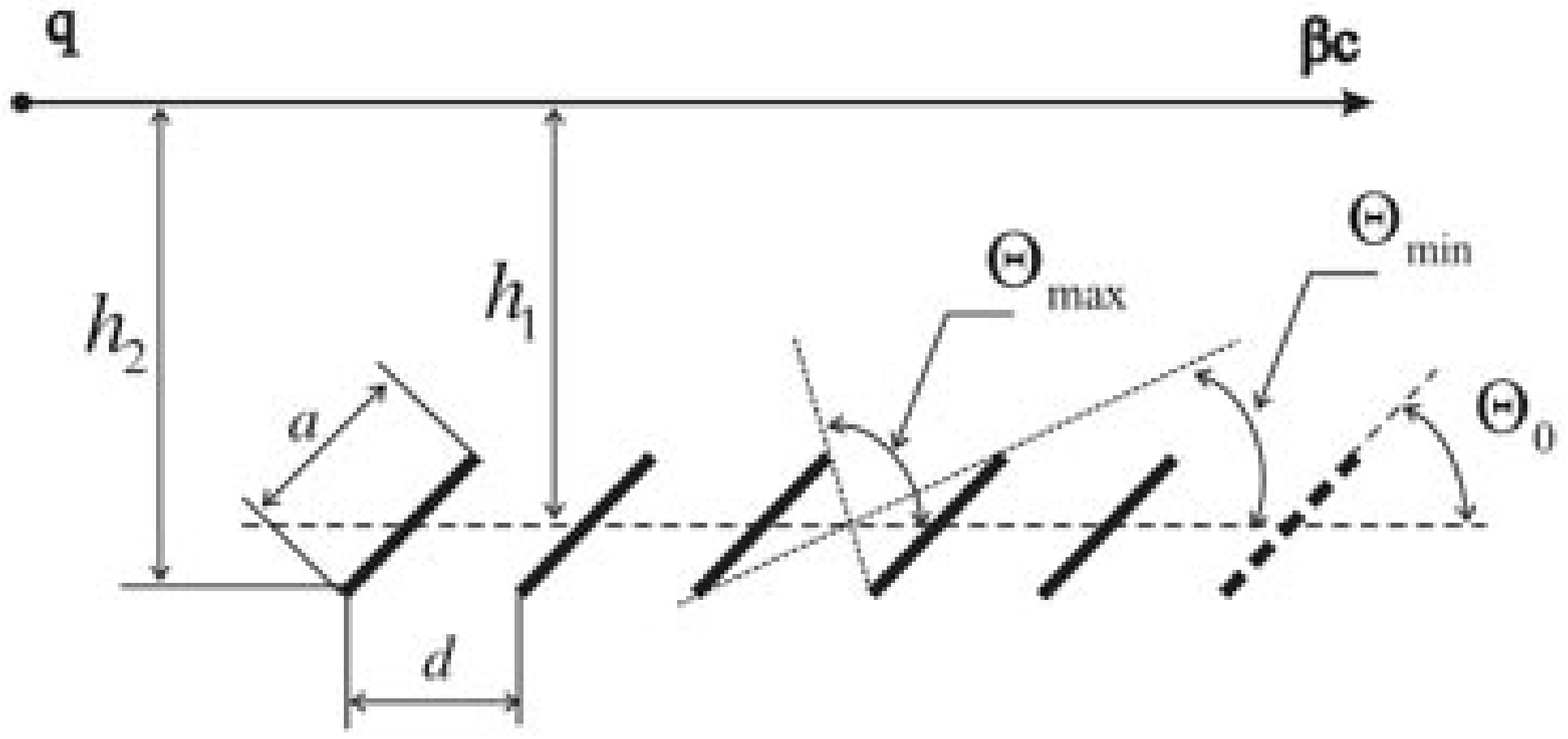,height=5.0cm}}
 \caption{Disposition of the grating, detector and bunch. Grating with inclined strips \cite{compsprKarlovetsPotylitsyn}}
 \label{fig:geom2}
 \end{center}
 \end{figure}

\section{Models}

The Surface Current (SC) and Resonant Diffraction Radiation (RDR)
models are presented in \cite{compsprKarlovetsPotylitsyn} as a
semi-analytical equations, which could be easily calculated, so
they are easy to use in the SP radiation simulations. However
those expressions are achieved with the assumption that the
detector is infinitely far from the source of radiation and
grating strips have infinite length.

Karlovets and Poptylitsyn in their work
\cite{sprpwKarlovetsPotylitsyn} proposed a basic model of SP
radiation. The idea was to use Backward Diffraction Radiation
(BDR), described in \cite{trVerzilov}, and calculate the intensity
on the detector point as interference of the waves from each point
of the grating. Using this approach automatically includes grating
sizes and source-detector distance. The problem of using this
method is long computation time, as the equations for intensity of
SP radiation are expressed through double integrals.

It is also must be mentioned that, unlike SC and RDR models, BDR
model proposed in \cite{sprpwKarlovetsPotylitsyn} does not include
blaze angle $\theta_0$ (see Fig. \ref{fig:geom2}). It could be
easily done, by changing every term of the Eq.~8 in
\cite{sprpwKarlovetsPotylitsyn}, that represents beam-grating
separation, by variable that depends on the position over the
grating. For example:

\begin{equation}
\label{eq:balzing_modifier} h \rightarrow h-\Delta Z_T\sin\theta_0
\end{equation}

\noindent where, $ h $ - impact parameter $ -a/2<\Delta Z_T<a/2 $
- position relative to the center of strip, that bunch currently
passing.

\subsection{Simplifications of BDR}

Numerical computation of BDR model requires considerable amount of
time, but there are cases, when it could be simplified. For the
"full" BDR model next equation was taken, with assumption, that
$h>>a/2 $:

\begin{equation}
 \label{eq:full_model}
 \begin{split}
 \left(
 \begin{array}{c}
   E_x^D\\
   E_z^D
 \end{array}
 \right)
 =
 Const\displaystyle\int\limits_{-M/2}^{M/2}dX_T\sum_{m=1}^{N}
 \displaystyle\int\limits_{-a/2}^{a/2}d\Delta Z_T
 \;A\frac{ K_1
 \left[
 \frac{2\pi}{\beta\gamma\lambda}
 \sqrt{X_T^2+h^2}
  \right]
  }
  {
 \sqrt{X_T^2+h^2}
   } \cos{\theta_0} \exp{\Delta\phi}
  \left(
  \begin{array}{c}
  X_T\\
  h
  \end{array}
  \right)
 \end{split}
 \end{equation}

\begin{equation}\label{eq:full_phi}
\begin{split}
\Delta\phi&=\frac{2\pi}{\lambda}\left(\Delta R+\frac{Z_{Tm} +\Delta Z_T \cos{\theta_0}}{\beta}\right)\\
                     \Delta R&=\bigg\{ (R_0\sin{\Theta}\sin{\Phi}-X_T)^2+(R_0\sin{\Theta}\cos{\Phi}-\Delta Z_T \sin{\theta_0})^2 \\
                     &+(R_0\cos{\Theta}-Z_{Tm}-\Delta Z_T \cos{\theta_0})^2\bigg\}^\frac{1}{2}
                     -R_0\\
                    Z_T&=Z_{Tm}+\Delta Z_T \cos{\theta_0}\\
                    Z_{Tm}&=\frac{d\left( 2m+1-N\right)}{2}
\end{split}
\end{equation}

Here, $X_T$ and $Z_T$ are the coordinates of the point on the
grating surface, $K_1$ is modified Bessel function of the second
kind, $\Delta \phi$ is the difference in phases of the waves
emitted from arbitrary point of the target and middle of the
grating, N is the number of grating strips, $\Delta Z_T$ is
coordinate of the source point relative to the center of the
closest strip, $h$ and $d$ are the same as in Eq.
(\ref{eq:diff_length}) and Eq. (\ref{eq:balzing_modifier}). In our
simulation waves were considered to interfere in one point so
$X_D$ and $Z_D$ from \cite{pwKarataev} had been considered 0.

\subsubsection{Far zone}

The far zone is the range of distances between the source and the
detector at which interference effects are not visible,
alternatively it could be said, that it is distances at which we
see our grating as single point of radiation. Criterion for the
far zone has been obtained by Karlovets and Potylitsyn in
\cite{sprpwKarlovetsPotylitsyn}:
\begin{equation}\label{eq:far_crit}
                R_0 \gg N^2 d (1+\cos{\theta})
\end{equation}
It was derived from the assumption, that difference in phases of
two waves coming from two furthest strips of grating is much
smaller than $2\pi$. Using this approximations it is easy to
integrals, which will significantly decrease computation time.

\subsubsection{Short grating strips}

When the detector is closer that the far zone limit different
assumption are possible if the following condition is met:
\begin{equation}\label{eq:short_crit}
                R_0 \gg \frac{d}{4(1-\cos{\theta})}
\end{equation}
With this simplification it is possible to obtain single integral
expression for the intensity of the SP radiation, but integrated
function is more complicated than in case of the
Eq.~(\ref{eq:far_crit}).

\subsection{Experimental parameters}

As parameters for our simulations it was decided to use real
experiments, as it would be possible to compare theory with
experiment later. The experiments are SPESO at Soleil Synchrotron
and E-203 at SLAC. For convenience the initial parameters are
presented in the table below.

\begin{center}
\begin{tabular}{|l|c|c|c|c|}
        \hline
        Parameter/Experiment & SPESO & E-203 & E-203 & E-203 \\ \hline
        $\gamma$           & 200 & 40000 & 40000 & 40000 \\
        $d, \mu m$         & 10000 & 1500 & 250 & 50 \\
        $h, mm$            & 1 & 1 & 1 & 1 \\
        $\theta_0,degree$  & 30 & 30 & 30 & 30 \\
        $M, mm$            & 100 & 20 & 20 & 20 \\
        $L, mm$            & 180 & 40 & 40 & 40 \\
        $R_0$              & 280 & 220 & 220 & 220 \\
\hline
\end{tabular}
\end{center}

Here $\gamma$ - lorentz factor, $d$ - period of the grating, $h$ -
inpact parameter, $\theta_0$ blaze angle, $M,L$ width,
length(along the beam direction) of the grating, $R_0$ -
source-detector distance. See Fig. \ref{fig:geom1},
\ref{fig:geom2}

\subsection{Comparison of the SC, RDR and BDR}

As both SC and RDR models are working in the far zone they could
be only compared to the far zone simplification of the BDR model.
For these several sets of data were created. For each model for
each $\theta$ from 40 to 140 degree with a step of 10 degree plots
of phi distribution of intensity per solid angle were created and
then combined into one figure for comparison. Also plot of ratio
between BDR and SC models was made.

\begin{figure}[htbp]
\begin{center}
\resizebox{0.45\textwidth}{!}{\includegraphics{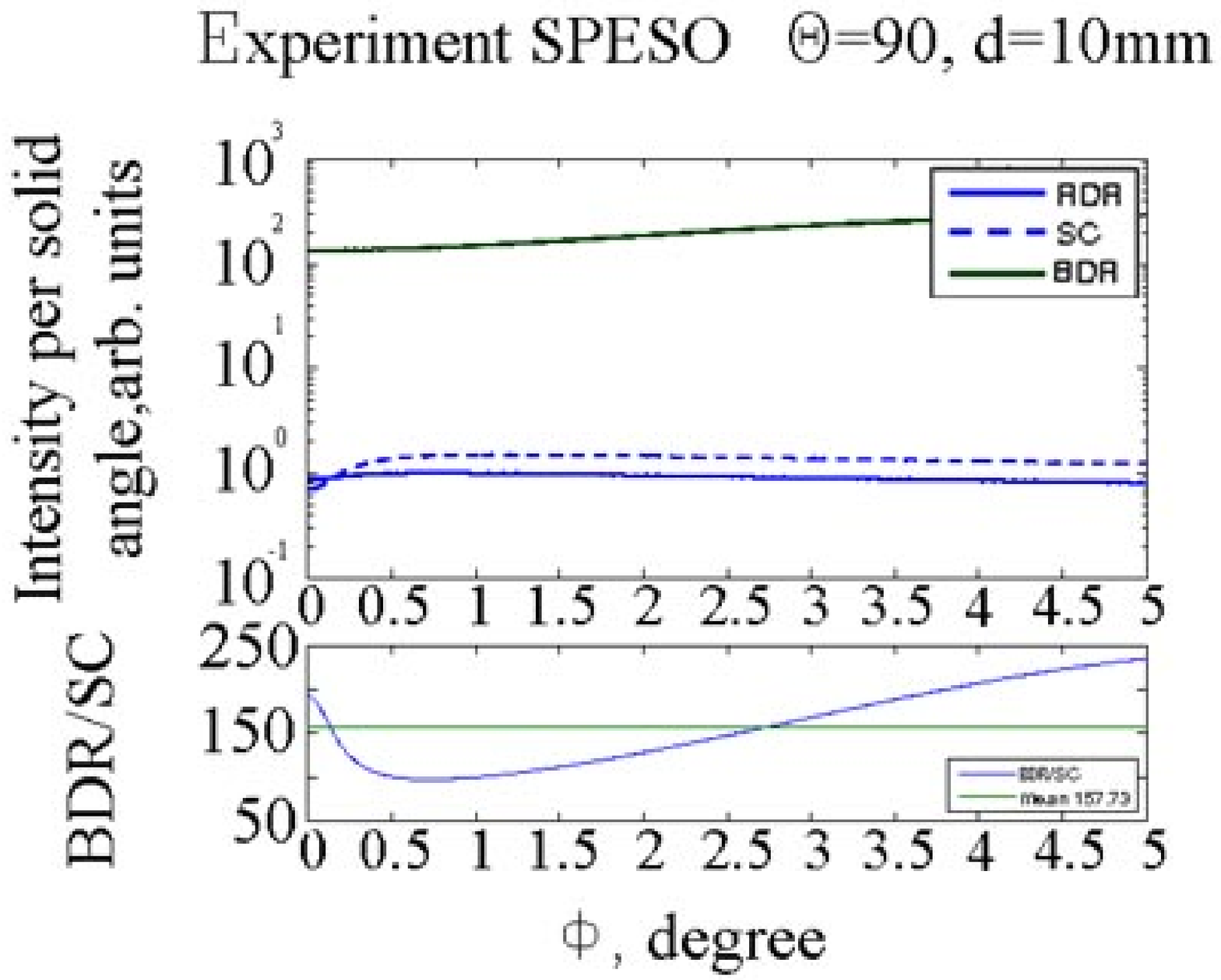}}
\resizebox{0.45\textwidth}{!}{\includegraphics{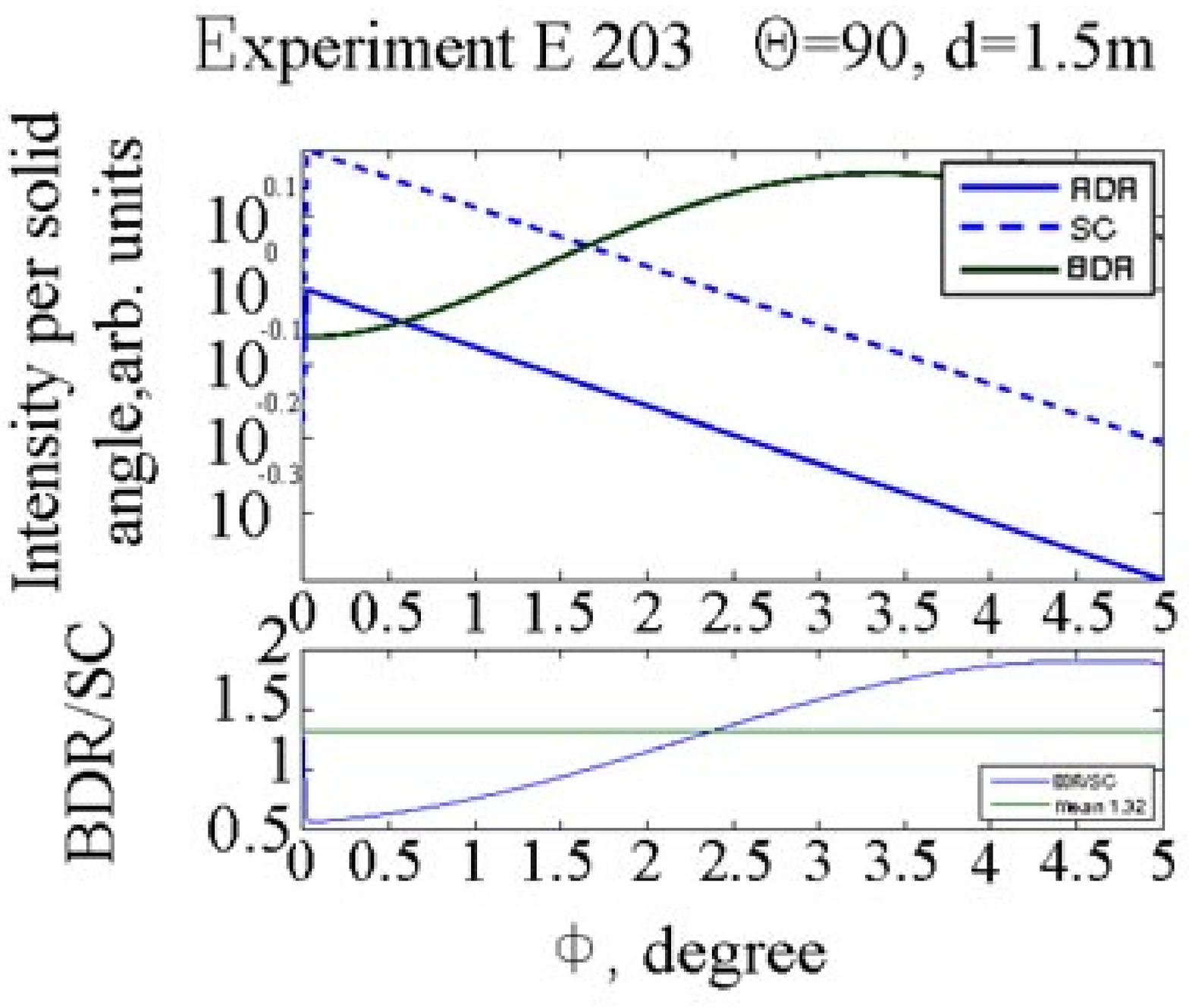}}
 \caption{Phi distribution for SPESO(left) and E 203 d=1.5mm(right).}
\end{center}
 \label{fig:Phi1}
\end{figure}

One can see that on the Fig. \ref{fig:Phi1} BDR model decreasing
slower than on \ref{fig:Phi2}, it could be easily explained. Even
if our source is point-like, it still can have interference at
large distances, and interference picture will be dependant on the
wavelength and the smaller the wavelength the faster intensity
will go down when  (in this case) azimuthal angle is changed.

Another problem is despite SC and RDR being close, they are much
different from BDR model. This happens because in BDR grating
sizes are accounted for. In fact if one start increasing length
(M) of the grating strips, simulated BDR model will be becoming
more like SC and RDR as shown on Fig. \ref{fig:Phi3}.

 \begin{figure}[htbp]
 \begin{center}
 \resizebox{0.45\textwidth}{!}{\includegraphics{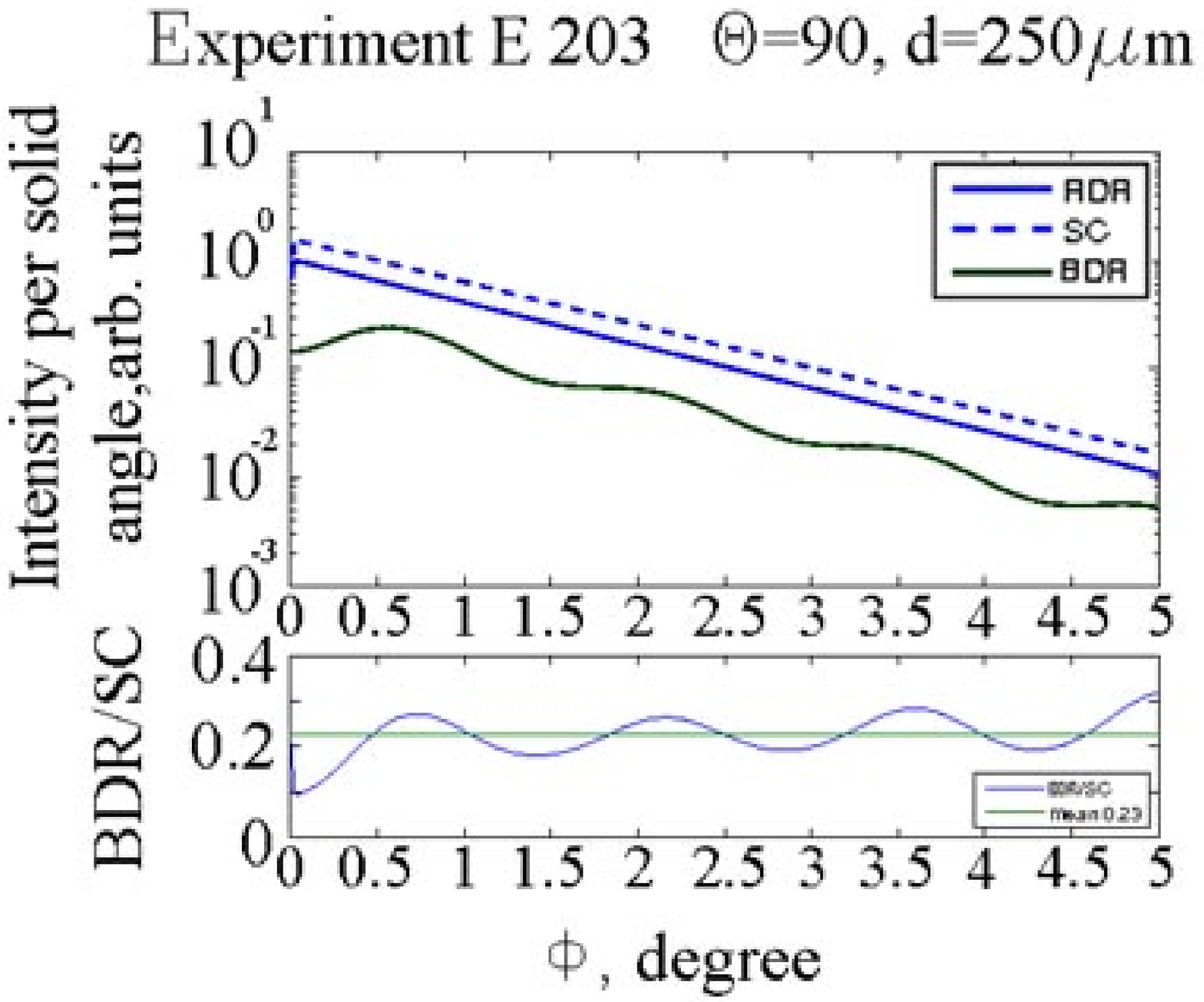}}
 \resizebox{0.45\textwidth}{!}{\includegraphics{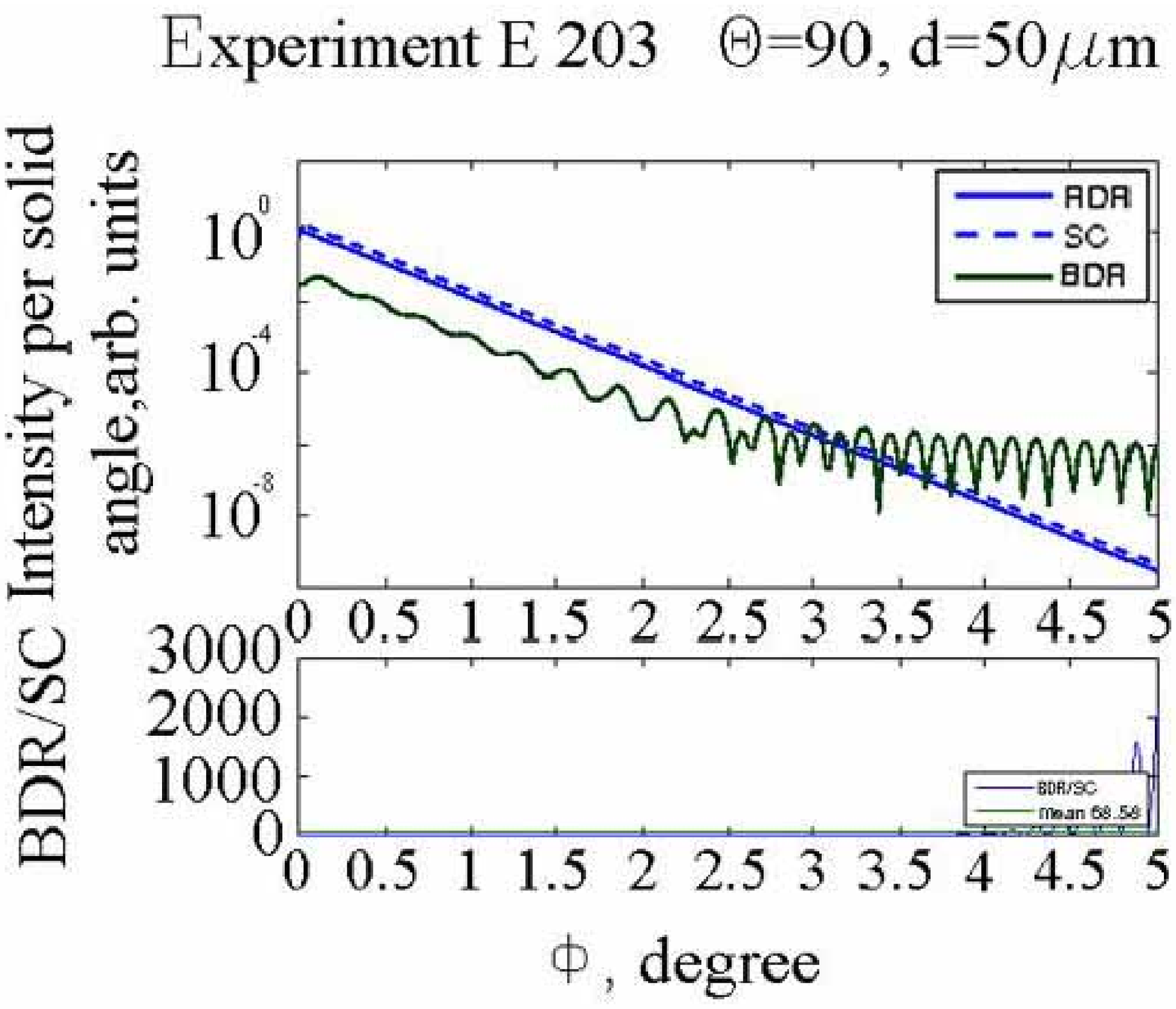}}
  \caption{Phi Distribution for E 203 d=250$\mu$m(left) and E 203 d=50$\mu$m(right)}
 \end{center}
 \label{fig:Phi2}
 \end{figure}

 \begin{figure}[htbp]
 \begin{center}
 \resizebox{0.45\textwidth}{!}{\includegraphics{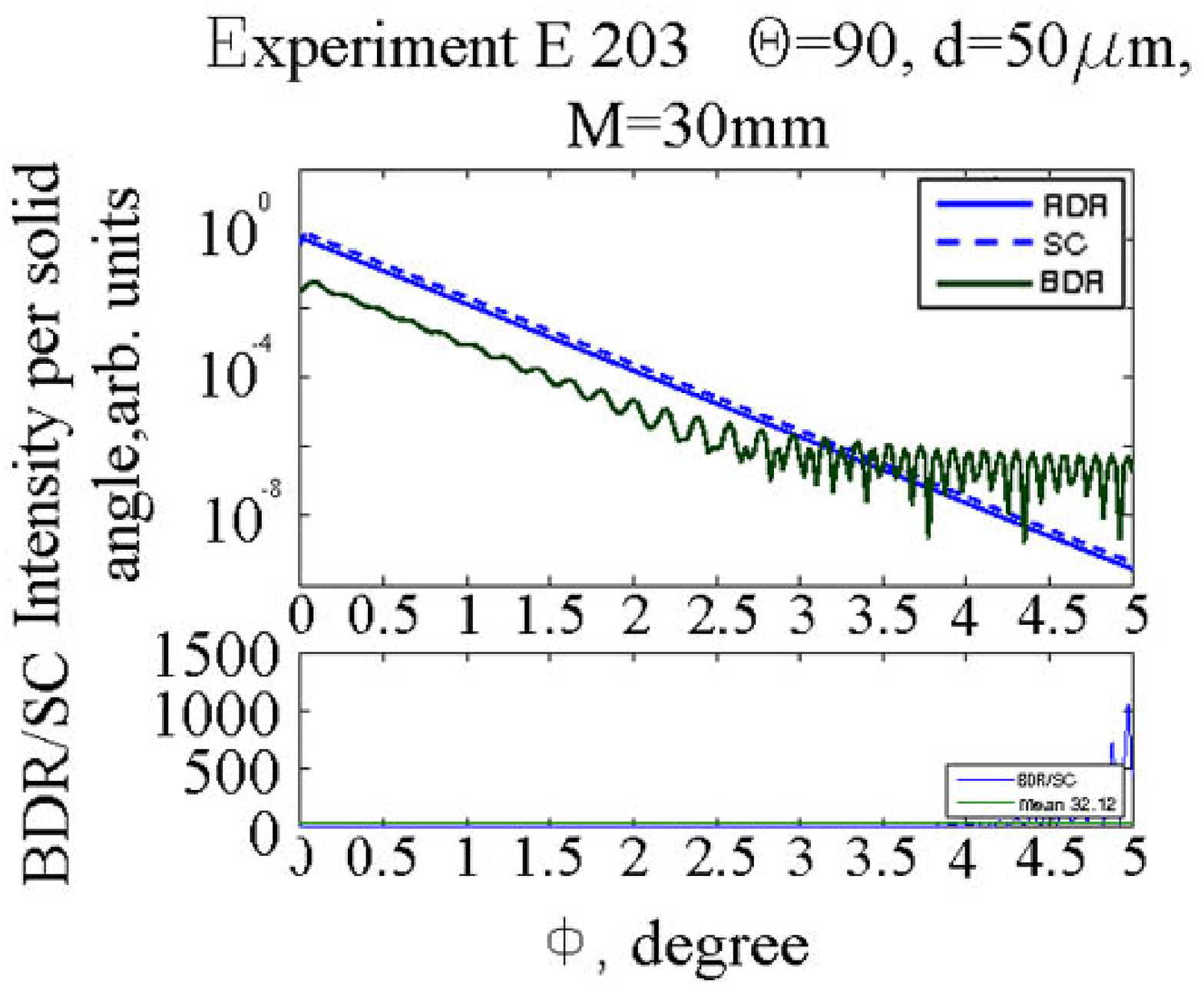}}
 \resizebox{0.45\textwidth}{!}{\includegraphics{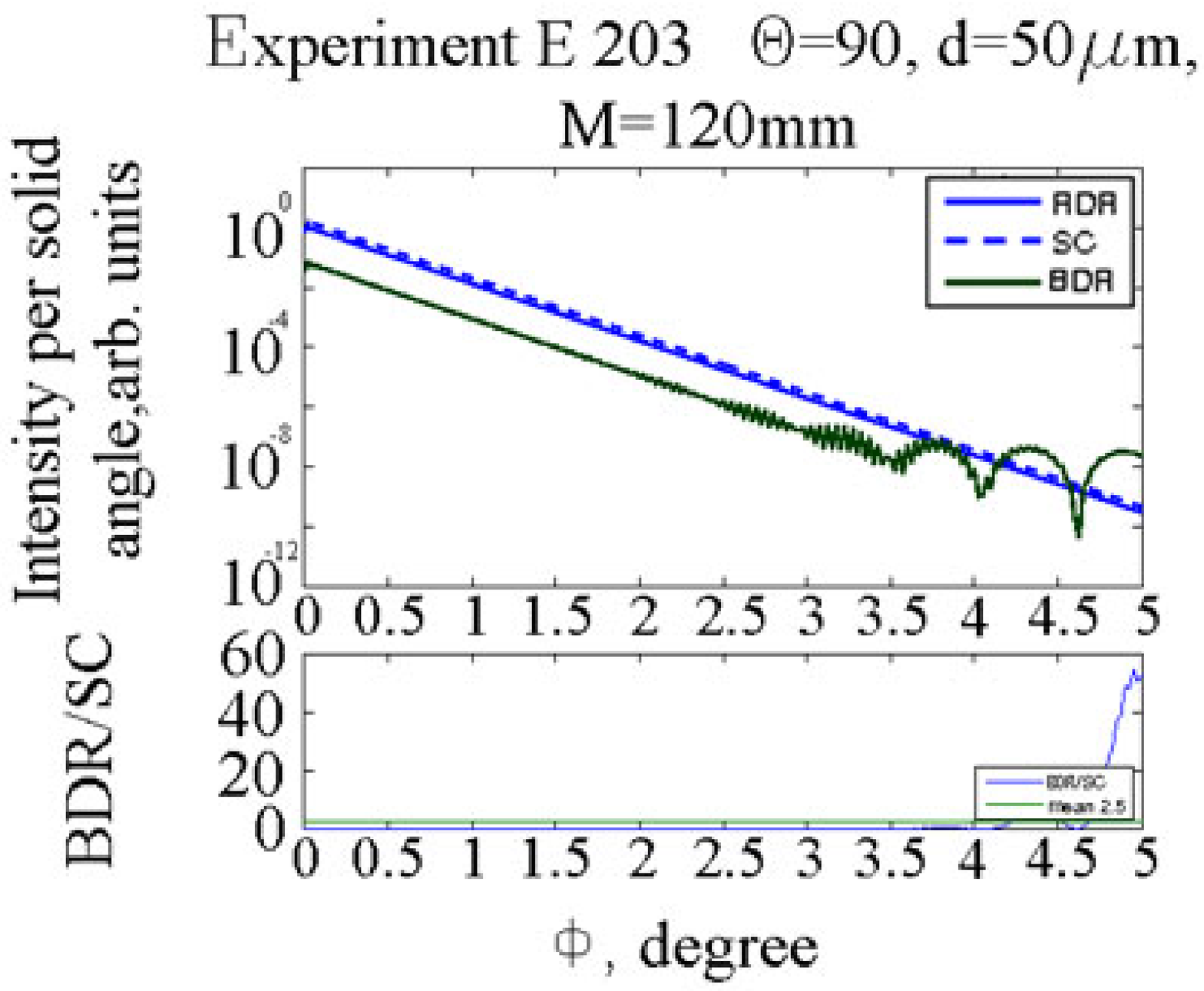}}
  \caption{Phi Distribution for E 203 d=50$\mu$m M=30mm(left) and M=120mm(right)}
 \end{center}
 \label{fig:Phi3}
 \end{figure}

As one can see on Fig.~\ref{fig:Phi2} and Fig.~\ref{fig:Phi3} the
RDR and SC models produce the same angular distribute and the
signal intensity are comparable (within experimental
uncertainties). It would there fore be difficult to discriminate
experimentally between  these models.

\section{Interference}

In far zone models the signal intensity is inversely proportional
to the square of  the distance between the source and the
detector. This changes when is closer closer to the source and one
must then use the full model for BDR. As can be seen on Fig.
\ref{fig:IntensR} the intensity per solid angle is constant at big
distances and begins to oscillate at small distances. This means
that the interference at near source distances depends not only on
the angles, but on the distance also. And as shown on the
\ref{fig:IntensR} the interference effect can be considerable.

\begin{figure}[htbp]
\begin{center}
\resizebox{0.3\textwidth}{!}{\includegraphics{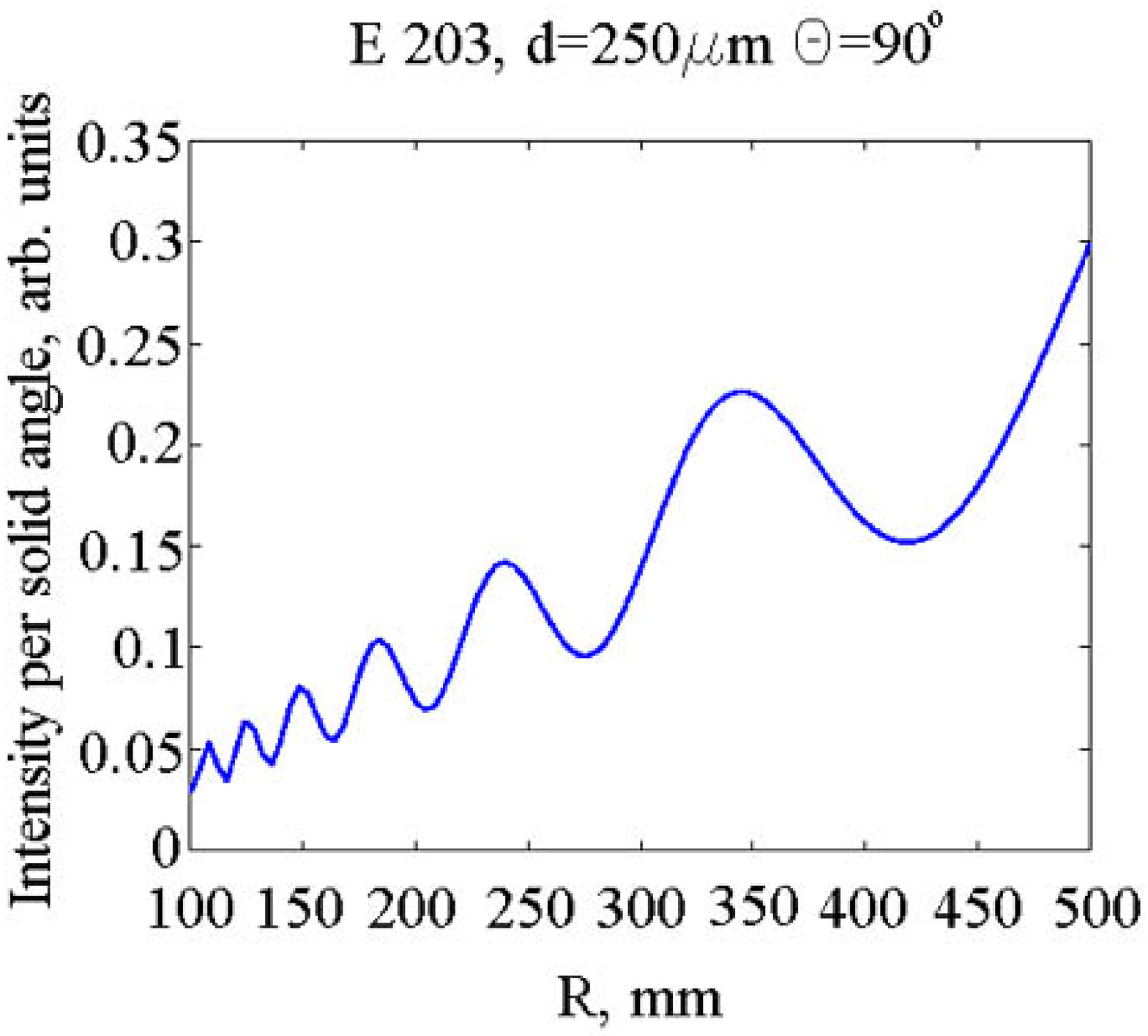}}
\resizebox{0.3\textwidth}{!}{\includegraphics{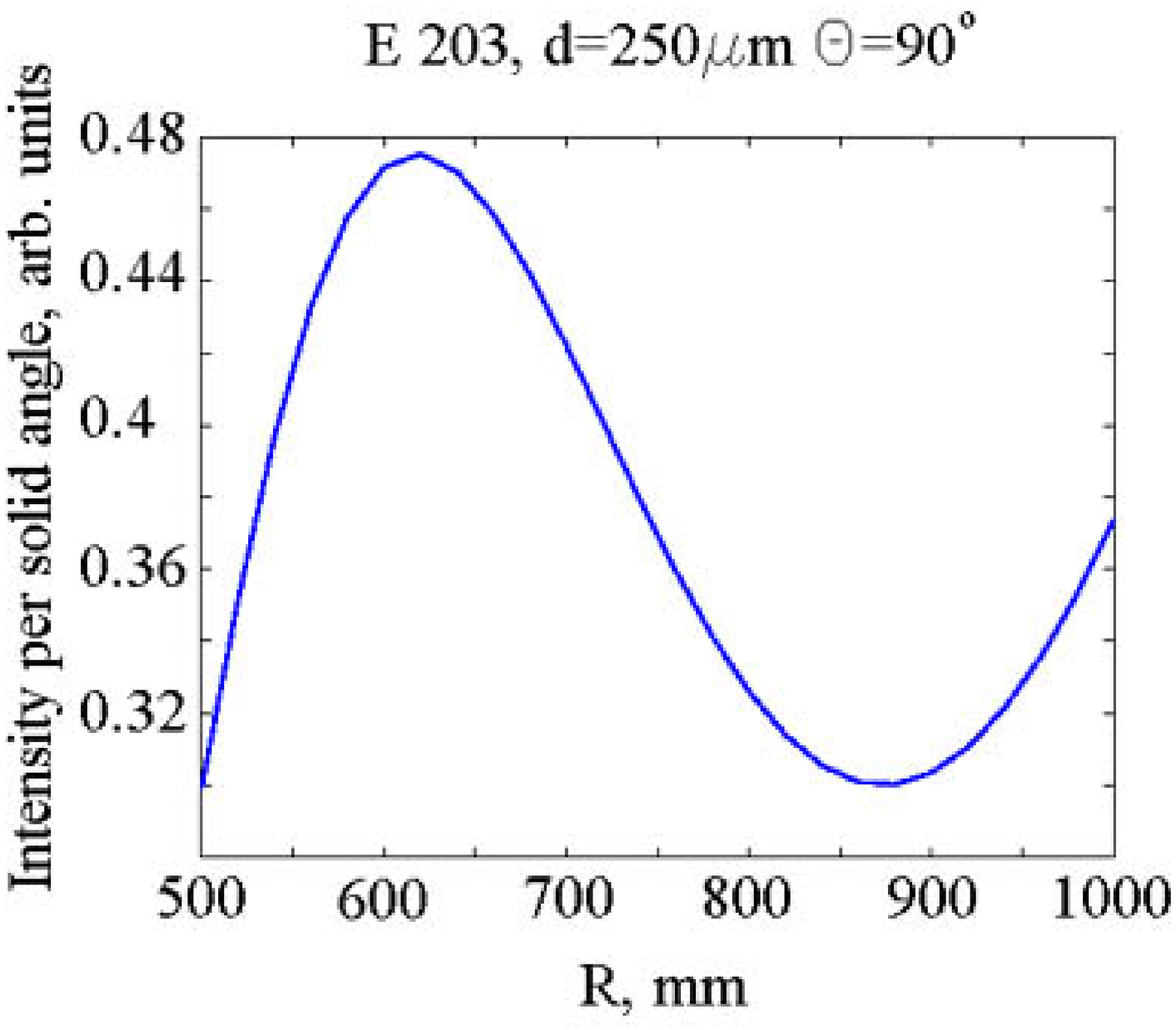}}
\resizebox{0.3\textwidth}{!}{\includegraphics{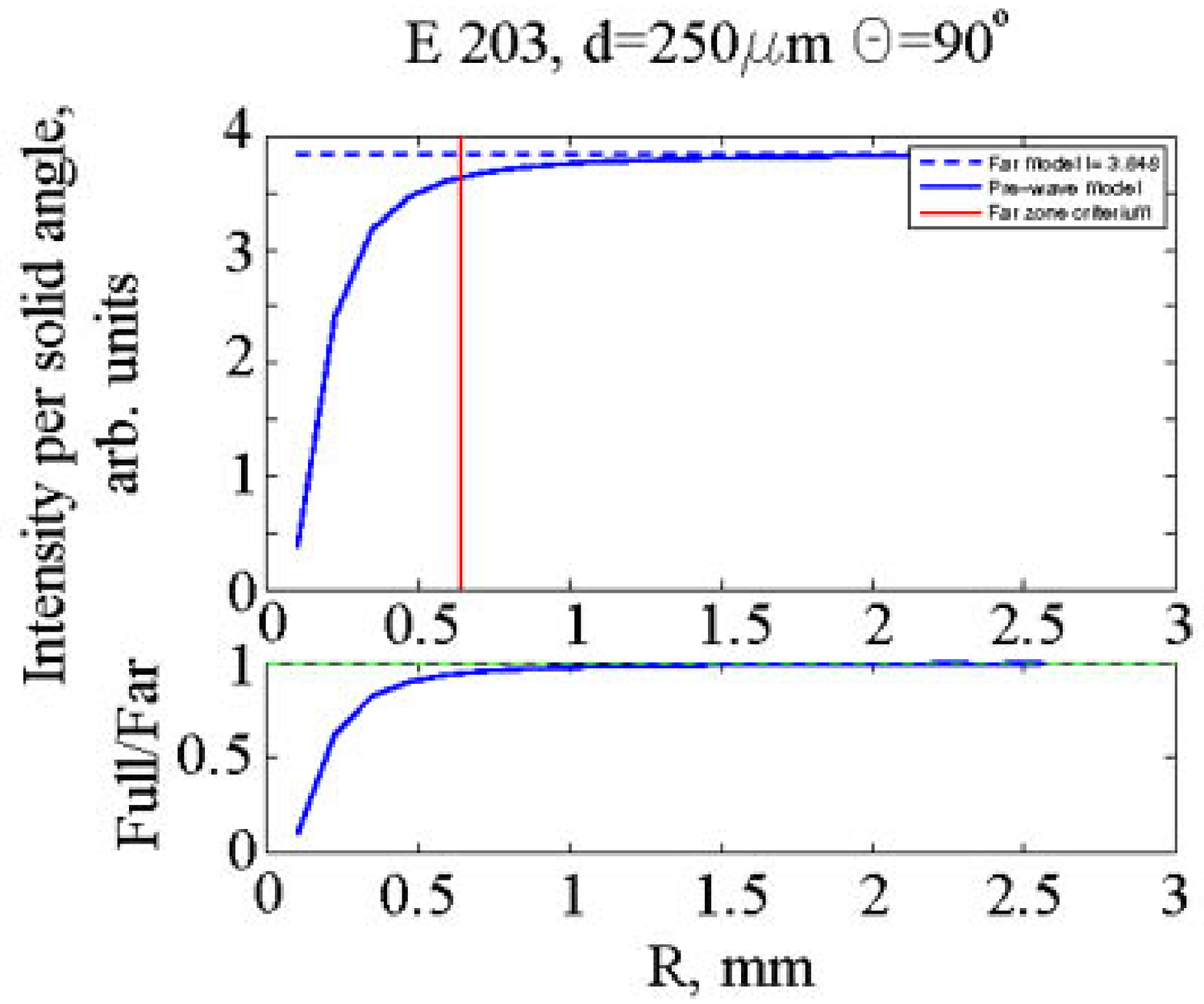}}
 \caption{Dependence of the intensity from R}
\end{center}
 \label{fig:IntensR}
\end{figure}

It is important to know if we should consider interference effects
in our calculations or not, for this we have criterion for
determining if we are in the far zone or not is Eq.
\ref{eq:far_crit}. For easier understanding it could be visualized
as curve in the two dimensional plot shown on  Fig.
\ref{fig:FarPlot}, where ZY plane is $X_T$$Z_T$ on Fig.
\ref{fig:geom1}.

\begin{figure}[htb]
\begin{center}
 \mbox{\epsfig{figure=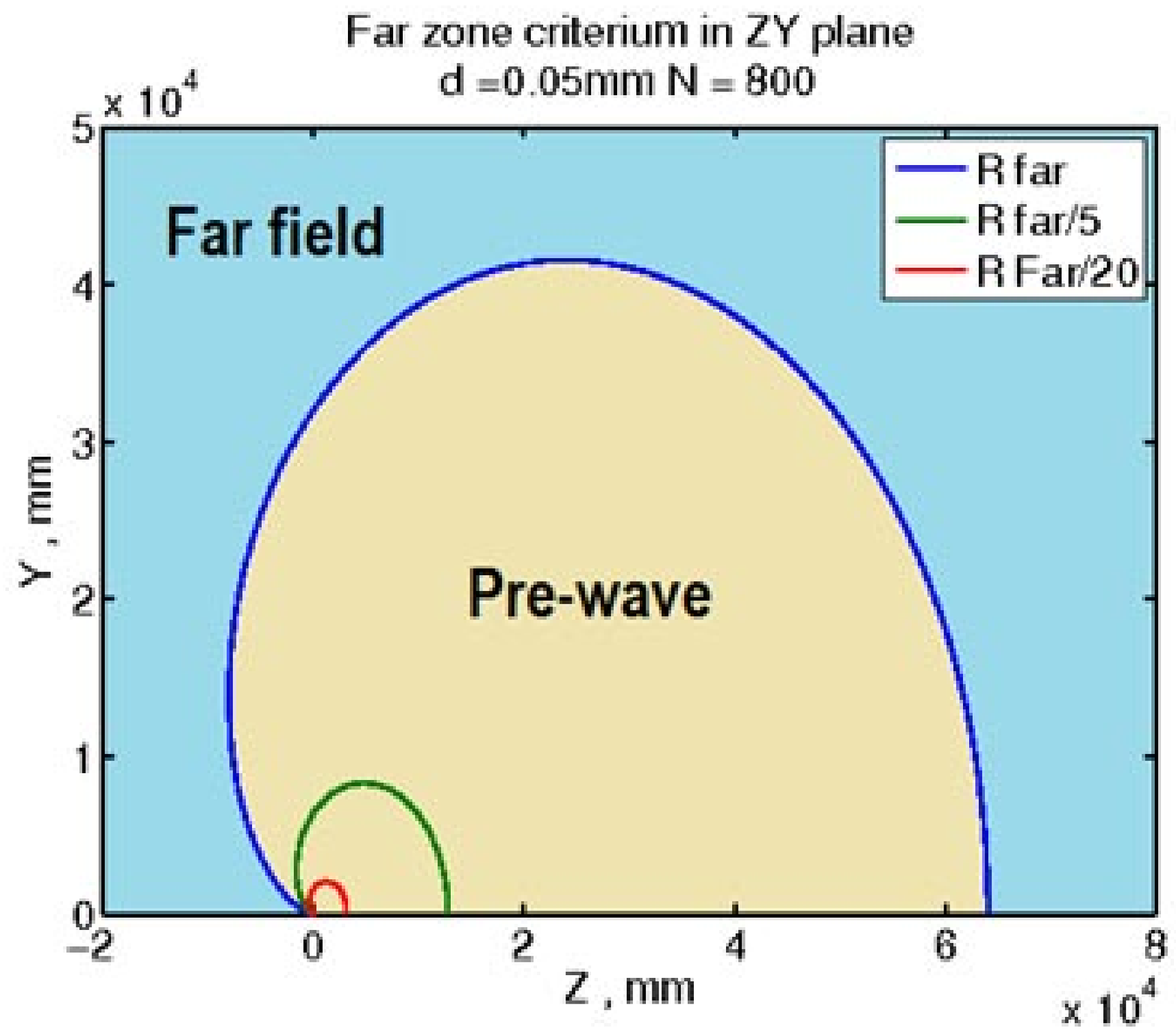,height=7.0cm}}
 \caption{Far zone criterion}
  \label{fig:FarPlot}
 \end{center}
 \end{figure}

\begin{figure}[htb]
\begin{center}
 \mbox{\epsfig{figure=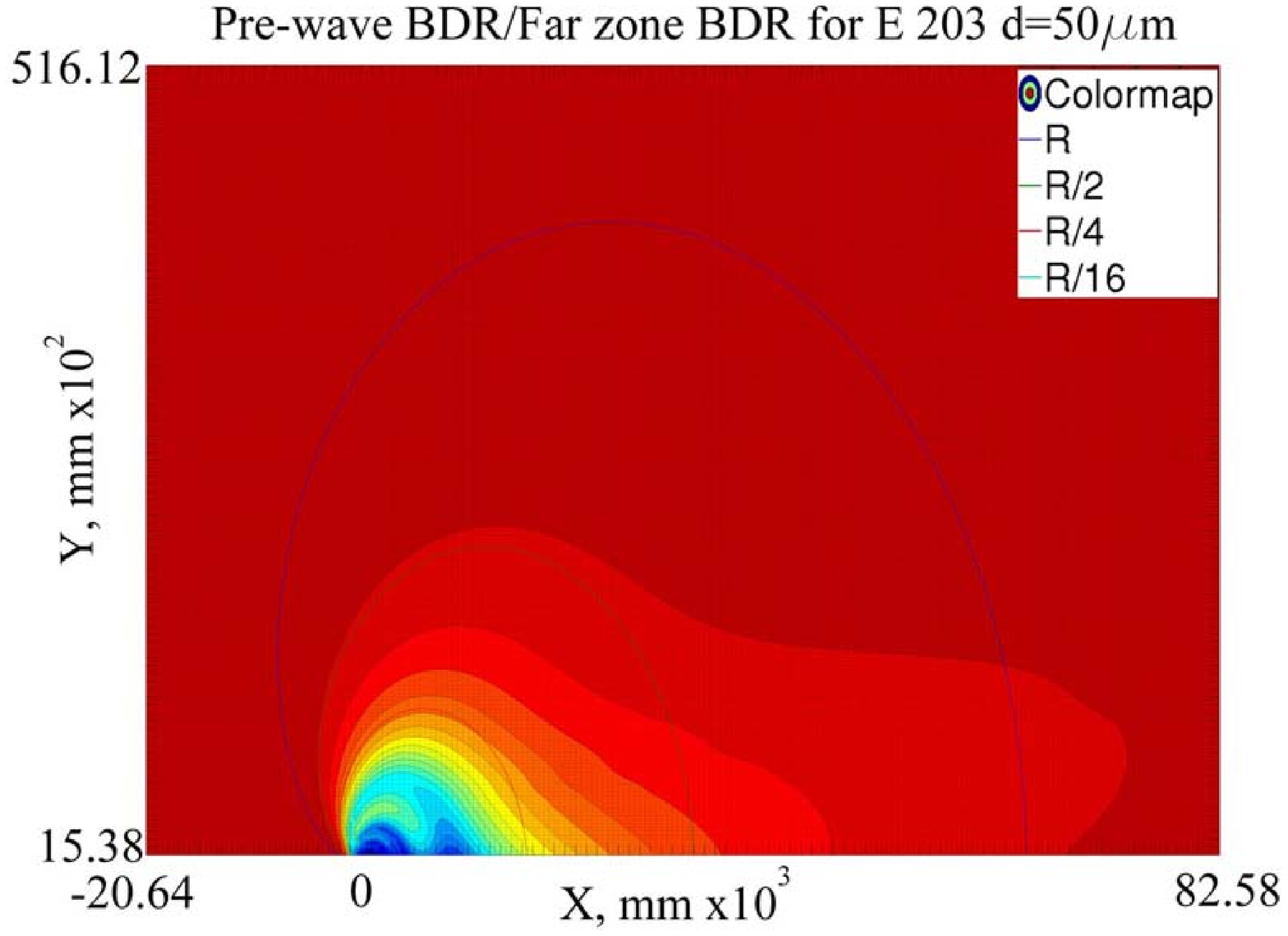,height=7.0cm}}
 \caption{Intensity color map}
  \label{fig:Colormap}
 \end{center}
 \end{figure}

\section{Summary}

We have studied two important effect that can change the intensity
of the Smith-Purcell Radiation and could thus affect its usability
as a longitudinal profile monitor. We have compared the two
leading models under experimental conditions and shown that it
would not be possible to discriminate them experimentally. We have
also shown that interferences in the near field zone have a
significant effect and must be taken into account to estimate the
correct intensity of SPR.

\coltoctitle{Production and applications of parametric X-rays emitted by relativistic particles}
\coltocauthor{A.V.~Shchagin, I.~Chaikovska, R.~Chehab}

\begin{center}

{\large \bf Production and applications of parametric X-rays emitted by relativistic particles}

\end{center}
\vspace*{3 mm}

\centerline{\bf A.V. Shchagin$^{a,\,b,\,c,}$\footnote{ Corresponding author: \texttt{shchagin@kipt.kharkov.ua}}, I. Chaikovska$^{c}$, R. Chehab$^{c}$}

\vspace*{3 mm}

\begin{center}
\textit{$^a$ Kharkiv Institute of Physics and Technology, Kharkiv 61108, Ukraine \\
$^b$ Belgorod State University, Belgorod 308015, Russia\\
$^c$ Laboratoire de l'Accelerateur Lineaire (LAL), Universite Paris-Sud, Bat. 200, 91898 Orsay, France
}
\end{center}

\vspace*{3 mm}

\textit{ Abstract:} The nature of parametric X-ray radiation (PXR) excited by relativistic charged particles passing through a family of crystallographic planes is discussed. The PXR is a quasi-monochromatic radiation with energy smoothly tunable in the range from few to hundreds keV. The main properties of parametric X-ray radiation, like yield, angular distributions, spectral peak energy and polarization are considered. Some results of theoretical calculations and experimental research of the PXR properties are presented. Areas of PXR applications like as calibration of X-ray space telescopes by the PXR beam, obtaining of a shadow and phase-contrast images, control of electron beam parameters are described. Some possibilities for generation of secondary X-ray beams from crystals at ThomX in LAL are discussed.

\vspace*{3 mm}

{\it PACS numbers:} 41.60.-m, 61.80.Cb

\vspace*{3 mm}

{\it Keywords:}   parametric X-ray radiation, virtual photons, crystal diffraction

\begin{flushright}{\it November 27, 2015}\end{flushright}

\section{Introduction}
Parametric X-ray radiation (PXR) arises when a relativistic charged particle crosses a family of crystallographic planes in a crystal. M.L. Ter-Mikaelian devised a kinematic theory to describe this radiation in a classical approximation (see equations (28.157 - 28.160) in book \cite{1} and also review \cite{2}). Ter-Mikaelian named this radiation ''resonant radiation'' \cite{1}, but in the later literature the term ''parametric X-ray radiation'' (PXR) is used by an analogy to optical parametric Cherenkov radiation \cite{3} which occurs when a charged particle crosses a medium consisting of periodically ordered layers with different dielectric constants. The results of classical calculations \cite{1} were confirmed by quantum calculations performed by H. Nitta \cite{4}. The fact that the kinematic theory can describe the basic properties of PXR emitted at large angles to the incident particle velocity vector is also confirmed by many experiments (see, for example, reviews \cite{5,6,7,8,9} of such research). Let us note that, unlike conventional Cherenkov radiation, PXR is emitted if the particle velocity is less than the phase velocity of the radiation propagation in the crystal. Below, we will consider some of main properties of the PXR, mainly following to Ref. \cite{9}, where the kinematic theory of the PXR has been developed with use of Fresnel coefficients, and also discuss some applications of the PXR.

\section{ The origin of the PXR}
The emission of the PXR can be considered as a transformation of virtual photons accompanying a relativistic charged particle. The transformation occurs due to interaction of the field of relativistic particle with a periodically arranged family of crystallographic planes in a crystal. As a result, the so-called PXR reflection is emitted from the crystal. The PXR reflection is emitted around of the Bragg direction relative to incident beam of particles. Typical diagram of generation of the PXR reflection is given in Fig.~\ref{shchag:fig1}.

Let us examine some properties of virtual photons that are associated with a charged particle which travels through a medium with constant velocity $\vec V$  and Lorentz factor $\gamma \gg 1$, where
 \begin{equation}\label{shchag:1}
 \gamma=\frac{1}{\sqrt{1-\left(\frac{V}{c}\right)^2}},\qquad  V=\left|\vec V \right|,
 \end{equation}
The angular distribution $J_v$  of normalized spectral density of virtual photons associated with a charged particle moving in a medium can be described by the formula (1.24) in book \cite{10}. In the approximation of small angular deviations from the direction of the particle velocity  $\vec V$ and in the conditions $\gamma \gg 1$, $\hbar \omega \ll \gamma \cdot mc^2$, where  $m$ is the particle mass, the formula for angular distribution $J_v$  has the form (see formula (4) in paper \cite{11}) 
 \begin{equation}\label{shchag:2}
        J_v=\left(\frac{dN_v}{d\Omega\frac{d\omega}{\omega}}\right)_{\bot}=
        \frac{\alpha \cdot z^2}{\pi^2}\frac{\delta_{v\bot}^2+\delta_{v\|}^2}{\left(\gamma_{eff}^{-2}+\delta_{v\bot}^2+\delta_{v\|}^2\right)^2},                          
      \end{equation}                             
where $dN_v$  is the number of virtual photons in the solid angle $d\Omega$; $\alpha=\frac{e^2}{\hbar\cdot c } \approx \frac{1}{137}$  is the fine-structure constant,   $z$ is the particle charge in units of the electron charge $-e$ ; $\hbar$  is the Planck constant,  $\delta_{v\bot}, \, \delta_{v\|} \ll 1$ are small angles between the direction of the radiation propagation and the direction of the particle velocity $\vec V$  in arbitrarily chosen perpendicular directions, $\sqrt{\delta_{v\bot}^2+\delta_{v\|}^2}$  is the angle between the vector $\vec V$  and the direction of the radiation propagation, $\gamma_{eff}=\left(\gamma^{-2}+\left|\chi_0\right|\right)^{-\frac{1}{2}}$ is the effective relativistic factor \cite{12} with due regard for Ter-Mikaelian longitudinal density effect \cite{1} (further, the density effect), $\chi_0$  is the dielectric susceptibility for the radiation frequency $\omega$  higher than the frequencies of the atomic transitions and off the resonance frequencies, $\left|\chi_0\right|=1-\varepsilon=\left(\frac{\omega_p}{\omega}\right)^2$  , where $\varepsilon$  is the mean dielectric constant, and $\omega_p$  is the plasma frequency of the medium. Let us also note that the phase velocity of the X-ray propagation in the medium is usually higher than the speed of light, $\frac{c}{\sqrt{\varepsilon}}>c$, because $\varepsilon<1$.

One can see from expression (\ref{shchag:2}) that the virtual photons are symmetrically distributed around the velocity vector of the particle  $\vec V$. The maximum of the distribution is resided at the angle  $\gamma_{eff}^{-1}$  to the particle's velocity vector $\vec V$, and there is a dip in the center of the distribution. The diagrams of the angular distributions of virtual photons are shown in Fig. \ref{shchag:fig1}. The spectral distribution of virtual photons does not have any peculiarities. One can see from formula(\ref{shchag:2})  that the number of virtual photons propagating in any direction per unit frequency interval monotonically decreases as the frequency increases, $\frac{dN_v}{d\Omega d\omega} \sim \frac{1}{\omega}$.
 
 \begin{figure}[ht]
\begin{center} 
\includegraphics[width=9 cm, height=4.5 cm]{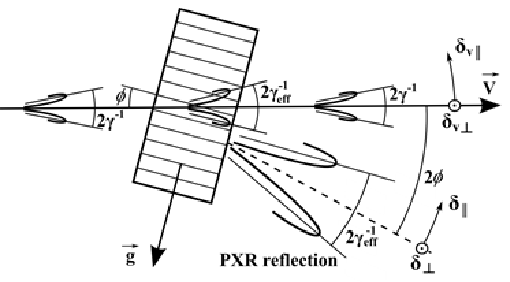}\\
\caption{Typical schematic of PXR reflection generation, shown in the diffraction plane. A charged particle moves with the constant velocity $\vec V$  in vacuum and crosses the crystalline plate. Angular coordinates $\delta_{v \bot}$, $\delta_\bot$, which are measured in the directions perpendicular to this plane, are marked by circles with dots. One of the crystallographic planes of the crystal with a nonzero structure factor, shown by the dashed line, is characterized by the reciprocal lattice vector  $\vec g$. Angular distributions for the number of virtual photons of the particle in a stationary state are depicted by diagrams at different sections of the particle trajectory: distribution in the crystal with a maximal number of virtual photons at angles $\delta_{v \bot}$, $\delta_{v \|}=\pm\gamma_{eff}^{-1}$, and distribution in the vacuum, before the particle enters and after it leaves the crystal, with the quantity $\gamma^{-1}$  instead of  $\gamma_{eff}^{-1}$  and a maximal number of virtual photons at angles  $\delta_{v \bot}$, $\delta_{v \|}=\pm\gamma^{-1}$;  $\phi$ is the angle between the crystallographic planes and the vector $\vec V$. The PXR reflection is emitted at the angle $2\phi$  to the vector $\vec V$. Angular distribution for the yield of the real photons in PXR reflection \cite{17} is shown by the diagram; the maximal yield in the PXR reflection is found in the directions at the angles $\delta_{\bot}$, $\delta_{ \|}=\pm\gamma_{eff}^{-1}$  to the direction of the PXR reflection center (Bragg direction), shown by the dashed line.}\label{shchag:fig1}
\end{center}
\end{figure}

\section{ The PXR yield}
Considering the reflection of the virtual photons (\ref{shchag:2}) from the crystal, one can find the formula for the differential (with respect to the angles) yield of PXR \cite{9,11}
 \begin{equation}\label{shchag:3}
 Y_{PXR}=\frac{dN_{PXR}}{d\Omega}=\frac{\alpha\cdot z^2\cdot M\cdot \left|
 \chi_g \left(\omega_{PXR}\right)\right|^2}{\left(\frac{c}{V\sqrt\varepsilon} - \cos \theta\right)^2}
 \frac{\delta_\bot^2+\delta_{\|}^2\cdot \cos^2 2\phi}{\left(
 \gamma_{eff}^{-2}+\delta_\bot^2+\delta_\|^2
 \right)^2},
 \end{equation}
where $\delta_{\bot}$, $\delta_{\|} \ll 1$  are the small angles that define the direction of the radiation propagation with respect to the direction of the PXR reflection center (Bragg direction). The angles $\delta_{\bot}$, $\delta_{\|}$  are measured in the directions perpendicular and parallel to the PXR diffraction plane, respectively. The PXR diffraction plane goes through the vectors $\vec V$  and $\vec g$. The angle between the velocity vector of the particles  $\vec V$ and the PXR reflection center equals $2\phi$  in the diffraction plane,  $\theta=2\phi-\delta_\|$, and $M$ is the number of crystallographic planes which contribute to the formation of PXR reflection. In a crystal which is transparent at frequency $\omega_{PXR}$, $M$ is simply the number of crystallographic planes crossed by the moving particle. In the case of an absorbing crystal, $M$ decreases due to the attenuation of the radiation in the crystal. The effective number of crystallographic planes which contribute to the formation of PXR reflection in this case can be found using Eqs. (13), (14) in paper \cite{13}: The expression for radiation frequency $\omega_{PXR}$   is described below by eq. (\ref{shchag:4}). Diagrams for angular distributions of the virtual photos accompanying the relativistic charged particle in the vacuum, and in the crystal, and of real photons in the PXR reflection are shown in Fig. \ref{shchag:fig1}. The formula (\ref{shchag:3}) has been obtained in kinematical approximation without account of secondary reflections of emitted radiation from crystallographic planes. Experimental research of the PXR differential yield and confirmation of the validity of kinematic PXR theory can be found, e.g. in papers \cite{5,6,9,12,13,14,15,16,17,18,19}. Calculations of the distribution of the PXR yield in the PXR reflections emitted at different angles can be found in refs \cite{20,21}.

The absolute differential yield in the maxima of the PXR reflection observed in experiments with a thin silicon crystal, is, for example, on the order of $2\cdot 10^{-4}$  quanta per electron per sr at an electron energy of 25.7 MeV \cite{12}, and on the order of $5\cdot 10^{-3}$  quanta per electron per sr at an electron energy of 855 MeV \cite{15}.

\section{ Frequency of the PXR spectral peak}
The PXR is quasi-monochromatic radiation. The frequency $\omega_{PXR}$  of the PXR first-order spectral peak is described by the formula
 \begin{equation}\label{shchag:4}
 \omega_{PXR}=\frac{g\cdot V\cdot \sin\phi}{1-\frac{V\sqrt \varepsilon}{c}\cos\theta}
 \end{equation}
Expression (\ref{shchag:4}) was first derived in monograph \cite{1} (see Eqs. (28.6) and (28.158) in \cite{1}) from the energy and momentum conservation laws. Moreover, frequency (\ref{shchag:4}) can be obtained from the Huygens scheme for PXR, proposed in book \cite{5}. One of the remarkable properties of PXR lies with the ability to gradually tune the frequency of the PXR spectral peak by simply rotating the crystal target placed in a beam of accelerated particles, i.e. correspondingly changing the angle $\phi$  in formula (\ref{shchag:4}). Gradual tuning of the PXR spectral peak frequency has been observed in various experiments in the energy range of PXR quanta from several keV \cite{12, 14} up to 400 keV \cite{22}. The measurements in cited articles experimentally confirmed the validity of expression (\ref{shchag:4}).

\section{The PXR spectral peak width}
The normalized FWHM (full width at half of maximum) of the PXR spectral peak is
  \begin{equation}\label{shchag:5}
  \frac{\Delta \omega}{\omega_{PXR}}=0.89\cdot M^{-1}.
 \end{equation}
This formula was obtained in Ref. ~\cite{23}, where PXR emitted from a finite transparent crystal was simply considered as a constant amplitude wave train. The number of oscillations in the train was equal to the number of crystallographic planes $M$  which the particle crossed. Note that the expression for the normalized FWHM of the PXR spectral peak (\ref{shchag:5}) was obtained for ideal conditions, without taking into account the effects of particle multiple scattering in the crystal, the radiation attenuation in the crystal itself, or the experimental angular resolution. Therefore, expression (\ref{shchag:5}) holds true only in the case of relatively thin transparent crystals and ideal angular resolution. Changing these conditions can lead to the broadening of the PXR spectral peak. The influence of the real experimental angular resolution, the experiment geometry, and the radiation attenuation on the width of the PXR spectral peak was studied in Refs.~\cite{14, 24}, and the influence of particle multiple scattering was looked at in paper \cite{25}. 

The normalized FWHM of the PXR spectral peak in experiments with a thin crystal is typically of the order of  $10^{-2}$  \cite{14}, and a minimal normalized width of up to $10^{-9}$ can be obtained by generating focused PXR from a particle moving in a channeling regime through a long bent crystal \cite{26} (see also Ref. ~\cite{20}). In this case, the length of the PXR wave train and coherent length can reach huge as for X-rays values of several centimeters \cite{26}. 

\section{ Linear polarization of the PXR}
The differential yield of the PXR is completely linearly polarized. But the directions of linear polarization are different in different parts of the PXR reflection. The polarization structure of the PXR reflection is described by curves that are tangent to the directions of linear polarization in every point. The curves can be represented by the equation 
 \begin{equation}\label{shchag:6}
\delta_{\bot}=B\cdot \left|\delta_{\|}\right|^{-\frac{1}{\cos (2\phi)}}
\end{equation}
with arbitrary constant  $B$. There is a significant difference in the structures of the linear polarization (\ref{shchag:6}) in the PXR reflections emitted in the front and rear hemispheres. In the case of PXR reflection emitted in the front hemisphere for $2\phi<\frac{\pi}{2}$, curves (\ref{shchag:6}), which are tangent to the directions of the linear polarization, have a hyperbolic form, while in the case of emission in the rear hemisphere, $2\phi>\frac{\pi}{2}$, these curves are parabolic. The calculated results for the structure of the linear polarization in PXR reflections emitted at various angles in the front and rear hemispheres are present in Fig.~2 in paper \cite{27}, as well as in figures plotted in papers \cite{20, 21}. The calculated results are in good agreement with the results of the lab measurements of the PXR polarization performed by in Ref. \cite{28}.

\section{ Applications of the PXR}

The PXR is a beam of linearly polarized monochromatic X-ray radiation. The PXR photon energies can be smoothly tuned in the range from a few to hundreds of keV. However, the sources of synchrotron or Compton backscattering radiation can be more powerful because of multiple emission of radiation by every electron that move in a storage ring. But PXR source can be easy installed at any linear accelerator without a storage ring and PXR reflection can be emitted at big angle to the electron beam. Besides, the PXR source can provide soft monochromatic tunable gamma-ray beam with energy of photons up to hundreds of keVs. Let us describe briefly some of possible applications of the PXR.

The experiment \cite{22} demonstrated that PXR can provide beam with photon energies no less than 400~keV. This fact encouraged research for application of the PXR for calibration of hard X-ray or soft gamma-ray space telescopes. It was found in Ref.~ \cite{29, 30} that PXR source can provide enough for calibration flux of monochromatic linearly polarized radiation at distance about 500~m with photon energies of hundreds keV. The distance is necessary to provide low enough divergence for whole telescope's aperture.

	Crystals have a number of crystallographic planes. The PXR reflections are produced in every of them. Therefore a number of PXR reflections are emitted simultaneously and coherently in different directions from only one crystal installed in a particle beam. The application of such multi-beam generation for obtaining several coherent X-ray beams simultaneously has been proposed in \cite{29}.

	The PXR beam with photon energy exceeding K-edges of elements U and Pu can provide K-shell ionization of these atoms. Observation of K-lines of characteristic X-ray radiation would allow identification of these elements. Possibilities of the X-ray locator based on the PXR source to control nuclear materials nonproliferation was analyzed in Refs.~\cite{31, 32}. It was found that the remote location of nuclear materials is possible up to distances up to about 20~m.
	
There were experiments for observation of shadow images of biological objects with use of the PXR beam \cite{33,34,35}. More recently, phase-contrast images in the PXR beam were obtained \cite{36}.

Since pioneering work \cite{37}, a series of experimental research about increasing of positrons production in crystalline targets were performed, see e. g. \cite{38,39,40}. The control of the production of coherent bremsstrahlung and positrons from a crystalline target by PXR emitted from the same target in backward direction has been proposed and considered in Ref. \cite{41}.

Possibilities for optical diagnostics of small intensive beams are restricted because of diffraction limitations and coherent emission from dense beams. In order to overcome these limitations, application of X-ray diagnostics based on PXR source has been proposed in Refs.~\cite{42, 43}. Recently, observation of the beam profile with use of the PXR source with pinhole has been demonstrated in experiment \cite{44}.

In Ref.~\cite{45}, it was proposed and analyzed the possibility for measurements of the size of crystalline grains of nanometers range with use of the spectral peak width of the PXR emitted in the backward direction.

Emission of wide PXR beam from the textured polycrystal was demonstrated in experiments described in Refs.~\cite{46, 47}. Such wide PXR beam can be used for imaging of large objects \cite{46, 48}. Besides, the application of rolled textured metal foils as a radiator of PXR would allow the increase of beam current and PXR beam power without problems connected with target damage. 

The application of the PXR for online diagnostics of the interaction of the beam with bent crystal was proposed in \cite{49}. The possibility to use PXR for control of the bent crystal degradation was analyzed \cite{50}. Thus, PXR can be used for online control of bent crystalline deflectors of high-energy beams. Experiments with PXR began in CERN \cite{51} and in Protvino \cite{52}.

Thus, there are a lot of fields for applications of the PXR. However, not all basic properties of the PXR are well known by now. For instance, properties of the PXR generated from bent crystal has been poorly studied by now, as well as coherent properties of the PXR beam.

\section{ Some possibilities for X-ray beam production based on radiation from crystals at Tho\-mX in LAL}

ThomX is a project proposed by a collaboration of French institutions and one company to build an accelerator based compact X-ray source in Orsay (France) \cite{53, 54}. The main goal of the project is to deliver a stable and a high energy X-ray beam (up to $\sim 90$ keV) with a flux of the orders of $ 10^{11}  - 10^{13}$ photons per second generated by the Compton backscattering process. At present, the ThomX machine is under construction. The ThomX accelerator facility is composed by the linac driven by 2998 MHz RF gun, a transfer line and a compact storage ring where the collisions between laser pulses and relativistic electron bunches result in the production of the X-rays. The energy of accelerated electron beam would be 50 -- 70 MeV. If there would be a possibility to install a goniometer under a vacuum on the electron beam, one can also provide another source of X-ray beam. A few kinds of radiation generated by electron beam passing through a crystal can be used, such as channeling radiation, coherent bremsstrahlung and PXR. This could extend the capabilities of the ThomX \cite{55}. As for the PXR, the X-ray beam can be used as it was described in previous section as well as for basic research.

\section*{ Acknowledgments}
This paper became possible due to supports from International Associated Laboratory on the field of high energy physics "Instrumentation Developments for Experiments at Accelerators facilities and accelerating Techniques" (LIA IDEATE), and LAL, and also due to travel financial support from the STCU.

\coltoctitle{Hybrid and Metal Microdetector Systems for measuring in real time spatial distribution of charged particles and X-rays beams}
\coltocauthor{V.~Pugatch, I.~Momot, O.~Kovalchuk, O.~Okhrimenko, Y.~Prezado}

\begin{center}
\large \bf
Hybrid and Metal Microdetector Systems for measuring in real time spatial distribution of charged particles and X-rays beams
\end{center}

\vskip 0.2cm

\begin{center}
{\bf  V.~Pugatch$^{a}$, I.~Momot$^{a}$, O.~Kovalchuk$^{a}$, O.~Okhrimenko$^{a}$, Y.~Prezado$^{b}$}
\end{center}

\vskip 0.2cm

\begin{center}
\textit{$^a$ Institute for Nuclear Research, National Academy of Sciences of Ukraine, Kyiv\\
$^b$ Laboratoire d'Imagerie et Mod\'elisation en Neurobiologie et Canc\'erologie 
(IMNC, CNRS), Orsay
}
\end{center}

\vskip 0.3cm

\textit{ Abstract:} Metal and hybrid micro-detector systems were tested at experimental facilities (HERA-B, LHCb, ESRF, KINR Tandem-generator) for measuring and imaging in real time spatial distribution of charged particles as well as X-ray beams. Monte-Carlo simulations as well as tests with low energy proton beam indicate a possibility to provide shaping and monitoring in real time of mini-beam structures for the purposes of fractionated hadron radiation therapy. 

\vskip 0.4cm
\noindent
{\it PACS numbers:} 29.40.Wk 

\vskip 0.4cm
\noindent
{\it Keywords:}  Monitoring of the charged particles and X-ray beams, Metal and hybrid micro-detectors, Real time systems.

\section{Introduction}
Various applications of the ionizing radiation beams require precise, long term reliable   monitoring of their spatial distribution in real time. In particular, the tissue-sparing effect of spatially fractionated beams was established in biological studies performed with mini-, micro-beams of synchrotron radiation \cite{1, 2}. The concept of hadrons minibeam radiation therapy has been proposed as well \cite{3}. Homogeneous distribution of the dose delivered by every beam as well as the highest possible Peak-to-Valley-Dose-Ratio (PVDR) for all beams in a multi-beam structure are required for reaching the best therapeutic result. To measure the dose profiles Gafchromic films are usually explored in conventional radiotherapy \cite{4}. Their off-line analysis provides excellent position accuracy (few micrometers) of the dose distribution. Yet, it is a time consuming procedure and it is impossible to have an online dose monitoring.

      In this talk the application of metal and hybrid micro-detector systems operating in real time is briefly presented for the X-rays beams monitoring while more detailed results are discussed for measuring charged particles beams also in multi-beam structures shaped by slit or matrix collimators. Studies were performed with prototypes of such systems exploring low energy protons at the KINR Tandem generator. 

\section{  Experimental studies}

      \subsection{ Metal Foil Detectors at high energy hadron beams}

           Physics and techniques principles of the Metal Foil Detector (MFD) have been developed at the Institute for Nuclear Research NASU (Kiev, KINR) \cite{5}. Incident on a metal foil sensor particles result in Secondary Electron Emission (SEE) from 10-50 nm of its surface. A positive charge appearing at isolated sensor is measured by sensitive Charge Integrator.  A single layer MFD had been explored for the multi-target steering at the HERA-B experiment \cite{6}. To provide unambiguous reconstruction of secondary vertices the interaction rate of 40 MHz had to be equally distributed over eight targets operated simultaneously in the halo of the 920 GeV proton beam. This has been realized by making targets as an MFD structure — thin metal strips connected to sensitive charge integrators. It has been proved that the SEE yield under the protons impinging onto a single target was strictly proportional to that target partial contribution into the total interaction rate. The equal sharing ($\sim 12,5 \%$) of the overall luminosity among eight operating metal targets-detectors has been demonstrated by means of the reconstructed primary vertices.

     The 12 sector MFD has been built and explored for the luminosity monitoring of the experiment HERA-B. The reliably detected flux of relativistic particles was in the range of $10^4$ particles/s per MFD sensor. Based on this technology Radiation Monitoring System (RMS) \cite{7} has been built at the Large Hadron Collider for the LHCb experiment for monitoring of the radiation load on the Silicon Tracker sensors. 
     
    \subsection{ Metal and Hybrid Micro-Detectors at X – rays beams}

     A reliable performance of the Metal Foil Detectors mentioned above as well growing demand for the low mass, radiation hard micro-detectors have evolved into the development of the Metal Micro-strip and Micro-pixel detectors (MMD). Such detectors have been successfully created at KINR in close collaboration with MPIfK (Heidelberg), DESY (Hamburg), Institute of Micro-devices NASU (Kiev) and CERN (Geneva) \cite{8,9,10}. To achieve the micrometer level position resolution and low thickness the silicon micro-strip detector technology combined with plasma-chemistry etching has been developed and used for the MMD design and production \cite{8,9}. Current technology allows for production of the $\sim  2$~μm thick metal strip sensors with width and pitch at the level of few micrometers. The sensitivity of the MMD to the radiation flux is determined by the physics conversion factor as well as by the readout electronics.  For instance, the MMD successfully applied for the 13 keV X-rays beam profile monitoring at HASYLAB (DESY, Hamburg) has had the conversion factor of $1.5 \times 10^4$ photons/electron \cite{9}.
     
     Characterization studies of the Metal Micro-detectors have been performed at the Mini-beam Radiation Therapy (MBRT) setup at ESRF (Grenoble) measuring in real time dose distribution of the synchrotron radiation \cite{11}. The biomedical features of that therapy requires high doses up to several kGy/sec. X-rays with peak energy ranging from 50 to 600 keV (peak energy of  
     100~keV)  were produced with intensity up to $5\times 10^{11}$ $\mbox{photons/(c}\times\mbox{mm}^2$). The spatially fractionated minibeam patterns were produced by multi slit collimator or utilizing a rotating chopper. Existing position sensitive gaseous or solid state detectors cannot operate reliably in real time at high radiation load exploited for the MBRT.

    TimePix detector designed at CERN \cite{12} has been used in these studies in a metal or hybrid mode for real time measurement of the dose distribution. TimePix detector in a metal mode  comprises a bare readout microchip ($256 \times 256$ pixels) and a metal grid for collecting SEE. Each readout pixel has a size of ($55 \times 55$) $\mu$m$^2$.  High spatial resolution as well as radiation hardness of the TimePix (up to 4.6 MGy) are suitable for measuring steep dose gradient areas and PVDR. The results obtained have demonstrated reliable performance of the TimePix in real time measurements at the radiation load of few kGy/s distributed over 10 beam-prints with a size of each $0.6 \times 10$ mm$^2$  and center-to-center distance of 1.2 mm. PVDR measured by the TimePix and gafhromic films agreed well.

     Data obtained for various modifications of mini- as well as micro- beam configurations will be used for creating a novel MMD-based radiation hard monitoring system for radiation therapy applications. In comparison with the latest developments in beam profile monitoring by various detector systems Metal Micro-strip Detectors have an advantage of being extremely thin and semi-transparent device, providing nearly non-destructive beam analysis in real time.  
     
     TimePix detector in a hybrid mode has been also applied for the X-rays diffraction studies of a dynamics of phase transitions in metal alloys under heating/cooling at the experimental setup of the Institute for Problems in Material Science NASU (Kiev) \cite{13}. Metal samples under heating/cooling transform their crystal structure resulting in a shift of corresponding diffraction peaks of scattered X-rays. TimePix detector allowed to get precise data on peaks position and their evolution  in real time exploring fast framing rate of up to 100 Hz with the USB interface FitPix \cite{14}. This is a key feature for dynamical phase transition studies. One may treat the TimePix detector as an ‘electronic plate’ imaging in real time a dynamics of phase transitions.

        \subsection{  Hybrid and Metal micropixel detectors measuring spatial distribution of low energy ions in a focal plane of the mass-spectrometer}

     The TimePix (Hybrid and Metal) detector response to low energy ions has been studied as a function of ion mass, charge, energy and beam position observed at the focal plane of the laser mass-spectrometer of the Institute of Applied Physics (Sumy, Ukraine) \cite{15}. Ion beams have been generated at the sample-target by the pulsed infrared (1064 nm) laser. Position of the TimePix detector alongside the focal plane, accelerating voltage and magnetic field were adjusted to observe few tens keV positively charged ions of the different isotopes (from carbon to lead). Real time observation of the mass-distribution of isotopes has proved usability of the TimePix detector for tuning and adjusting mass-spectrometer. This provides an improvement in a mass-resolution as well as introduces a new possibility of a simultaneous mass-spectrometry of few target samples, prepared, for instance, by different technologies. It was established that pixels response is more uniform for the TimePix (Metal) mode operation. Results of the tests show the ability of both detector modes to be used for building electronic focal plane of the mass-spectrometer to determine the chemical composition of a complex target samples. Mapping isotope population over the sample area by scanning it by a laser or charged particle micro-beam might be useful for applications in material studies, microbiology, medical diagnostics etc. 

 \subsection{ Feasibility studies of Metal and Hybrid microdetectors for Monitoring of the Hadron Beams for the purposes of fractionated radiation therapy}

     Spatially fractionated hadron beams (several hundred MeV/nucleon) are considered as promising tool to get better therapeutic results due to dose localization, and tissue-sparing effect observed with mini-beams of synchrotron radiation \cite{3}. We report here some results of the experiment performed for optimization of the characteristics of the setup for carrying out feasibility studies at Heidelberger Ionenstrahl-Therapiezentrum (HIT, Heidelberg) related to the design of collimators and their effectiveness for the purposes of the fractionated mini-beam hadron radiation therapy. Monte Carlo simulations (Gate$\_$v.6.2 and Geant4 versions) of dose distribution resulting from spatially fractionated irradiations have been explored for the design of collimators shaping multi-beam structure. As figure of merit, peak and valley depth dose curves, penumbras, and central Peak-to-Valley Dose Ratios (PVDR) have been calculated for various collimators, incident ions and their energies.

 \begin{figure}[ht]
\includegraphics[width=0.5\textwidth, height=4.5 cm]{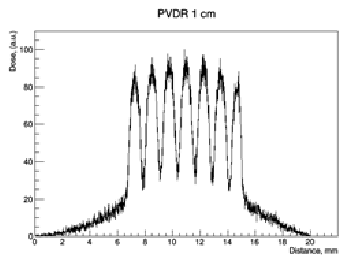}
\includegraphics[width=0.5\textwidth, height=4.5 cm]{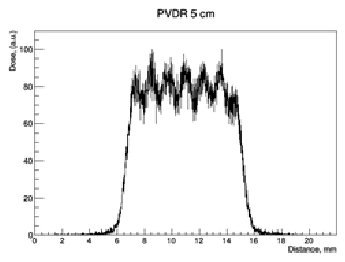}
\\
\caption{ Monte-Carlo simulation of the dose profile (projection onto X-axis) from the  $7\times 7$  holes with diameter of 0.6 mm in matrix aluminum collimator with c-t-c distance of 1.2 mm. Protons energy: 87 MeV. Left: depth in a water phantom -- 1 cm. Right -- 5 cm.}\label{06VP:fig1}
\end{figure}

An example of such calculations is shown in Fig.~\ref{06VP:fig1} presenting Monte-Carlo simulation of the dose profile expected at different depth in a water phantom irradiated by the 87 MeV proton multi-beam structure, shaped by the  matrix aluminum collimator ($7\times 7$ holes with diameter of 0.6 mm and c-t-c distance of 1.2 mm). One may see clear separation of the beams (projection onto X-axis) at the depth of 1 cm (left part of the Fig.~\ref{06VP:fig1}) with PVDR value about 5, while at the depth of 5 cm the fractionation of the beam becomes negligible (due to scattering of protons on their way through phantom). This means that the normal tissue would benefit from the spatial fractionation of the dose while a quasi-homogeneous dose distribution is achieved in the tumor position. It is clear that 1D- or 2D-scanning by the hadron pencil beam \cite{15} over the tumor area would provide the best result, keeping in mind, otherwise, an additional radiation load on patient from the irradiated collimator. It is the task for the further bio-medical studies to find out the best therapeutic approach in shaping the multi-beam structure and evolution of the fractionation over the depth of tissue. One could assume that it might be in providing well separated beams on the way through the healthy tissue and making them merged in a tumor.

    Test of the prototypes of the experimental setup for shaping and monitoring of the hadron beam for the purposes of the fractionated radiation therapy has been performed at the KINR Tandem generator. 
 
\begin{figure}
\begin{center} 
\includegraphics[width=0.5\textwidth, height=4.5 cm]{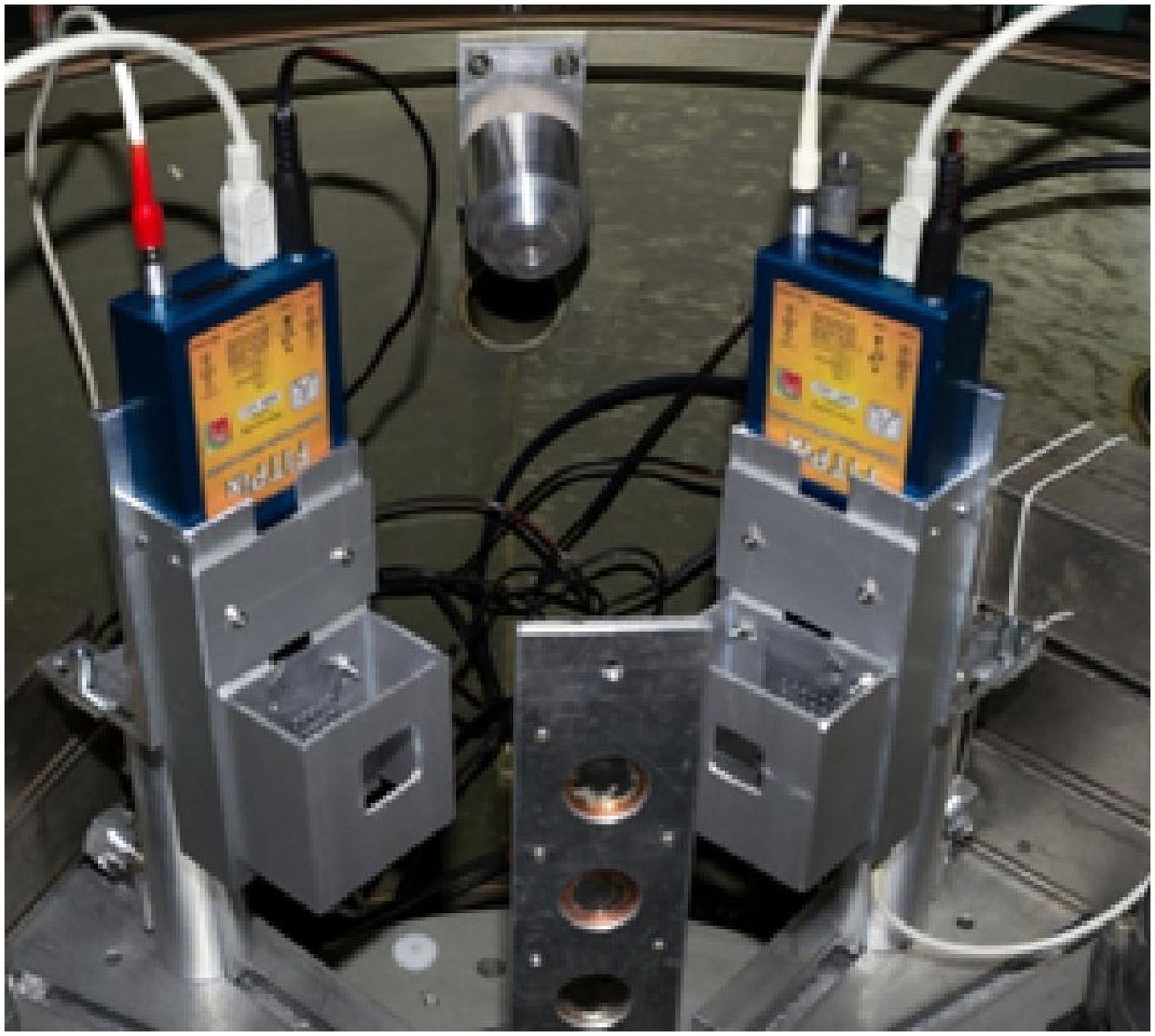}
\\
\caption{Photo of the reaction chamber at the KINR Tandem generator. Two TimePix detectors and matrix collimators in front of them were mounted inside of the chamber.}\label{06VP:fig2}
\end{center}
\end{figure}

     Protons with the energy of 3 MeV were incident on the polyethylene target installed inside the reaction chamber (Fig.~\ref{06VP:fig2}). For shaping multi-beam structure from the scattered protons collimators of two types were explored,  slit and matrix.  Both were manufactured out of 2 mm thick Al plates: SLIT - with 5 slits; MATRIX – with $5 \times 5$ = 25 holes, covering active area of $14 \times 14$ $\mbox{mm}^2$ of the TimePix detector. Protons scattered by the target were registered by two TimePix detectors installed at 100 mm ddistance from the target center, at 45 degrees from both sides of the proton beam axis. One of the detectors was operated in a metal mode while another one in a hybrid mode with 300 $\mu$m thick silicon micropixel sensor. As an example Fig.~\ref{06VP:fig3} illustrates 2D image of the protons intensity distribution measured by metal TimePix detector with the SLIT collimator installed in front of it. Perfectly separated `beams'  were observed. We point out in Fig.~\ref{06VP:fig3} (Left) vertical and horizontal narrow regions with zero intensity inside the slits area. Those regions correspond to the image of the grid (metal wires, 100  $\mu$m diameter) installed in close vicinity to the TimePix readout microchip for providing its metal mode of operation with positive 20 V applied. Projection of the 2D data onto X-axis also exhibits the drop of the intensity inside the slit area at position of the grid wires. This illustrates a nice performance of the TimePix detector in reflecting details of the 2D beam intensity distribution with an accuracy of 55 $\mu$m. 

\begin{figure}
\begin{center} 
\includegraphics[width=0.5\textwidth, height=4.5 cm]{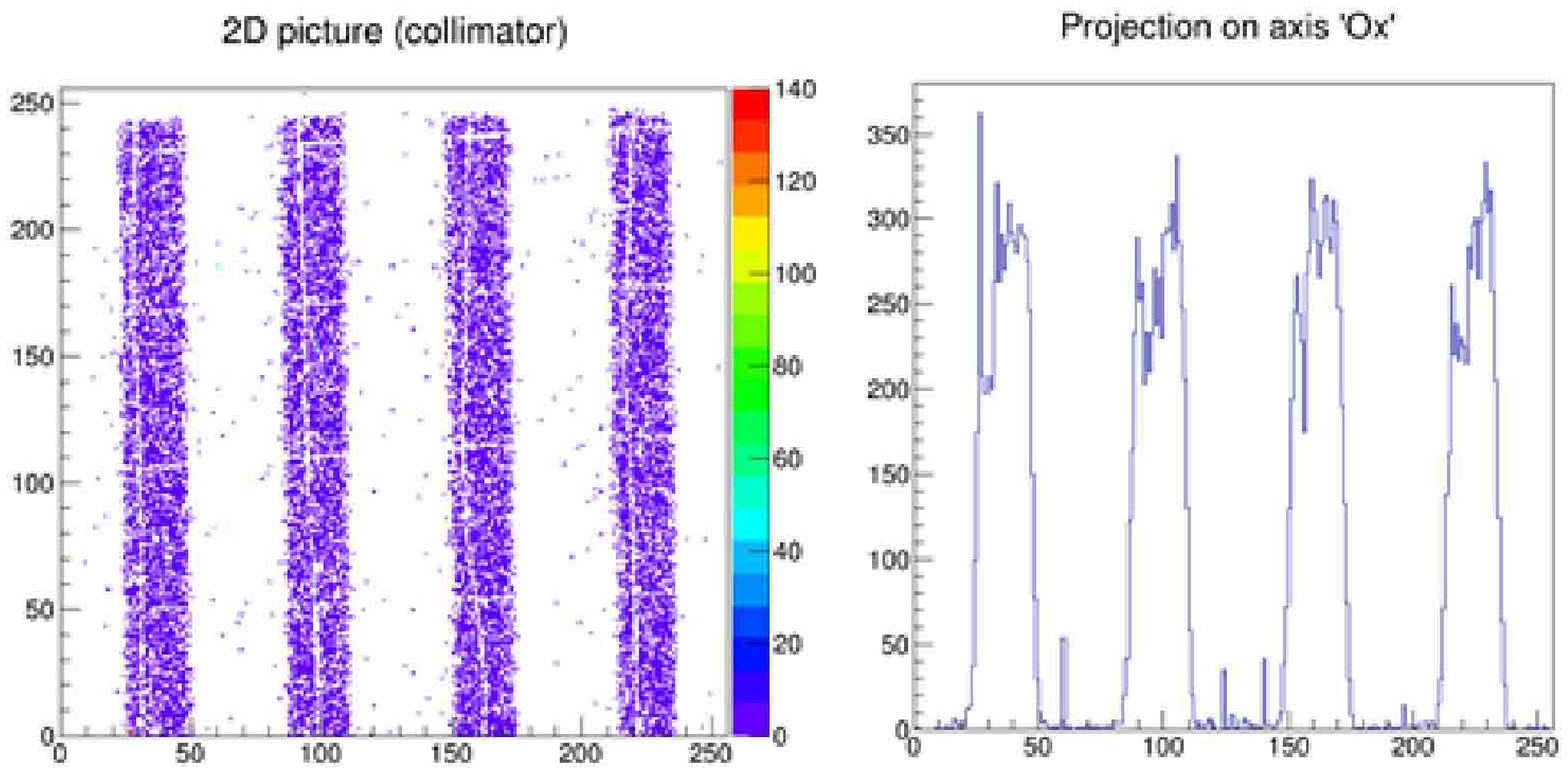}
\\
\caption{Left: 2D distribution of the of the direct proton beam intensity (color scale at the right side) measured by metal TimePix detector (X- and Y-axis -- pixel number).  Slit collimator was installed in front of the TimePix. Right: Projection of the data onto X-axis.}\label{06VP:fig3}
\end{center}
\end{figure}

     The results obtained with a MATRIX collimator installed at the entrance window of the detector TimePix are illustrated in Fig.~\ref{06VP:fig4}.

  \begin{figure}[ht]
\begin{center} 
\includegraphics[width=0.8\textwidth, height=4.5 cm]{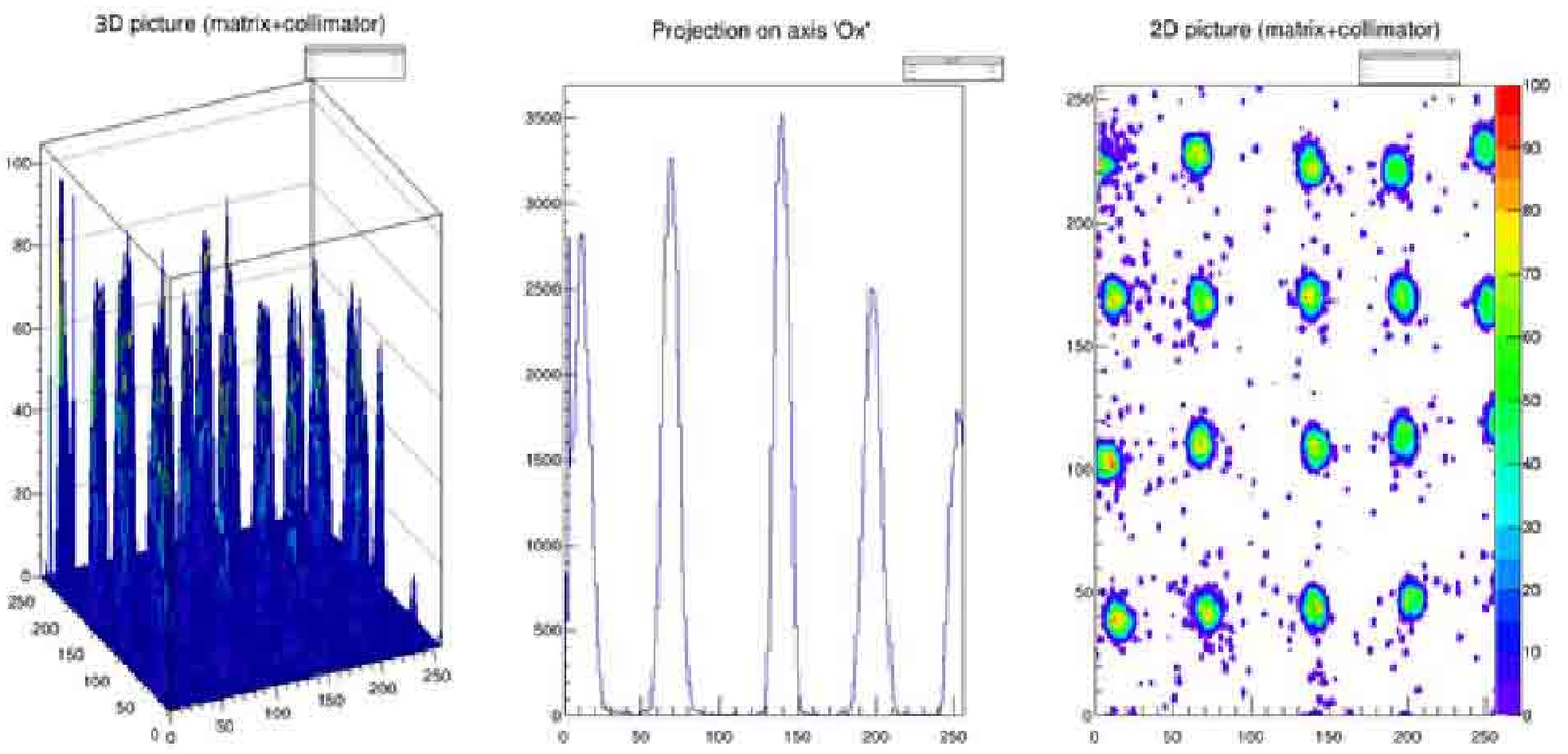}\\
\caption{Right: 2D distribution of the of the scattered proton beam intensity (color scale at the right side) measured by hybrid TimePix detector (X- and Y-axis -- pixel number) at 45 degrees.  Matrix collimator was installed in front of the TimePix detector. Middle: Projection of the data onto X-axis. Left: 3D-view of the proton beam intensity distribution over the TimePix detector area.}\label{06VP:fig4}
\end{center}
\end{figure}  

 The important feature of the multi-channel detecting system is the uniformity of its response. For testing this feature of the TimePix detector we have measured the angular distribution of the protons scattered by the carbon target within the TimePix acceptance ($41\,^\circ<\theta<49\,^\circ$) and compared it with the corresponding calculations. From that study we have concluded that the tested sample of the TimePix detector had non-uniform response of its 65 536 pixels  varying within +/-- 15 $\%$.

\section{	Summary and Outlook}

     Metal and hybrid micro-detector systems tested at different experimental facilities for measuring spatial distribution of charged particles as well as X-rays beams have demonstrated reliable performance also in real time imaging of the measured data. The results obtained in frames of Monte-Carlo simulations as well as in the tests with low energy proton beam indicate a possibility to provide shaping and monitoring of mini-beam structures for the purposes of fractionated hadron radiation therapy. Based on these studies similar metal and hybrid micro-detector systems as well as slit and matrix collimators have been recently used in feasibility studies with high energy hadron beams at Heidelberger Ionenstrahl-Therapiezentrum (HIT, Heidelberg). We plan to publish their positive results, soon.

\section*{ Acknowledgments}
This research was partially conducted in the scope of the IDEATE International Associated Laboratory (LIA). We thank MEDIPIX Collaboration, in particular, M. Campbell, X. Llopart, S. Pospisil, C. Granja for introducing an opportunity to use TimePix detectors for these studies. We express our gratitude to the personnel of the KINR Tandem generator for providing stable beam and frank collaboration atmosphere. This work has been supported by the grants CO-4-1/2015 (NASU) and Ф58/382-2013 (DFFD).

\coltoctitle{Studies of sensitive area for a single InGrid detector}
\coltocauthor{A.~Chaus, M.Titov, O.Bezshyyko, O.Fedorchuk}

\begin{center}
{\Large {\bf Studies of sensitive area for a single InGrid detector}}

\vskip 0.5cm

{\bf A.~Chaus$^{a,b}$, M.Titov$^{b}$, O.Bezshyyko$^{c}$, O.Fedorchuk$^{c}$}

\vskip 0.3cm

$^{a}${\it Kyiv Institute for Nuclear Research}

$^{b}${\it CEA, Saclay}

$^{c}${\it Taras Shevchenko National University of Kyiv}

\end{center}

\vskip 0.4cm

\begin{abstract}
\noindent
The novel structures where Micromegas or GEM are directly coupled to the CMOS multi-pixel readout represent an exciting field and
allow to reconstruct fine-granularity, two-dimensional images of physics events. One of such structures that has become subject
of this article is a single InGrid detector.
In order to study sensitive area of a single InGrid detector, an InGrid chip has been
operated in Mini-TPC.  Full simulation of experimental setup and  electric field were performed.
 Experimental data were compared with simulation in several different configurations.

\end{abstract}

\vskip 0.4cm

\noindent  Keywords: Micromegas, InGrid, Gas detectors, TPC

\section{Introduction}

The availability of highly integrated amplification and readout electronics allows
for the design of gas-detector systems with channel densities comparable
to that of modern silicon detectors. The fine granularity and high-rate capability
of GEM and Micromegas devices can be fully exploited by using high-density
pixel readout with a size corresponding to the intrinsic width of the detected avalanche charge.
However, for a pixel pitch of the order of 100 $\mu m$, technological
constraints severely limit the maximum number of channels that can be brought
to the external front-end electronics. While the standard approach to readout the signals is a segmented
strip or pad-plane with front-end electronics attached through connectors from the
backside, an attractive alternative is to place CMOS chip in the gas volume (without
 bump-bonded semiconductor sensor), with GEM or Micromegas amplification
structure directly above it.
With this arrangement signals are induced at the
input gate of a charge-sensitive preamplifier (top metal layer of the CMOS chip).
Every pixel is then directly connected to the amplification and digitization circuits,
integrated in the underlying active layers of the CMOS technology.
 The proof-of-principle of this concept has been demonstrated in the past by several groups~\cite{Bellazzini:2004hr,Campbell:2004ib,Bamberger:2006xp}.

\section{Pixel readout of Micro-Pattern Gas Detectors. The ``InGrid'' Concept}

The original motivation of combining a Micro-Pattern Gas Detector (MPGD) with Medipix2~\cite{Llopart1046904}
and Timepix~\cite{Llopart2007485} chips was the development of a new readout system for a
large TPC at the ILC. The digital Medipix2 chip was originally designed for
single-photon counting by means of a semiconductor X-ray sensor coupled to the chip. In gas
detector applications, the chip is placed in the gas volume without any semiconductor sensor,
with a GEM or Micromegas amplification structure above it~\cite{Campbell:2004ib,Bamberger:2006xp,Colas:2004ks}.
Approximately 75~$\%$ of each pixel is covered with an insulating passivation layer; therefore,
avalanche electrons are collected on the metalized bump-bonding pads exposed to the gas.
Figure~\ref{Fig:InGrid}(Right) and (Left) shows an enlarged photo of the
Medipix2 pixel cells.

\begin{figure}[!ht]
\includegraphics[width=0.45\textwidth,height=5cm]{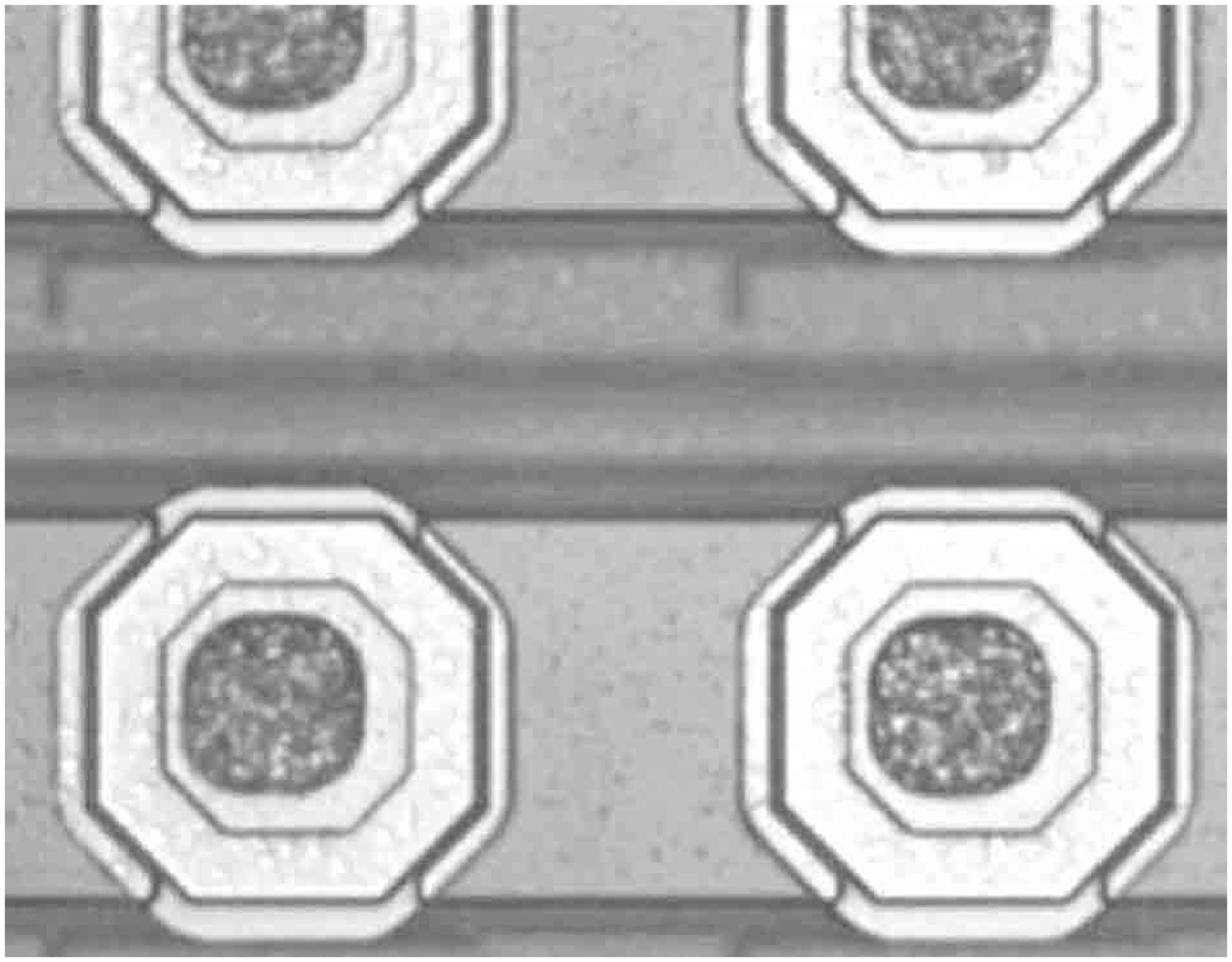}
\includegraphics[width=0.45\textwidth,height=5cm]{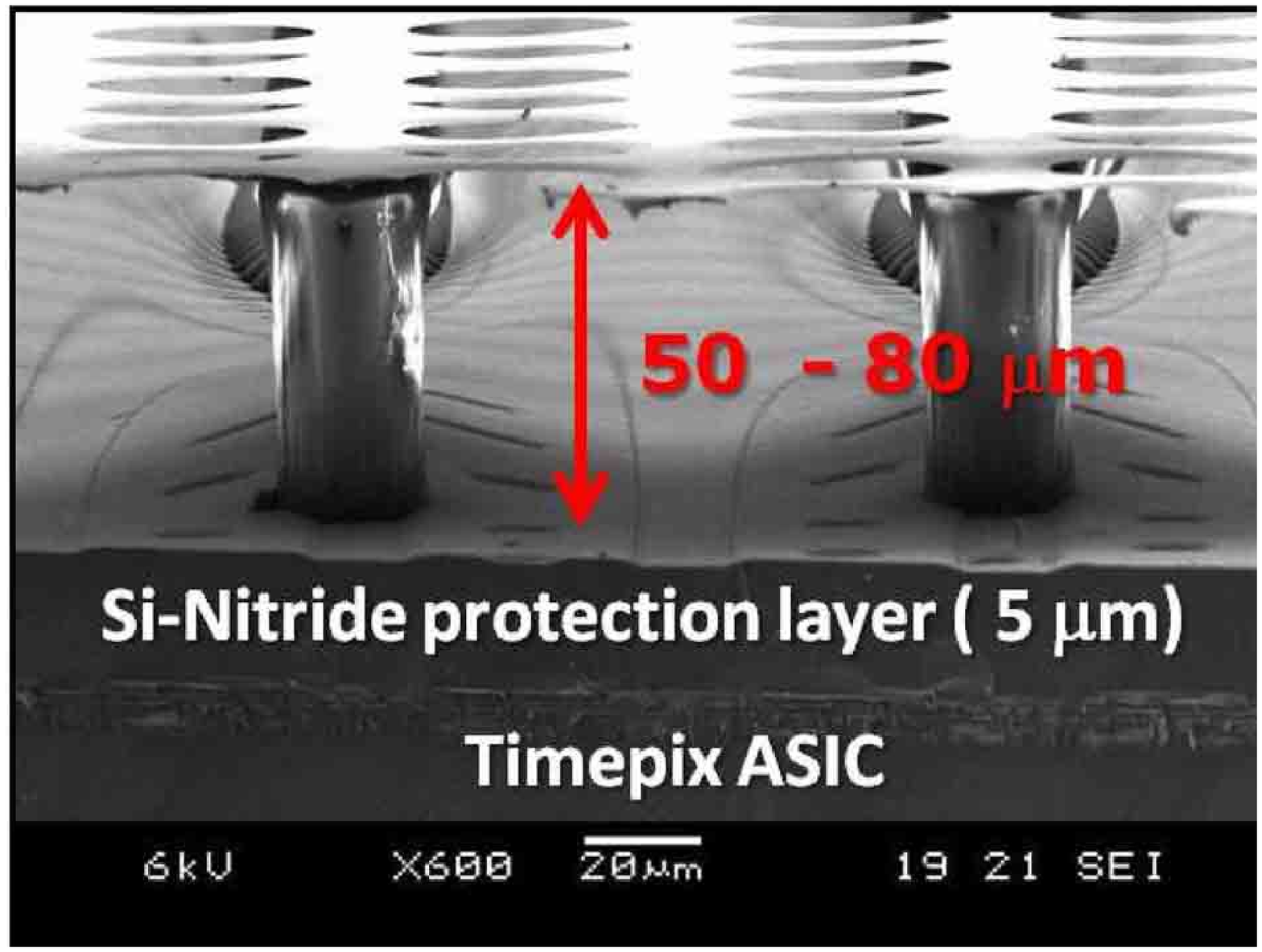}
\caption{Left: Magnified view of bump-bonding openings on top of
the Medipix/Timepix chip. Right: SEM image of the ``InGrid''
structure with ``SiProt'' layer. } \label{Fig:InGrid}
\end{figure}

\section{Measurement setup}

 A special mini-TPC was designed, developed and constructed in Saclay; a schematical
drawing of the chamber with the 10 cm field cage is shown in Figure~\ref{Fig:MiniTPCShema}.
This chamber consists of a printed circuit board where the chip was mounted and
wire bonded.
   The box is built of aluminium with a volume of $\sim  2  l$
and has  three high voltage connectors and input and output gas pipes.
 A transparent window for the radioactive source (Fe$^{55}$ photons)
was made from a 12~$\mu m$ Mylar foil.
 A 10$\,\text{cm}$ gap field cage was also installed inside the box.
The field uniformity is created with the help of 25 voltage degrader rings.
Each ring has a height of 3$\,\text{mm}$ and an
isolation of 1$\,\text{mm}$ between two ring segments.
The segments are connected by 1 M$\varOmega$ resistors and last (bottom) segment is connected to the
ground by 10 M$\varOmega$ resistor.
The top and bottom segments  are connected via high voltage connectors to the power supplies ($U_{cathode}$ and $U_{first}$).
Remaining HV connector is used to supply grid voltage  ($U_{mesh}$).
All voltage parameters were chosen in order to have uniform
electric field 200 V/cm inside the field cage.
This optimal value was chosen from Magboltz simulation for Ar/Iso (95/5) gas mixture.
This experimental setup gives us full sharing of primary electrons that were created in Ar=Iso (95:5) from $^{55}$Fe X-rays.
The distance between the end of the field cage and the surface of the ``InGrid'' chip inside
the box is 8 mm.

\nopagebreak
\begin{figure}[htb]
\centering
\includegraphics[width=0.6\textwidth,height=6.0cm]{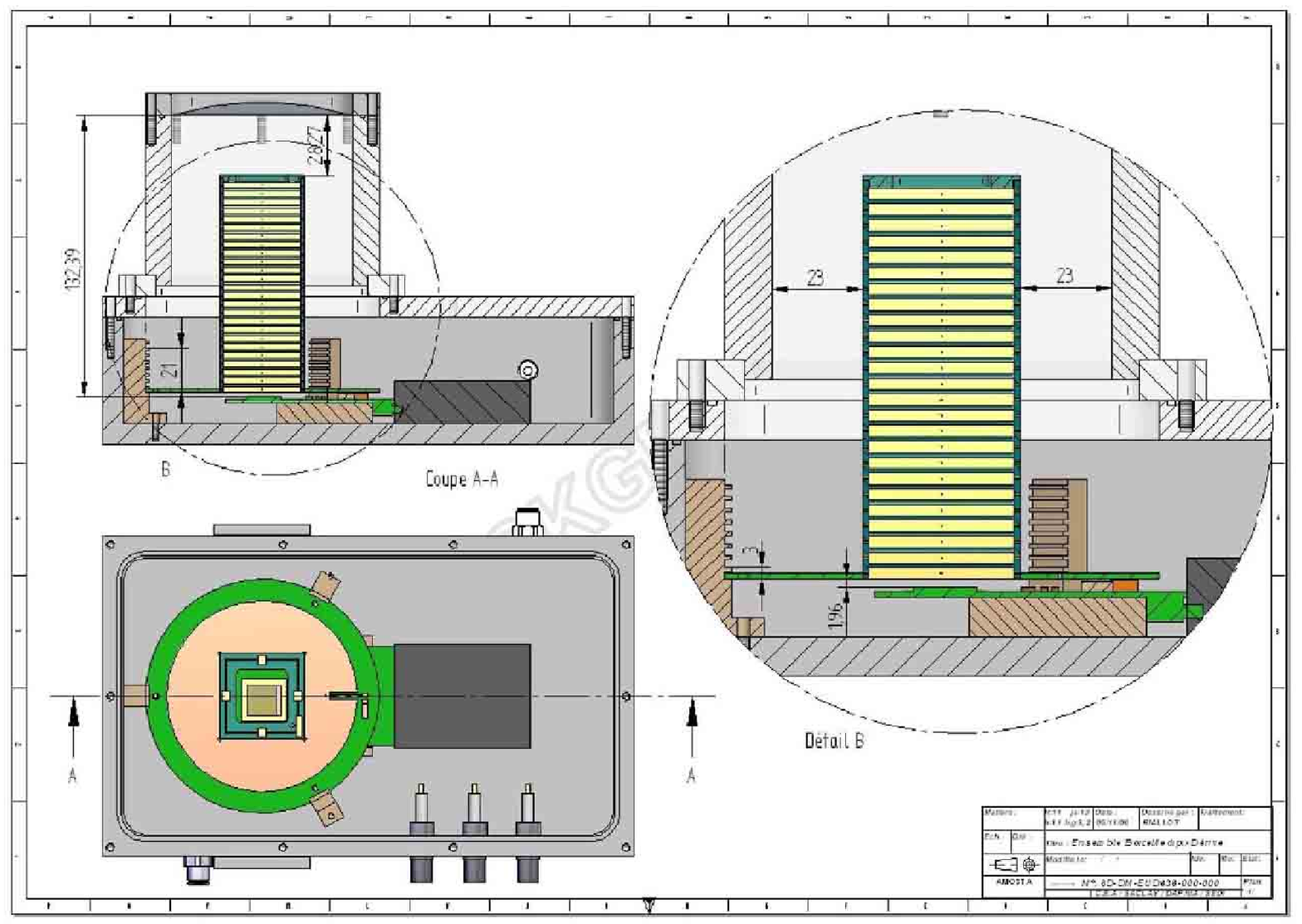}
\caption[Schematic of the Saclay gas box with 10 cm drift cage
]{Schematic of the Saclay gas box with 10 cm drift cage}
\label{Fig:MiniTPCShema}
\end{figure}

Most of the studies in this analysis were performed using D7-W0056
chip from IZM-3 badge (W0056 plate); measurements were done with
and without guard ring using gas P5 (Ar/Isobutan 95/5).

\section{Results and discussion}

\subsection{Data analyses and selection criteria}

All data were saved in ASCII files. Format of saving is (x,y,c),
where "x,y" are the 2D coordinate of the active pixel and "c" is
number of counts for respected counting mode (TOT of Time). For
the analysis the ASCII files  Medipix Analysis Framework (MAFalda)
framework~\cite{website:MAFalda}, which allows the recognition of
particle/tracks, was used. {\it MAFalda} is a set of algorithms
written in the C++ language and based on the Root framework and is
dedicated to data analysis of any device of the Medipix family.
For different purposes several algorithms were written. Further in
the following, criteria used by ``SingleIngrid'' algorithm, will
be explained.

To reject unusable events like cosmics or clusters that were coming from a conversion that
was too close to the chip to provide enough separation of the primary electrons by diffusion, some cuts were used.
 The analyses used the following cuts:
\begin{itemize}
 \item minimal size of electron cloud
 \item position of the geometrical center of electron cloud
 \item circularity of electron cloud
\end{itemize}

{\it Minimal cluster size}  needed to skip the events coming from a conversion that
was too close to the chip.
We suppose that for escape peak for the chosen gas mixture the minimal number of the
primary electrons will not be less than 25 electrons (active pixels) in the electron cloud.

{\it Position of the geometrical center} was used to avoid the clusters which are
registered close to chip border and where part of electrons from the cloud can be lost.
The clusters with the geometrical center ($\overline{x},\overline{y}$) close to the middle of the chip are accepted.
 The range of 100 central pixels in the middle of the chip in x and y-direction was accepted.
If geometrical center is in the range for x-direction
\begin{equation}
     75\leq\overline{x}\leq175
\end{equation}
and the same range is chosen for y-direction, then this cluster is used in analyses. There were chosen 100$\times$100 pixels in the center of the chip.
Figure~\ref{Fig:CentralRegion} shows the central region acceptable by the geometrical center cut.

\begin{figure}[!ht]
\centering
\includegraphics[width=0.5\textwidth,height=5cm]{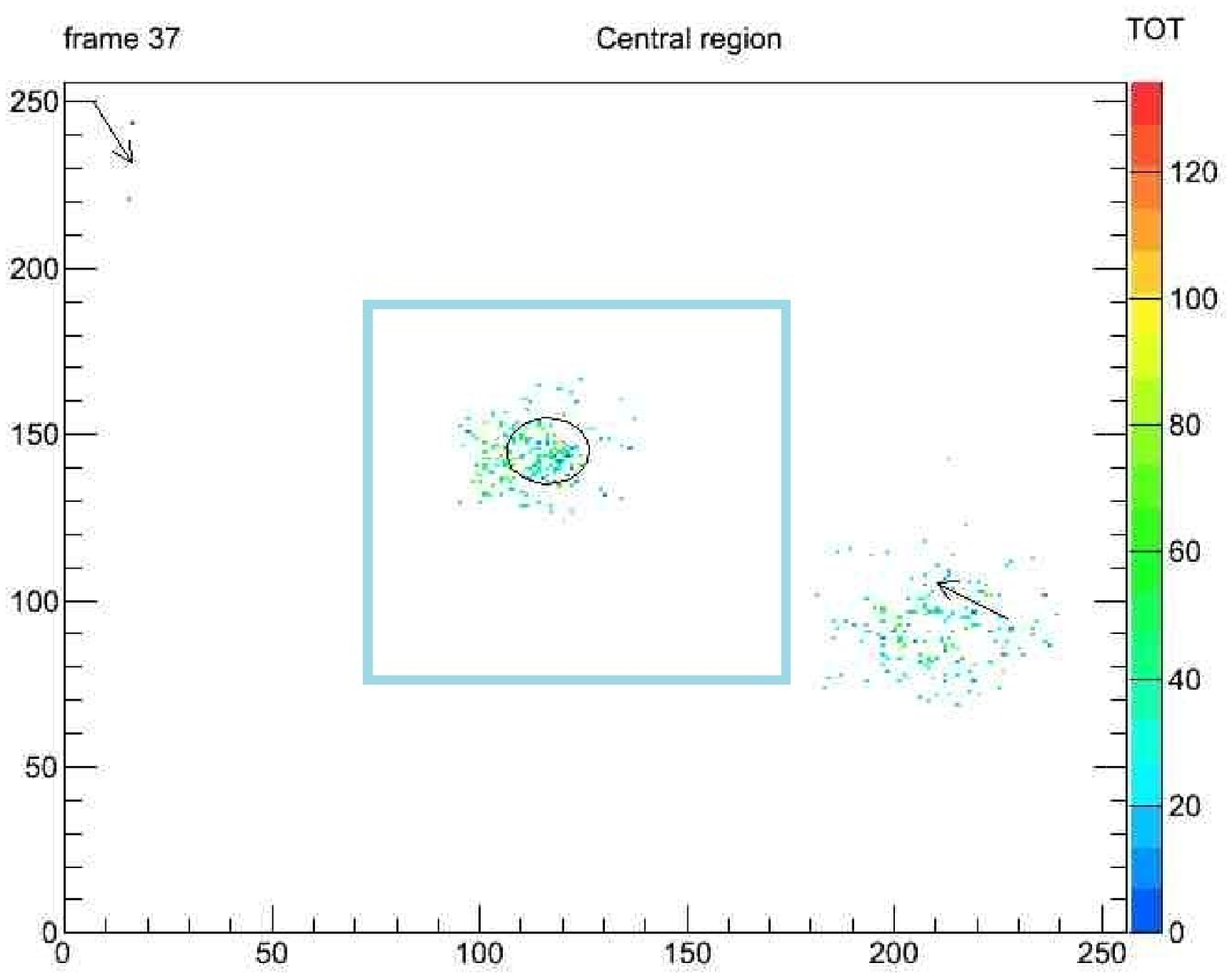}
\caption{MAFalda event display with central area is highlighted.
Only electron clouds with centers in this area are accepted by a
geometry center cut. } \label{Fig:CentralRegion}
\end{figure}

{\it Circularity of the cluster} was used to exclude several electron clouds in TOT mode.
The cut is chosen so that circularity is defined as $R_a/R_b$, where the $R_a$ and  $R_b$ are
ellipse radii calculated for the cluster.
Only electron cloud with
\begin{equation}
     0.75\leq R_a/R_b\leq1.25
\end{equation}

is accepted (Figure~\ref{Fig:CiclingOne}). This cut helps to reject event where  two photon conversions consist in one frame.

\begin{figure}[!ht]
\centering
\includegraphics[width=0.5\textwidth,height=5cm]{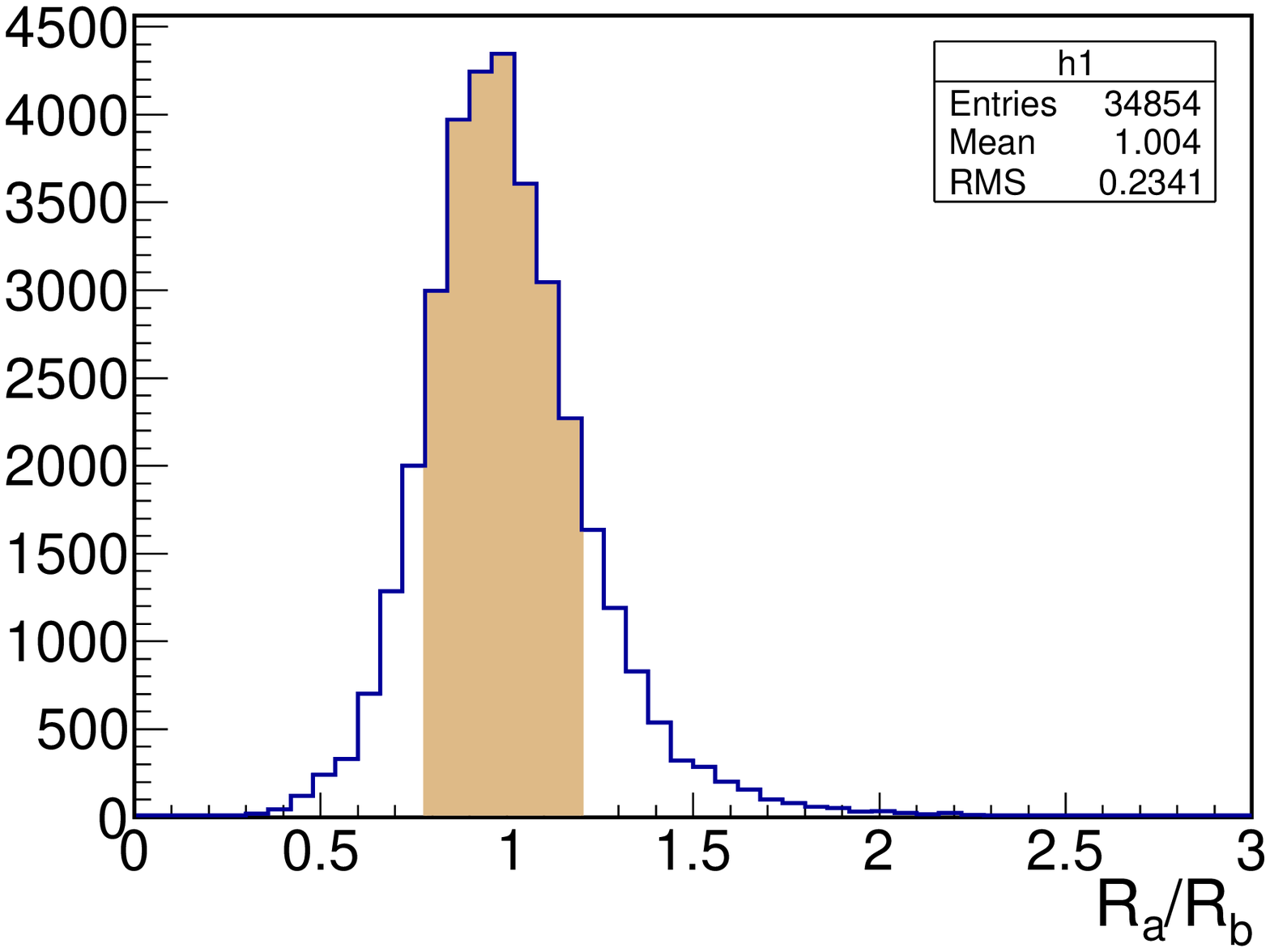}
\caption{Histogram of the $R_{a}/R_{b}$ of the electron cloud in
one data sample. The filled area is the data which were used after
cut. } \label{Fig:CiclingOne}
\end{figure}

\subsection{Occupancy results}

The optimal value was taken for electric field for current gas
mixture from Magboltz simulation. To have the homogeneous electric
field $E=200$ V/cm inside the drift cage the cathode voltage
$U_{cathode}$ = 3100 V (top of drift cage)
 and $U_{first}$ = 1100 V (bottom of drift cage) was chosen. One can see that field distortions on the border of the chip
decrease the sensitive area to $\sim$~35\%, Figure~\ref{Fig:OccupancyTwo} (Left). Figure~\ref{Fig:OccupancyTwo} (Right) demonstrates
 projection on X-axis for this occupancy plot,  but the size of sensitive area can't be enough.

\begin{figure}[!ht]
\includegraphics[width=0.45\textwidth,height=6cm]{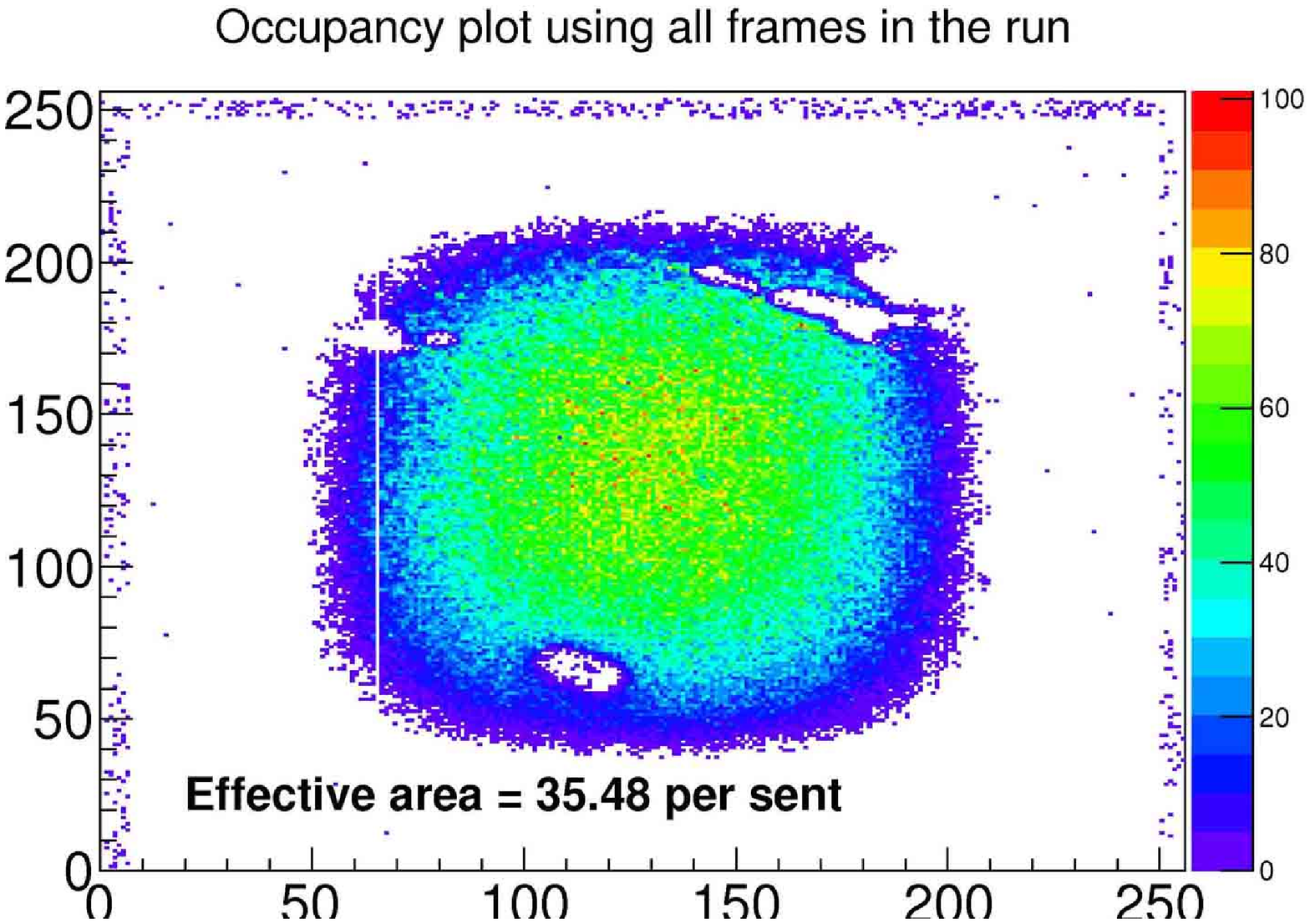}
\includegraphics[width=0.45\textwidth,height=6cm]{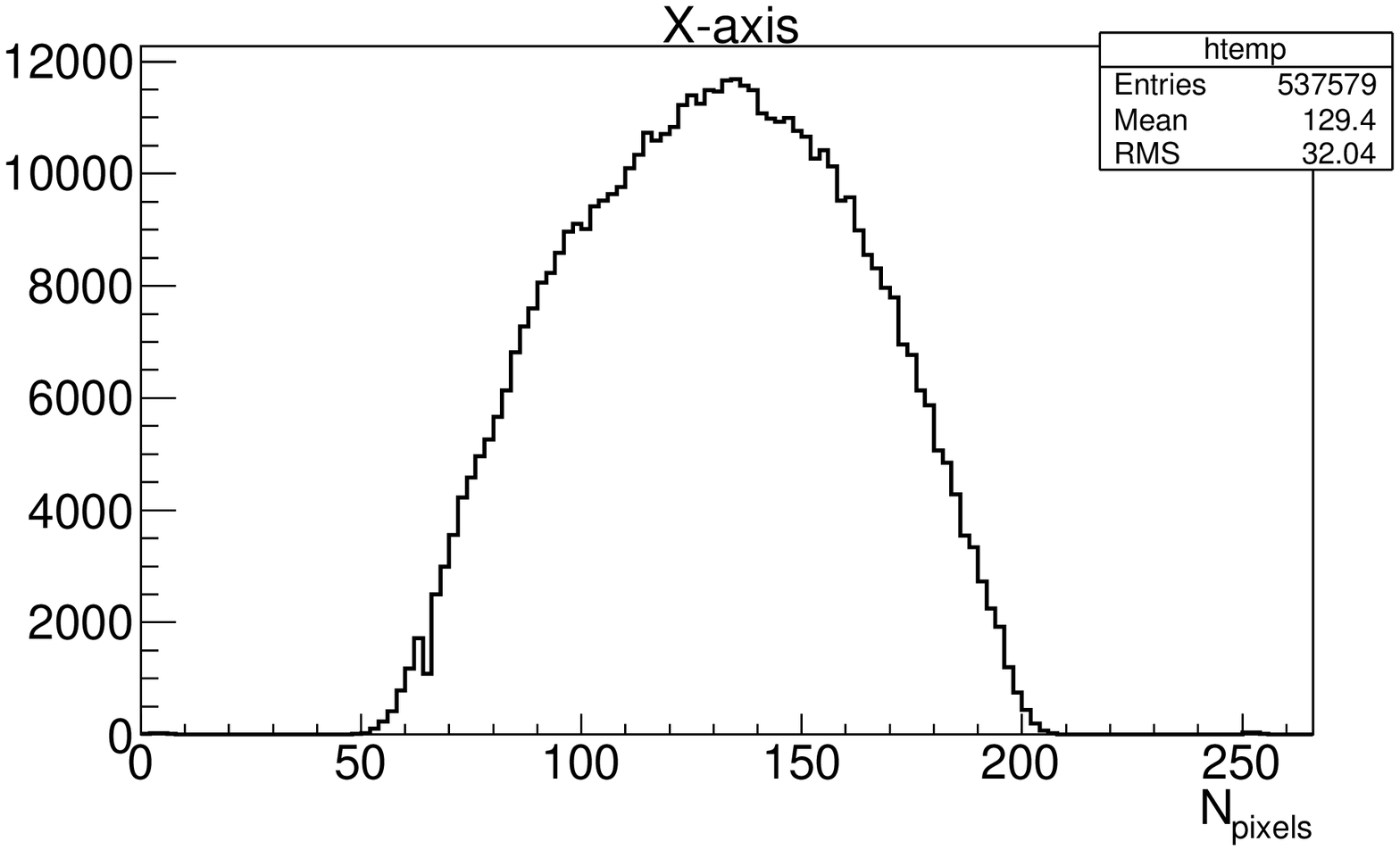}
\caption[Occupancy plot for the chip D07-W0058 is shown: Left: for
the following parameters $U_{mesh}$ = 350 V, $U_{first}$ = 1100 V
and $U_{cathode}$ = 3100 V. Right: projection on X-axis ]{
Occupancy plot for the chip D07-W0058 is shown:
Left: for the following parameters $U_{mesh}$ = 350 V, $U_{first}$ = 1100 V  and $U_{cathode}$ = 3100 V. \\
Right: projection on X-axis.
}
\label{Fig:OccupancyTwo}
\end{figure}

The decision has been made to manufacture and install a guard ring to the setup. The guard ring was installed
on the top on chip board in such a manner that the distance between mesh and guard ring surface is 1$\,\text{mm}$.
The voltage on guard ring was the same as on the mesh $U_{mesh}$ = $U_{guard}$. By applying the guard ring
the  sensitive area can be increased significantly.

\begin{figure}[!ht]
\includegraphics[width=0.45\textwidth,height=6cm]{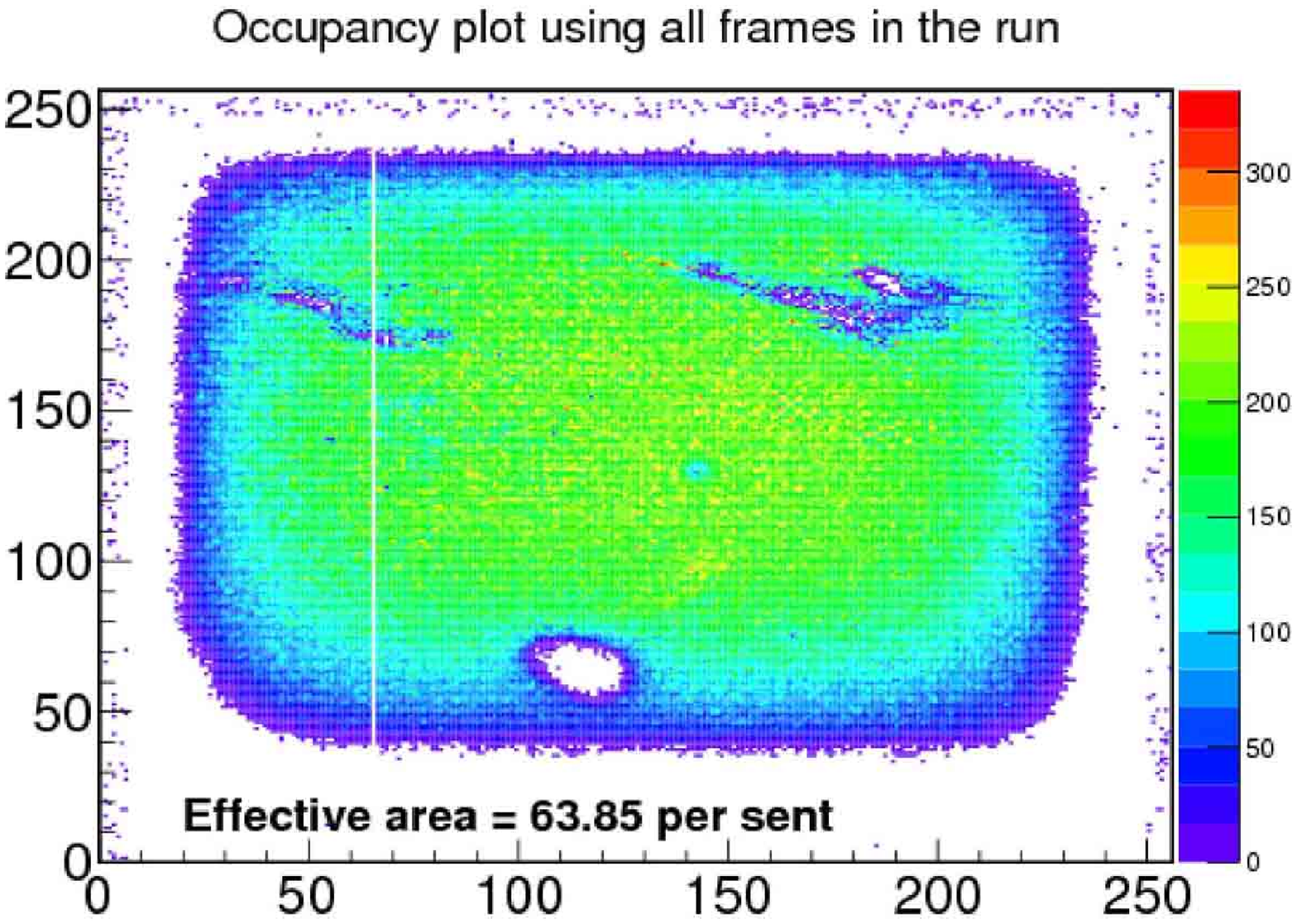}
\includegraphics[width=0.45\textwidth,height=6cm]{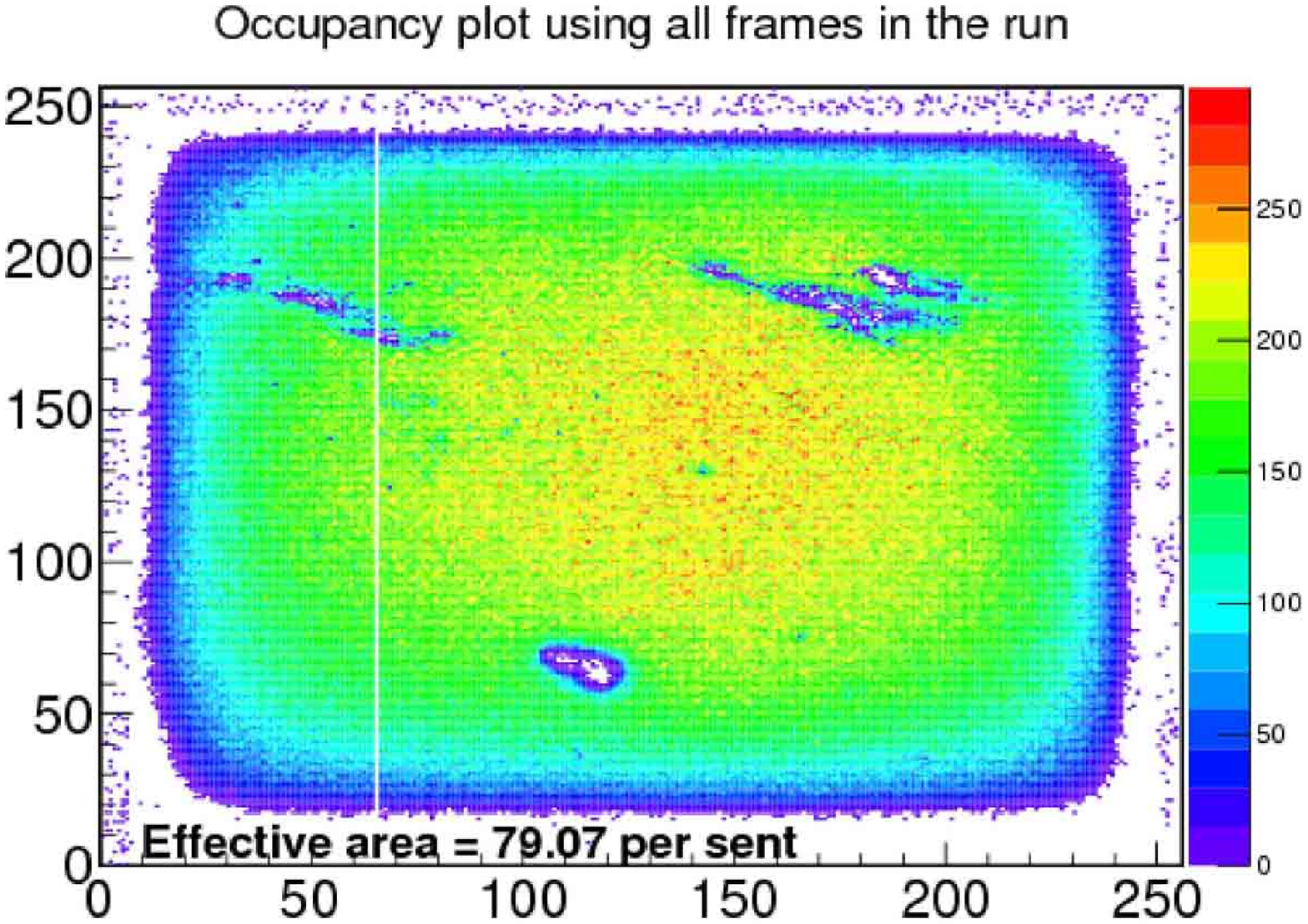}
\caption[Occupancy plot for the chip D07-W0058 with guard ring is shown:
Left: for the following parameters $U_{mesh}$ = $U_{guard}$ = 350 V, $U_{first}$ = 600 V  and $U_{cathode}$ = 2600 V.
Right: for the following parameters $U_{mesh}$ = $U_{guard}$ = 350 V, $U_{first}$ = 1100 V  and $U_{cathode}$ = 3100 V]
{ Occupancy plot for the chip D07-W0058 with guard ring is shown:  \\
Left: for the following parameters $U_{mesh}$ = $U_{guard}$ = 350 V, $U_{first}$ = 600 V  and $U_{cathode}$ = 2600 V.
Right: for the following parameters $U_{mesh}$ = $U_{guard}$ = 350 V, $U_{first}$ = 1100 V  and $U_{cathode}$ = 3100 V.
}
\label{Fig:OccupancyThree}
\end{figure}

Using the guard ring with the same potential as the mesh gives us the possibility to decrease field distortions on the edge of the
InGrid chip.
However, in Figure~\ref{Fig:OccupancyThree} (Right) the sensitive area is $\sim$80\% with guard ring and
highest potential between drift cage and mesh ($U_{first}$ = 1100 V), while in Figure~\ref{Fig:OccupancyThree} (Left) occupancy
plot for $U_{first}$ = 600 V and gave the sensitive area  $\sim$64\% is presented.

\subsection{Simulation}

For the more proper studies of the field distortions on the chips borders, the simulations were done.
Geometry of experimental setup  was reconstructed. It consisted of a drift cage and an InGrid chip.
The hole in the guard ring is not square-shaped.
  Such shape was chosen, because one side of chip has wire bonding. To be compared with real experiment,
  distance between chip and guard ring is 1 mm.  All geometry was created in CST EM Studio~\cite{website:CST}.
Using CST EM Studio electric field simulation was obtained.
To compare the MC with measurements simulation without guard ring was done.
When field map was obtained it was implemented to Garfield++ by using special tools\cite{Zenker}.
Trajectories for every primary electron were obtained using Garfield++.
Several simulations for different size of ground around chip (0 mm, 0.5 mm, 1 mm, 2 mm, full grounding of PCB) were performed to find
admissible accordance between simulation and experimental data.

\subsubsection {Comparison of simulation and experimental data}

In this work comparison of experimental data with simulation was done.
The realization of  simulation  was described.
Best agreement is shown for a setup with guard ring  and without ground around chip.

Statistics of simulation results is poor compared to experimental data, because simulation needs a lot of CPU time.
But already at this stage we can see that it is possible to know status of future experiment  using this type of simulation.
 Best agreement is shown for 800 V  on bottom side of drift cage for experimental
 setup in comparison with simulation when ground is 0 mm around the chip.

In Figure \ref{Fig:SimulatinThree}  comparison of two sets of experimental setup with simulations is shown.
One can see that geometrical forms of occupancy for simulation and experiment are similar.

\begin{figure}[!ht]
\begin{center}
\subfigure[MC, occupancy plot in case without guard ring]{
\includegraphics[width=0.48\textwidth,height=5cm]{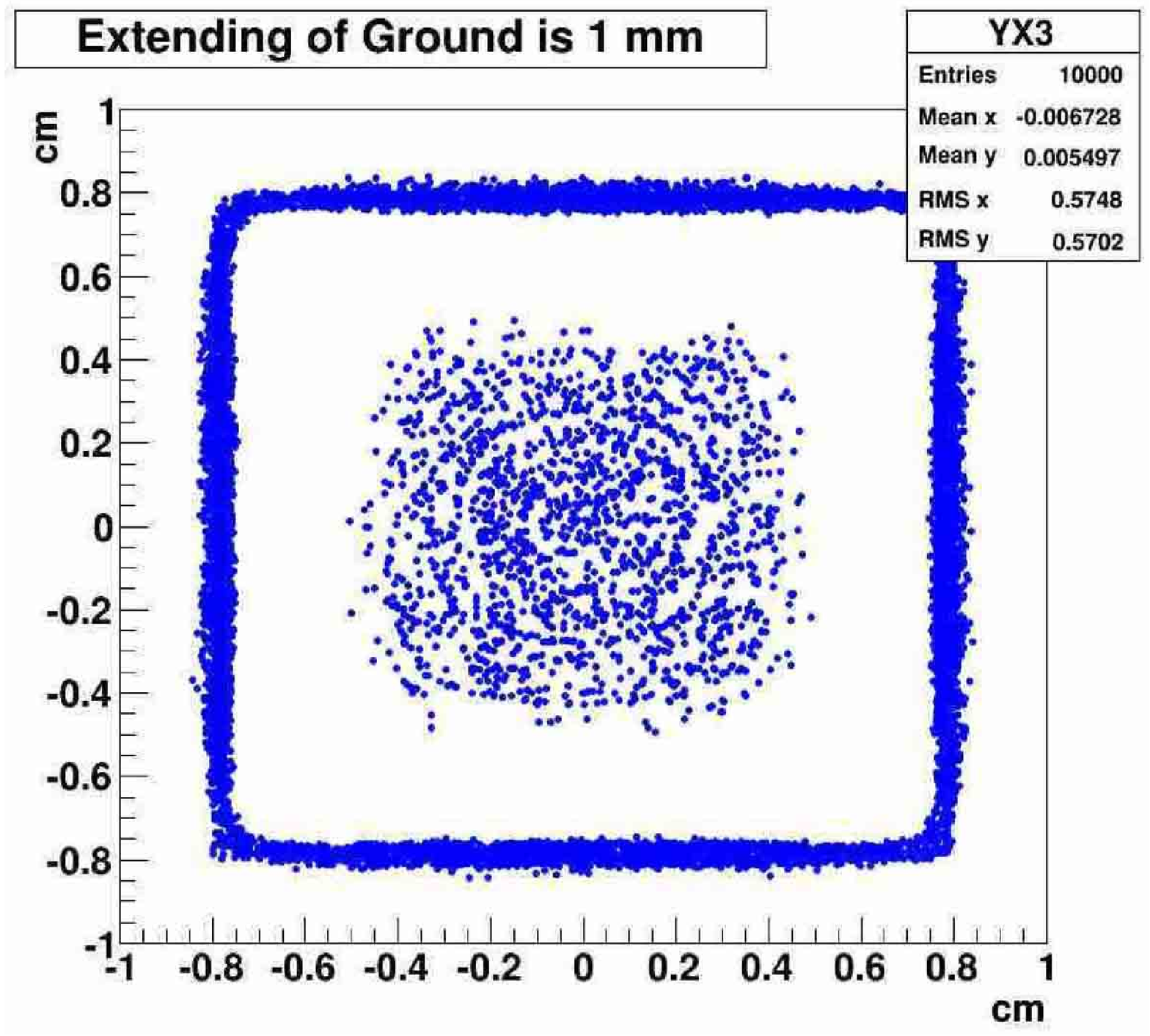}
}
\hfill
\subfigure[MC, occupancy plot in case with guard ring.]{\includegraphics[width=0.48\textwidth,height=5cm]{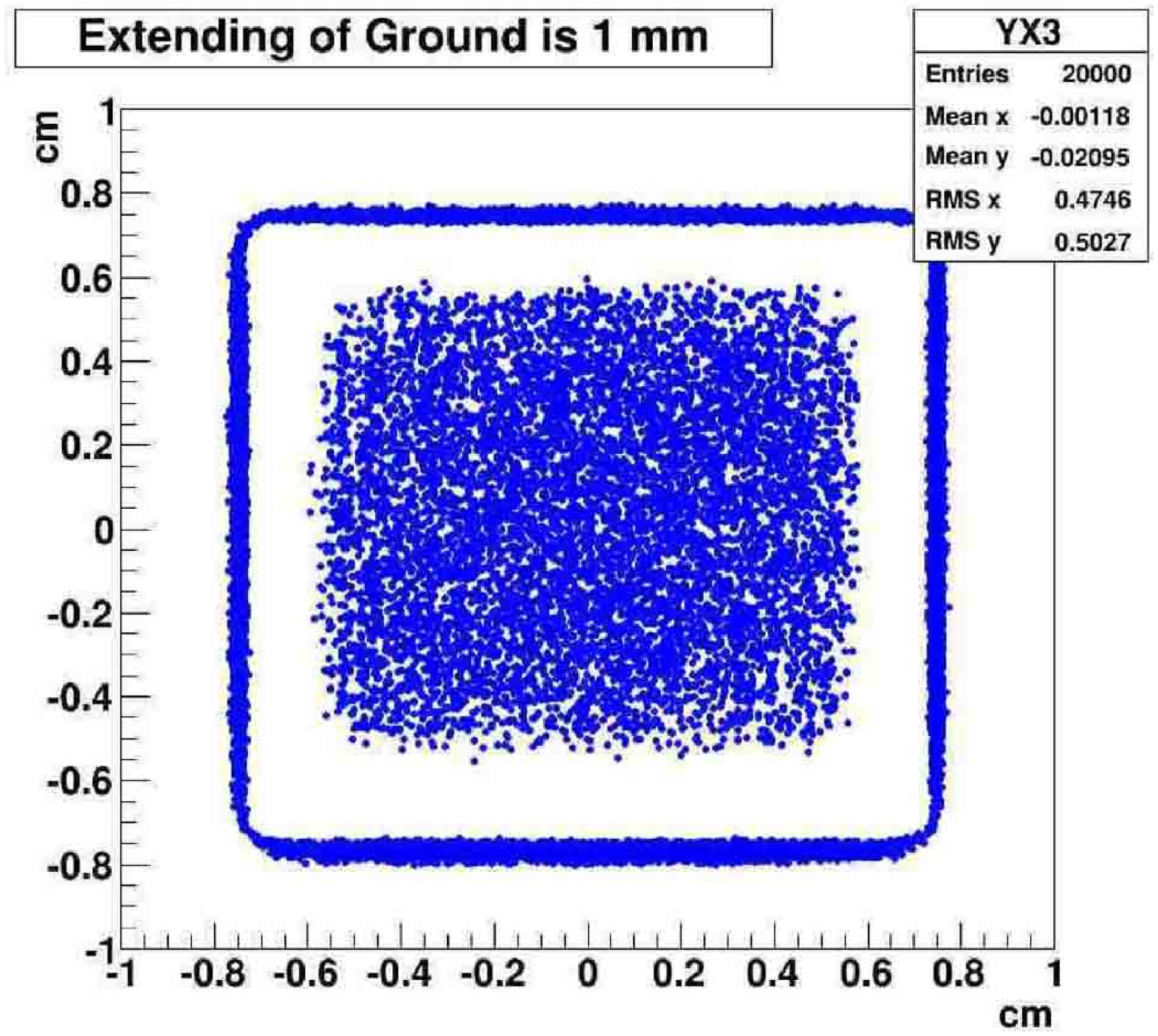}}\\ 

\vspace{0.5cm}

\subfigure[Measured, occupancy plot in case without guard ring.]{
\includegraphics[width=0.45\textwidth,height=5cm]{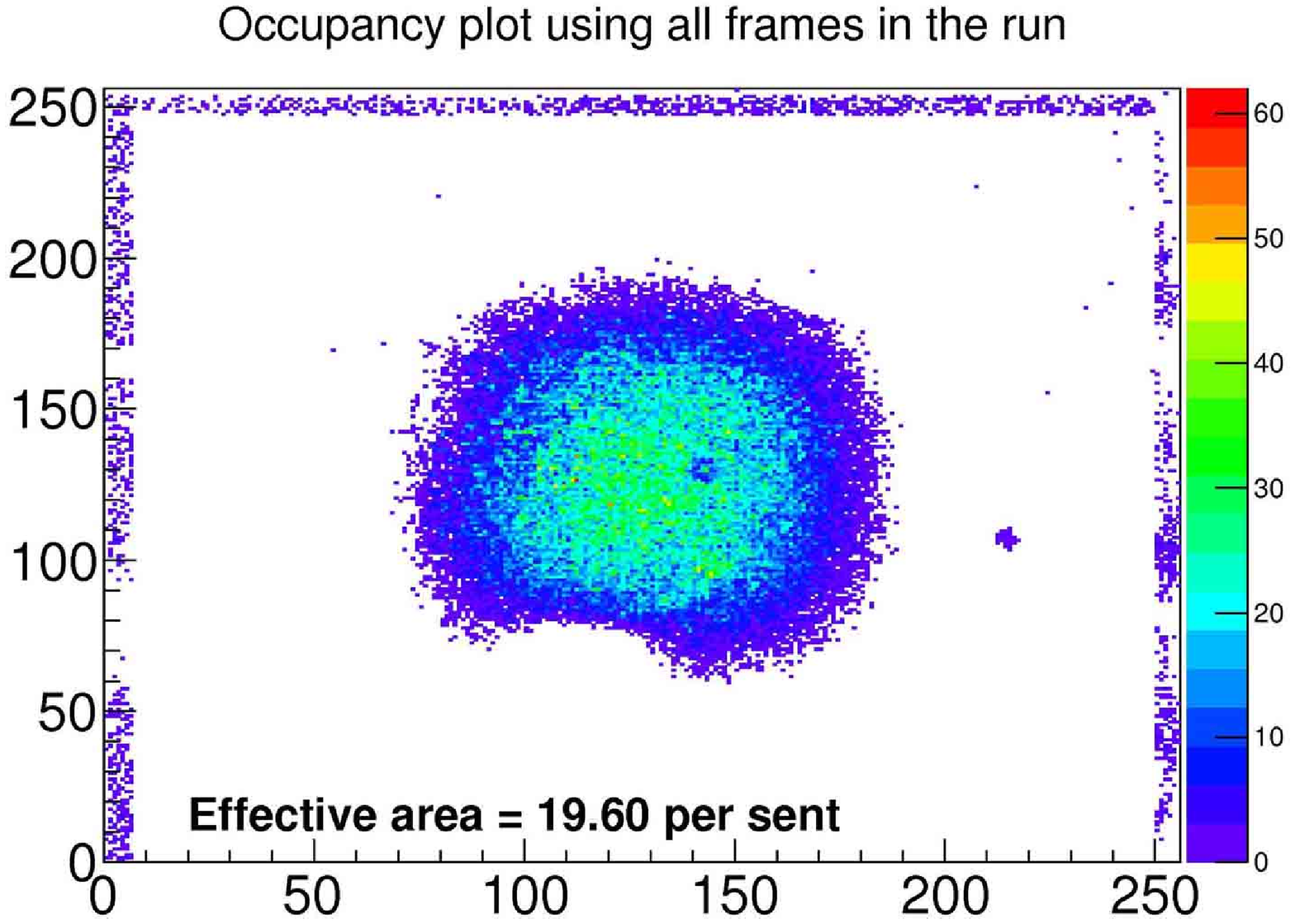}
    }
\hfill
\subfigure[Measured, occupancy plot in case with guard ring.]{
    \includegraphics[width=0.45\textwidth,height=5cm]{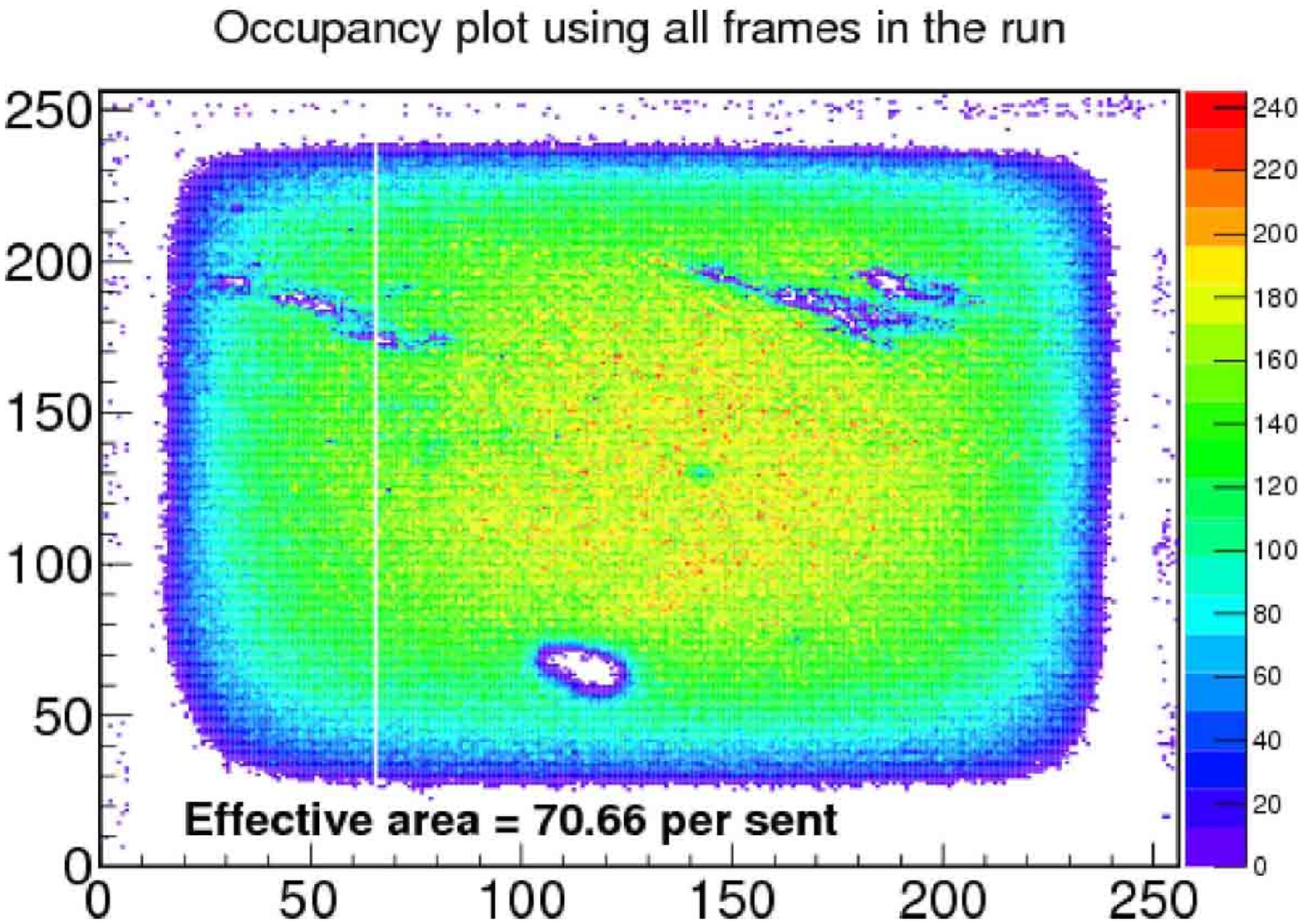}}

  \end{center}
    \caption{%
    Monte Carlo simulation. Occupancy plots.
     }%
   \label{Fig:SimulatinThree}
\end{figure}

\section{Conclusions}

The Micromegas and GEM detectors became a wide-spread tool for
high-rate tracking of sensitive areas, precision reconstruction of
charged particles in the TPC, X-ray etc. In its turn InGrid
detectors combine advantages of Micromegas and CMOS chips with
high granularity. Thus InGrid detectors could become one of the
readout options for ILD TPC.

In this article the possibility to increase the size of sensitive
area for a single  detector was described. In order to select and
analyse measurement data from the radioactive source (Fe$^{55}$
photons), special algorithm and selection criteria were developed.
In the studies it was shown that it is possible to reach almost
80\% of sensitive area. These results well agreed with simulation.
Still, there is a room for optimization when using different
voltage on guard ring and on mesh (as was shown in simulation).

\coltoctitle{LHCb RMS status and operation at 13 TeV}
\coltocauthor{O.~Okhrimenko, S.~Barsuk, F.~Alessio, V.~Pugatch}
\label{art8}

\begin{center}
{\Large {\bf LHCb RMS status and operation at 13 TeV}}

\vskip 0.5cm

{\bf O.~Okhrimenko$^{a}$, S.~Barsuk$^{b}$, F.~Alessio$^{c}$, V.~Pugatch$^{a}$}

\vskip 0.3cm

$^{a}${\it Institute for Nuclear Research NAS of Ukraine, Kyiv, Ukraine}

$^{b}${\it LAL, Universit\'{e} Paris-Sud, CNRS/IN2P3, Orsay, France}

$^{c}${\it European Organization for Nuclear Research (CERN), Geneva, Switzerland}

\end{center}

\vskip 0.4cm

\begin{abstract}
\noindent In this paper, results of the LHCb Radiation Monitoring System operation
in 2015 are presented. Comparison of the RMS baselines behaviour during collisions
at 8 and 13~TeV is shown. Integrated luminosity measurements are described.
\end{abstract}

\noindent

\vskip 0.4cm

\noindent PACS numbers: 29.40.Mc; 23.40.-s; 23.60.+e; 95.35.+d

\vskip 0.4cm

\noindent  Keywords: LHCb, Radiation monitoring, Beam and background control

\section{Introduction}

By the summer of 2015, the Large Hadron Collider (LHC) has started its
operation at a new energy of 13~TeV center-of-mass proton-proton
collisions. This is 5~TeV higher than the value of energy explored at CERN
during Run1 in the years of 2011--2012. For the LHCb
experiment~\cite{lhcb}, the Run2 period of data taking (2015--2017 years)
will be marked by an increase of the instantaneous luminosity up to
4$\cdot$10$^{32}$~cm$^{-2}$s$^{-1}$ which is 2 times higher than the initial
design value, while keeping the average number of proton-proton
interactions per bunch crossing ($\mu$) at a constant value of 1.1, which
is between 2 and 3 times higher than the initial design value. These
actions are essential steps towards reaching the main goal of the LHCb
experiment in precision measurements of the CP-violation phenomena as well
as in searching for the rare decay modes of heavy flavored hadron probes.
The new domain of the collision energy as well as the increase of the
luminosity are good arguments for observation of NP signals. On the other
hand, there is an obvious challenge to the experimental techniques for
keeping reliable performance in these new and harsh environment.
In particular, the silicon microstrip detectors of the LHCb Silicon Tracker
(ST)~\cite{st} are very sensitive to the radiation load. The Radiation
Monitoring System (RMS) installed at the ST IT-2 station in close vicinity
to the beam pipe provides real time measurements of the charged particle
flux spatial distribution allowing for evaluation of the dose absorbed
by the silicon sensors and related increase of the leakage current. The
data measured by the RMS are applied to evaluate instantaneous and
integrated luminosity of the LHCb experiment relative.

\section{The LHCb Radiation Monitoring System}

The LHCb experiment is the forward spectrometer located at the interaction
point IP8 of the LHC (CERN). The LHCb detector consists of following
parts: Vertex Locator (VELO), Silicon (IT, TT) and Outer Trackers
to reconstruct tracks of charged particles and their decay vertices; Magnet
to measure charged particle momentum; Cherenkov Detectors (RICH1, RICH2)
to separate kaons and pions; Hadronic and Electromagnet Calorimeters (HCAL,
ECAL) to measure the particles energy; Muon detector to detect the muons.
The level of charged hadron fluxes at the location of the silicon sensors
of the ST IT-2 station varies from $10^4$ to $10^5$~cm$^{-2}$s$^{-1}$
at nominal LHCb luminosity of 2$\cdot$10$^{32}$~cm$^{-2}$s$^{-1}$~\cite{radenv}.
These fluxes are high enough to make a significant impact onto the performance
of the IT silicon sensors and their frond-end electronics caused by the
radiation load. The main goal of the RMS is a measurement of that radiation
dose load to exclude their damage as a result of an unexpected radiation
incident, i.e. change of the beam trajectory, partial beam loss in the
region of the detector etc~\cite{rmsdes1, rmsdes2}.

The RMS is based on the Metal Foil Detector (MFD) technology. The
principle of the MFD operation is Secondary Electron Emission
(SEE) from the metal foil surface (emission layer
$\sim$~10--50~nm) caused by impinging charged particles. SEE
causes positive charge in isolated metal foil read out by
sensitive Charge Integrator (ChI). The ChI is equipped by a
current-to-frequency converter allowing to achieve high dynamic
range (up to $10^6$). A current from the stable external source
(250~pA) is injected to the ChI's inputs to make baselines
($\sim$~25~kHz).The MFD is a 5-layer structure manufactured out of
50~$\mu$m thick Al foils supported by insulating epoxy frames. The
central sensitive layer is connected to the readout electronics,
while two neighboring (from both sides) accelerating layers are
biased by positive voltage (24~V) to reduce recombination after
SEE. The additional two outer shielding layers are grounded. RMS
sensor and accelerating layers are divided into 7 parts ($110
\times 75$~mm, with a layout which is similar to IT silicon
sensors size). The RMS consists of 4 modules (Top, Cryo, Bottom,
Access) containing 7 sensors each (in total 28 sensors), which are
located at IT-2 station ($\sim$~8.4~m from the LHCb interaction
point) around the beam-pipe.

\section{Results}

\subsection{RMS in 2011--2012}

The performance of the RMS system was evaluated during the period
2011--2012 while the initial calibration of the RMS was performed
in September--October 2010~\cite{rmsres1, rmsres2}. In this period, the RMS was exposed
to about 29~pb$^{-1}$ of luminosity. The calibration was done during
high-intensity beams ($>10^{11}$ protons-per-bunch).

During the period of data taking 2011--2012, the LHCb experiment took data
in these conditions: center-of-mass pp collision energy of 7--8~TeV;
average luminosity of 4$\cdot$10$^{32}$~cm$^{-2}$s$^{-1}$; average number
of visible collisions per bunch crossing ($\mu$) 1.7; number of colliding
bunches per fill up to 1262; bunch spacing 50~ns (corresponding to a total
of $\sim$~20~MHz of bunch crossing rates). A total of about 3.3~fb$^{-1}$
of integrated luminosity was accumulated.

The RMS measures in real time fluxes of charged particles from the LHCb
interaction point IP8 as well as radiation induced background. The data
accumulated by the RMS were explored to calculate the absorbed dose
distribution and related leakage currents increase in the silicon
microstrip sensors of the LHCb Inner Tracker. For the operational years
2011--2012, the values are in the range from 0.3 to 1.5~kGy and from 50 to 400~$\mu$A,
respectively, depending on the sensor position and strip length. The
accuracy of the RMS measurements was about 10\% , and obtained data were
in good agreement with other measurements and Monte-Carlo simulations. More
details were described  in~\cite{rmsres3}.

\subsection{RMS first data in 2015}

After the first long shutdown (LS1), LHCb has started collecting data
in 2015 under following conditions: proton-proton collisions c.m.s. energy
13~TeV; average luminosity up to 3.8$\cdot$10$^{32}$~cm$^{-2}$s$^{-1}$; average number
of visible collisions per bunch crossing ($\mu$) 1.1; number of colliding
bunches per fill up to 1866; bunch spacing 25~ns ($\sim$~40~MHz of bunch
crossing rates). In total, about 0.4~fb$^{-1}$ of integrated luminosity has been
currently (October 2015) delivered. Data taking will resume at these
conditions in 2016 and LHCb plans to record $>5$~fb$^{-1}$ up to the end of 2018,
when a second long shutdown will take place.

During LS1, the Top and Access RMS modules were replaced to improve the
operational performance (the drift of baselines has become there comparable
with modules response to pp collisions). As an example of the upgraded RMS
performance the baselines evolution of the Cryo 2 and Access 5 sensors
is shown in Figure~\ref{fig:rms} (top plots) during pp collisions at LHCb
in the year 2015 (fill 4449). The fill duration was $\sim$~15h, 1278
colliding bunches at 25~ns spacing, an average instantaneous
luminosity---2.34$\cdot$10$^{32}$~cm$^{-2}$s$^{-1}$ and average $\mu = 1.08$.
Different colors indicate a status of the LHC. Blue one
corresponds to ``NO BEAM'' conditions. The baselines are measured here
as `signal zero' level to be subtracted later to obtain the RMS response
to pp collisions. The yellow, green and magenta colors corresponds
to ``INJECTION'', ``RAMP'' and ``ADJUST'' LHC status, respectively. During
this time proton beams are injected in LHC, ramped to the energy
of 6.5~TeV, focused and made collided by applying local corrections on the
beams orbits. Finally, the status of ``STABLE BEAMS'' is declared when the
LHCb experiment starts recording data with the full detector on and the DAQ
system ready. The red color corresponds to this state of ``PHYSICS''. For
the sake of comparison with the data of the year 2012, the same plot
is generated for fill 3318 for the Cryo 2 sensor.

From the Figure~\ref{fig:rms}, it can be seen that the baselines/responses of the new
Access module has no large drift anymore but the baselines/responses of all modules
are affected by some super-imposing signal during beams circulating at the
LHC---from the INJECTION state all the way through the physics data taking
period. The two sets of three plots at the bottom in Figure~\ref{fig:rms}
show the baselines/responses parameters (mean and standard deviation) during the
various states of a fill. The left plot corresponds to the NO BEAM status
before beams injection (blue histogram) and INJECTION\,+\,RAMP states
(yellow-green histograms). It can be seen that the baseline width has grown
up between NO BEAM and INJECTION states. The plot in the middle of the
Figure~\ref{fig:rms} corresponds to the PHYSICS status during the pp
collisions. The right plot corresponds to the NO BEAM state after beams
were dumped to show that the baselines returned back the nominal ones (i.e.
as before injection). The baselines widening has an impact on the accuracy
of the RMS fluxes measurements (10--30\% depending on the sensor position).

Using data from the Cryo and Bottom modules the integrated luminosity
of fill 4449 (2015 year) was estimated using the RMS calibration. The
obtained value of $16 \pm 6$~pb$^{-1}$ agrees well with the official LHCb
luminosity of $13 \pm 2$~pb$^{-1}$. The Top and Access modules will
be calibrated with 2015 data later this year.
\pagebreak
\begin{figure}[h!]
\begin{center}
 \mbox{\epsfig{figure=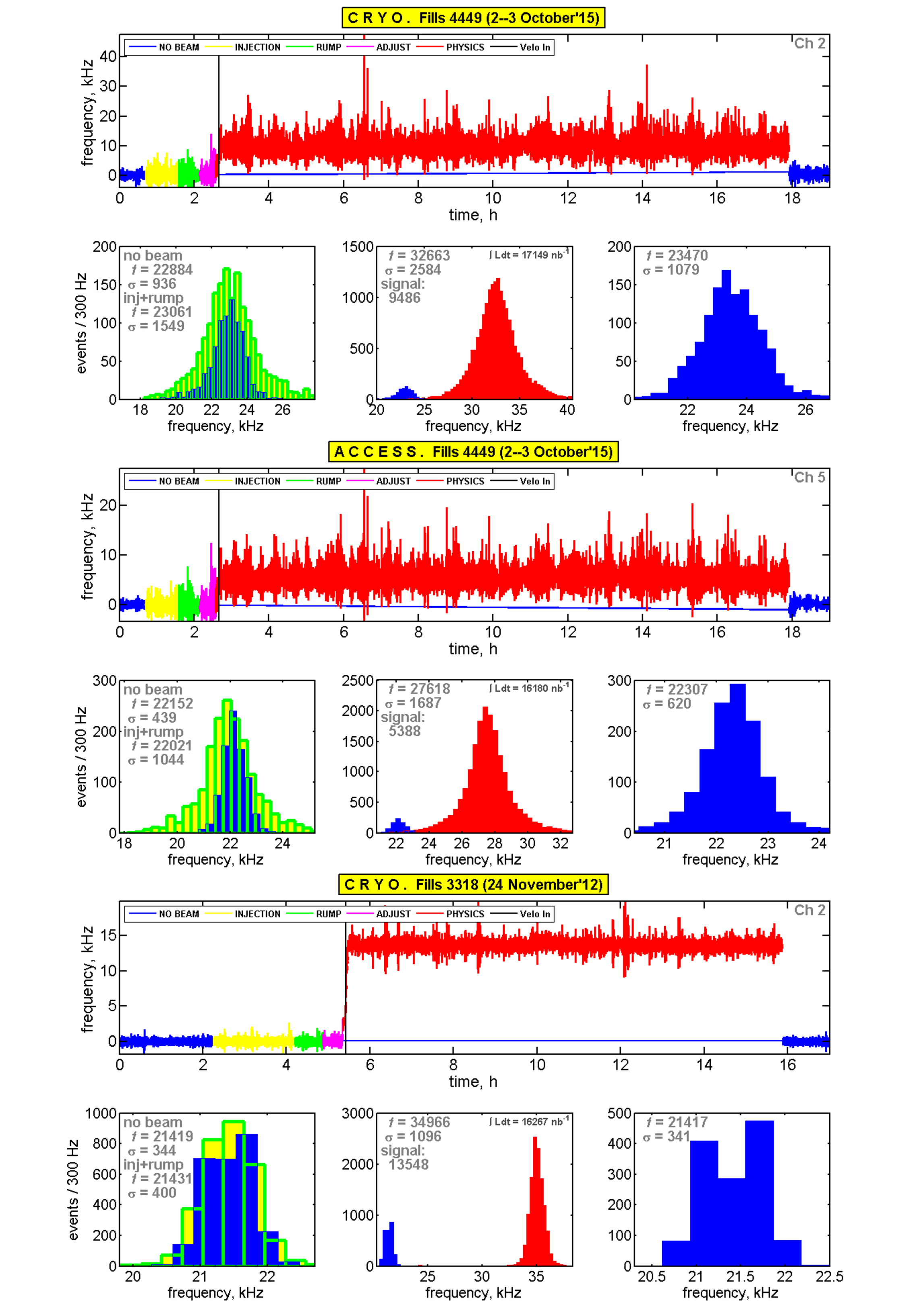,height=20.0cm}}
 \caption{Trend plots for baselines behaviour of Cryo 2 and Access
 5 sensors during $pp$ collisions at fill 4449 and fill 3318 (top plots) and
 histograms of these baselines (bottom plots). Different colours
 corresponds to different LHCb states: blue---NO BEAM; yellow---INJECTION;
 green---RAMP; magenta---ADJUST; red---PHYSICS.}
\label{fig:rms}
\end{center}
\end{figure}

\section{Conclusions}

During the years 2011--2012 the Radiation Monitoring System has successfully
provided monitoring of the radiation load on Si-sensors of the LHCb Inner
Tracker. RMS calibrated data allowed to evaluate the LHCb integrated
luminosity. The RMS data are in good agreement with other detectors and
Monte-Carlo simulations. The first data taken in 2015 year show that
in spite of the baselines widening due to some currently unknown reason RMS
allows to monitor the radiation load with an accuracy of 10--30\% depending
upon the sensor position. The understanding of this problem is under way.

\section{Acknowledgement}

This research was partially conducted in the scope of the IDEATE
International Associated Laboratory (LIA). This work has been supported
by the grants CO-4-1/2015 (NASU) and F58/382-2013 (DFFD). We thank the LHCb
ST and B\&B groups and the LHCb Collaboration. Special thanks to Fred,
Helge, and Richard.

\coltoctitle{Focusing of relativistic electron Gaussian bunches by nonresonant wakefield excited in plasma}
\coltocauthor{V.I.~Maslov, N.~Delerue, I.P.~Levchuk, I.N.~Onishchenko}

\begin{center}
\large \bf
Focusing of relativistic electron Gaussian bunches
by nonresonant wakefield excited in plasma
\end{center}
\vspace*{1 mm}
\begin{center}
\bf
V.I. Maslov$^{a}$\footnote{Corresponding author: \texttt{vmaslov@kipt.kharkov.ua}}, N. Delerue$^{b}$, I.P. Levchuk$^{a}$, I.N. Onishchenko$^{a}$
\end{center}
\vspace*{1 mm}

\begin{center}
\textit{$^a$ NSC Kharkov Institute of Physics and Technology, Kharkov, Ukraine\\
$^b$ Laboratory of Linear Accelerator, Orsay Science Centre, Orsay, France
}
\end{center}

\vspace*{3 mm}

\textit{ Abstract:} Focusing of relativistic electron Gaussian bunches by nonresonant wakefield, excited by them in plasma, is investigated by numerical simulation. It has been shown that in the case where the electron plasma frequency is larger than the repetition frequency of bunches, all bunches are in focusing wakefield of excited beatings except bunches, which are located at the fronts of beatings where they are not focused.

\vskip 0.3cm
\noindent
{\it PACS numbers:} 29.17.+w; 41.75.Lx

\vskip 0.3cm
\noindent
{\it Keywords:}  train of relativistic electron bunches, wakefield, plasma lens, focusing

\section{Introduction}
Focusing of bunches by radial wakefield is an important problem. The intensity of this focusing is larger on a few orders in comparison with usual magnetic focusing \cite{1}. However focusing, which occurs in the plasma at space charge compensation of bunches, is also not enough intense. The intensity of focusing can be increased significantly by using an excited transverse wakefield. Focusing by excited resonant wakefield was studied in \cite{2, 3}. Also a uniform focusing by excited wakefield has been studied in \cite{3, 4} for long bunches and in \cite{5} for short bunches. Because it is difficult to maintain in an experiment a uniform and stationary plasma density, resonant for train of electron bunches, in this paper focusing of train of bunches of relativistic electrons by excited non-resonant wakefield is considered.

\section{Nonresonant wakefield plasma lens for short train of bunches}

Numerical simulation has been performed using 2d3v-code lcode \cite{6}. For the numerical simulations the following parameters are selected: $n_{res} =10^{11}$ cm$^{-3}$ is the resonant plasma density which corresponds to $\omega_{pe}= \omega_m = 2\pi \cdot 2.8\cdot 10^9$, relativistic factor of bunches equals $\gamma_b=5$. $\omega_m$ is the repetition frequency of bunches, $\omega_{pe}=(4\pi n_{res}e^2/m_e)^{1/2}$ is the electron plasma frequency. The density of bunches $n_b=6\times 10^8$ cm$^{-3}$  is distributed in the transverse direction according to Gaussian distribution,  $\sigma_f=0.5$ cm, $\lambda=10.55$ cm  is the wavelength, $\xi=V_b t-z$, $V_b$ is the velocity of bunches. Time is normalized on $\omega_{pe}^{-1}$, distance -- on $c/\omega_{pe}$, density -- on $n_{res}$, current $I_b$ -- on $I_{cr}=\pi m c^3/4e$, fields -- on $(4\pi n_{res} c^2 m_e)^{1/2}$.

As it has been shown in \cite{3}, at the resonant excitation of wakefield the shorter first fronts of the bunches are defocused by smaller fields, and longer back fronts of the bunches are focused by larger fields (see Fig. 1) i.e. focusing by resonant wakefield is inhomogeneous.

\begin{figure}
\centering
  \includegraphics[width=.4\linewidth, height=.4\linewidth]{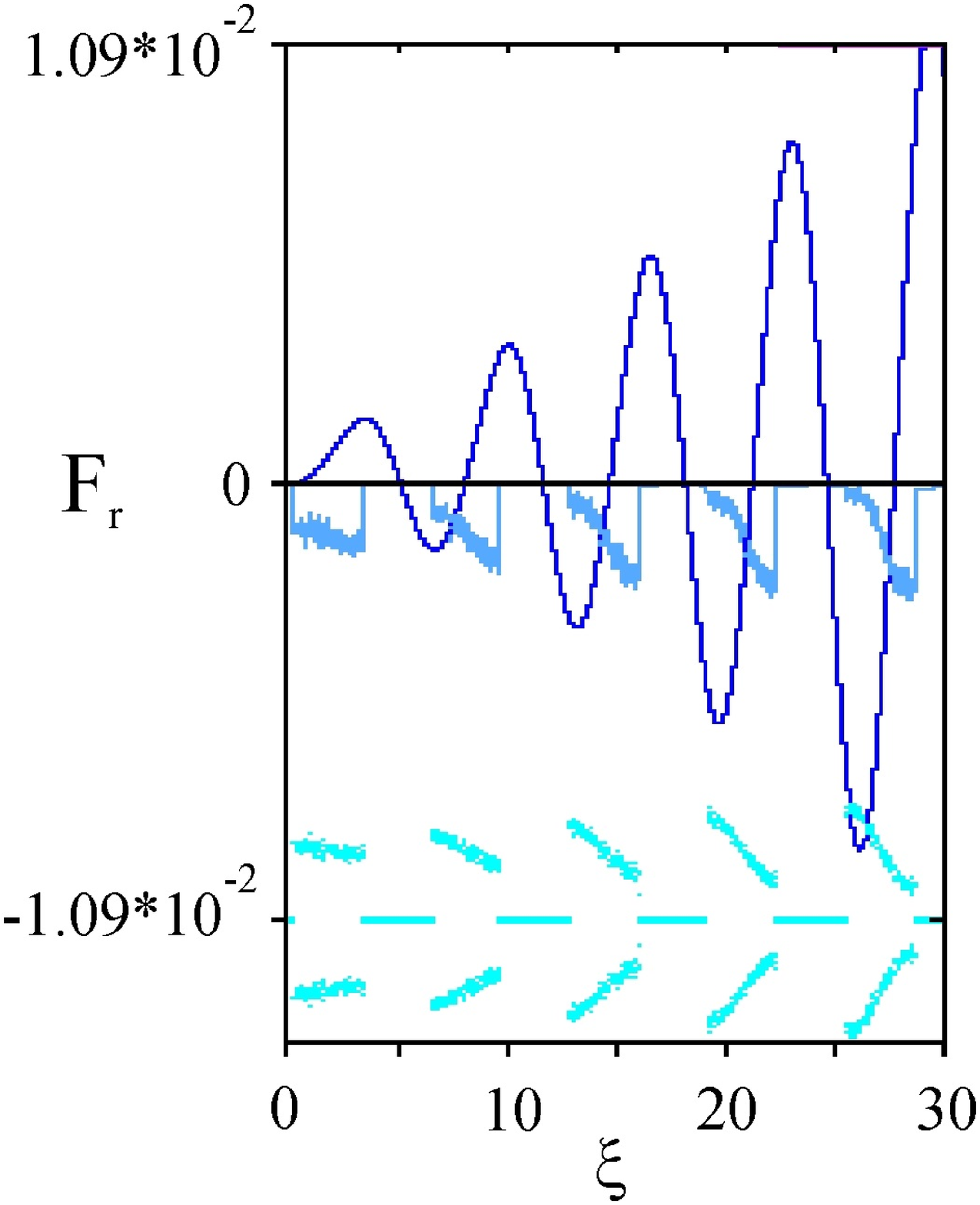}
  \caption{Longitudinal, along train, distribution of radius $r_b$ (bunches-horns in cyan) and density $n_b$ (trapezoids in blue) of train of resonant rectangular bunches and of radial wakefield $F_r$ (oscillating line in dark blue), after their focusing/defocusing at the distance $z=33$ cm from the boundary of injection at $\xi_b=\lambda/2=5.275$ cm, $I_b=0.45\times 10^{-3}=6$~A. Normalized length of 30 for the train of bunches in the Fig.~\ref{08France:fig1} corresponds to a length of 50.6~cm or to a duration of 1.68~ns. The achieved normalized radial wakefield $F_r=1.09\times10^{-2}$ after five bunches corresponds to 3.27~kV/cm. }
  \label{08France:fig1}
 \vspace{0.05\textwidth}
  \includegraphics[width=.4\linewidth, height=.4\linewidth]{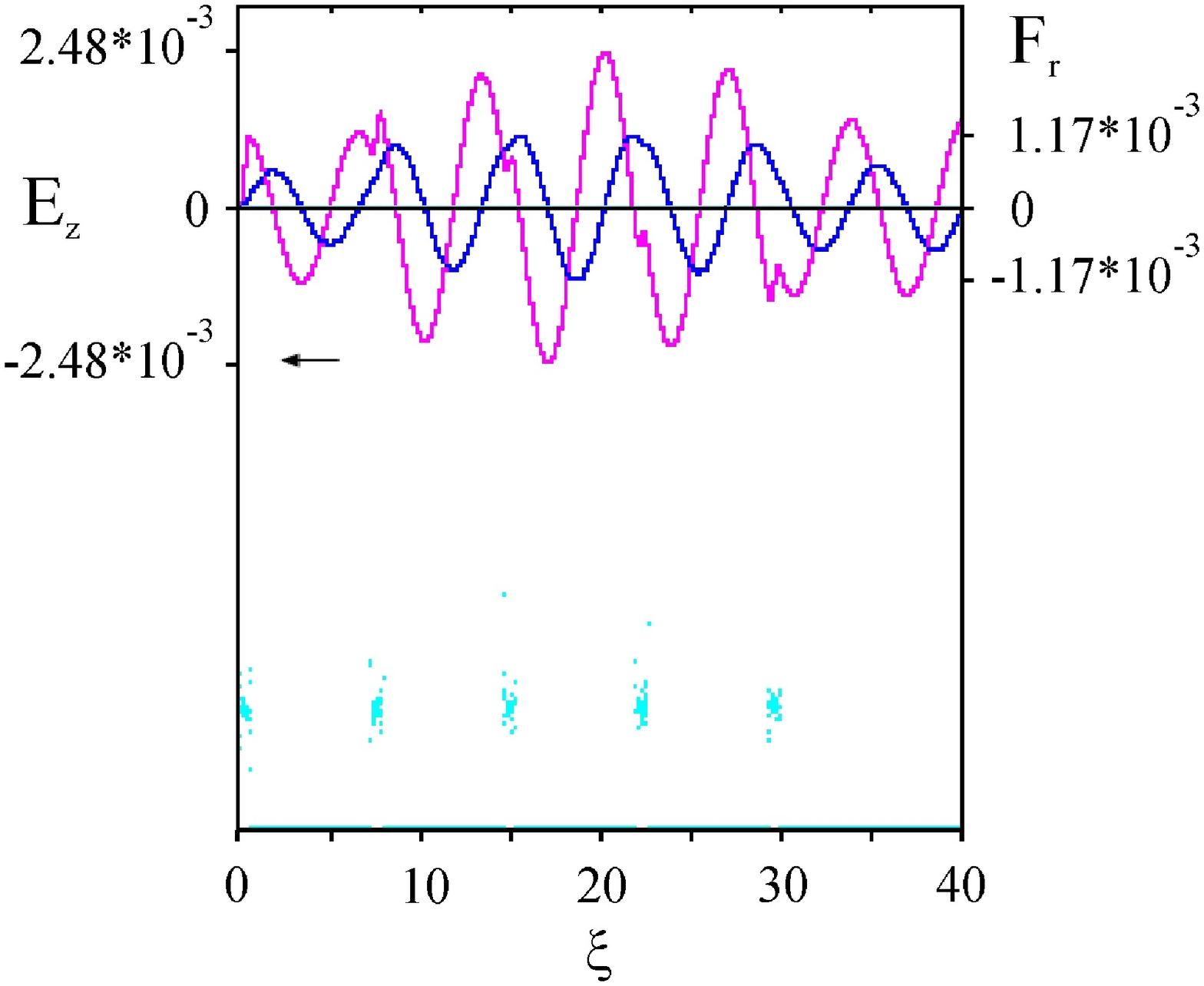}
  \caption{ Longitudinal, along train, distribution of radius $r_b$ (points in cyan) of train of very short Gaussian bunches, of radial wakefield $F_r$ (oscillating line of smaller amplitude in blue) and of longitudinal wakefield $E_z$ (oscillating line of larger amplitude in magenta) near the boundary of injection at $n_e=1.35 n_{res}$, $\xi_b=0.1\lambda =1.055$~cm, $I_b=1.56\times 10^{-3} = 20.9$~A. The arrow shows the direction of the bunch motion.}
  \label{08France:fig2}
\end{figure}

\begin{figure}[p]
\centering
\includegraphics[width=.5\linewidth]{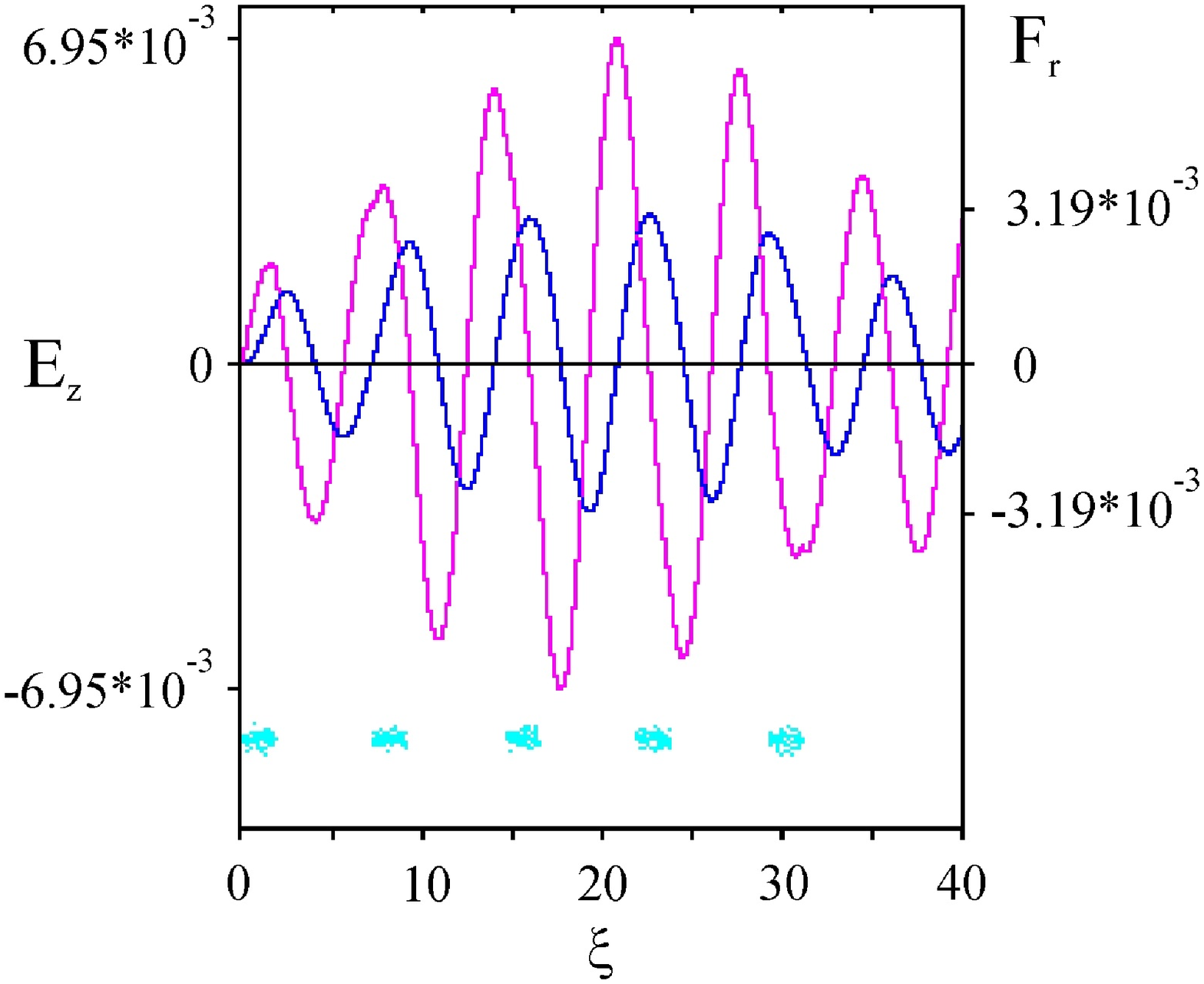}
\caption{Longitudinal distribution, along the train, of radius
$r_b$ (short linear segments) of train of rectangular bunches of
length, $\xi_b=\lambda/4=2.6375$~cm, of radial wake force $F_r$
(oscillating line of smaller amplitude) and of longitudinal
wakefield $E_z$ (oscillating line of larger amplitude) near
boundary of injection at $n_e=1.35n_{res}$, $I_b=10^{-3}=13.4$~A.}
\label{08France:fig3} \vspace{0.05\textwidth}
\includegraphics[width=.5\linewidth]{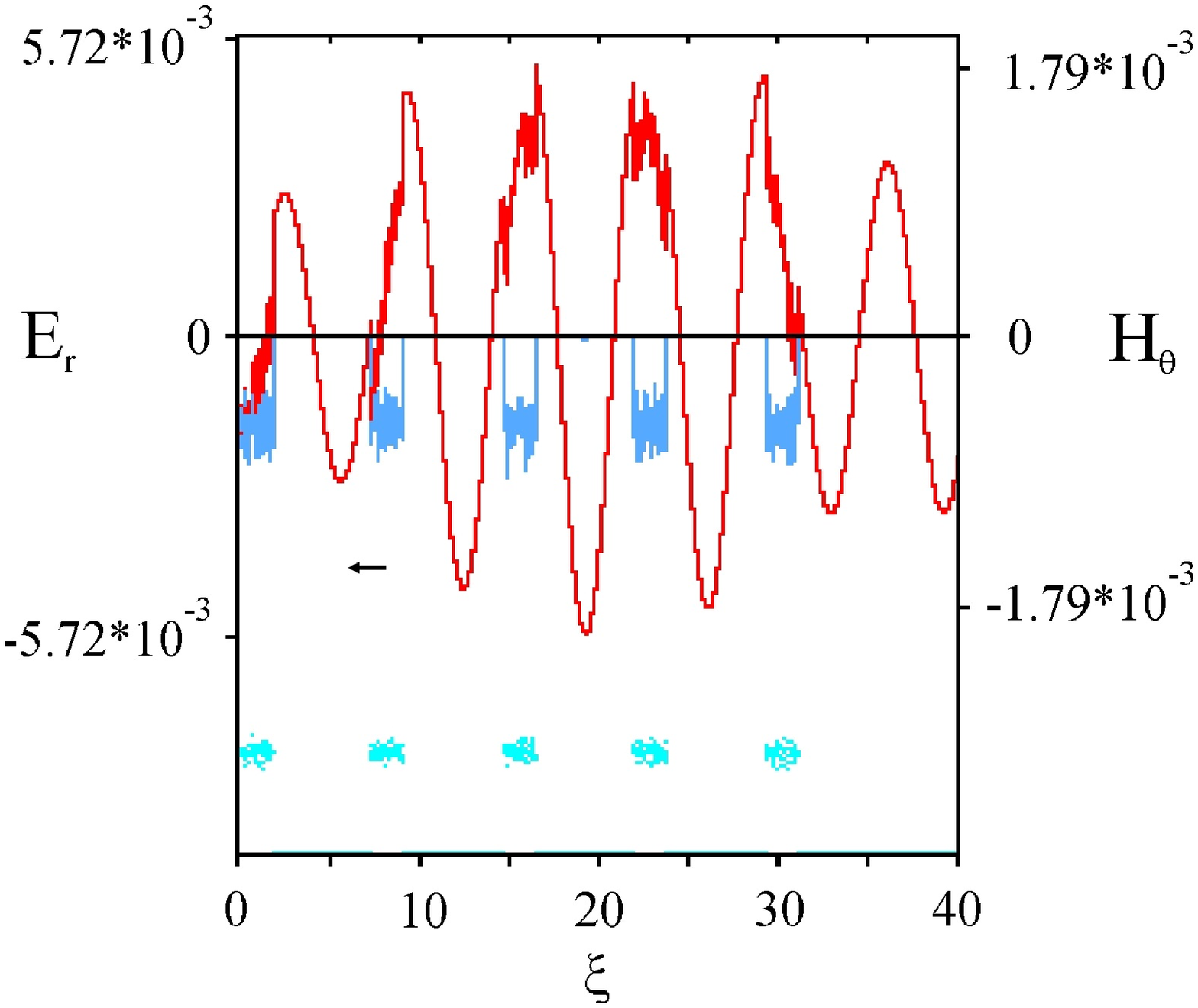}
\caption{ Longitudinal, along train, distribution of radius $r_b$
(short linear segments) of train of rectangular bunches of length
$\xi_b=\lambda/4$, of radial wakefield $E_r$ (oscillating line)
and of magnetic wakefield $H_\theta$ (trapezoids) near boundary of
injection at $n_e=1.35n_{res}$, $I_b=10^{-3}$.}
\label{08France:fig4}
\end{figure}

Let us consider the optimum parameters for the case of nonresonant wakefield plasma lens for short train of identical bunches of relativistic electrons. I.e. we will show that for selected: the length of bunches, less than half the wavelength $\xi_b<\lambda/2$, their number $N_b$ and repetition frequency of bunches $\omega_m$ there is a range of suitable electron plasma frequency $\omega_{pe}$ such that all bunches are in focusing wakefields $F_r$. As it will be demonstrated below, that in the case $\omega_{pe}>\omega_m$ all point (very short) bunches are in focusing or in a zero radial field, we use the range of parameters when $\omega_{pe}>\omega_m$. For determination the optimal parameters we use two conditions. Namely, that all $N_b$ bunches are placed on the length of a beating, it is necessary $0<\omega_{pe}- \omega_m<\omega_{cr}$. $\omega_{cr}$ is some critical frequency, associated with $N_b$. At the same time, for all electrons of all bunches are in focusing wakefields, it is necessary $\xi_b<xi_{cr}$.

In the case of point $\xi_b\rightarrow 0$  bunches one restriction is removed, and the relative position of bunches and $F_r$ at $n_e/n_{res} - 1=0.35$ ($n_{res}$ is determined from $(4\pi n_{res}e^2/m_e)^{1/2}=\omega_m)$ has the form shown in Fig.~\ref{08France:fig2}. One can see that $N_b=5$ bunches are in focusing wakefields $F_r$. In the case of bunches of finite length, $\xi_b=\lambda/4$  at the plasma density, equal to $n_e=1.35n_{res}$, the relative position of bunches and $F_r$ has the form shown in Figs.~\ref{08France:fig3}, ~\ref{08France:fig4}. $F_r$ is the total, i.e. radial field of the space charge of the bunch, wakefield and its own magnetic field of the bunch current $H_\theta$. $E_r$ is the total, i.e. radial field of the space charge of the bunch and wakefield. As one can see, for each frequency difference $\omega_{pe}-\omega_m$ there exist the length of train and the length of the bunches, when all the electrons of all bunches are in focusing fields.

Let us consider the distribution of longer train of short relativistic electron bunches (Fig.~\ref{08France:fig5}) relative to excited wakefield beatings at $\omega_{pe}>\omega_m$ (Fig.~\ref{08France:fig6}).

\begin{figure}
\centering
  \includegraphics[width=.6\linewidth, height=.2\linewidth]{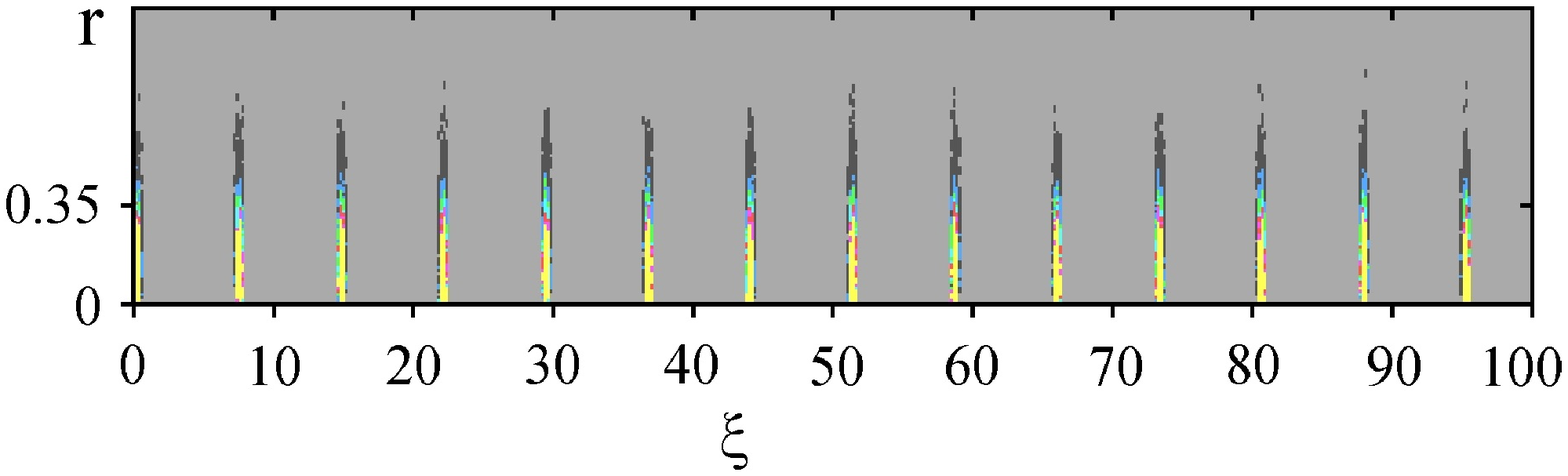}
  \caption{Spatial in $(r, \xi)$ distribution of density $n_b$ of train of very short approximately Gaussian bunches $\xi_b=0.1\lambda$ near boundary of injection at $n_e=1.35n_{res}$, $I_b=1.56\times 10^{-3}$. }
  \label{08France:fig5}
\end{figure}

\begin{figure}
\centering
  \includegraphics[width=.5\linewidth, height=.33\linewidth]{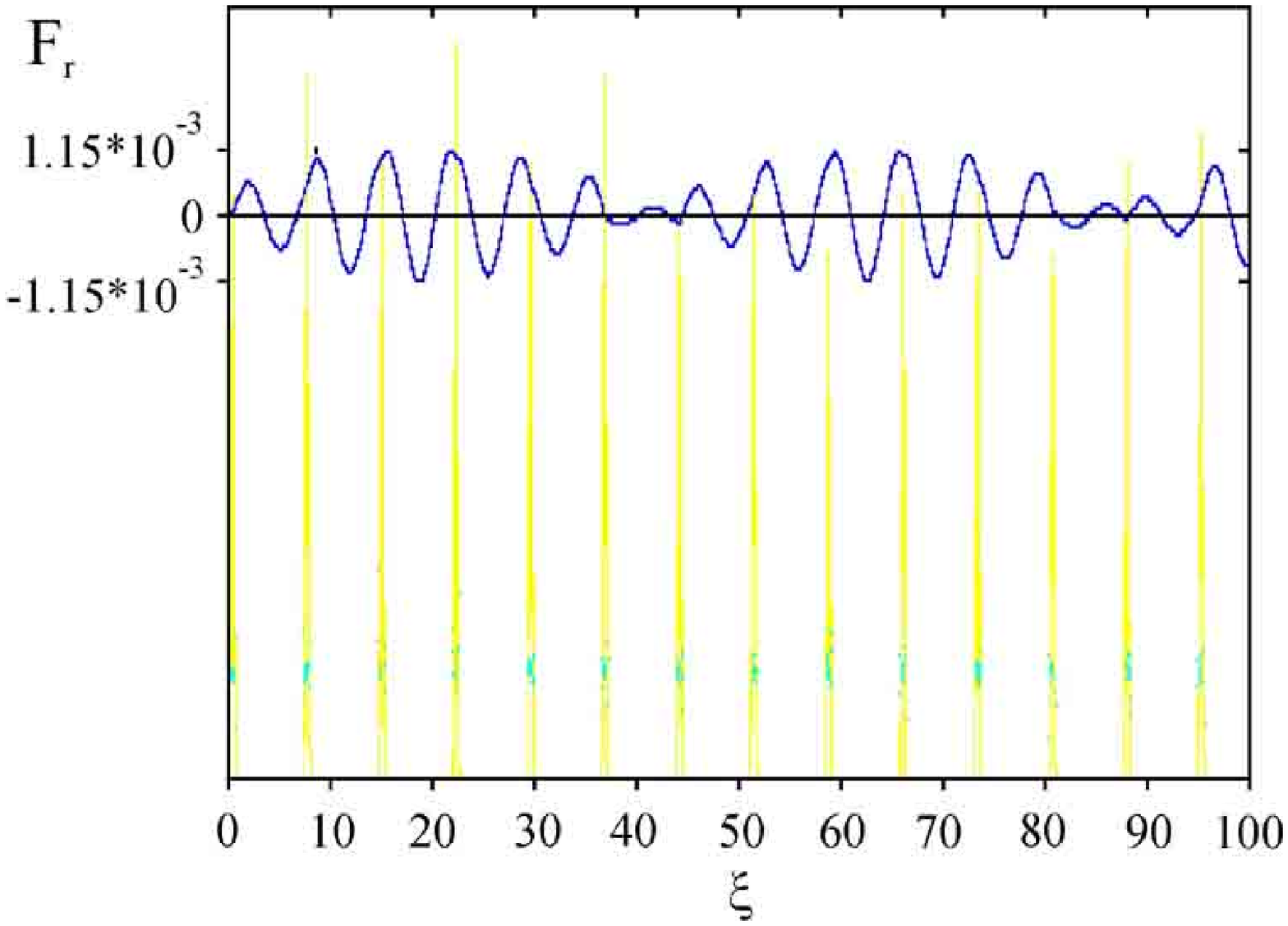}
  \caption{Longitudinal, along train, distribution of radius rb (points), of density nb (vertical lines) of train of very short approximately Gaussian bunches $\xi_b=0.1\lambda$ and of radial wake force $F_r$ (oscillating line) near boundary of injection at $n_e=1.35n_{res}$, $I_b=1.56\times 10^{-3}$.}
  \label{08France:fig6}
\end{figure}

\begin{figure}
\centering
  \includegraphics[width=.6\linewidth, height=.2\linewidth]{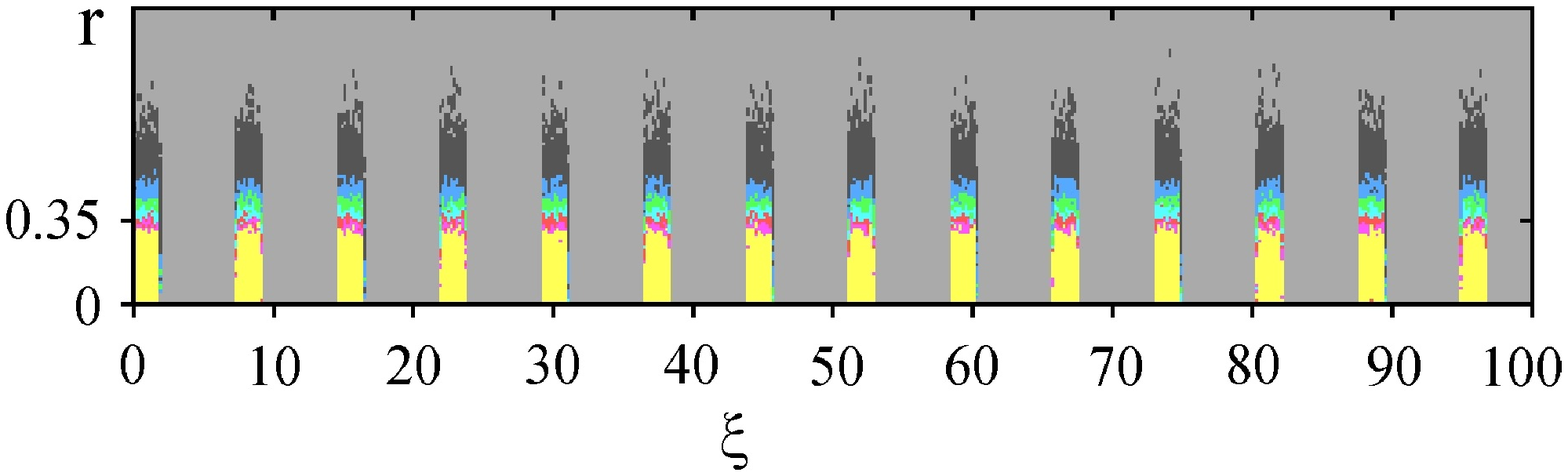}
  \caption{Spatial in $(r,\xi)$ distribution of density nb of train of rectangular bunches of length $\xi_b=\lambda/4$  near boundary of injection at  $n_e=1.35n_{res}$, $I_b=10^{-3}$.}
  \label{08France:fig7}
\end{figure}

\begin{figure}
\centering
  \includegraphics[width=.5\linewidth, height=.33\linewidth]{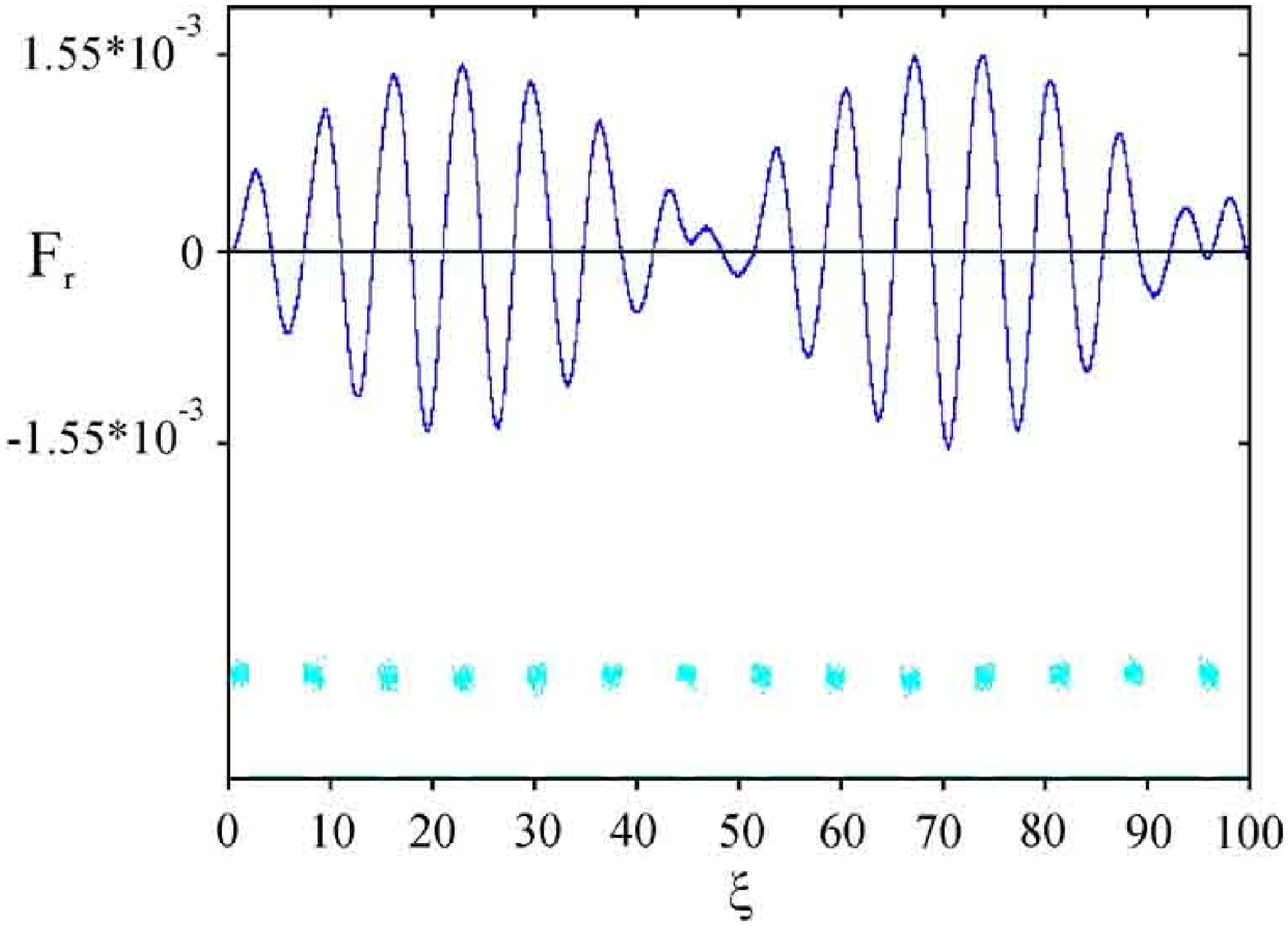}
  \caption{Longitudinal, along train, distribution of radius $r_b$ (short linear segments) of train of rectangular bunches of length $\xi_b=\lambda/4$ and of radial wake force $F_r$ (oscillating line) near boundary of injection at  $n_e=1.35n_{res}$, $I_b=10^{-3}$.}
  \label{08France:fig8}
\end{figure}

At $\omega_m<\omega_{pe}$ beatings are excited. All bunches are in focusing fields of beatings except at fronts of beatings, where they are not focused.

In the case of bunches of length $\xi_b=\lambda/4$, $\lambda=2\pi V_b/\omega_p$ one can see Fig.~\ref{08France:fig7} and Fig.~\ref{08France:fig8}.

All bunches are in focusing fields except at fronts of beatings, where they are not focused. Let us compare focusing in nonresonant (Fig.~\ref{08France:fig9}) $\omega_m<\omega_{pe}$  and in resonant (Fig.~\ref{08France:fig10}) $\omega_m = \omega_{pe}$  cases.

One can see that in nonresonant case all bunches are focused except at fronts of beatings, where they are not focused.

Now we consider the long train of short Gaussian bunches. The train is shaped according to linear dependence. The space interval between bunches equals to the wavelength (see Fig.~\ref{08France:fig11}). One can see that all bunches are in maxima of focusing field and thus they are decelerated slowly, as they are in zero decelerating field, excited by previous bunches.

 \begin{figure}
\centering
  \includegraphics[width=.6\linewidth, height=.2\linewidth]{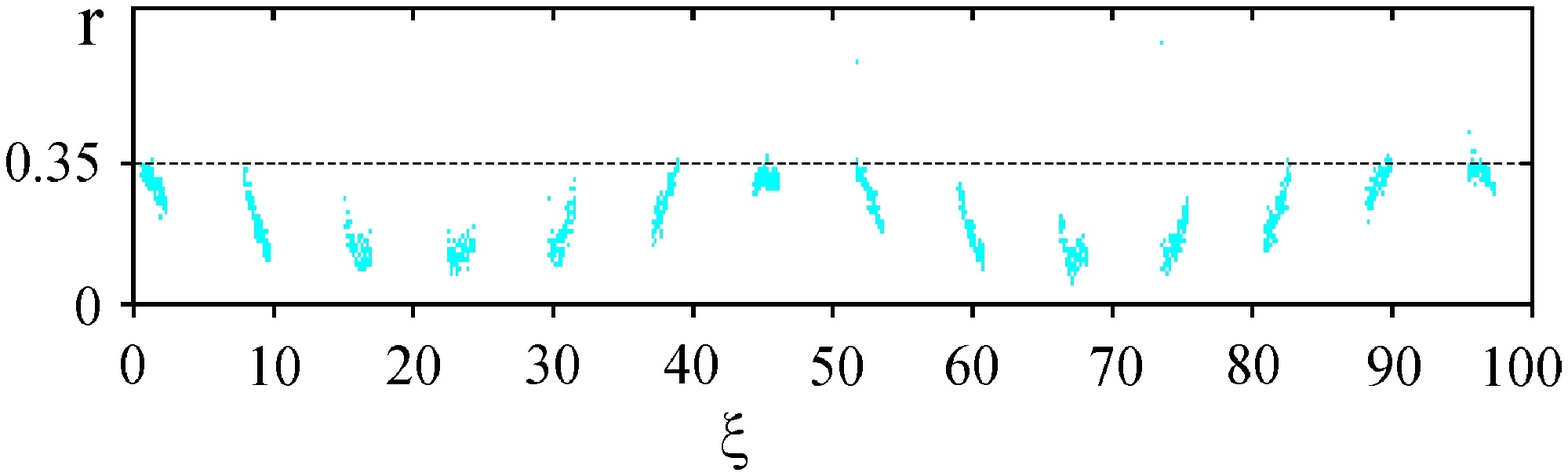}
  \caption{ Longitudinal, along train, distribution of radius $r_b$ of train of rectangular bunches of length $\xi_b=\lambda/4$ at $n_e=1.35n_{res}$, $I_b=10^{-3}$ after their focusing at the distance z=50cm from the boundary of injection.}
  \label{08France:fig9}
\end{figure}

\begin{figure}
\centering
  \includegraphics[width=.6\linewidth, height=.2\linewidth]{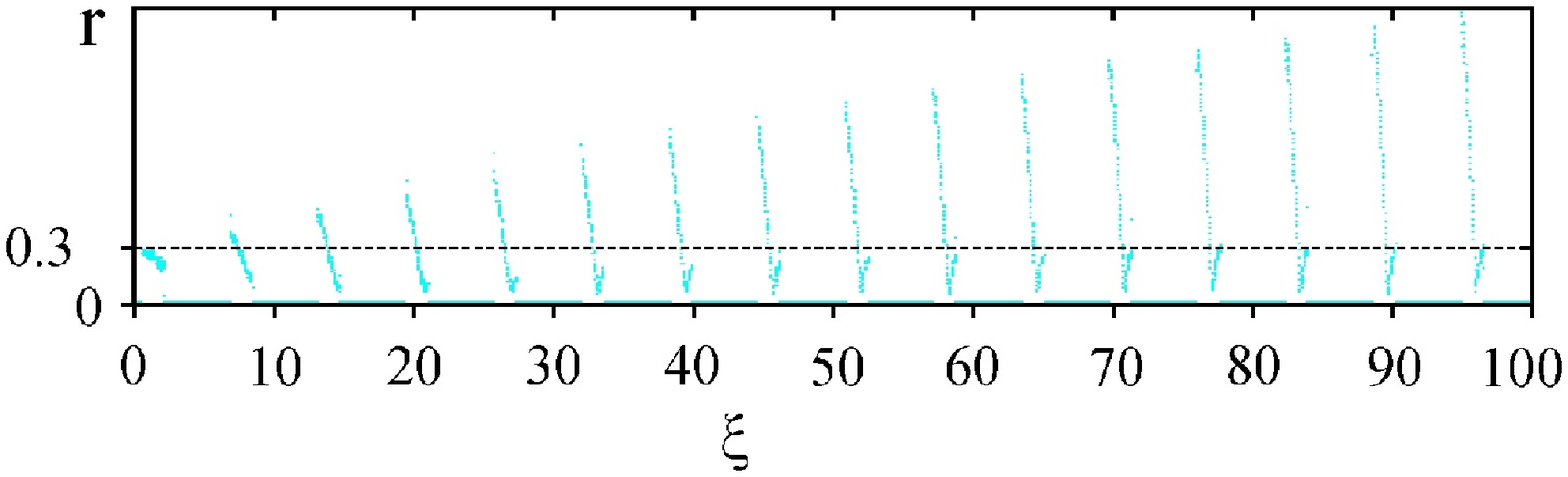}
  \caption{Longitudinal, along train, distribution of radius $r_b$ of resonant train of rectangular bunches of length $\xi_b=\lambda/4$ at $I_b=10^{-3}$ after their focusing/defocusing at the distance $z=50$~cm from the boundary of injection.}
  \label{08France:fig10}
\end{figure}

\begin{figure}
\centering
  \includegraphics[width=.8\linewidth, height=.27\linewidth]{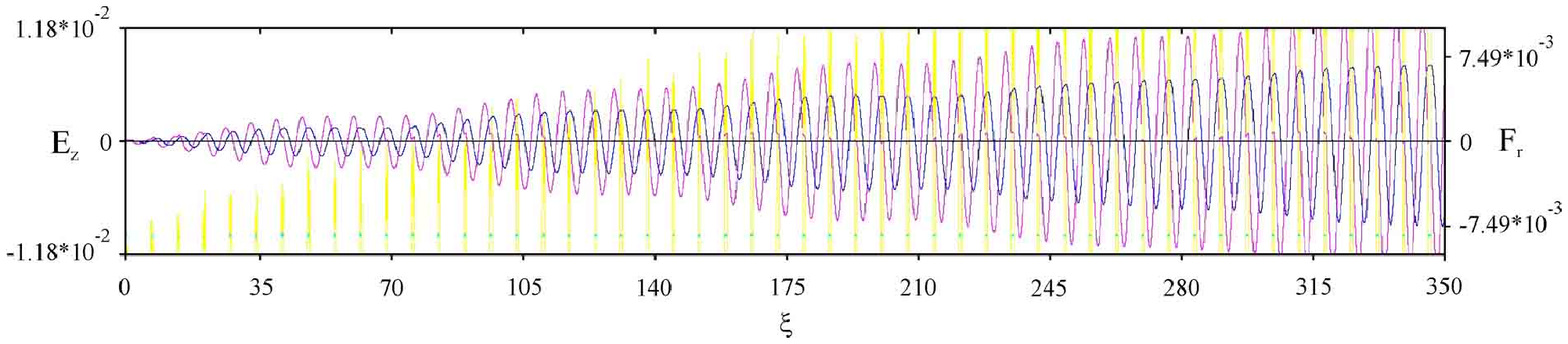}
  \caption{Longitudinal, along train, distribution of radius $r_b$ (points) and density $n_b$ (vertical lines) of long shaped according to linear dependence train of very short bunches, of radial wake force $F_r$ (oscillating line of smaller amplitude) and of longitudinal wakefield $E_z$ (oscillating line of larger amplitude) near boundary of injection at $I_b=10^{-3}$.}
  \label{08France:fig11}
\end{figure}

\begin{figure}
\centering
  \includegraphics[width=.8\linewidth, height=.27\linewidth]{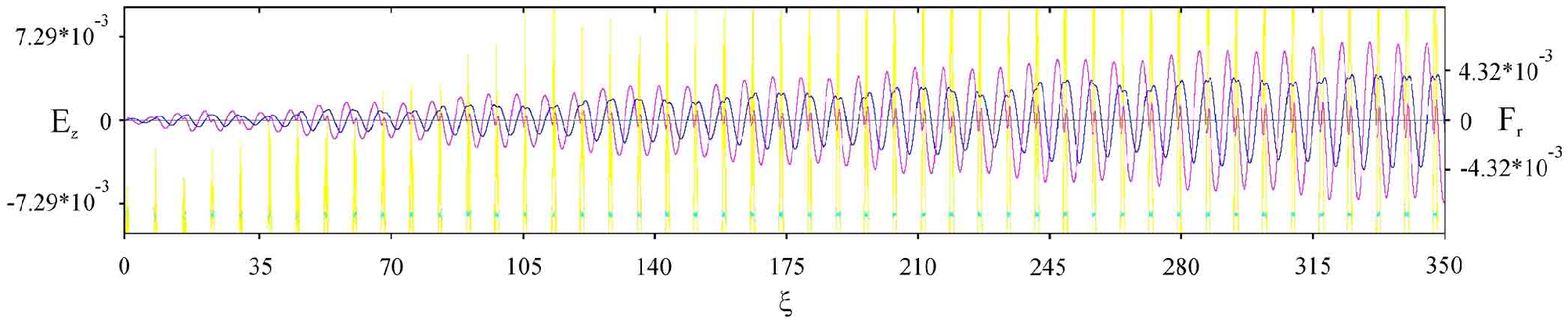}
  \caption{Longitudinal, along train, distribution of radius $r_b$ (points) and density $n_b$ (vertical lines) of long shaped according to linear dependence train of approximately Gaussian bunches, the length of which equals $\xi_b=\lambda/5=2.11$~cm, of radial wake force $F_r$ (oscillating line of smaller amplitude) and of longitudinal wakefield $E_z$ (oscillating line of larger amplitude) near boundary of injection at $I_b=2.5\cdot 10^{-3} = 33.5$~A.}
  \label{08France:fig12}
\end{figure}

\begin{figure}
\centering
  \includegraphics[width=.8\linewidth, height=.27\linewidth]{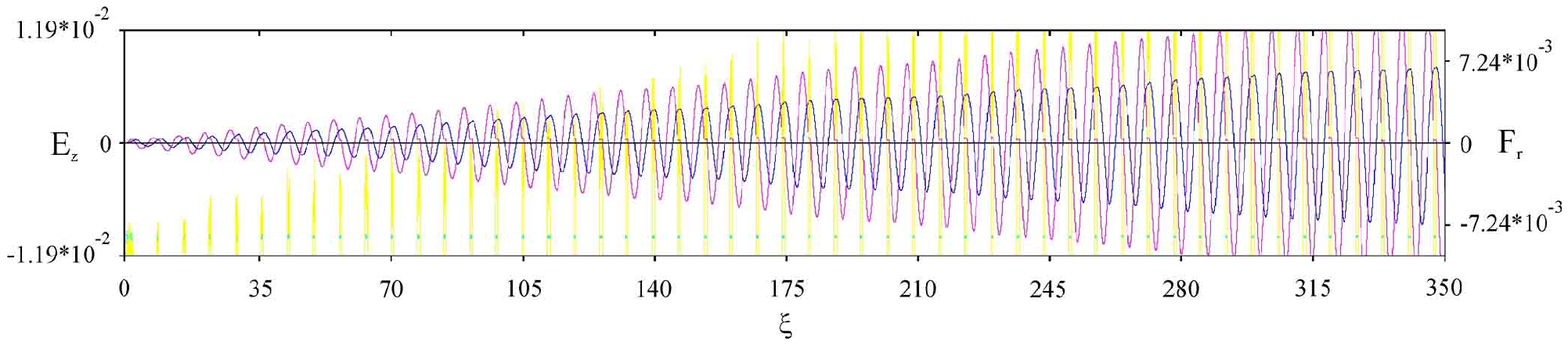}
  \caption{ Longitudinal, along train, distribution of radius $r_b$  (points) and density $n_b$ (vertical lines) of long train with precursor shaped according to linear dependence along train as well as along each bunch, of radial wake force $F_r$ (oscillating line of smaller amplitude) and of longitudinal wakefield $E_z$ (oscillating line of larger amplitude) near boundary of injection at $I_b=10^{-3}$.}
  \label{08France:fig13}
\end{figure}

Now we consider the long train of Gaussian bunches, shaped according to linear dependence. The space interval between bunches is equal to the wavelength, and the bunch length equals $\xi_b=\lambda/5$ (Fig.~\ref{08France:fig12}). Also we consider a long train of short bunches with precursor, shaped according to linear dependence along the train and along each bunch (Fig.~\ref{08France:fig13}). The space interval between bunches equals wavelength. One can see that in both cases all bunches are in maximal focusing fields and in a small $E_z$.

Thus, bunches of sequence, shaped according to linear dependence, and bunches of sequence, shaped according to linear dependence with precursor, are in maximal focusing fields.

\section{Conclusions}

Focusing of relativistic electron bunches by nonresonant wakefield, excited by them in a plasma, has been investigated by numerical simulation. It has been shown that in the case where the electron plasma frequency is larger than the repetition frequency of bunches, all bunches are in focusing wakefield of excited beatings except bunches, which are located at the fronts of beatings which are not focused.

\coltoctitle{Modelling point defects dynamics in irradiated foils: patterning and pattern selection}
\coltocauthor{D.~Kharchenko, V.~Kharchenko, I.~Lysenko}

\begin{center}
{\Large {\bf Modelling point defects dynamics in
irradiated foils: patterning and pattern selection}}

\vskip 0.5cm

{\bf D.~Kharchenko, V.~Kharchenko, I.~Lysenko}

\vskip 0.3cm

{\it Institute of Applied Physics NAS of Ukraine}

\end{center}

\vskip 0.4cm

\begin{abstract}
\noindent Dynamics of nano-sized vacancy clusters formation in a
prototype systems of an irradiated foils is studied by using  a dynamical model
describing spatio-temporal behavior of vacancy concentration and temperature of
irradiated sample. It is shown that pattern selection processes
and morphology of defect structures can be controlled by damage rate.
Oscillatory dynamics of  both vacancy
concentration and sample temperature is analyzed. It is shown that the
mean size of vacancy clusters (from 30nm up to 300nm) evolves in oscillatory
manner due to pattern selection processes.
\end{abstract}

\noindent

\vskip 0.4cm

\noindent PACS numbers: 05.40.-a, 89.75.Kd, 65.80.-g, 81.16.Rf

\vskip 0.4cm

\noindent  Keywords: defects, pattern formation, nano-structure, thermal
effects

\section{Introduction}

It is well known that irradiated materials are nonequilibrium systems
manifesting self-organization of nano-structures of point defects due to laser
or particle irradiation. These processes are results of rearrangement of mobile
point defects (vacancies ($v$), interstitials ($i$, doped atoms) produced by an
irradiation influence. It leads to formation of surface structures like dots
having sizes of nano- or micrometers on semiconductors and metals
\cite{Walgraef}. Depending on irradiation conditions (displacement damage rate
and temperature) point defects can arrange into objects of higher dimension
such as clusters (di-, tri-, tetra-vacancy clusters), defect walls with vacancy
and interstitial loops, voids, precipitates and bubble lattices. Fabricated in
such a way nano-structured thin films or foils can be exploited in developing
novel electronic devices, memory storage, detectors, etc. A rearrangement of
mobile nonequilibrium point defects leads to emergence of mechanical stresses
in irradiated foils. This results to self-organization of nano-structures
leading to compensation of such stresses \cite{QE2006}.

Usually, considering dynamics of defects in a bulk  one admits that the temperature of the irradiated sample remains constant due to high thermal conductivity (for
example, for metals) and one can control the surface morphology varying damage
rate (defects production rate) only. This effects is possible for massive
samples. In most of studies it is assumed that the system attains the thermal
state realized under stationary external conditions. At the same time it is
well known that in thin foils (with thickness around $0.3\div 0.5\mu m$) a
nonequilibrium defect concentration and a temperature of a sample are able to
reach much higher values than stationary ones and oscillate around them.
In thin foils  local temperature variations of a sample due to irradiation and point defect
rearrangement can be realized. Considering  foils as a sample in
an environment characterized by a constant temperature one deals with processes
of local changes in the sample temperature due to its heating, production of
defects and their annihilation (a range of ionizing particles is smaller then
thickness of the sample). A local increase in the sample temperature results in
defects annealing. It leads to defect energy release into a heat increasing the
temperature of the sample. As a result a number of defects decreases with
increasing the heat transfer. Next, the sample cools, heat transfer decreases
and new defects accumulate. A repetition of this scenario leads to
self-oscillations of the sample temperature and point defects concentration. An
emergence of such self-oscillations in homogeneous systems were reported in
Ref.\cite{SS92}. There are many experimental observations of
self-organization of point defects in solids with spatio-temporal oscillations
of point defect concentration (see, for example,
Refs.\cite{Krishan,Carpenter}). Among them one can note: void size
oscillations were observed in irradiated nickel samples \cite{Steel};
temperature oscillations of crystals $CH_4$ were reported in
Ref.\cite{Carpenter}; periodic variations of microhardness of Nimonic 90 with
$\gamma$-precipitates with radiation dose growth were discussed in
Ref.\cite{Varatharajan}.

From the theoretical and practical viewpoints one of the interesting problem relates to study surface layer self-organization of point defects due to irradiation influence with
redistribution of the conjugated temperature field. In this work we study
spatio-temporal evolution of point defects and local temperature in a prototype model of irradiated sample following
the approach developed in Refs.\cite{QE2006,EPJB2012}, where local
temperature field in irradiated layers evolves according to main mechanisms
proposed in Ref.\cite{SS92}. We take into account fluctuations in damage
rate related to the fact that bombarding particles have dispersion in energy.
In our consideration we assume Gaussian fluctuations in defect production rate
and study an influence of these fluctuations onto stationary surface patterns
realization. It should be noted that we study patterning effects due to point
defects interaction in a prototype theoretical model excluding  sputtering
and erosion of the surface. The main attention is paid to describe
formation of clusters of point defects and patterning of the conjugated
temperature field caused by agglomeration of defects. We consider a development of spatio-temporal patterns manifesting pattern selection processes.

\section{Prototype model of irradiated foils}

Considering a system (irradiated metallic foil), dynamics of concentration of point defects (interstitials and
vacancies, $c_{i,v}=c_{i,v}(\mathbf{r},t)$, $\mathbf{r}=\{x,y,z\}$) and local temperature $T(\mathbf{r},t)$ of a sample  can be described  by following  equations:

\begin{equation}\label{eq1_def}
\begin{split}
\partial_t c_v&=\mathcal{K}-\beta_v (c_v-c_v^0)-\alpha c_ic_v-\nabla\cdot\mathbf{J}_v,\\
\partial_t c_i&=\mathcal{K}-\beta_i c_i-\alpha c_ic_v-\nabla\cdot\mathbf{J}_i,\\
C\rho\partial_t T&=\chi\Delta T-\frac{\gamma_0}{h}(T-T_0)+\zeta E_{f} \mathcal{K}+E_{f}[\beta_v (c_v-c_v^0)+\alpha c_ic_v],
\end{split}
\end{equation}

\noindent
where $\mathcal{K}$ is the defect production rate; $\beta_{i,v}=\rho_dD_{i,v}$ is the inverse lifetime of defects of $i/v$ type defined through the dislocation density $\rho_d$ and diffusion coefficient $D_{i,v}$, $c_v^0$ is the equilibrium
vacancy concentration, $\alpha$ is the recombination
rate. In equation for the local temperature $T$ (measured in energetic units) of a sample $C$ is the specific thermal capacity, $\chi$ is the thermal conductivity,
 $\gamma_0$ is the heat transfer coefficient, $h$ is the foil thickness,  $T_0$ is the environment temperature; $\zeta\gg 1$ is the ratio of energy of irradiation which transforms into heating and energy of irradiation which transforms into defect generation; $E_{f}$ is the energy of defect formation; the last term defines energy release when defects are captured by dislocations and due to recombination with interstitials. As far as we consider thin foils one can put $z=0$ and study planar problem only.  Here we
assume that a substrate for the layer location does not affect the irradiation
influence, or its influence on substrate is negligible small, and is not
considered. Formally, by using the well known relation between diffusivities
 $D_i/D_v\gg1$,
one can exclude the fast variable $c_i$, setting $\partial_t c_i\simeq0$.
The diffusion flux of vacancies as mobile species is defined as follows
$
\mathbf{J}_v=-D_v\nabla c_v+ c_v\mathbf{v},
$
where the first term is responsible for the free diffusion of defects, the
second term relates to the thermodynamical (chemical) force $\mathbf{f}=
-\nabla (U/T)$ influence inducing a speed $\mathbf{v} = D_v\mathbf{f}$; here $U$ is the interaction potential between defects. The lateral force exerting
on the defect from the deformed elastic continuum is $
-\nabla U=\left.\theta_d\nabla \vartheta\right|_{z=0}$,
where $\vartheta(\mathbf{r})=\nabla\cdot \mathbf{u}$,
$\mathbf{u}=\mathbf{u}(\mathbf{r},t)$ is the vector of material displacement in
the layer, $\theta_d\equiv\Omega K$ is the strain potential of defect, $K$ is
the elasticity modulus. As was shown in previous studies (see
Ref.\cite{QE2006}) the deformation of the surface layer depends nonlocally on
the defect concentration on the surface as follows:
$\vartheta(z=0)=\frac{\nu\theta_d}{\rho c_{||}^2}\mathcal{L}c_v,\quad
\mathcal{L}=(1+l_{||}^2\Delta)$;
here $\nu=(1-2\sigma_P)/(1-\sigma_P)$, $\sigma_P$ is the Poisson coefficient of
the layer; $c_{||}^2=E/\rho(1-\sigma^2_P)=\sigma_{||}/\rho$ is the  bending
stiffness; $E$ is the Young modulus; $\rho$ is the density; $\sigma_{||}$ is
the tensile isotropic stress in the defect-enriched surface layer of the height
$h$ (thickness of radiation induced defect enriched surface layer);
$l_{||}=h(\rho c_{||}^2/12\sigma_{||})^{1/2}$.

Next, it is convenient to use  rescaled concentration $x\equiv \mu c_v$ with  $\mu\equiv \alpha/\beta_i$. Measuring time in unites  $\tau_d\equiv (\rho_dD_v^0)^{-1}$, spatial coordinate in
units $L_D\equiv (\rho_d)^{-1/2}$, one can introduce dimensionless quantities
$t'\equiv t/\tau_d$, $\mathbf{r}'\equiv \mathbf{r}/L_D$, $\ell\equiv
l_{||}/L_D$, $\epsilon\equiv \nu\theta_d^2/\mu\rho c^2_{||}T_0$, $\Theta\equiv
T/T_0$,  $\eta\equiv C\rho h/\tau_d\gamma_0$,
$\kappa\equiv\chi h/L_D^2\gamma_0$, $\varpi\equiv E_f h/T_0\mu\tau_d\gamma_0$,
$\varepsilon\equiv E_v/T_0$. Generally, in order to make a statistical
description one assumes that defects can be produced in stochastic manner due
to fluctuations in damage rate. Hence, by taking $\mathcal{K}\to \mathcal{K}(t)=\mathcal{K}_0+\xi(t)$, where
$\xi(t)$ is a white Gaussian noise with properties $\langle\xi(t)\rangle=0$,
$\langle\xi(t)\xi(t')\rangle=2\mathcal{K}_0\sigma^2\delta(t-t')$ one admits that
fluctuations in the damage rate are possible at nonzero flux described by
$\mathcal{K}_0$; $\sigma^2$ denotes fluctuations intensity.

\section{Results and discussion}

Considering $\mathcal{K}$ and $T_0$ as independent parameters
and using material constants for pure $Ni$
one can obtain phase diagram illustrating domain of oscillatory dynamics (see
Fig.\ref{phd_osc}a). The typical set of the system parameters is:
$\mathcal{K}\sim 10^{-3}dpa/sec$, $T_0\in[300,900]K$, $L_D\simeq 10^{-7}m$,
$\tau_d\simeq 10^{-6}sec$, $E_f=1.6eV$, $h=0.5\mu m$, $\eta=100$, $\kappa=1$, $\ell=0.7$,
$\zeta=7$, $\varpi=0.1$, $\mu\simeq 10^8$.
According to obtained diagram one finds that in the deterministic limit
($\sigma^2=0$) oscillatory dynamics is realized at values of $\mathcal{K}$ and
$T_0$ lying above the solid line. In the stochastic case ($\sigma^2\ne 0$)
fluctuations do not change lower values of $\mathcal{K}$ and $T_0$ belonging to
the solid line. The noise leads to bounding the domain of oscillatory dynamics
limiting upper values of damage rate and the temperature $T_0$. Hence,
oscillations are possible inside the bounded domain for $\mathcal{K}$ and $T_0$
in the stochastic system ($\sigma^2\ne 0$).
\begin{figure}[!t]
 \centering
 a)\mbox{\epsfig{figure=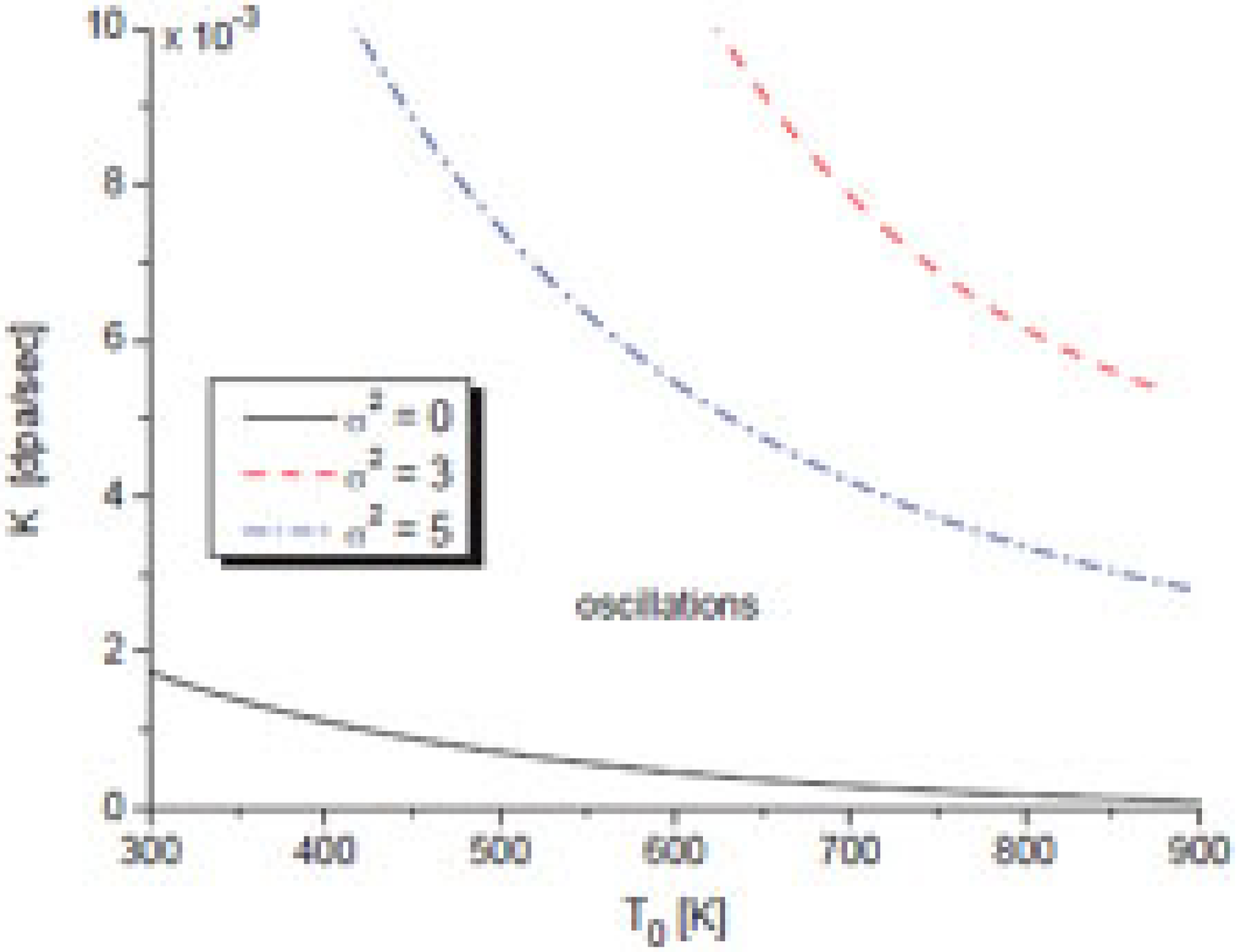,height=5.3cm}}
 b)\mbox{\epsfig{figure=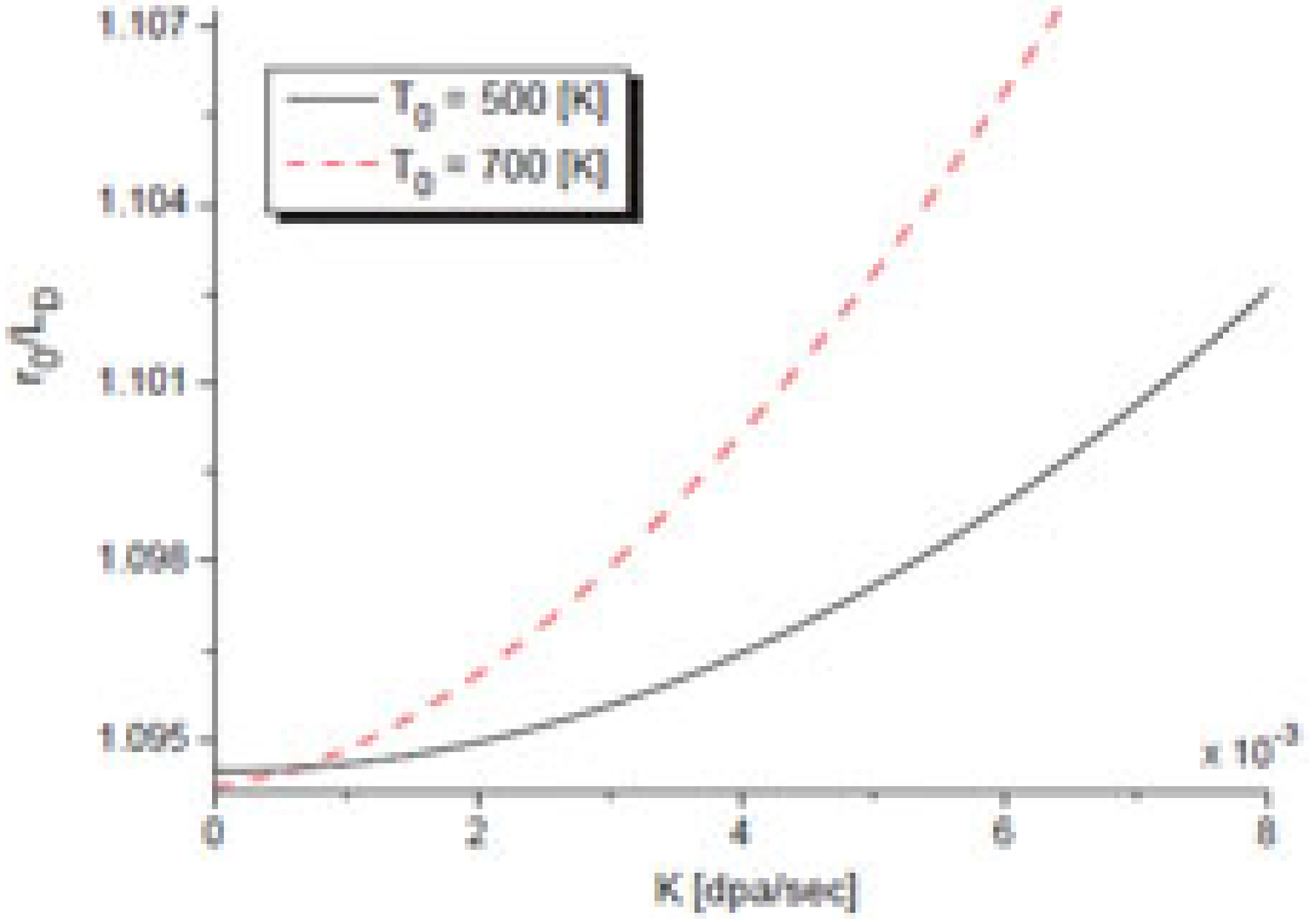,height=5.3cm}}
\caption{(a)Phase diagram for linear stability analysis. Domain of oscillatory
dynamics is bounded by solid curve. (b) Dependencies of
period of patterns  at different $T_0$
 \label{phd_osc}}
\end{figure}

From the linear stability analysis one can obtain a period  $r_0$ of spatial patterns realized during irradiation (see Fig.\ref{phd_osc}b). It is seen that with an increase in the damage rate $\mathcal{K}$ the period of patterns slightly increases. The noise does not affect the quantity $r_0$ essentially (not shown here). Exposing the target at elevated environment temperature $T_0$ one can generate patterns with larger period.

To make a quantitative analysis we solve numerically the system (\ref{eq1_def}) on a quadratic grid with linear  size $9.2L_D$.
Typical patterns of both defect concentration field and temperature of the
target at different times are shown in Figs.\ref{snap}. It is well seen that during the system evolution formation of vacancy patterns with nonuniform distribution of the temperature field is realized. The local temperature is larger in domains with vacancy clusters.
Here  one gets dot-like patterns, where vacancies are concentrated in
spherical clusters forming voids/holes on the surface of the target. Increasing
the damage rate one induces a formation of nonequilibrium vacancies able to
form elongated clusters transformed into ripples.

\begin{figure*}[!t]
\centering
 \includegraphics[width=0.8\textwidth]{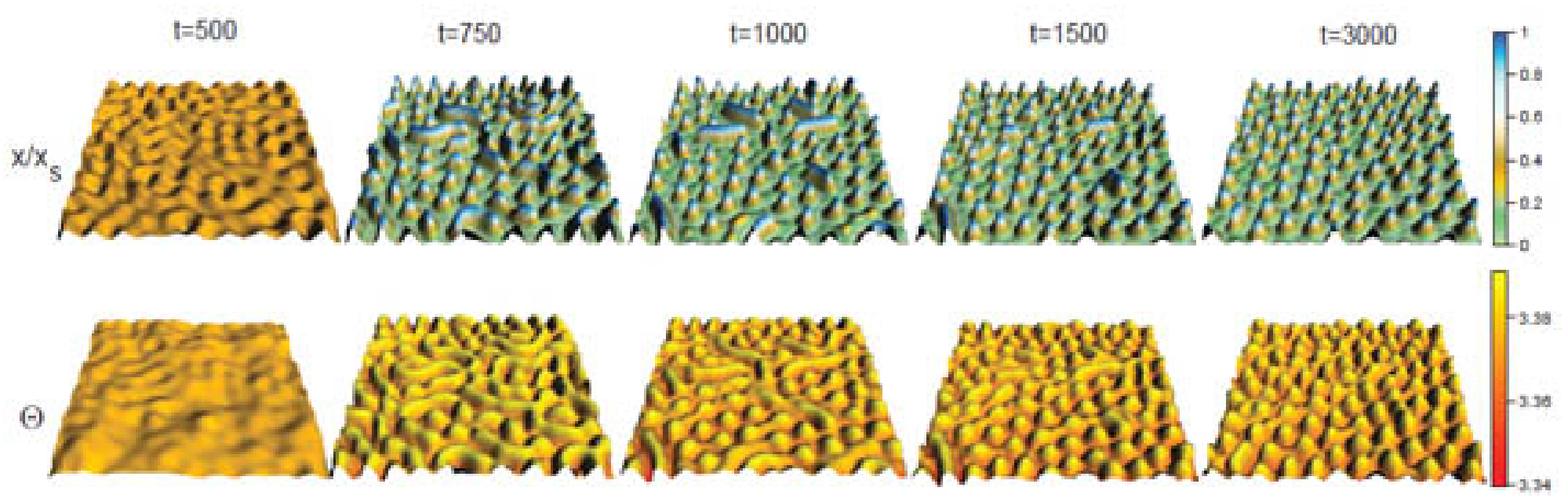}
 \caption{Snapshots of vacancy concentration field (top panels) and temperature field (bottom panels) at $T=500K$:
 $\mathcal{K}=3\cdot10^{-3} dpa/sec$ in the deterministic case; $x_s$ denotes the stationary homogenous vacancy concentration\label{snap}}
\end{figure*}

From Fig.\ref{time}a it follows  that both $\langle
x\rangle$ and $\langle\Theta\rangle$ manifest small decaying oscillations near
the steady states.  Such oscillatory behaviour is responsible for pattern selection
processes. At this stage one can observe formation of patterns with spherical
and elongated structures. During exposing the system selects one type of
spatial structures. As far as temperature variations are strongly related to
behaviour of vacancies, therefore, temperature oscillations become possible
at this stage.
\begin{figure*}[!t]
\centering
 a) \includegraphics[width=0.45\textwidth]{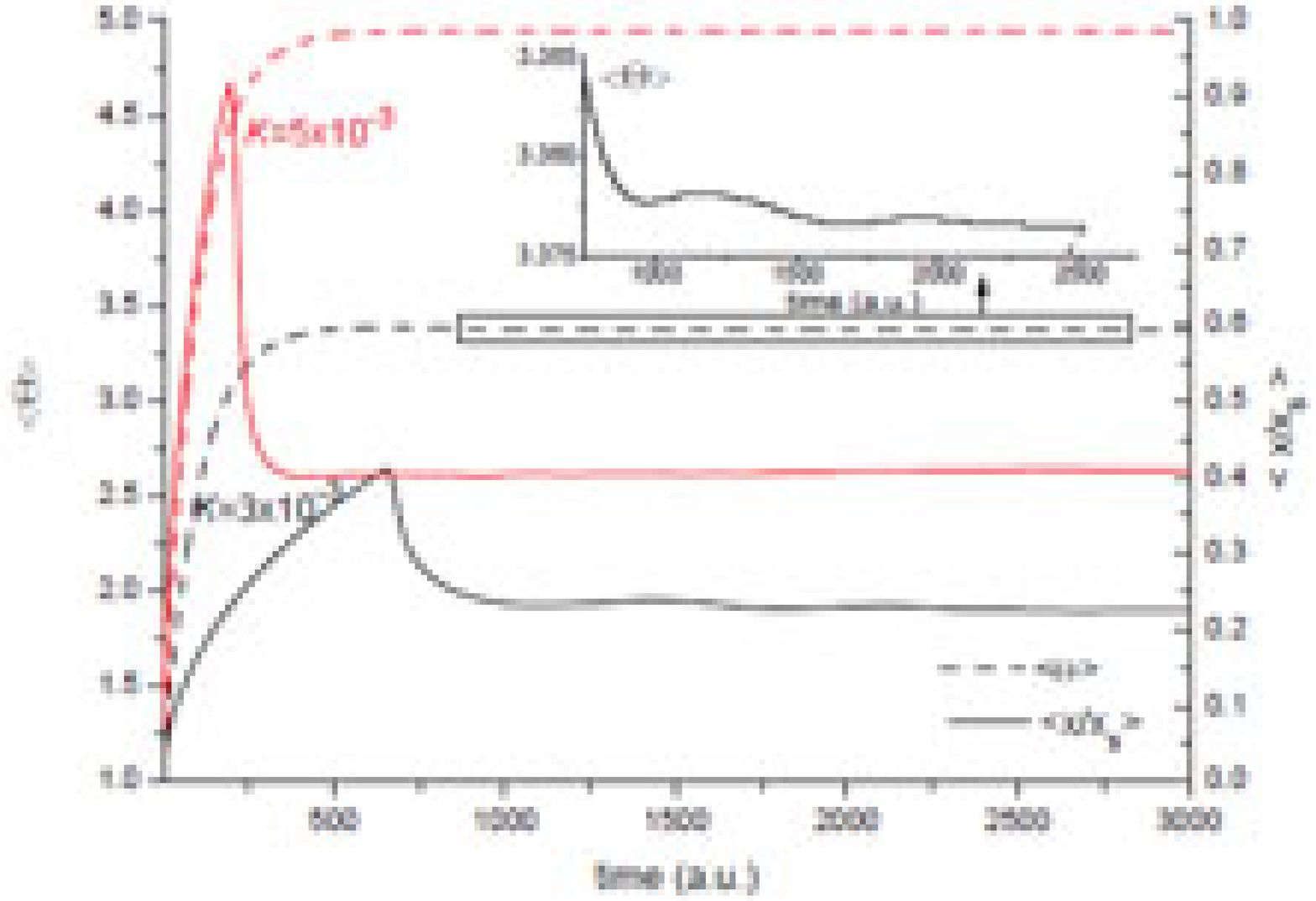}
 b) \includegraphics[width=0.45\textwidth]{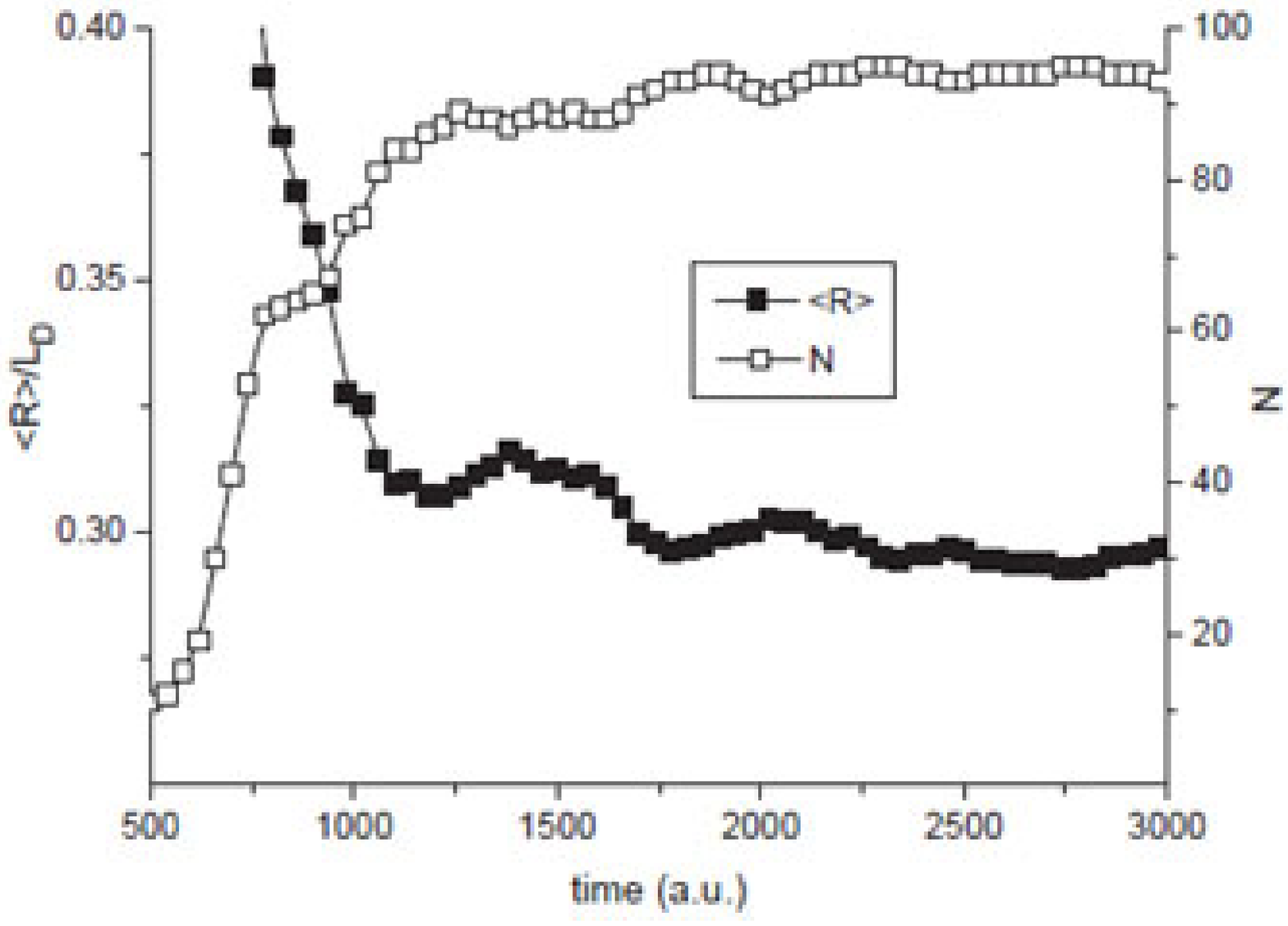}
 \caption{Evolution of   averaged temperature of a sample and  concentration of defects (panel a) and mean radius $\langle R\rangle$ of spherical vacancy
clusters (filled squares) with the corresponding number of clusters (empty
 squares) (panel b)   at $\mathcal{K}=3\cdot10^{-3} dpa/sec$, $T_0=500K$\label{time}}
\end{figure*}
Oscillatory dynamics of a mean radius of vacancy islands $\langle
R(t)\rangle$ and their number $N(t)$ (see Fig.\ref{time}b) illustrate
that pattern formation starts from organization of small amount of clusters
having large characteristic lengths.  During the system evolution  some of
them dissolves, some new clusters can emerge due to interactions of
supersaturated vacancies. With an increase in the exposition time most of
clusters become identical and are characterized by the constant averaged radius
at large time scales.  To estimate $\langle R\rangle$ we take $L_D\simeq(\rho_d)^{-1/2}$
with $\rho_d\simeq 10^{14}\div 10^{12} m^{-2}$. It gives typical size of
vacancy clusters $\langle R\rangle\simeq 0.3L_D\simeq 30\div 300 nm$; the
related distance between them is around diffusion length. Dynamics of $\langle
R \rangle$ is accompanied by an increase in the number of islands toward the
corresponding stationary value.

\section{Conclusions}
It was shown that pattern selection processes are realized due coupling between
defect concentration and sample temperature fields.
We have shown that
stochastic contribution can reduce the domain of pattern selection processes.
Oscillatory dynamics of both vacancy
concentration and temperature of the sample averaged over the whole system is discussed. It
was found that pattern selection processes are accompanied by decaying
oscillatory dynamics of mean size of vacancy islands and their number.

\section{Acknowledgement}

The support  of the LIA project is greatly acknowledged.

\coltoctitle{ALERT: A Low Energy Recoil Detector}
\coltocauthor{G.~Charles}

\begin{center}
{\Large {\bf ALERT: A Low Energy Recoil Detector}}

\vskip 0.5cm

{\bf G.~Charles$^{a}$}

\vskip 0.3cm

$^{a}${\it Institut de Physique Nucl\'{e}aire d'Orsay, CNRS-IN2P3, Universit\'{e} Paris-Sud, Universit\'{e} Paris-Saclay}

\end{center}

\vskip 0.4cm

\begin{abstract}
\noindent We present here a preliminary study of a tracker that could be used to reconstruct recoil nuclei fragments at CLAS. The main characteristics of this tracker are its ability to reconstruct protons with a minimum momentum of about 70 MeV/c and to identify all particles with a mass between the proton and the alpha. The detectors have also been selected in order to be included in the trigger to ensure a fast event selection. This paper focused on the state of our research on the project as well as on the expected performances obtained from simulations.
\end{abstract}

\noindent

\vskip 0.4cm

\noindent  Keywords: Tracker, Drift chamber, stereo-angle, hadronic physics

\section{Introduction}

The ALERT tracker is intended to be used at Jefferson Laboratory (JLab) is located in Virginia (USA). It is an electron accelerator facility, with a beam of energy up to 12 GeV now accessible after the recent three years long upgrade. The beam will be distributed to four halls and in particular in Hall B where the CLAS12 experiment will start taking data at the end of 2016. With the predecessor of CLAS12, CLAS \cite{CLAS}, a whole new physics program has emerged based on the use of detectors dedicated to low energy nuclear recoils (kinetic energy of few MeV). The recoils are produced during high energy reactions such as deep inelastic scattering (DIS) eA $\rightarrow$ eX or deep virtual Compton Scattering (DVCS) eA $\rightarrow$ eA. In particular, radial time projection chambers (rTPC) were successfully used in CLAS to measure the structure function of the free neutron by tagging slow protons out of deuterium targets \cite{TagSlowProt} and to measure coherent DVCS off helium nuclei \cite{DVCSOffHelium}. An international group of scientists from the CLAS collaboration has formed around the project of creating a new nuclear recoil detector for the upgraded CLAS12. We seek to develop a new detector offering better timing and spatial resolution as well as better particle identification than the previous rTPCs \cite{PrevRTPC}.

\section{Detector setup}

The construction of a low energy recoil detector for CLAS12 is challenging in many aspects. Indeed such detector will have to run in a high rate region (several MHz of protons in particular) in a very strong magnetic field (5T) due to the magnetic shield of the CLAS12 spectrometer. At the same time it needs to provide fast and precise response ($<$~2~$\mu$s) in order to be used by the trigger, in coincidence with the normal CLAS12 trigger. Finally, in order to detect the recoil before they are absorbed in any materials the chamber needs to be right around the target. We envision a 30 cm long detector with a radius of 10 cm surrounding a 30 cm long gaseous target at 3~bars, the layout can be seen figure \ref{ALERT_scheme}.


\nopagebreak
\begin{figure}[htb]
\begin{center}
 \mbox{\epsfig{figure=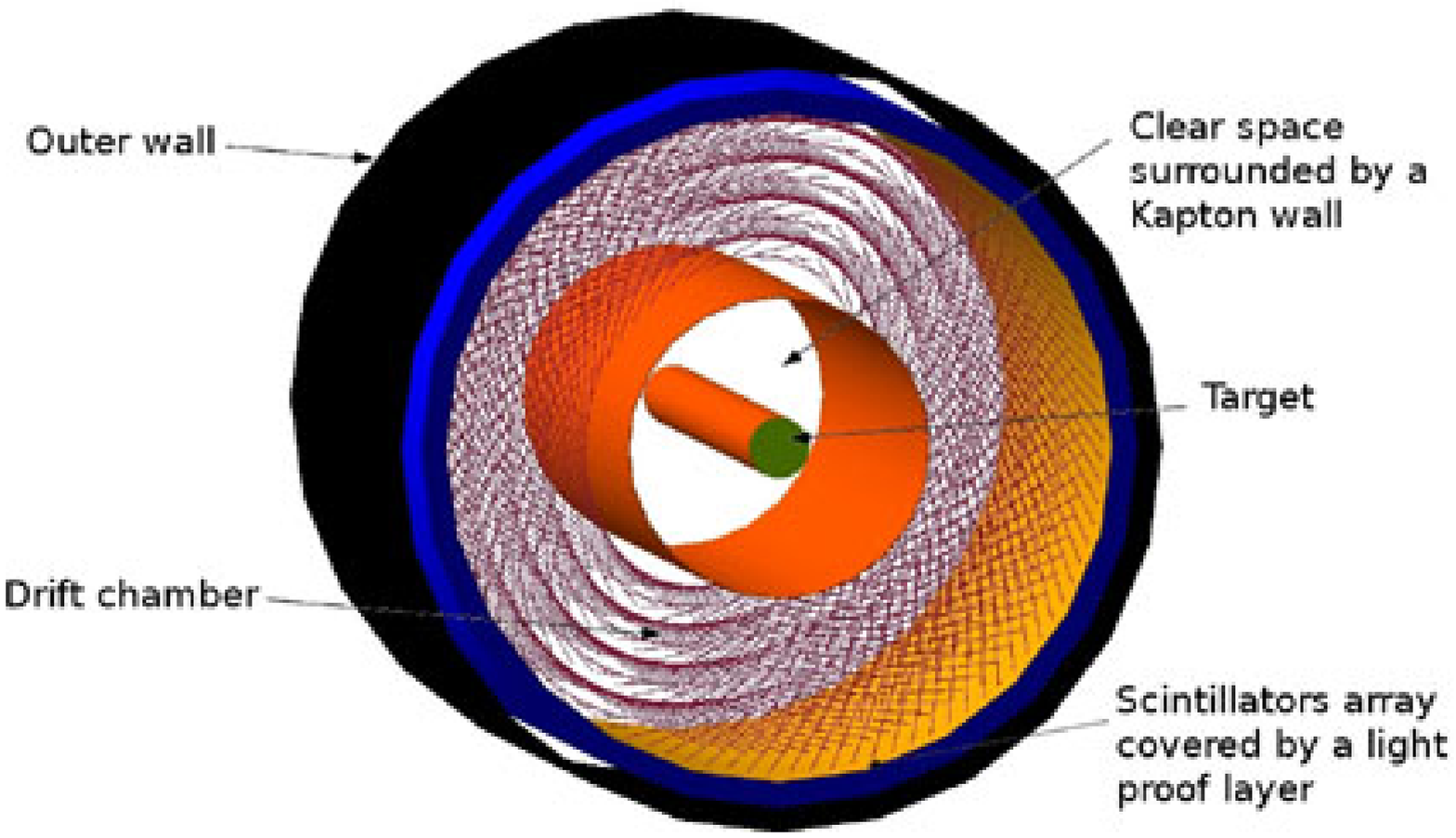,height=7.0cm}}
 \caption{Schematic layout of the detector, viewed from the beam direction.}
 \label{ALERT_scheme}
 \end{center}
 \end{figure}

\subsection{Drift chamber}
The drift chamber is composed of 8 layers of sense wires. The later are each surrounded by field wires to create the electric field. The geometry of the basic cell, composed of one sense wire and a certain number of field wires, needs to be determined. The cell must ensure a good uniformity of the electric field around the sense wire but the total number of wires should be as low as possible to reduce multiple scattering, the total tension on the forward end plate and the number of particles stopped by the wires. The quantity of material in the detector can also be limited by using lighter wires, such as aluminium wires or any other light conductive material.

As mentionned earlier, the thickness of the forward end plate is critical as the scattered electrons will go through it. In order to keep it thin, the wires will be readout only on one side of the chamber. The position along the beam axis is thus determined by the stero-angle given to the wires. This angle will probably be between 10$^{\circ}$ and 15$^{\circ}$. Its value will be tuned by simulations.

The last key parameter of the drift chamber is the space between the wires. Indeed, it influences the speed and the precision of the detector.

To optimize the parameters, tests on a prototype will be carried out. It will be an 8 layer drift chamber. Each layer will have either a different cell geometry or different spacing between wires (from 1.5 mm to 3 mm). We will also test different gas mixtures to study its influence on the efficiency and the speed of the detector.

\subsection{Scintillators}
The scintillators are used to measure the time of flight for particle identification. The main aspects considered to determine their size and material are the following ones:

\begin{itemize}
\item Simulations have shown that a resolution of 200 ps or less may be necessary to perform separation at an acceptable level (see next section)
\item They should have a granularity that will allow a matching with the drift chamber without ambiguity
\item Precise light detection in a 5~T magnetic field when coupled to a silicon based photo-detectors
\item Number of channels and cost
\end{itemize}

The preliminary design for the scintillators is based on a multi-layer array of which the granularity still remains to be determined. The layer closest to the beam would be thinner to detect alpha and low momentum particles and give a good time resolution. The second layer would stop the particles to determine their energy. The use of several thin layers of scintillators will also help to differentiate the different nuclear species which are penetrating the scintillators differently. In particular time of flight has difficulties differentiating nuclei that have same mass/charge ratio.

\subsection{Electronics}

The main requirements for the electronics is to have a 10 ns time resolution or better while keeping an energy resolution that allows particle identification through energy deposition in the detector. In this project, we will perform tests on electronics initially developed for other detectors and eventually propose modifications. First, we will investigate the possibility to use stand-alone preamplifiers. Based on the gain in the drift chamber and the number of primary ionizations, it should be possible to use a design similar to the one developed for the Heavy Photon Search \cite{HPS} experiment installed in the Hall B. The main challenge is to adapt the board and elements to the higher voltages (up to 2~kV for the drift chamber). The time resolution has already been shown to be around 2~ns \cite{HPS}, so well below our 10~ns requirement. However, more studies will be needed to evaluate how the gains of the chamber and the preamplifier can be tuned to ensure a signal over noise ratio that allows to discriminate electrons from protons and light nuclei between each other (p, 2H, 3H, 3He and 4He). Second, we will study the possible use of DREAM electronics \cite{DREAM} which was developed for the Micromegas detectors of CLAS12. A charge simulator of drift chamber wire signal will be built and used with existing electronics in order to optimize its parameters for our application: sampling frequency, peaking time and gain. Like the previous solution, it has the advantage to be already compatible with the CLAS12 data acquisition system.

\section{Expected performances}

\subsection{Implementation}
A simulation of ALERT has been developed using Geant4 \cite{Geant4}. In order to include multiple scattering and energy loss effects, all the different layers and elements, in particular the wires, have been included. A Kalman filter is under development but for now the fitting algorithm is using a global helix fit. The point coordinates sent to the fit are given by using the particle path returned by Geant4 and smearing its position each time it crosses a signal wire layer by the resolutions expected. The resolutions are given by the product of the time resolution (10~ns) by the drift speed ($\approx 2 \cdot 10^{4}$~m$\cdot$s$^{-1}$) expected in the $^{4}$He (90\%)- iC$_{4}$H$_{10}$ mixture in a plan perpendicular to the wire. The resolution along the wire also takes into account the stereo-angle. \\

\subsection{Acceptance and resolutions}
Figures \ref{Resos_Protons} and  \ref{Resos_Alphas} show from top left to bottom right, the angular resolutions, the $z$ resolution, where $z$ is the coordinate along the beam axis, and the transverse momentum resolution at the vertex for protons and for alphas. To understand the acceptance, it is important to notice that a particle is considered detected when it reaches the scintillators. \\


\begin{figure}[htb]
\begin{center}
 \mbox{\epsfig{figure=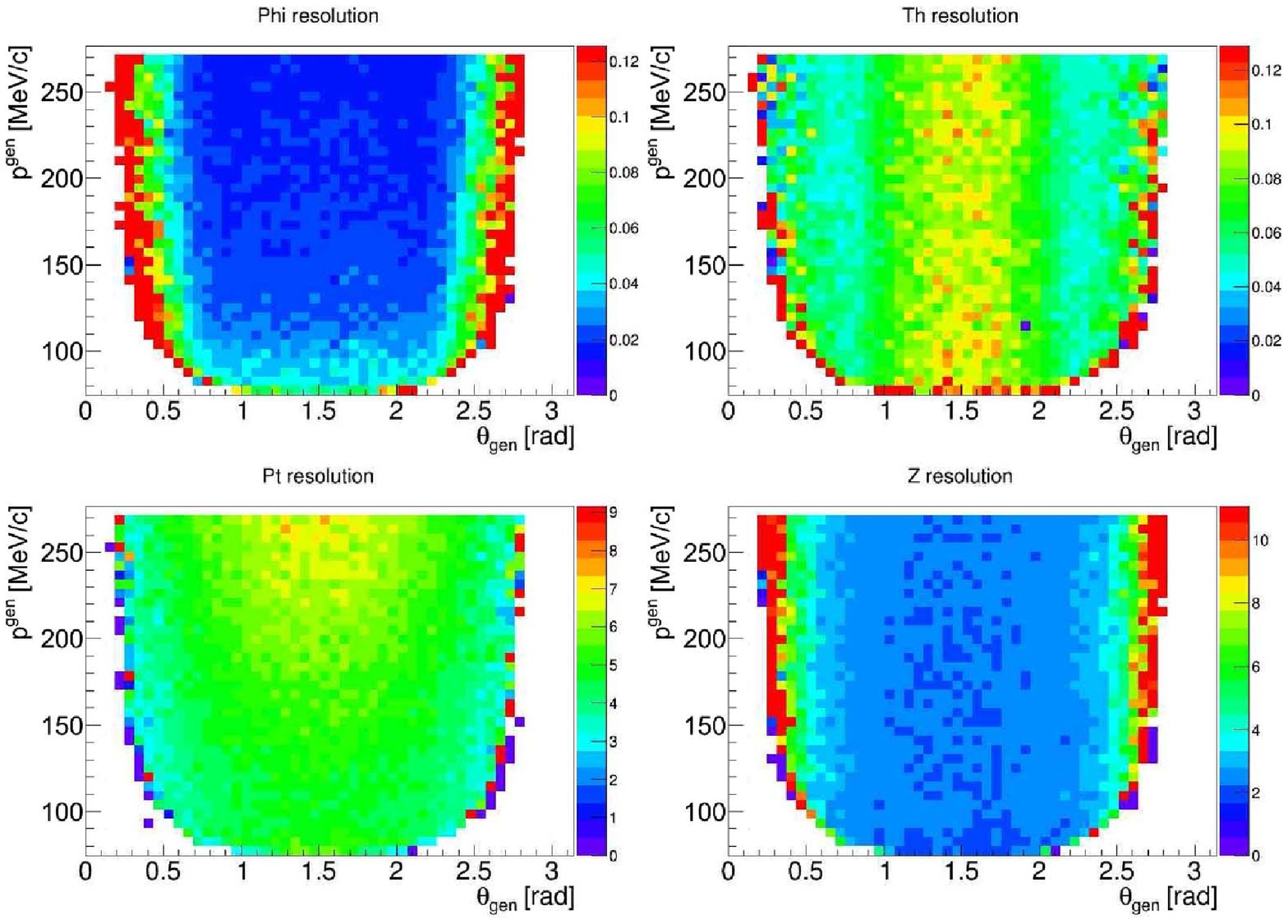,height=10.0cm}}
 \caption{Expected time of arrival as a function of the radius of the trajectory for protons.}
 \label{Resos_Protons}
 \end{center}
 \end{figure}
 
\begin{figure}[htb]
\begin{center}
 \mbox{\epsfig{figure=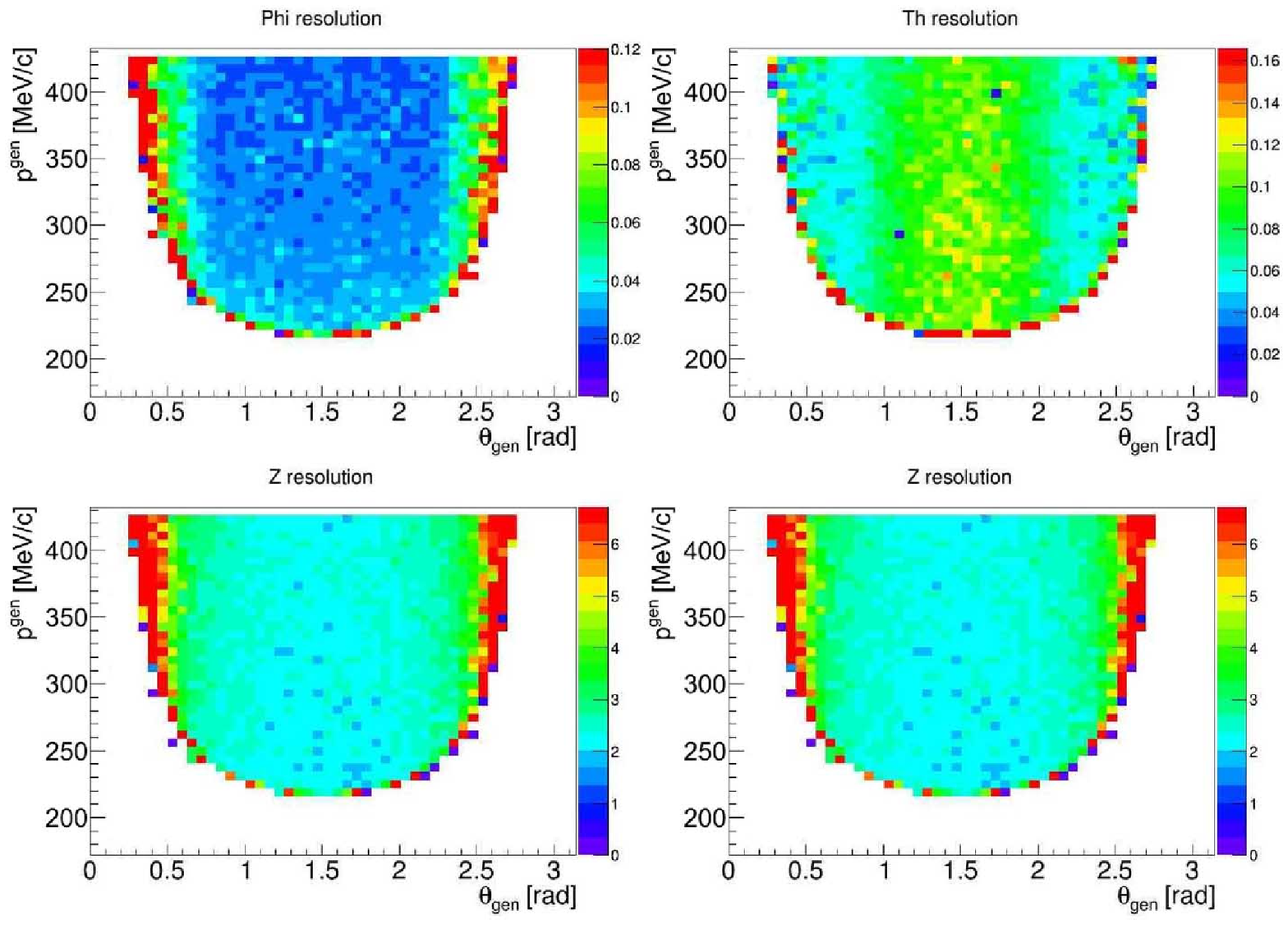,height=10.0cm}}
 \caption{Expected time of arrival as a function of the radius of the trajectory for alpha particles.}
 \label{Resos_Alphas}
 \end{center}
 \end{figure}

As expected the acceptance is larger near $\pi$/2 as the particle goes through less material when emitted at this angle. Also the resolutions for alphas are generally better than for protons due to the fact that the curvature is larger making the fit easier. A fast Monte Carlo using the acceptances and energy resolutions has been made available for collaborators. Conclusions concerning the match between our design and their needs will then be made.

\subsection{Particle identification}
One of the key point of this detector is its ability to differenciate different species. The differentiation relies upon the time of arrival in the scintillators, the initial time being given by another particle reconstructed by the rest of CLAS12, and the reconstructed radius in the drift chamber (fig. \ref{Diff}).

\begin{figure}[htb]
\begin{center}
 \mbox{\epsfig{figure=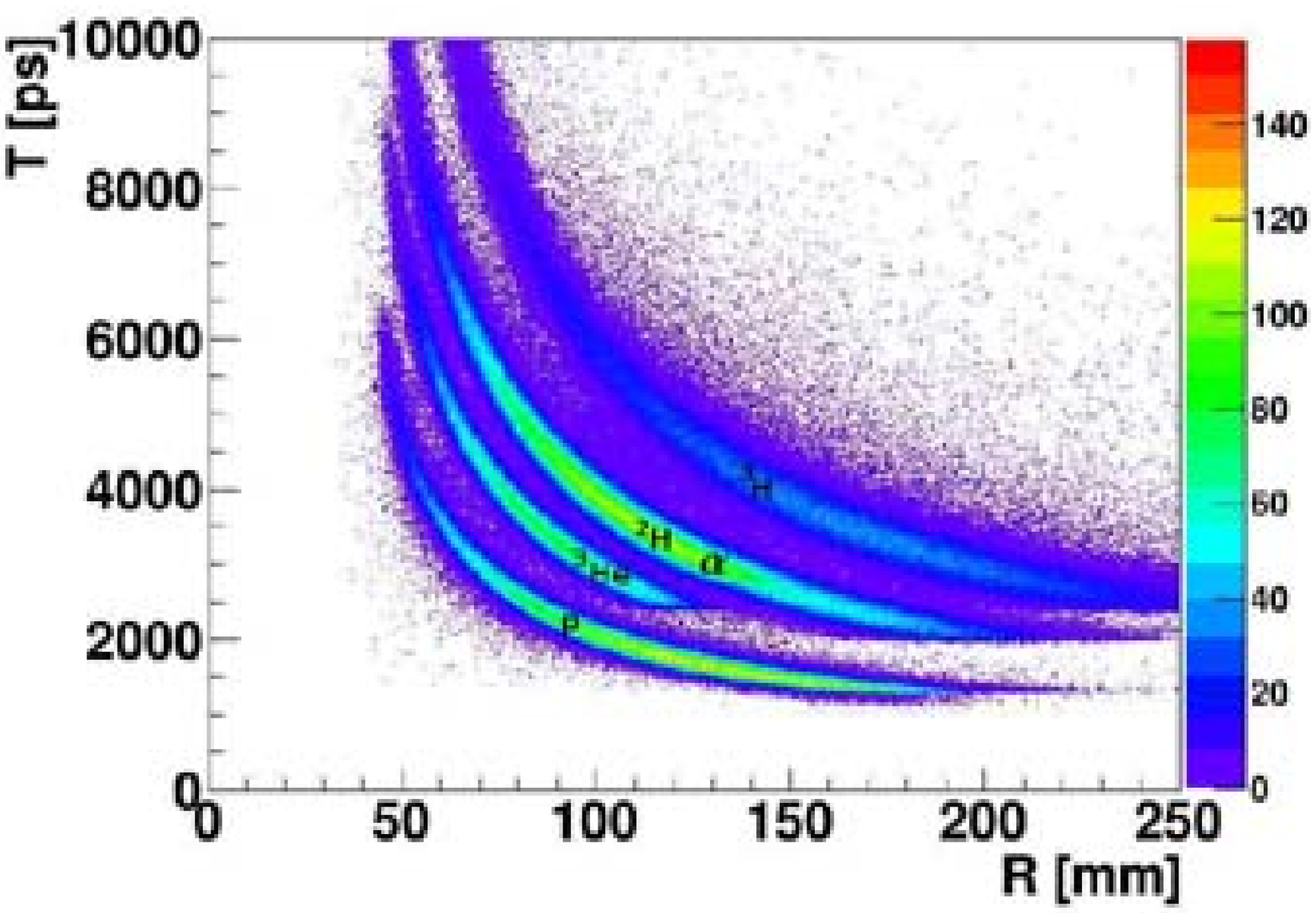,height=8.0cm}}
 \caption{Expected time of arrival as a function of the radius of the trajectory for different particles.}
 \label{Diff}
 \end{center}
 \end{figure}

As one can see, this method allows to separate clearly almost all kind of species we are interested in. To discriminate $^{2}$H from $\alpha$ a more traditional method based on the energy deposited in the drift chamber and/or the scintillators will be used.

\section{Conclusions}

The preliminary design shows that the speed, precision as well as separation power of a drift chamber coupled with an array of scintillators fullfills the needs of the second generation of experiments measuring recoil nuclei fragments. Two prototypes will be build in the next years. These prototypes will permit to finalize the design of the detector. In the mean time, collaborators will use the available fast Monte Carlo to check if it meets their needs. Finally, depending on the results of the two previous steps, the ALERT collaboration will submit a proposal to conduct experiment at CLAS12 using the detector described here.

\section{Acknowledgement}

The support of the LIA project is greatly acknowledged. This work was supported by a grant from the French National Research Agency (ANR) as part of the “Jeunes chercheuses et jeunes chercheurs” Programme (ANR-13-JS05-0001).

\coltoctitle{Molybdenum containing scintillating bolometers for double-beta decay search (LUMINEU program)}
\coltocauthor{D.V.~Poda (the LUMINEU Collaboration)}
\label{art12}

\begin{center}
{\Large {\bf Molybdenum containing scintillating bolometers for double-beta decay search (LUMINEU program)}}

\vskip 0.5cm

{\bf D.V.~Poda$^{a,b}$, L.~Berg\'e$^{a}$, R.S.~Boiko$^{b}$, M.~Chapellier$^{a}$, D.M.~Chernyak$^{b}$, N.~Coron$^{c}$,
F.A.~Danevich$^{b}$, L.~Devoyon$^{d}$, A.-A.~Drillien$^{a}$, L.~Dumoulin$^{a}$, C.~Enss$^{e}$, A.~Fleischmann$^{e}$,
L.~Gastaldo$^{e}$, A.~Giuliani$^{a,f,g}$, D.~Gray$^{h}$, M.~Gros$^{h}$, S.~Herv\mbox{\'e}$^{h}$, V.~Humbert$^{a}$,
I.M.~Ivanov$^{i}$, A.~Juillard$^{j}$, V.V.~Kobychev$^{b}$, F.~Koskas$^{d}$, M.~Loidl$^{k}$, P.~Magnier$^{h}$,
E.P.~Makarov$^{i}$, M.~Mancuso$^{a,f}$, P.~de~Marcillac$^{a}$, S.~Marnieros$^{a}$, C.~Marrache-Kikuchi$^{a}$,
X.-F.~Navick$^{h}$, C.~Nones$^{h}$, E.~Olivieri$^{a}$, B.~Paul$^{h}$, Y.~Penichot$^{h}$, G.~Pessina$^{g,l}$,
O.~Plantevin$^{a}$, T.~Redon$^{c}$, M.~Rodrigues$^{k}$, V.N.~Shlegel$^{i}$, O.~Strazzer$^{d}$, M.~Tenconi$^{a}$,
L.~Torres$^{c}$, V.I.~Tretyak$^{b,m}$, Ya.V.~Vasiliev$^{i}$, M.~Velazquez$^{n}$ and O.~Viraphong$^{n}$
(LUMINEU Collaboration)}

\vskip 0.3cm

$^{a}${\it CSNSM, Univ. Paris-Sud, CNRS/IN2P3, Universit\'e Paris-Saclay, 91405 Orsay, France}

$^{b}${\it Institute for Nuclear Research, MSP 03680 Kyiv, Ukraine}

$^{c}${\it IAS, CNRS, Univ. Paris-Sud, 91405 Orsay, France}

$^{d}${\it CEA, Centre d'Etudes Saclay, Orph\'{e}e, 91191 Gif-Sur-Yvette Cedex, France}

$^{e}${\it Institut f\"{u}r Angewandte Physik, Universit\"{a}t Heidelberg, D-69120 Heidelberg, Germany}

$^{f}${\it Dipartimento di Scienza e Alta Tecnologia dell'Universit\`{a} dell'Insubria, I-22100 Como, Italy}

$^{g}${\it INFN, Sezione di Milano-Bicocca, I-20126 Milano, Italy}

$^{h}${\it CEA, Centre d'Etudes Saclay, IRFU, 91191 Gif-Sur-Yvette Cedex, France}

$^{i}${\it Nikolaev Institute of Inorganic Chemistry, 630090 Novosibirsk, Russia}

$^{j}${\it IPNL, Universit\'{e} de Lyon, Universit\'{e} Lyon 1, CNRS/IN2P3, 69622 Villeurbanne Cedex, France}

$^{k}${\it CEA, LIST, Laboratoire National Henri Becquerel, 91191 Gif-Sur-Yvette Cedex, France}

$^{l}${\it Dipartimento di Fisica, Universit\`{a} di Milano-Bicocca, I-20126 Milano, Italy}

$^{m}${\it INFN, Sezione di Roma, I-00185 Roma, Italy}

$^{n}${\it ICMCB, CNRS, Universit\'{e} de Bordeaux, 33608 Pessac Cedex, France}

\end{center}

\vskip 0.4cm

\begin{abstract}
\noindent A scintillating bolometer technology, promising to be used in a next-generation cryogenic experiment to search
for neutrinoless double-beta decay, is currently under development within the LUMINEU (Luminescent Underground Molybdenum
Investigation for NEUtrino mass and nature) program. The recent results about the R\&D of high quality large volume molybdenum
containing crystal scintillators (zinc and lithium molybdates), including ones produced from $^{100}$Mo-enriched powder, and
aboveground / underground tests of cryogenic detectors based on these crystals are presented here.
\end{abstract}

\noindent



\vskip 0.4cm

\noindent  Keywords: Double-beta decay, Scintillating bolometer, ZnMoO$_4$, Li$_2$MoO$_4$,
Low background experiment, Radiopurity

\section{Introduction}
The recent discovery of neutrino oscillations (the Nobel prize award in physics 2015), phenomena which
demonstrate that neutrinos have mass \cite{RPP14}, stirs up an additional great interest in the searches for
neutrinoless double-beta ($0\nu2\beta$) decay, a nuclear transformation of the type $(A, Z) \rightarrow (A, Z+2) + 2e^-$.
This never observed process is beyond the Standard Model (SM) and definitely requires finite value of neutrino mass,
equivalence between neutrino and anti-neutrino (Majorana nature of neutrinos), and violation of the total lepton number
by two units (see details in the recent review \cite{Bil15} and references herein).

The LUMINEU (Luminescent Underground Molybdenum Investigation for NEUtrino mass and nature) program \cite{Ten15,Ber14}
is going to bridge the gap for high-sensitivity next-generation $0\nu2\beta$ studies by the development of a technology
based on Mo-containing scintillating bolometers, heat-light double read-out cryogenic calorimeters, and the demonstration
of its feasibility by the successful realization of a small-scale experiment to search for $0\nu2\beta$ decay of $^{100}$Mo
with zero background in the range of interest (around $Q_{2\beta}$ = 3034 keV, the total energy of the two electrons emitted
in $0\nu2\beta$ decay of $^{100}$Mo). It will play an important role in the selection for the technology to be adopted for
a tonne-scale cryogenic $0\nu2\beta$ experiment within the CUPID (CUORE Upgrade with Particle IDentification) project \cite{CUPID}.
The main R\&D results in the framework of the LUMINEU program are briefly presented here.

\section{R\&D of scintillating bolometers within LUMINEU}

\subsection{Tasks of the LUMINEU program}

LUMINEU is performing an extensive R\&D of Mo-containing scintillating bolometers which involves several general
tasks related with: I) crystal scintillators; II) light detectors; III) temperature sensors; IV) prototypes of
scintillating bolometers; V) final detectors for a pilot $0\nu2\beta$ experiment. Below we describe briefly the program
and the main requirements of these tasks.

A high detection efficiency for $0\nu2\beta$ decay can be obtained in an experiment with the so-called ``active-source''
technique for which a $0\nu2\beta$ source (e.g. $^{100}$Mo) is embedded into a detector. The baseline detector material for
LUMINEU is a zinc molybdate (ZnMoO$_4$) scintillator, however another promising Mo-based material, lithium molybdate
(Li$_2$MoO$_4$), is also under consideration. Therefore, task I is devoted to the development of purification and
crystallization procedures for producing large volume, high optical quality, radiopure scintillators both from natural
and $^{100}$Mo-enriched molybdenum. In view of the high cost of enriched $^{100}$Mo and the prospect to grow $\sim$ 1000
crystals (for a large-scale project), one needs to optimize a production line in order to get high throughput and low
irrecoverable losses of the enriched material. Specifying these requirements, the main objectives are scintillators
with a mass of $\sim$ 1 kg (scintillation elements up to 0.4 kg) produced with more than 70\% crystal yield, a few \% of  irrecoverable losses, and with a level of internal radioactive contamination by $^{228}$Th and $^{226}$Ra $\leq$ 0.01 mBq/kg
(the total alpha activity of radionuclides from U/Th, except that of $^{210}$Po, is below 1~mBq/kg).

The main background source above 2615 keV in a bolometric $0\nu2\beta$ experiment is caused by $\alpha$ decays of natural
radioactivity. Usually, a scintillator produces less light for $\alpha$ particles with respect to $\gamma$($\beta$)'s of the
same energy (this phenomenon is known as quenching). Therefore, by using a light detector complementary to a bolometer which
scintillates at low temperature, one can efficiently discriminate an $\alpha$-induced background. It is convenient to make
this light detector as a thin bolometer and mount it close to a scintillator-based bolometer, which all together constitute
a scintillating bolometer. Therefore, task II includes the fabrication of optical bolometers from high purity Ge wafers, widely
used for this purpose. An important part of the activity within this task is dedicated to the optimization of light detectors
with the aim to get an acceptable performance. Since both ZnMoO$_4$ and Li$_2$MoO$_4$ scintillators are characterized by very
modest light yield (of the order of $\sim$ 1 keV per 1 MeV deposited energy), efficient light collection and absorption are
important parameters. The energy resolution (FWHM) of the photodetector should be good enough to provide a 99.9\% discrimination
between light signals caused by $\sim$ 3 MeV $\gamma$($\beta$) and $\alpha$ particles impinging in the Mo-based scintillator.
In particular, an $\alpha$/$\gamma$ separation at the level of 5$\sigma$ at $\sim$ 3 MeV can be achieved with a light detector
characterized by FWHM $\sim$ 15\% at 5.9 keV X-ray of $^{55}$Fe (typically used for calibration of these devices) \cite{Ten15a}.
Finally, random coincidences of two neutrino double-beta decay of $^{100}$Mo, allowed in the SM process and registered with a
half-life $\sim$ 10$^{18}$ yr \cite{RPP14}, can constitute a major background for a bolometric $0\nu2\beta$ experiment
with Mo-containing scintillators, as it was pointed out for the first time in Ref. \cite{Bee12}. Therefore a light detector
should have a fast response (e.g. a rise time $\sim$ 1 ms or less) and an as high as possible signal-to-noise ratio to provide
an efficient discrimination of random coincidences.

A small temperature rise appeared after particle interaction inside a bolometer can be measured by a dedicated thermometer.
Within task III, LUMINEU is developing three technologies of temperature sensors: Neutron Transmutation Doped (NTD) Ge
thermistors \cite{Nav15}, superconducting Transition-Edge Sensors (TES) \cite{Non12}, and Magnetic Metallic Calorimeters
(MMC) \cite{Loi14}. The first two types exploit the dependence of resistivity on temperature, while the last one that of
magnetization on temperature. All results given below were obtained with scintillating bolometers instrumented with NTD Ge-based thermometers.

Task IV involves the construction of small / middle size prototypes of scintillating bolometers and their low temperature test,
below tens mK, both at aboveground (at CSNSM) and underground (at Modane in France and Gran Sasso in Italy) cryogenic facilities.
The crucial point of this activity is achieving excellent performance of Mo-containing scintillating bolometers in terms of energy
resolution (FWHM $\leq$ 10 keV at 3 MeV) and particle identification ($\alpha/\gamma$ separation more than 99.9\%), as well as
proving low bulk radioactive contamination of crystals.

Task V requires the fabrication of at least two complete single modules based on massive ZnMoO$_4$ crystals (with mass up
to 0.4 kg; one produced from natural molybdenum and another one enriched in $^{100}$Mo), and a realization of a pilot
experiment with the aim to demonstrate the abovementioned key performance and radiopurity, which are associated with a
zero-background $0\nu2\beta$ experiment. Current status of the LUMINEU tasks are briefly summarized below (task III will
not be highlighted here in more detail than it was done above).

\subsection{R\&D and preformance of Mo-containing scintillating bolometers}

A technology of fabrication of large area light detectors (with diameter up to 50 mm and thickness around 250 $\mu$m)
from high purity germanium wafers (Umicore, Belgium) was developed at CSNSM (Orsay, France) \cite{Ten15a}.
The adopted LUMINEU standard is a $\oslash$44-mm Ge slab coated by SiO to improve light absorption \cite{Man14}.
In addition to the batch of the LUMINEU NTD Ge-based light detectors, photodetectors produced within the LUCIFER $0\nu2\beta$
project \cite{Bee13}, as well as state-of-the-art extra thin ($\sim$ 30--50 $\mu$m) germanium optical bolometers developed at
IAS (Orsay, France) \cite{Cor04} were used to build and test LUMINEU scintillating bolometers. First test of MMC-based light
detectors developed within LUMINEU \cite{Gra15} demonstrate their potential to get $\sim$ 100 times faster response than that
typical for NTD Ge-based photodetectors.

A dedicated protocol for molybdenum purification (double sublimation of molybdenum oxide in vacuum and double recrystallization
of ammonium molybdate from aqueous solutions) has been adopted by LUMINEU to get high purity initial compound for
crystal growth \cite{Ber14}. The chosen advanced directional solidification method developed at NIIC (Novosibirsk, Russia),
low-thermal-gradient Czochralski (LTG Cz) technique, demonstrates the possibility to grow large (up to 1.5 kg) ZnMoO$_4$
crystals with a high crystal yield ($\sim$ 80\% of charge) \cite{Ber14}. The aboveground low temperature test of first LUMINEU
crystals (scintillation elements with size of $\oslash$20$\times$40~mm and $\oslash$35$\times$40~mm and masses of 55 g and 160 g
respectively), produced from ZnMoO$_4$ boules grown from deeply purified materials, shows excellent signal-to-noise ratio,
high signal amplitude, expected values of light-to-heat ratio and light quenching factor for $\alpha$ particles \cite{Ber14}.
Only internal $^{210}$Po contaminant was clearly observed over about two weeks of data taking which indicates high crystals' radiopurity and effectiveness of the purification procedure.

Advances in large volume ZnMoO$_4$ growth have been achieved by
applying double crystallization. In particular, by using this
technique an improved quality $\sim$ 1.0 kg boule was grown from
molybdenum purified by double recrystallization \cite{Arm15}. Two
ZnMoO$_4$ optical elements produced with a size expected for a
pilot LUMINEU $0\nu2\beta$ experiment ($\oslash$50$\times$40~mm;
masses 336 and 334 g) were used to construct identical
scintillating bolometers \cite{Pod15} according to a new special
design compatible with the EDELWEISS set-up at the Modane
undeground laboratory (LSM, France). Results of long term (about
3000 h) low background measurements with these devices
\cite{Arm15,Pod15a} demonstrate an excellent energy resolution
(e.g. FWHM $\sim$ 9 keV at 2615 keV) and an efficient
$\alpha/\gamma$ separation (15$\sigma$ above 2.5 MeV) achieved
with ZnMoO$_4$-based scintillating bolometers, as well as very
high internal radiopurity of ZnMoO$_4$ crystals (e.g. activity of
$^{228}$Th and $^{226}$Ra $\leq$ 0.004 mBq/kg), which completely
satisfy the LUMINEU requirements and those of a future large-scale
experiment. An illustration of the main performance of a 334-g
ZnMoO$_4$ detector is shown in Fig.~1. Taking into account that
the additional crystallization improves quality of a crystal boule
and could also enhance bulk radiopurity thanks to segregation of
radionuclides during growing process, double crystallization can
be used for the production of large LUMINEU crystals.

\nopagebreak
\begin{figure}[ht]
\begin{center}
\mbox{\epsfig{figure=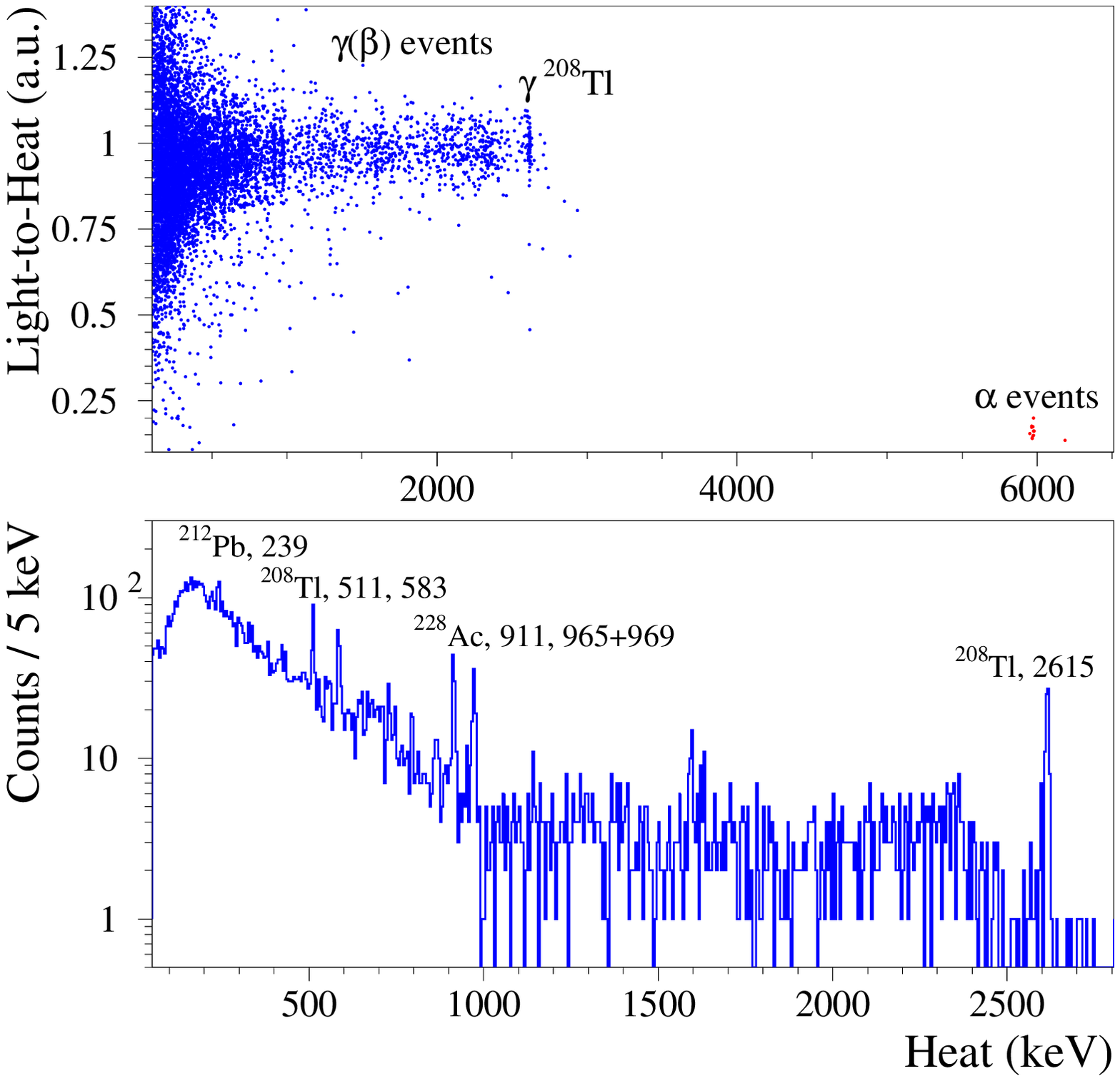,height=9.0cm}}
\caption{(Top) A 2-D histogram showing light-to-heat ratio versus
heat for the data of the 14-h $^{232}$Th calibration measurements
with the 334-g ZnMoO$_4$-based scintillating bolometer operated in
the EDELWEISS set-up at LSM. A high potential of full separation
between $\alpha$ events and $\gamma$($\beta$) band (red and blue
dots respectively) is clearly evident. The region, where events
caused by decays of $\alpha$ radionuclides from U/Th (above 4 MeV)
are expected, is mainly populated by $^{210}$Po, which
demonstrates encouraging internal radiopurity. (Bottom) 1-D
projection of the 2-D histogram shown at the figure top. Energies
of $\gamma$ quanta are given in keV.}
\end{center}
\label{fig:cali}
\end{figure}

A first $^{100}$Mo-enriched Zn$^{100}$MoO$_4$ crystal with a mass
of $\sim$ 0.17 kg has been successfully developed with 84\%
crystal yield and $\sim$ 4\% irrecoverable losses of the enriched
material \cite{Bar14}. The used $^{100}$MoO$_3$ (99.5\% enrichment
in $^{100}$Mo) was purified by sublimation in vacuum and double
recrystallization from aqueous solutions. Molybdenum purification
is related with the major part of losses (85\%) and the minor part
with the crystal growth. The Zn$^{100}$MoO$_4$ boule had a
non-regular shape and yellow coloration unlike similar size
natural ZnMoO$_4$ crystals. It is caused by still existing
difficulties during the solidification, in particular by the
effect of a second phase formation. A Zn$^{100}$MoO$_4$
scintillating bolometer array, built from two produced small
Zn$^{100}$MoO$_4$ elements (59 g and 63 g) and tested aboveground,
shows bolometric properties similar to non-enriched ZnMoO$_4$
detectors \cite{Bar14}. The energy resolution of both bolometers
was reasonably good taking into account a high counting rate due
to the background conditions of the sea-level laboratory. In
particular, 5 and 10 keV FWHM at 609 keV $\gamma$'s of $^{214}$Bi
from environmental radioactivity was obtained with the 59-g and
63-g Zn$^{100}$MoO$_4$ bolometers respectively \cite{Bar14}. After
42 h of data taking only a hint associated with $^{210}$Po was
observed in the $\alpha$ region, which implies encouraging
radiopurity of the crystals \cite{Bar14}. During an underground
test at LSM, the Zn$^{100}$MoO$_4$ array which had a non-standard
holder was strongly affected by a microphonic noise deteriorating
the detector's energy resolution. It is caused by the absence of a
suspension system in the EDELWEISS set-up (to be implemented
soon), which can reduce effect of microphonics, in contrast to the
cryostat \cite{Man14a} used for the aboveground measurements.
Anyway, a full particle identification by Zn$^{100}$MoO$_4$
detectors was clearly demonstrated in the underground test.

The ZnMoO$_4$ solidification procedure was further improved by introducing a small amount of tungsten oxide (up to
$\sim$ 0.5 wt\%) to stabilize the melt and several high quality $\sim$ 0.15-kg crystal boules were grown. No effect of
W-doping on bolometric properties of ZnMoO$_4$ was evidenced by the aboveground measurements with a scintillating bolometer
array constructed from stoichiometric and W-doped ZnMoO$_4$ crystals ($\oslash$20$\times$40~mm each) \cite{Che15}.

Recently, a first large volume colorless Zn$^{100}$MoO$_4$ crystal
($\sim$ 1.4 kg) has been developed \cite{Pod15a,Dan15}. Two large
scintillation elements were produced with a shape of hexagonal
prism (with diagonal 60 mm and height 40 mm, masses 379 g and 382
g) in order to have an improved light output according to studies
\cite{Dan14}. Massive Zn$^{100}$MoO$_4$-based scintillating
bolometers have been tested in the CUORE R\&D test cryostat at
Gran Sasso National Laboratories (LNGS, Italy) and the data
analysis is in progress. A new test in the EDELWEISS set-up at
Modane is foreseen. Meanwhile, background conditions of a
middle-scale $0\nu2\beta$ experiment with 48 Zn$^{100}$MoO$_4$
scintillating bolometers ($\oslash$60$\times$40 mm, 495 g each)
installed in the EDELWEISS set-up have been simulated with
GEANT4-based code \cite{Dan15}. Considering the already achieved
performance, radiopurity, and pulse-shape discrimination of random
coincidences developed by LUMINEU \cite{Che14}, a background
counting rate $\approx$ 5$\times$10$^{-4}$ counts/yr/kg/keV at
$Q_{2\beta}$ of $^{100}$Mo, corresponding to zero-background
conditions, can be reached \cite{Dan15}.

In spite of the achieved progress in the development of large Zn$^{100}$MoO$_4$, growing a regular cylindrical shape
crystal boule with similar quality along its length is still a difficult task (e.g. see Fig.~1 in \cite{Pod15a}).
This problem can be resolved, without further R\&D, just by pulling a short crystal boule. Another solution could be
related with a choice of another Mo-based material, Li$_2$MoO$_4$, which is characterized by comparatively easy crystal
growth process and has bolometric properties similar to ZnMoO$_4$. In particular, several perfect
quality Li$_2$MoO$_4$ crystal boules with masses 0.1--0.4 kg were grown by LTG Cz method from deeply purified Mo and
commercial Li$_2$CO$_3$ (99.99\% purity grade) \cite{Bek16}. A first large Li$_2$MoO$_4$-based ($\oslash$40$\times$40 mm,
151 g) scintillating bolometer was tested aboveground with quite encouraging results \cite{Bek16}, which were reinforced
by subsequent $\sim$ 350-h underground measurements at LNGS \cite{Pat14}. In particular, an excellent spectrometric
properties (FWHM $\sim$ 4 keV at 2615 keV), an efficient particle identification and high radiopurity for U/Th nuclides
($^{228}$Th and $^{226}$Ra $\leq$ 0.02 mBq/kg) were obtained \cite{Pat14}. There is only an issue caused by considerably
high activity of $^{40}$K inside the tested crystal ($\sim$60 mBq/kg), which can be a problem due to random coincidences in
a future $0\nu2\beta$ experiment. Fortunately, this issue can be easily resolved by selection of a Li-containing raw material
with low concentration of $^{40}$K and by recrystallization of Li$_2$MoO$_4$ crystals. It is worth to note that two scintillating
bolometers based on large Li$_2$MoO$_4$ crystals ($\oslash$50$\times$40 mm, 242 g each) produced from different Li-based
powder with low $^{40}$K content (tested by HPGe $\gamma$ spectroscopy) are under investigation in order to prove low bulk
activity of $^{40}$K. A Li-powder with the lowest $^{40}$K content will be used to produce a first Li$_2$$^{100}$MoO$_4$ crystal.
A dedicated low temperature test will be performed in order to completely demonstrate that Li$_2$MoO$_4$-based scintillating
bolometer is a viable detector for a large-scale cryogenic $0\nu2\beta$ experiment.

\section{Conclusions}

A protocol for producing high quality large volume Mo-containing
crystal scintillators from deeply purified molybdenum with both
natural isotopic composition and enriched in $^{100}$Mo has been
developed within the LUMINEU program. A technology for the
development of high performance LUMINEU single modules based on
radiopure large ZnMoO$_4$ and Li$_2$MoO$_4$ crystals is well
established providing high energy resolution (better than 10 keV
FWHM at 2615 keV), efficient $\alpha$/$\gamma$ separation (more
than 5$\sigma$) and required low level of bulk radioactivity (e.g.
$^{228}$Th and $^{226}$Ra $\sim$ 0.01 mBq/kg). A first large
Zn$^{100}$MoO$_4$ crystal ($\sim$ 1.4 kg) was successfully grown
and two 0.4-kg Zn$^{100}$MoO$_4$-based scintillating bolometers
have been preliminary tested at LNGS (Italy). The development of a
first large Li$_2$$^{100}$MoO$_4$ crystal scintillator is in
progress and a pilot LUMINEU test of massive Zn$^{100}$MoO$_4$-
and Li$_2$$^{100}$MoO$_4$-based scintillating bolometers in the
EDELWEISS set-up at LSM (France) is foreseen at the beginning of
2016. After completing this test we will make a final choice about
the Mo-containing crystals to be produced from $\sim$ 10 kg of
$^{100}$Mo (enriched isotope is already available) for a
middle-scale $0\nu2\beta$ experiment (LUCINEU project) based on
the LUMINEU technology. A Monte Carlo simulation of 48 0.5-kg
Zn$^{100}$MoO$_4$-based scintillating bolometers installed in the
EDELWEISS set-up demonstrates the possibility to get a
satisfactory low background counting rate in the range of interest
(at $\sim$ 3 MeV). The LUMINEU activity is now part of the CUPID
project, a proposed bolometric tonne-scale $0\nu2\beta$ experiment
experiment to be built as a follow-up to CUORE and exploiting as
much as possible the CUORE infrastructures.

\section{Acknowledgements}

This work is part of the LUMINEU project funded by the Agence Nationale de la Recherche (ANR, France).
The Li$_2$MoO$_4$ aboveground test was performed with the determinant contribution of ISOTTA, a R\&D ASPERA common call.
The Ukrainian group was supported in part by the IDEATE International Associated Laboratory (LIA).
DVP was supported by the P2IO LabEx (ANR-10-LABX-0038) in the framework ``Investissements d'Avenir''
(ANR-11-IDEX-0003-01) managed by the ANR (France).

\end{papers} 
\clearpage

\end{document}